\documentclass[12pt]{pbook}
\usepackage{a4,amsmath,amssymb}
\usepackage{epsfig,epsf,psfrag,epic}
\usepackage{makeidx}
\usepackage{mystyle}

\def\hilbert{\mbox{$\mathcal{H}$}}
\def\fock{\mbox{$\mathcal{F}$}}

\makeindex

\begin{document}

\pagenumbering{roman}

\thispagestyle{empty}

\begin{center}
\begin{tabular}{|p{14cm}|}
\hline \phantom{K} \\ \phantom{K} \\
\centerline{\sc\Huge State Preparation}\bigskip
\centerline{\sc\large and}\bigskip
\centerline{\sc\Large Some Applications}\bigskip
\centerline{\sc\large in}\bigskip
\centerline{\sc\Huge Quantum Optics}\bigskip
\centerline{\sc\large within}\bigskip
\centerline{\sc\large the context}\bigskip
\centerline{\sc\large of}\bigskip
\centerline{\sc\Huge Quantum Information}\bigskip
\centerline{\sc\Huge Theory}

\vspace{2cm}

\centerline{\sc\Large by Pieter Kok}

\vspace{6.2cm}

\centerline{PhD thesis, University of Wales, Bangor.} \bigskip
\\ \hline 
\end{tabular}
\end{center}

%\centerline{\sc\Huge State Preparation}\bigskip
%\centerline{\sc\large and}\bigskip
%\centerline{\sc\Large Some Applications}\bigskip
%\centerline{\sc\large in}\bigskip
%\centerline{\sc\Huge Quantum Optics}\bigskip
%\centerline{\sc\large within}\bigskip
%\centerline{\sc\large the context}\bigskip
%\centerline{\sc\large of}\bigskip
%\centerline{\sc\Huge Quantum Information}\bigskip
%\centerline{\sc\Huge Theory}
%
%\vspace{2cm}
%
%\centerline{\sc\Large by Pieter Kok}
%
%\vfill
%
%\centerline{PhD thesis, University of Wales, Bangor.}

\newpage\thispagestyle{empty}

\phantom{lege bladzijde}

\newpage\thispagestyle{empty}

$\qquad\qquad${\sl How I need a drink, alcoholic of course, after}

$\qquad\qquad${\sl the heavy chapters involving quantum mechanics.}

\medskip

{\hfill ---Mnemonic for the first fifteen digits of $\pi$.}

\newpage\thispagestyle{empty}

\tableofcontents
\newpage\thispagestyle{empty}

\newpage\addcontentsline{toc}{chapter}{$\quad\,\,$List of Figures}
\listoffigures
\newpage\thispagestyle{empty}

\chapter*{Summary}\addcontentsline{toc}{chapter}{$\quad\,\,$Summary}

Entanglement is perhaps the single-most important resource of quantum 
information theory. The first part of this thesis deals with the creation of 
optical event-ready entanglement with a specific class of optical 
circuits. These circuits include passive components such as beam-splitters 
and phase-shifters, and active components such as optical parametric 
down-converters and optical squeezers. Furthermore, the entangled-state 
preparation may be conditioned on one or more detector outcomes. 
In this context, I discuss the statistics of down-converters and give a 
quantitative comparison between realistic detectors and detector cascades, 
using the {\em confidence} of the detection. The outgoing states of the 
optical circuits can be expressed in terms of multi-dimensional Hermite 
polynomials. Event-ready entanglement cannot be created when the outgoing 
state is conditioned on two detected photons. For six detected photons using 
ideal photo-detectors a scheme is known to exist. 

Part two of this thesis includes two applications of optical 
entanglement. First, I discuss quantum teleportation and entanglement 
swapping using down-conversion. It is shown that higher-order photon-pair 
production degrades the fidelity of the teleported (or swapped) states. 
The interpretation of these states proved controversial, and I have attempted 
to settle this controversy. As a second application, quantum lithography 
uses optical (`which-way') entanglement of multiple photons to 
beat the classical diffraction limit. Given a suitable photo-resist, 
this technique results in sub-wavelength optical resolution and can be used 
to write features much smaller than is possible with classical 
lithography. I present classes of states which can be used to create patterns 
in one and two dimensions with sub-wavelength resolution. 

\newpage 
\thispagestyle{empty}

\chapter*{Acknowledgements}
\addcontentsline{toc}{chapter}{$\quad\,\,$Acknowledgements}

I could not have written this thesis without the help of many people. I would
like to express my gratitude towards Samuel L.\ Braunstein for his excellent 
supervision, and my colleagues Peter van Loock and Arun Pati for the many 
discussions we've had over the years. Special thanks to prof.\ Rajiah Simon, 
who helped me understand the maths of operators and maps, Jonathan Dowling for 
inviting us to work on quantum lithography and for offering me a job, and Apy 
Vourdas for the fruitful discussions on multi-dimensional Hermite polynomials. 
I also thank all the others whom I have had the pleasure to meet and who have 
generally made me wiser.

During the past three years I have often found consolation and playful 
adversity from my friends in the Netherlands, especially from Jasper for his 
long-distance moral and immoral support, and Alex, Gijs and Maarten. The 
RISK-club rules! Many thanks also to Ang\`ele (mooi, die middeleeuwse kastelen 
van voor de Renaissance), Jeroen (Slaatje kameraadje), Joep (d'r zit 'n haar 
in m'n glas) and Klaas-Jan (``mobile construction yard deployed'').

I have had an unforgettable time in Bangor, the responsibility of which can 
be traced almost entirely to my friends here. Many thanks to Barbara (my own 
drama-queen), Mireia (fly, my pretty), Consuelo (\`e chica!), Ana (when will 
we dance again?), Claudia (Mel G.\ meets tequila), Carlotta (close that 
cupboard), Martin ({\em you} are the best man), Marc (I'll see you in 
Cambridge), Ross (our man in Havana) and Johnny (watch out for that bottle!): 
pray I don't publish my memoirs\ldots

Finally, I thank my parents and my brother Joost (architectuur = kunst).

\bigskip

\hfill Pieter Kok,$\qquad$

\hfill December 2000.$\qquad$

\newpage
\thispagestyle{empty}

\phantom{x}

\newpage
\pagenumbering{arabic}

\chapter{Introduction}

The closing decade of the twentieth century has witnessed the coming-of-age of
a new field, called {\em quantum information 
theory}.\index{quantum!information} This field includes 
the development of quantum computation\index{quantum!computation} and quantum 
communication.\index{quantum!communication} At this point
a fully scalable quantum computer has not been built, but there are numerous 
experimental and theoretical
proposals to achieve this \cite{braunstein00}. At the same time, the quest
for quantum algorithms continues. So far, we have Shor's algorithm to factor 
large numbers into primes \cite{shor97}, the Deutsch-Jozsa algorithm 
\cite{deutsch85,deutsch92} and Grovers search algorithm \cite{grover96}. The 
possibility of quantum error correction was discovered \cite{steane95}, which 
is very important to any practical application of quantum computation.

Considerable progress has also been made in quantum communication. It is a 
generic term for communication protocols based on quantum mechanical principles
and includes cryptography \cite{bennett84,ekert91}, 
teleportation \cite{bennett93}, entanglement swapping \cite{zukowski93},
dense coding \cite{bennett92}, quantum clock synchronisation \cite{jozsa00},
entanglement purification \cite{bennett96a} and quantum networks 
\cite{grover97,ekert98}. Another recent application of quantum mechanics is 
quantum lithography \cite{boto00}. The common divisor of nearly all elements 
of quantum information is quantum {\em entanglement}\index{entanglement} 
\cite{braunstein}. In 
this thesis I study the creation of entanglement in quantum optics, and 
some of its applications.

This introduction will provide the motivation and physical background for the 
thesis. I discuss entanglement, teleportation and lithography. It will be 
largely non-mathematical and aimed at an audience of non-specialists. The 
subsequent chapters will then develop these issues in a rigorous 
mathematical way.

\section{Quantum entanglement}\label{sec1}

In order to explain what quantum entanglement is about, I will first discuss
the double slit experiment\index{experiment!double-slit} as presented by 
Richard Feynman\index{Feynman, R.P.} \cite{feynman65}. 
Suppose we have a gun firing bullets at a screen with two holes which are 
close to each other. Most of the bullets will hit the screen and fall on the 
floor, but some of them will pass through the holes and hit a wall of clay. In 
effect, this wall records the position of impact of the bullets which passed 
through the holes.

\begin{figure}[t]
\begin{center}
  \begin{psfrags}
     \psfrag{a}{a)}
     \psfrag{b}{b)}
     \epsfxsize=8in
     \epsfbox[-50 20 650 130]{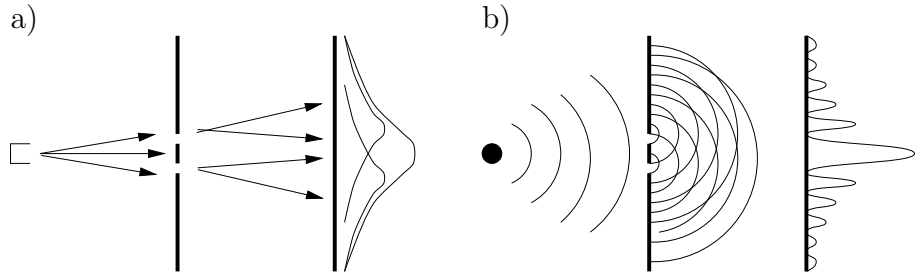}
  \end{psfrags}
  \end{center}
  \caption{The double slit experiment a) with particles and b) with waves.} 
\label{fig:1}
\end{figure}

When we inspect the wall, we will see that the bullets are spread around
the centre of the clay wall in a straight line behind the gun and the holes 
in the screen. Each bullet must have passed through either hole to make it to 
the wall. When we record the process with a high-speed camera we can see the 
bullets going through the holes. In fact, we can mount paint sprayers next to 
the holes, colouring the bullets which pass through the left hole red, and the
bullets which pass through the right hole blue. The clay wall will be peppered 
with red and blue bullets, with the red bullets shifted slightly to the left 
and the blue bullets slightly to the right (see figure \ref{fig:1}a).

Let us now repeat this experiment with waves instead of bullets. Suppose we 
have a shallow tray of water with a screen containing two narrow openings
close to each other at the waterline. On one side of the screen a pin is moving
up and down in the water, creating a wave which spreads out in all directions. 
When the wave reaches the screen, the two slits start to act as if they were 
vertically moving pins {\em themselves}! The slits thus create {\em two} waves 
which spread out in all directions behind the screen. 

These two waves will soon start to interfere: when a wave-crest meets another
crest, the result will be a crest twice as high; when a trough meets another 
trough, the result will be a trough twice as deep. And finally, when a crest
meets a trough they cancel each other. When we record the vertical displacement
of the water at the far end of the tray of water, we will find 
an {\em interference} pattern\footnote{More precisely, the 
interference\index{interference} pattern
is given by the square of the displacement: the {\em intensity}.} of peaks 
and troughs (see figure \ref{fig:1}b).

The difference with bullets is obvious: the bullets arrive in a spread area
with its bullet density falling off uniformly with the distance from the 
centre, whereas waves will show an intensity pattern which rises and falls in 
alternation with increasing distance from the centre. The simple (classical) 
picture is: waves give interference and particles (bullets) don't.

Now let's take a look at light. Suppose we have again a screen 
with two slits, and a laser which is aimed at the slits. The light which passes
through the slits is recorded on a photographic plate. After development, we
will see an interference pattern on the photographic plate: light seems to be 
a wave.

When we weaken the intensity of the laser enough, we will see (using 
very sensitive equipment) that the light is no longer a continuous stream, but 
that instead, it is a succession of small `bursts'. We call these bursts {\em 
photons},\index{photon} and they are described by quantum theory. It thus 
seems that light consists of particles which interact to give an 
interference\index{interference} pattern just like waves.

We now attenuate the laser so that we fire individual photons at the double 
slit. This way, the photons cannot interact to give an interference pattern 
since at any time there is only one photon travelling between the laser and the
photographic plate. The photons which make it past the double slit will give a 
dot on the photographic plate, analogous to the bullets in the clay wall. If 
the photons are truly classical particles, they should pass through either 
slit, just like the bullets, and they should {\em not} make an interference 
pattern. 

However, when we develop the photographic plate after a long exposure time 
we {\em do} find an interference pattern! We have set up this experiment in 
such a way that the photons, which seem to behave like particles (indivisible,
giving dots on a screen), are passing the slits one at a time so they don't 
interact with each other. The only way to get an interference pattern is thus 
when the photon somehow interferes with {\em itself}. Has the photon gone 
through both slits simultaneously? Let's test this.

Again, we fire individual photons at a double slit and record the pattern 
on a photographic plate. But this time we place a detector behind both slits.
These detectors tell us through which slit the photon passes. While running
this experiment, the detectors are clicking when a photon passes through 
its corresponding slit, giving us information about the paths of the 
successive photons. They really go through one slit at a time.

But when we now develop the photographic plate, the interference pattern has
gone! Instead, we have a concentration of dots, its density decreasing with 
increasing distance from the centre. This is the bullet pattern. Apparently, 
when we 
{\em know} through which slit the photons pass, we do not get an interference 
pattern. When we do not look, it is {\em meaningless} to ask through which
slit the photons pass. In describing the path of the photons without detection,
we need to include both possibilities: the path is a {\em 
superposition}\index{superposition} of 
going through the left and the right slit.

Feynman elevated this to a general principle: when an event can occur in 
several different ways, we need to describe the event in terms of a 
superposition of these ways \cite{feynman65}. The superposition principle 
is responsible for many of the counterintuitive aspects of quantum mechanics. 
This simple thought-experiment 
thus takes us straight into the heart of the theory. 

\medskip

Let's now consider entanglement. Photons have an extra internal property called
{\em polarisation}.\index{polarisation} A photon which reflects off this paper 
towards your eye 
(which can be represented graphically as $\odot$) vibrates in the plane of the 
paper perpendicular to the direction of travelling ($\updownarrow$ or 
$\leftrightarrow$, or a combination of these two. Technically, we also have 
circular polarisation). The polarisation of the photon is determined by the 
angle of this vibration direction. 

When we want to {\em measure} the polarisation of a photon, we place a 
polarising beam-splitter,\index{beam-splitter} or {\em polariser} in the path 
of the photon. This 
is essentially a piece of glass which reflects horizontally polarised photons 
and transmits vertically polarised photons. When we place photo-detectors in 
the paths of reflected and transmitted photons, a detector click will tell us 
the polarisation of that photon. When a horizontally (vertically) polarised 
photon encounters the polariser, it will always be reflected (transmitted). 
But what if the photon has a diagonal polarisation?

When a diagonally polarised photon encounters the polariser it will be either
reflected or transmitted. We can only make a probabilistic
prediction as to which path the photon will take. When we {\em rotate} the 
polariser so that its horizontal orientation is turned parallel to the 
(diagonal) polarisation of the photon, the photon will be reflected {\em with 
certainty}. We now consider two polarised photons.
 
Suppose we have two photons originating from a common source and heading off 
in opposite directions. One photon is received by Alice, and the other by Bob.
Furthermore, Alice and Bob are far away from each other, possibly in different
galaxies. 

First, we consider the case where both photons are horizontally polarised 
($\leftrightarrow$). When Alice and Bob measure the polarisation of the photon 
in the horizontal and vertical direction using polarisers, both will find with 
certainty that the photons have horizontal polarisation. When Alice rotates 
her polariser by 45 degrees, the probability that either detector signals the
detection of a photon is one half. This situation is similar to the measurement
of a single photon since the photons received by Alice and Bob behave 
completely independent from each other.

Now suppose that the two photons are prepared in the following way: either
Alice's photon is horizontally polarised and Bob's photon is vertically 
polarised, {\em or} Alice's photon is vertically polarised and Bob's photon is 
horizontally polarised. Furthermore, the photons are prepared in a 
{\em superposition}\index{superposition} of these two possibilities. When 
Alice and Bob measure 
the polarisation of these photons they will find that their photons always 
have opposite polarisations: when Alice detects a horizontally polarised 
photon, Bob will find a vertically polarised photon and vice versa. This means 
that given a measurement outcome, Bob knows what Alice's measurement outcome 
will be, even though she might be light years away. The measurement results 
are said to be correlated.

So far, nothing strange has happened. We know these correlations from 
classical physics. Suppose Alice and Bob meet in Amsterdam. They blindly draw
a marble from a vase containing only one black and one white marble. Alice
travels to New York and Bob travels to Tokyo. When Alice looks at her marble
and finds that it is white, she immediately knows that Bob's marble is black.
These outcomes are also correlated.

There is, however, a difference in the case of polarised photons. Suppose 
Alice and Bob both rotate their polariser over 45 degrees. According to the
classical picture, both photons have a 50:50 chance to end up in either 
detector. That means that with 50\% probability the photons have equal 
polarisation. But this is not what they find: Alice and Bob {\em always} find
that they have opposite polarisations! Clearly, this is not just a classical 
correlation. The two photons are said to be {\em entangled}.

\medskip

\begin{figure}[t]
\begin{center}
  \begin{psfrags}
     \psfrag{p}{laser}
     \psfrag{c}{crystal}
     \psfrag{r}{$\updownarrow$}
     \psfrag{l}{$\leftrightarrow$}
     \epsfxsize=8in
     \epsfbox[-70 20 470 120]{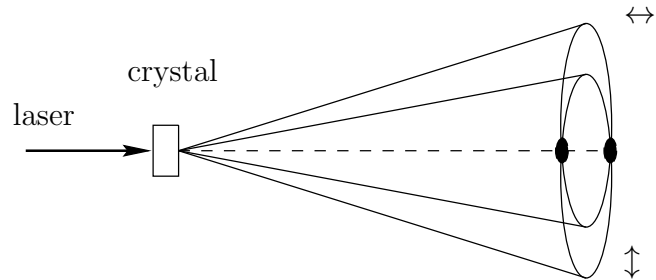}
  \end{psfrags}
  \end{center}
  \caption{A down-converter.} 
\label{fig:2}
\end{figure}

The question is now: how can we make entangled photons? One way of doing it 
is to use a so-called down-converter.\index{down-converter} In a 
down-converter, a high-powered laser
is sent into a special crystal. A photon of the laser interacts with the 
crystal and breaks up into two photons with half the energy. The photons will
travel away from the central axis (defined by the path of the laser light)
under a fixed angle. The photons thus travel on the surface of a {\em cone}
originating from the crystal (see figure \ref{fig:2}).

Furthermore, we can set up the down-converter in such a way that the photons 
have opposite polarisations. This is where the crystal performs its special 
trick: the refraction index\index{refraction index} of the crystal is 
different for horizontally and 
vertically polarised photons. This means that the cone corresponding to the 
possible paths of horizontally polarised photons is tilted upwards from the 
central axis. Similarly, the cone for vertically polarised photons is tilted 
slightly downwards. 

Due to momentum conservation, the two photons are always travel on different 
cones along lines opposite of each other with respect to the central axis. The 
cones intersect each other at two opposite lines, and as a 
consequence, we find a photon in one of those lines {\em if and only if} there
is a photon in the other line. Furthermore, we cannot tell to which cone the
photons on the intersecting lines belong. Either the left photon belongs to 
the upper cone and the right photon to the lower, or the other way around. By 
virtue of Feynman's principle we have to take the superposition of these two 
possibilities.

The down-converter only produces two entangled photons probabilistically; not
every laser pulse results in two down-converted photons. Furthermore, since we 
select only the intersection of the two cones, we lose all the instances where
photons were not produced along the intersecting lines. This means that most 
of the time we fire the laser into the crystal we do not produce entanglement.
In this thesis I study whether and how we can minimise the number of cases 
where no photons are produced.

\section{Teleportation}

Another subject of this thesis is quantum 
teleportation,\index{quantum teleportation} in particular the 
teleportation of a photon. In this procedure, the (unknown) polarisation of one
photon is transferred to another photon far away. It is not true that the 
photon itself is magically transported from Alice to Bob, only the polarisation
direction (or, more generally, the {\em state} of the photon) is transferred.

The same is true for other types of matter: we can teleport atoms, but that 
does not mean we can make an atom appear somewhere in the distance. I will now
present the general protocol, using polarised photons.

Suppose Alice received a photon with a polarisation direction which is unknown 
to her. We assume that she has some way of storing it without disturbance. In 
other words, she has a device called a `quantum memory'.\index{quantum!memory} 
Now she wants to 
transfer the polarisation direction of the photon to Bob. When she has only 
measurements and a telephone at her disposal to tell Bob the results, she has 
a problem. Since she doesn't know what the polarisation direction is, she 
cannot choose her polariser to be parallel to this direction. Therefore, when 
she measures the polarisation of the photon in a chosen direction and tells 
Bob the result, his reconstruction of the polarisation direction will generally
be off by a certain angle. Faithful teleportation cannot be performed this way.

However, the story changes when Alice and Bob share entanglement (produced, 
for example, by the down-converter of the previous section). Alice and Bob
both hold one part of an entangled photon pair. Remember that these photons 
are correlated: whatever the polarisation direction measured by Alice, Bob 
will always find the opposite polarisation. Alice proceeds by making a {\em 
joint} measurement of her part of the entanglement\index{entanglement} and 
the the incoming 
photon with unknown polarisation. Such a measurement does not give any 
information about the individual photons, but determines the relation of the 
photons relative to each other. It is a carefully chosen measurement which 
will correlate the incoming photon with unknown polarisation to Alice's part 
of the entangled photon-pair. In the case of polarisation we consider here, 
the outcome of Alice's measurement has four possible outcomes. These outcomes 
correspond to four different ways the two photons can be correlated 
(technically, we have four {\em orthogonal} ways). 

Let's pause for a second to contemplate the current state of affairs. Alice 
has just correlated the unknown incoming photon with her half of the entangled 
photon-pair. She cannot choose or predict how she correlates them, every one 
of the four possibilities is equally likely. But her half of the entangled 
photon-pair is already correlated with the other half. It then follows that 
the unknown incoming photon is now correlated with Bob's half of the entangled 
pair.

The only thing Bob does not know is the exact nature of the correlation. 
Every one of the four possible correlations will give a different polarisation
direction in Bob's photon. That is why Alice has to tell him. She picks up the 
phone and gives one of four possibilities (i.e., she sends two classical bits)
corresponding to her measurement outcome. The beautiful thing about 
teleportation is now that Bob has to perform a polarisation rotation 
corresponding to the measurement outcome, which is {\em independent of the 
unknown polarisation direction of the incoming photon!} His photon now has the 
same polarisation direction as the incoming photon and teleportation is 
complete.

We have to note three things. First of all, neither Alice, nor Bob gains any
information about the direction of the polarisation of the incoming photon.
Secondly, no photon magically appears at Bob's site; he already held a photon 
during the whole procedure. In this respect, quantum teleportation is quite 
unlike the Star Trek version. And finally, teleportation cannot be used for 
superluminal signalling. If Alice does not tell Bob her measurement outcome
(which is a classical message and thus restricted by the speed of light),
teleportation will fail since Bob does not know what polarisation rotation
he has to perform.

The entanglement shared by Alice and Bob is typically produced with a
down-converter. In this thesis I study the effects of the down-conversion 
characteristics on the quality of teleportation.

\section{Lithography}

The second application of optical quantum entanglement I study in this thesis
is quantum lithography.\index{quantum lithography} This technique may be used 
to write components on 
micro-chips which are smaller than possible with classical optical lithography.
It works as follows.

Consider again the double slit experiment\index{experiment!double-slit} with 
photons, given in section 
\ref{sec1}. When photons are fired at the slits one at a time without looking
through which slit they pass, we obtain an interference pattern on the 
photographic plate. In this case the photon can travel along two possible 
paths: either through the left slit or through the right.

Suppose we now fire two photons per shot at the slits. If we assume that all
photons pass the slits we now have three possible paths: both photons may pass
through the left slit; one may pass through the left and the other though the
right; or both may pass through the right slit. In this case the interference 
pattern will be twice as bright, because we use twice as much light.

But now we can ask what happens if we suppress one of these three possible 
paths. What will the interference pattern look like when the photons {\em do
not separate}, that is, what happens when the photons either both pass through 
the left slit, or both through the right slit? The answer is that the 
interference pattern, which is an array of bright and dark lines, will become 
twice as narrow: the distance between two bright lines is halved. The reason 
why this happens is because the photons `stick together', thus effectively 
acting as a single particle with twice the momentum. The De Broglie wavelength 
\index{wavelength!DeBroglie}
(which determines the line spacing of the interference pattern) is inversely 
proportional to the momentum of the particle. The higher the momentum, the 
shorter this wavelength and the narrower the interference pattern. Since 
photons also have momentum, this means that the more photons we can make 
acting as a single particle, the smaller the interference pattern. Note that 
in general, we need a special surface which is sensitive to two photons (a 
`two-photon resist') to record these patterns.

Classically, light cannot resolve features which are much smaller than its 
wavelength. As a consequence, classical optical lithography, in which light is 
used to etch a surface, cannot write features much smaller that its wavelength.
This is Rayleigh's diffraction\index{Rayleigh criterion} limit. It is derived 
from the interference 
between two waves. We have seen that we can narrow this interference pattern 
using the quantum properties of light, which means that the Rayleigh limit is
a {\em classical} limit. Quantum lithography can therefore be used to create 
sub-wavelength patterns, to be used in, for example, the micro-chip industry.

So far, quantum lithography is still a theoretical method. 
Only the two-photon case
described above has been experimentally tested. It is not easy to see how more
exotic patterns may be produced, and what the requirements for the surface are.
Nevertheless, it gives us a new insight in the nature of light. In this thesis
I study how we can create arbitrary sub-wavelength patterns in one and two 
dimensions.

\section{Thesis outline}

This thesis is organised in two parts. The first part, called `Quantum State 
Preparation' is divided in four chapters. Chapter \ref{chap2} gives the 
general quantum mechanical background. It includes the postulates of quantum 
mechanics, the quantisation of the electro-magnetic field and some topics 
from quantum information theory such as the Von Neumann entropy and the 
fidelity.

In chapter \ref{chap3}, I study a limited set of optical circuits for creating 
near maximal polarisation entanglement {\em without} the usual large vacuum 
contribution. The optical circuits I consider involve passive interferometers, 
feed-forward detection, down-converters and squeezers. For input vacuum fields 
the creation of maximal entanglement using such circuits is impossible when 
conditioned on two detected auxiliary photons. Furthermore, I derive the 
statistical properties of down-converters and show that coincidences between 
photon-pairs from parametric down-conversion automatically probe the 
non-Poissonian structure of these sources.

So far, the photo-detectors I considered are ideal. In chapter \ref{chap4} I
study the use of detection devices in entanglement-based state preparation.
In particular I consider realistic optical detection devices such as 
single-photon sensitivity detectors, single-photon resolution detectors and 
detector cascades (with a limited efficiency). I develop an extensive theory 
for the use of these devices. In entanglement-based state preparation we 
perform measurements on subsystems, and we therefore need precise bounds on 
the distinguishability of these measurements. To this end, I introduce the 
{\em confidence} of preparation, which may also be used to quantify the 
performance of detection devices in entanglement-based preparation. I give a 
general expression for detector cascades of arbitrary size for the detection 
up to two photons. I show that, contrary to the general belief, cascading does 
not give a practical advantage over detectors with single-photon resolution in 
entanglement-based state preparation.

Finally, in chapter \ref{chap5}, I study a special class of optical circuits 
and show that the {\em outgoing state} leaving the optical circuit can be 
expressed in terms of so-called multi-dimensional Hermite polynomials and give 
their recursion and orthogonality relations. I show how quantum teleportation 
of photon polarisation can be modelled using this description.

The second part is called `Some Applications' and covers two chapters. In 
chapter \ref{teleportation}, I study the experimental realisation of quantum 
teleportation as performed by Bouwmeester {\em et al}.\ \cite{bouwmeester97}
and the adjustments to it suggested by Braunstein and Kimble 
\cite{braunstein98}. These suggestions include the employment of a detector 
cascade and a relative slow-down of one of the two down-converters. 
Furthermore, I discuss entanglement swapping and the creation of GHZ states
within this context.

Chapter \ref{lithography} gives the theory of quantum lithography. I generalise
the lithography procedure in order to create patterns in one and two 
dimensions. This renders quantum lithography a potentially useful tool in 
nano-technology. 

\newpage
\thispagestyle{empty}

\part*{Quantum State Preparation}\addcontentsline{toc}{part}
	{$\quad\,\,$Quantum State Preparation}

\chapter{Quantum Theory}\label{chap2}

This chapter presents the mathematical background theory for the understanding 
of this thesis. It does not contain new results. First, I present quantum 
mechanics in the Hilbert space formalism. The second section discusses the
quantisation of the electro-magnetic field and quantum optics, and in the last
section I treat some aspects of quantum information theory, such as entropy,
fidelity and non-locality.  

\section{Quantum mechanics in a nutshell}

What is quantum mechanics\index{quantum mechanics} all about? Initially, 
the theory was developed
to describe the physical world of atoms, that is, to explain the observed 
spectral lines in spectrometers. In 1913, Niels Bohr,\index{Bohr, N.} then at 
the Cavendish
laboratory in Cambridge, developed what is now called the `old quantum theory',
in which he presented a model explaining the spectral lines of hydrogen
\cite{bohr13,heilbron77}. As the theory was developed further (culminating in 
the work of Schr\"odinger\index{Schr\"odinger, E.} and subsequently Heisenberg)
\index{Heisenberg, W.} it became clear that 
quantum mechanics is a mathematical theory which describes {\em measurement 
outcomes},\index{measurement!outcome} rather than the underlying physical 
processes \cite{peres95}.

Von Neumann\index{Von Neumann, J.} proved the equivalence of Schr\"odingers 
wave mechanics\index{wave!mechanics} and 
Heisenberg's matrix mechanics\index{matrix!mechanics} in {\em Mathematical 
foundations of quantum 
mechanics} \cite{vonneumann55} and introduced the Hilbert 
space formalism\index{Hilbert space!formalism} 
still in use today. Dirac\index{Dirac, P.A.M.} \cite{dirac30} developed his 
own version of the 
theory (of which, incidentally, Von Neumann did not approve\footnote{The 
preface of his book makes very enjoyable reading.}), and his bracket notation 
has become the standard. In accordance with the convention, I will follow Von 
Neumann's framework and use Dirac's bracket notation.

This section is organised as follows: first I will give the postulates for 
quantum mechanics. Then I discuss mixed states and composite systems. Finally, 
this section ends with measurement theory according to Von Neumann and its 
generalisation to projection operator valued measures (POVM's).

\subsection{The postulates of quantum 
mechanics}\index{quantum mechanics!postulates}

Using some properties of complex vector spaces\index{vector space!complex} 
(see appendix 
\ref{app:vectorspaces}), we can formulate the postulates of quantum mechanics 
\cite{bransden89,cohen77,hilgevoord93}. More properties of operators on 
Hilbert spaces can be found in appendix \ref{app:povm}.
 
\begin{description}
 \item[Postulate 1] For every physical system\index{physical system} there is 
	a corresponding Hilbert\index{Hilbert space}
	space $\hilbert$. The accessible (pure) states\index{state} of the 
	system are completely determined by rays with unit length 
	in $\hilbert$.  
\end{description}

A ray\index{ray} in Hilbert space is a set of unit vectors which differ only 
by an 
arbitrary (complex) phase. A physical state corresponds to a ray. Thus, a 
state $|\psi\rangle$ is physically equivalent to $e^{i\varphi}|\psi\rangle$ 
with $0\leq\varphi\leq 2\pi$. A complete set of orthonormal states (rays) form 
a basis of $\hilbert$. I will use the terms `ray', `vector' and `state' 
interchangeably, while remembering that an overall phase does not change the 
physical state. Later, in section {\ref{sec:mixed}}, this class of states is
extended to {\em mixed} states.\index{state!mixed}

Since the Hilbert space $\hilbert$ is a (complex) 
vector space,\index{vector space!complex} if any two 
normalised rays $|\psi\rangle$ and $|\phi\rangle$ in $\hilbert$ are accessible
states to the system, then their superposition\index{superposition} 
$\alpha|\psi\rangle + 
\beta|\phi\rangle$ is also an accessible state to the system. Normalisation
\index{normalisation}
then requires $|\alpha|^2 + |\beta|^2=1$. A superposition of this type is 
sometimes called a {\em coherent} superposition.\index{superposition!coherent} 
Note that in this case the 
phase of a ray {\em does} have a physical meaning. Consider two orthonormal 
states\index{state!orthonormal}
 $|\psi\rangle$ and $|\phi\rangle$ (i.e., $\langle\psi|\psi\rangle = 
\langle\phi|\phi\rangle = 1$ and $\langle\psi|\phi\rangle=0$), and consider 
the two superpositions
\begin{eqnarray}
 |\upsilon_1\rangle &=& 
	\frac{1}{\sqrt{2}}\left(|\psi\rangle + |\phi\rangle\right)\; ,\cr
 |\upsilon_2\rangle &=& 
	\frac{1}{\sqrt{2}}\left(|\psi\rangle - |\phi\rangle\right)\; ,
\end{eqnarray}
then it is easy to verify that $\langle\upsilon_1|\upsilon_2\rangle=0$, i.e.,
they are orthogonal\index{orthogonality} (in fact, they are orthonormal). The 
two superpositions 
differ only in a {\em relative} phase, but they yield two physically distinct
(orthonormal) states.

\begin{description}
 \item[Postulate 2] For every physical observable\index{observable} of the 
	system there is a unique corresponding self-adjoint (Hermitian) 
	operator\index{operator!self-adjoint}\index{operator!Hermitian} $A$ in 
	$\hilbert$.
\end{description}

An operator $A$ is called self-adjoint if and only if $A^{\dagger}=A$. An 
operator is Hermitian if and only if its eigenvalues\index{eigen!-values} are 
real. I will now 
prove that (for finite-dimensional Hilbert spaces) any operator is 
self-adjoint if and only if it is Hermitian.

To show that self-adjointness implies Hermiticity, observe that according to
the eigenvalue equation\index{eigen!-value!equation} 
$A|\psi\rangle=\alpha|\psi\rangle$ in Eq.~(\ref{eigeneq}) we have
\begin{equation}
 \langle\psi|A|\psi\rangle = \alpha \quad\mbox{and}\quad 
 \langle\psi|A^{\dagger}|\psi\rangle = \alpha^*\; .
\end{equation}
Substituting $A^{\dagger}=A$ immediately yields $\alpha^*=\alpha$, i.e., a real
eigenvalue. The second implication is proved by running the argument backwards.
In this thesis I will use the terms Hermitian and self-adjoint interchangeably.
Also, I will use the convention that Greek letters ($\alpha,\beta$\ldots) 
denote complex numbers and Roman letters ($a,b$\ldots) denote real numbers. 
The fact that self-adjoint operators have real eigenvalues lead to the next 
postulate.

\begin{description}
 \item[Postulate 3] The only possible measurement 
	outcomes\index{measurement!outcome} obtainable from the
	measurement of an observable\index{observable} are the eigenvalues 
	\index{eigen!-values}of its corresponding
	self-adjoint operator $A$. If the 
	state of the system is $|\psi\rangle$, then the probability $p(a_i)$ 
	of finding the $d_{a_i}$-fold degenerate eigenvalue $a_i$ of the 
	observable $A$ is equal to the probability of finding the system in 
	the corresponding eigenspace: \index{eigen!-space}
	\begin{equation}\label{postulate3}
	 p(a_i) = \sum_{j=1}^{d_{a_i}} |\langle\alpha_{ij}|\psi\rangle|^2\; ,
	\end{equation}
	where $|\alpha_{ij}\rangle$ are the eigenvectors\index{eigen!-vectors} 
	corresponding to the $d_{a_i}$-fold degenerate eigenvalue $a_i$. 
\end{description}

The outcome of a measurement in the laboratory can only yield a real number, 
and since the measurement outcomes are the eigenvalues of operators, these 
operators must be Hermitian.\index{operator!Hermitian} Postulate 2 ensures 
that there is a one-to-one correspondence between self-adjoint operators
\index{operator!self-adjoint} and physical observables, and 
postulate 3 determines the possible measurement outcomes for these observables.
Eq.~(\ref{postulate3}) is the so-called {\em Born rule}\index{Born rule} 
\cite{hilgevoord93}.

\begin{description}
 \item[Postulate 4] The evolution\index{evolution} of a system is governed by 
	a unitary transformation\index{transformation!unitary} $U$:
 	\begin{equation}\label{postulate4}
 	 |\psi(\theta')\rangle = U(\theta,\theta') |\psi(\theta)\rangle\; ,
	\end{equation}
	where $\theta$ and $\theta'$ are (vectors of) real parameters.
\end{description}

Any operator $U$ on a Hilbert space\index{Hilbert space} $\hilbert$ for which 
$U^{\dagger} = U^{-1}$ is called a {\em unitary} operator on $\hilbert$. In 
general, every unitary operator\index{operator!unitary} can be written as 
\begin{equation}\label{unipos4}
 U(\theta) = \exp\left( iA\theta \right)\; ,
\end{equation}
with $A$ a self-adjoint operator on $\hilbert$. To prove this statement, note 
that $U^{\dagger} = \exp(-iA^{\dagger}\theta) = \exp(-iA\theta)$. Thus 
$U^{\dagger}U = UU^{\dagger}= {\unity}$, and $U^{\dagger}=U^{-1}$. Unitary 
operators in matrix representation\index{matrix!representation} always have 
determinant 1, and they can be viewed as rotations in a complex vector 
space.\index{vector space!complex}

A special choice for $U(\theta,\theta')$ in Eq.~(\ref{postulate4}) is the 
infinitesimal time evolution\index{time evolution} where $\theta=t$ and 
$\theta'=t+dt$:
\begin{equation}
 U(t,t+dt) = \exp\left[iHt/\hbar-iH(t+dt)/\hbar\right] = \exp(-iHdt/\hbar)\; ,
\end{equation}
and $H$ the {\em Hamiltonian}\index{Hamiltonian} of the system. It is the 
observable associated 
with the total energy of the system. Substituting this evolution into
Eq.~(\ref{postulate4}) and neglecting higher-order powers of $dt$, we obtain
in the Taylor expansion\index{Taylor expansion}
\begin{eqnarray}
 |\psi(t+dt)\rangle &=& e^{-iHdt/\hbar} |\psi(t)\rangle \quad\Leftrightarrow\cr
 |\psi(t)\rangle + d|\psi(t)\rangle &=& \left( 1-\frac{i}{\hbar} H dt \right) 
 |\psi(t)\rangle \quad\Leftrightarrow\cr
 i\hbar \frac{d}{dt} |\psi(t)\rangle &=& H |\psi(t)\rangle\; .
\end{eqnarray}
This is the famous Schr\"odinger equation.\index{Schr\"odinger!equation} 

In this thesis I will not use the Schr\"odinger equation. Instead, I will use
the fact that the self-adjoint operator\index{operator!self-adjoint} $A$ in 
Eq.~(\ref{unipos4}) acts as a {\em generator}\index{group!generator} of the 
group of unitary evolutions\index{unitary evolution} parametrised by $\theta$.
This approach will have great benefits in chapter \ref{chap3}. The theory of 
(Lie) groups\index{group} and their generators is treated in appendix 
\ref{app:group}.

Unitary transformations\index{transformation!unitary} not only govern the 
evolution of quantum states, they also constitute basis transformations. 
\index{transformation!basis}If $A$ and $U$ are a linear 
operator\index{operator!linear} 
and a unitary transformation on $\hilbert$ respectively, then there exist 
another linear operator $A'$ on $\hilbert$ such that
\begin{equation}
 A' = U^{\dagger} A U\; .
\end{equation}
In particular, if $A$ is self-adjoint there always exist a unitary 
transformation such that $A'$ is diagonal.

\begin{description}
 \item[Postulate 5] When a measurement\index{measurement} of an observable $A$ 
	yields the 
	(non-degenerate) eigenvalue $a_i$, the state of the system immediately
	after the measurement will be the eigenstate $|\alpha_i\rangle$
	corresponding to $a_i$. 
\end{description}

This is the so-called {\em projection} postulate.\index{projection!postulate} 
It is often referred to as state collapse,\index{state!collapse} since a 
measurement can induce a discontinuous jump from a
superposition\index{superposition} to an eigenstate of the measured 
observable.\index{observable} This postulate 
has caused severe problems for interpretations of quantum mechanics which 
assign some form of `reality' to the state. Such interpretations suffer from 
what has become generically known as the `measurement 
problem'\index{measurement!problem} 
\cite{redhead87,hilgevoord93}. In this thesis I will ignore this problem, 
since it does not seem to have any effect on the experimental success of 
quantum mechanics\footnote{The reader should note that, although I will not
discuss the measurement problem, this does not imply that there {\em is} no
measurement problem. This is still very much open to debate \cite{fuchs00}.}.
The general theory of measurements is discussed in section {\ref{sec:meas}}.

So far, I have presented quantum mechanics in the so-called {\em Schr\"odinger
picture}.\index{Schr\"odinger!picture} In this picture the time dependence is 
captured in the {\em state}: $|\psi(t)\rangle$. Alternatively, we can choose 
the states to be time independent, and have all the time dependence in the 
{\em operators}. This is called the {\em Heisenberg 
picture}:\index{Heisenberg picture}
\begin{equation}
 A_{\rm H} = U^{\dagger}(t) A_{\rm S} U(t)\; ,
\end{equation}
where $A_{\rm H}$ denotes the operator in the Heisenberg 
picture\index{Heisenberg picture} and $A_{\rm S}$ the operator in the 
Schr\"odinger picture.\index{Schr\"odinger!picture} When part of the time 
dependence is in the states and part is in the operators, we speak of the 
{\em interaction picture}.\index{interaction picture}

Another alternative formulation of quantum mechanics is Feynman's {\em path 
integral formalism}. This is particularly useful in the formulation of quantum 
field theories, but I will not discuss it here.

\subsection{The linear harmonic oscillator}\index{harmonic oscillator}

One application of quantum mechanics which deserves attention in the context 
of this thesis is the description of the linear harmonic oscillator. I will 
treat this in a telegraphic manner, since this is a well known example. For
a full derivation see, for example, Merzbacher \cite{merzbacher98}.

We start by defining a quadratic potential $V(x)$ for a classical particle 
with mass $m$, position $x$ and momentum $p$:
\begin{equation}
 V(x) = \frac{1}{2} m\omega^2 x^2\; ,
\end{equation}
where $\omega$ is, loosely speaking, the classical frequency of the oscillator.
The classical Hamiltonian\index{Hamiltonian} is then given by the sum of the 
kinetic and potential energy:
\begin{equation}
 H_{\rm classical} = \frac{p^2}{2m} + \frac{m\omega^2 x^2}{2}\; .
\end{equation}
In quantum mechanics the observables $x$ and $p$ have to be replaced by
self-adjoint operators.\index{operator!self-adjoint} This procedure is called 
\index{quantisation}`quantisation'\footnote{Or {\em first} quantisation. 
\index{quantisation!first} Indeed, there is something called `second' 
quantisation,\index{quantisation!second} in which the fields are written in 
the operators formalism.
We will encounter this in section 2, where I introduce quantum optics and the
quantisation of the electro-magnetic field.}. The quantum mechanical 
Hamiltonian thus becomes
\begin{equation}
 H_{\rm quantum} = -\frac{\hbar^2}{2m}\frac{d^2}{dx^2} + 
 \frac{m\omega^2 x^2}{2}\; ,
\end{equation}
where $p\rightarrow\hat{p} = -i\hbar d/dx$ and $x\rightarrow\hat{x}=x$ with 
$[\hat{x},\hat{p}]=i\hbar$. When the quantum 
mechanical state of the harmonic oscillator is denoted by $\psi(x) =
\langle x|\psi\rangle$, with $|x\rangle$ the position eigenvector corresponding
to the position $x$, then we obtain the differential (Schr\"odinger) equation
(see postulate 4):
\begin{equation}
 \frac{d^2 \psi(x)}{dx^2} -\left(\frac{m\omega}{\hbar}\right)^2 x^2\, 
 \psi(x) = E\psi(x)\; .
\end{equation}
This equation is satisfied by the following class of wave-functions:
\index{wave!function}
\begin{equation}\label{lho}
 \psi_n (x) = 2^{-n/2} (n!)^{-1/2} \left(\frac{m\omega}{\hbar\pi}\right)^{1/4}
 \exp\left(-\frac{m\omega}{2\hbar}x^2\right) H_n \left(
 \sqrt{\frac{m\omega}{\hbar}}x\right)\; ,
\end{equation}
corresponding to energies
\begin{equation}
 E_n = \hbar\omega \left( n+\frac{1}{2}\right)\; .
\end{equation}
The $H_n(x)$ are the so-called {\em Hermite} polynomials\index{Hermite 
polynomial} (see appendix \ref{app:mdhp}).

The Hamiltonian\index{Hamiltonian} of the harmonic oscillator\index{harmonic 
oscillator} can also be expressed in terms of so-called {\em raising} and {\em 
lowering} operators\index{operator!raising}\index{operator!lowering} 
$\hat{a}^{\dagger}$ and $\hat{a}$ respectively:
\begin{equation}
 \hat{a}^{\dagger} \equiv \sqrt{\frac{m\omega}{2\hbar}} \left( \hat{x} - i
 \frac{\hat{p}}{m\omega} \right) \qquad\mbox{and}\qquad
 \hat{a} \equiv \sqrt{\frac{m\omega}{2\hbar}} \left( \hat{x} + i
 \frac{\hat{p}}{m\omega} \right)\; 
\end{equation}
(remember that $\hat{x}^{\dagger}=\hat{x}$ and $\hat{p}^{\dagger}=\hat{p}$
since position and momentum are physical observables).\index{observable} It is 
easily found that $[\hat{a},\hat{a}^{\dagger}]=1$. The 
eigenstate\index{eigen!-states} corresponding to the energy 
$E_n$ of the linear harmonic oscillator is now symbolically denoted by 
$|n\rangle$, and we have
\begin{equation}
 \hat{a}^{\dagger} |n\rangle = \sqrt{n+1}|n+1\rangle\; , \quad
 \hat{a} |n\rangle = \sqrt{n} |n-1\rangle \quad\mbox{and}\quad 
 \hat{a}^{\dagger} \hat{a} |n\rangle = n |n\rangle\; .
\end{equation}
The operator $\hat{a}^{\dagger}\hat{a}$ in the last equation is also called 
the {\em number} operator\index{operator!number} $\hat{n}$. The Hamiltonian of 
the linear harmonic oscillator in terms of the raising and lowering operators 
is then given by
\begin{equation}
 H_{\rm LHO} = \hbar\omega\left(\hat{a}^{\dagger}\hat{a}+\frac{1}{2}\right)\; .
\end{equation}
The raising and lowering operators will return in section \ref{2quant} as 
creation and annihilation operators.

\subsection{Composite and mixed states}\label{sec:mixed}\index{state!composite}

After this brief, but necessary digression I now return to the definition 
of states of {\em composite} systems. Postulate 1 tells us that with every
physical system corresponds a Hilbert space.\index{Hilbert space} Two systems, 
1 and 2, therefore 
have two Hilbert spaces ${\hilbert}_1$ and ${\hilbert}_2$. However, the 
composite system $1+2$ is also a physical system. The question is thus which
Hilbert space corresponds to system $1+2$.

Let $\{ |\psi_i\rangle_1 \}$ be an orthonormal basis for ${\hilbert}_1$ and 
let $\{ |\phi_j\rangle_2 \}$ be an orthonormal basis for ${\hilbert}_2$. When
the two systems are independent of each other, every basis vector in 
${\hilbert}_1$ can be paired with every basis vector in ${\hilbert}_2$ and 
still give a mathematically legitimate description of the composite system.
Therefore, one possible orthonormal basis for the Hilbert space of the 
composite system is given by the set of ordered pairs $\{ |\psi_i\rangle_1,
|\phi_j\rangle_2\}$. This is a basis of the {\em tensor 
product},\index{tensor product} or {\em direct} product\index{direct!product} 
of the two Hilbert spaces of the subsystems:
\begin{equation}
 {\hilbert}_{1+2} = {\hilbert}_1 \otimes {\hilbert}_2\; .
\end{equation}
An orthonormal basis\index{orthonormal basis} is given by 
$\{ |\psi_i\rangle_1\otimes|\phi_j\rangle_2\}$.

From postulate 1 and the fact that a Hilbert space\index{Hilbert space} is a 
complex vector space\index{vector space!complex}
we immediately see that any tensor product of two superpositions is again a
superposition\index{superposition} of tensor product states (see appendix 
\ref{app:vectorspaces}):
\begin{equation}\label{compstate}
 \sum_i \alpha_i |\psi_i\rangle_1\otimes \sum_j \beta_j |\phi_j\rangle_2 = 
 \sum_{ij} \alpha_i \beta_j |\psi_i\rangle_1\otimes|\phi_j\rangle_2\; ,
\end{equation}
i.e., the tensor product is linear. Note that the right-hand side can in
general {\em not} be written as a state 
$|\zeta_k\rangle_1\otimes|\xi_l\rangle_2$. This exemplifies the fact that two 
systems need not be independent of each other. This property, called {\em 
entanglement}\index{entanglement} is crucial to quantum information 
theory.\index{quantum!information} It will be discussed 
in detail later on in chapter \ref{chap3}.

The total state of two systems can always be written in a special form, called 
the {\em Schmidt decomposition}.\index{Schmidt decomposition} The most general 
composite state is given by 
Eq.~(\ref{compstate}), which involves a double sum over the indices $i$ and 
$j$. In the Schmidt decomposition the state is written as a single 
sum \cite{vonneumann55}:
\begin{equation}\label{smitdec}
 |\Psi\rangle_{12} = \sum_i c_i |\psi'_i,\phi'_i\rangle_{12}\; ,
\end{equation}
where $c_i$ can be chosen real and $\{ |\psi'_i\rangle_1\}$ and 
$\{ |\phi'_i\rangle_2\}$ are two orthonormal bases for the two subsystems.
The bases of the subsystems in eqs.~(\ref{compstate}) and (\ref{smitdec}) are
transformed into each other by a unitary 
transformation:\index{transformation!unitary}
\begin{equation}
 |\psi'_i\rangle_1 = \sum_k U_{ik} |\psi_k\rangle_1 \quad\mbox{and}\quad
 |\phi'_j\rangle_2 = \sum_l U_{jl} |\psi_l\rangle_2\; . 
\end{equation}

The Schmidt decomposition is unique (up to phase factors) if and only if the 
$c_i$ are non-degenerate. If the dimensions of the Hilbert spaces of the two 
subsystems are $d_1$ and $d_2$ respectively, the sum in Eq.~(\ref{smitdec}) 
the index $i$ runs up to the dimension of the smallest Hilbert space 
\cite{vonneumann55,peres95}. A Schmidt decomposition of the state of three or 
more subsystems exists only in special circumstances \cite{peres95b}.

When a state is in a superposition\index{superposition} $|\Psi\rangle = \sum_i 
\alpha_i|\psi_i\rangle$ with $\sum_i|\alpha_i|^2=1$, the operator 
$|\Psi\rangle\langle\Psi|$ is a so-called projection operator (see appendix 
\ref{app:vectorspaces}):
\begin{eqnarray}
 \left(|\Psi\rangle\langle\Psi|\right)^2 &=& \left( \sum_{ij}\alpha_i\alpha_j^*
 |\psi_i\rangle\langle\psi_j| \right)^2 = \sum_{ijkl} \alpha_i\alpha_j^*
 \alpha_k\alpha_l^* |\psi_i\rangle\langle\psi_j|\psi_k\rangle\langle\psi_l|\cr
 &=& \sum_{ijkl} \alpha_i\alpha_j^* \alpha_k\alpha_l^* |\psi_i\rangle
 \langle\psi_l| \delta_{jk} = \sum_j |\alpha_j|^2 \sum_{il}\alpha_i\alpha_l^*
 |\psi_i\rangle\langle\psi_l| \cr &=& \sum_{il}\alpha_i\alpha_l^*
 |\psi_i\rangle\langle\psi_l| = |\Psi\rangle\langle\Psi|\; .
\end{eqnarray}
We can now extend our notion of states for a system. In particular, suppose 
that we have a classical probability 
distribution\index{distribution!probability} over a set of 
states. We write this as 
\begin{equation}
 \rho = \sum_i p_i |\psi_i\rangle\langle\psi_i|\; ,
\end{equation}
where $p_i$ is the probability to find the system in state $|\psi_i\rangle$.
This is sometimes called an {\em incoherent} superposition.
Since the $p_i$ are probabilities, we have $\sum_i p_i =1$. The operator 
$\rho$ is called the {\em density operator}\index{operator!density} of the 
system. It is also referred to as a {\em mixed} state.\index{state!mixed} It 
has the following properties:
\begin{enumerate}
 \item $\rho^{\dagger} = \rho$;
 \item $\langle\psi|\rho|\psi\rangle \geq 0$ for all $|\psi\rangle$;
 \item ${\rm Tr}\rho=1$. 
\end{enumerate}
In a complex Hilbert space,\index{Hilbert space} properties 1 and 2 are 
equivalent.

\subsection{Measurements}\label{sec:meas}\index{measurement}

I will now consider the effect of a measurement on the state of a system. 
According to the projection postulate,\index{projection!postulate} immediately 
after the measurement of an observable\index{observable} $A$, the state of the 
system is in the eigenstate\index{eigen!-states} $|\psi_i\rangle$ corresponding
to the eigenvalue $a_i$ found in the measurement outcome. With the knowledge 
of projection operators\index{operator!projection} given in appendix
\ref{app:vectorspaces} we can now model this as follows.

Suppose that the system under consideration is in a mixed state $\rho$. Let
the eigenvalues of $A$ be given by $\{ a_i\}$. Then the probability $p(a_i)$ 
that we obtain outcome $a_i$ in a measurement of $A$ is given by 
\begin{equation}\label{mixedborn}
 p(a_i) = \langle\psi_i|\rho|\psi_i\rangle = 
 {\rm Tr}\left( \rho |\psi_i\rangle\langle\psi_i| \right)\; .
\end{equation}
The right-hand side can be shown to equal the centre term by using the cyclic 
property of the trace. This type of measurement is called a {\em Von Neumann} 
measurement\index{measurement!Von Neumann} or {\em ideal} measurement 
\index{measurement!ideal}\cite{hilgevoord93,merzbacher98,vonneumann55}. The 
underlying assumption in this model is that the measurement outcome faithfully 
identifies the state of the system immediately after the measuring process. 

In practice, this is of course not always the case. Instead, due to the 
imperfections of the measurement apparatus there might be a whole family of
projectors $\{|\psi_k\rangle\langle\psi_k|\}$ which, with some probability
$\eta_k>0$, give rise to the measurement outcome $p(a_i)$. Rather than a 
projection operator $|\psi_i\rangle\langle\psi_i|$ in Eq.~(\ref{mixedborn}) 
we include a {\em projection operator valued measure}, or POVM:\index{POVM}
\begin{equation}
 P_{|\psi_i\rangle} = |\psi_i\rangle\langle\psi_i| ~\longrightarrow~
 E_{\mu} = \sum_k \eta_k^{\mu} |\psi_k\rangle\langle\psi_k|\; ,
\end{equation}
with $\sum_{\mu} E_{\mu} = \unity$. When $\{|\psi_k\rangle\}$ is an 
orthonormal basis,\index{orthonormal basis} this implies 
$\sum_{\mu}\eta_k^{\mu}=1$. In chapters 
\ref{chap4} and \ref{teleportation}, I will use these POVM's to model non-ideal
measurements\index{measurement!non-ideal} in the context of quantum optics. A 
more formal presentation of POVM's is given in appendix \ref{app:povm}.

\section{Quantum optics}\label{2quant}

In this section I present quantum optics;\index{quantum optics} the quantum 
theory of light. First, the electro-magnetic 
field\index{field!electro-magnetic} is quantised and given a particle 
interpretation, yielding the concept of photons.\index{photon} Then I describe 
various optical components in terms of unitary evolutions\index{unitary 
evolution} and their generators.

\subsection{Quantisation of the electro-magnetic field}

Quantum mechanics, as presented in the previous section, can describe a 
particle in an electro-magnetic field given by a vector potential\index{vector 
potential} ${\mathbf A}({\mathbf r},t)$ by making the following substitution:
\begin{equation}
 \hat{\mathbf p} ~\longrightarrow~ \hat{\mathbf p} - e \hat{\mathbf A}
 (\hat{\mathbf r},t)\; ,
\end{equation}
where $e$ is the charge of the particle and $\hat{\mathbf A}$ is the vector 
potential operator obtained by replacing the coordinates ${\mathbf r}$ by their
corresponding operator $\hat{\mathbf r}$. 

Alternatively, the quantisation of the electro-magnetic field can be derived 
from the Maxwell equations\index{Maxwell equations} for the electric and 
magnetic field ${\mathbf E}$ and ${\mathbf B}$ respectively\footnote{I avoid 
the notation ${\mathbf H}$, 
because its components may be confused with the Hamiltonian $H$ later on.}:
\begin{eqnarray}\label{maxwell}
 \nabla\times{\mathbf B} - \frac{1}{c^2}\frac{\partial{\mathbf E}}{\partial 
 t} = 0 \quad &,&\quad \nabla\cdot{\mathbf B} = 0 \; , \cr && \cr
 \nabla\times{\mathbf E} + \frac{\partial{\mathbf B}}{\partial t} = 0
 \quad &,&\quad \nabla\cdot{\mathbf E} = 0\; ,
\end{eqnarray}
with $c$ the velocity of light in free space.
For the fully quantum mechanical description of the electro-magnetic field
\index{field!electro-magnetic} in free space, I will follow the derivation of 
Scully and Zubairy \cite{scully97}. Other books on quantum optics include 
Loudon \cite{loudon83} and Walls and Milburn \cite{walls94}.

Suppose we want to quantise the electro-magnetic field in a cavity with length
$L$ and volume $V$. Classically, we can describe the electric field in terms 
of the transverse modes\index{mode!transverse} in the $x$-direction:
\begin{equation}\label{electric}
 E_x (z,t) = \sum_j A_j q_j(t) \sin k_j z\; ,
\end{equation}
where $z$ is the propagation direction, $q_j(t)$ the mode amplitude, 
$k_j=j\pi/L$ the wave number\index{wave!number} and $A_j$ a proportionality 
constant:
\begin{equation}
 A_j = \sqrt{\frac{2\nu_j^2 m_j}{V\epsilon_0}}\; .
\end{equation}
The $\nu_j$ are the eigenfrequencies\index{eigen!-frequencies} of the cavity. 
The constant $m_j$ is a dummy mass, included to make the subsequent argument 
more suggestive \cite{scully97}. 

From the Maxwell equations\index{Maxwell equations} (Eq.~(\ref{maxwell})), we 
can derive the magnetic field (which is only non-zero in the $y$-direction):
\begin{equation}\label{magnetic}
 B_y (z,t) = \sum_j A_j \dot{q}_j(t) \frac{\epsilon_0}{\mu_0 k_j} \cos k_jz\; ,
\end{equation}
where $\dot{q}_j(t)$ denotes the time derivative of the mode amplitude $q_j$.
The classical Hamiltonian\index{Hamiltonian} then reads
\begin{equation}
 H_{\rm classical} = \frac{1}{2} \int_V dv \left( \epsilon_0 E_x^2 + 
 \mu_0^{-1}B_y^2\right)\; .
\end{equation}
After substitution of Eqs.~(\ref{electric}) and (\ref{magnetic}) in the 
classical Hamiltonian and integrating over the cavity volume we obtain
\begin{equation}\label{hclassical}
 H_{\rm classical} = \frac{1}{2} \sum_j \left( m_j \nu_j^2 q_j^2(t)
 + m_j \dot{q}_j^2(t) \right)\; .
\end{equation}
When we write $p_j=m_j \dot{q}_j$, this has exactly the same form as the 
classical Hamiltonian of the harmonic oscillator.\index{harmonic oscillator} 
Therefore, when we want to quantise the
electro-magnetic field we proceed in a similar fashion as in section 1.{\sl c}.
We replace the variables $q_j$ and $p_j$ by their respective operators 
$\hat{q}_j$ and $\hat{p}_j=i\hbar\partial/\partial q_j$.

There are, however, several subtleties. The variables $q_j$ are {\em 
amplitudes} of the field modes, and {\em not} coordinates, as is the case in 
the linear harmonic oscillator.\index{harmonic oscillator} The variables 
$\dot{q}_j$ are their 
corresponding conjugate variables, which facilitate the position-momentum 
interpretation since $[\hat{q},\dot{\hat{q}}]=1$. But this is really {\em 
field quantisation},\index{field!quantisation} or second 
quantisation.\index{quantisation!second} Secondly, the masses $m_j$ do 
{\em not} have any physical meaning. They are removed by changing our 
description from $q_j$ and $\dot{q}_j$ to creation\index{operator!creation} 
and annihilation operators\index{operator!annihilation} 
$\hat{a}_j^{\dagger}$ and $\hat{a}_j$, as I shall now show.

\subsection{Creation and annihilation operators}

Starting with the classical Hamiltonian of the free field in 
Eq.~(\ref{hclassical}) and replacing the variables $q_j$ and $p_j$ with the
quantum mechanical operators $\hat{q}_j$ and $\hat{p}_j$, we obtain the 
quantum mechanical Hamiltonian\index{Hamiltonian}
\begin{equation}
 H = \frac{1}{2} \sum_j \left( m_j \nu_j^2 \hat{q}_j^2(t) +
 \frac{\hat{p}_j^2}{m_j} \right)\; ,
\end{equation}
where 
\begin{equation}\label{cancomrel}
 [\hat{q}_j,\hat{p}_{j'}] = i\hbar \delta_{jj'} \qquad\mbox{and}\qquad
 [\hat{q}_j,\hat{q}_{j'}] =  [\hat{p}_j,\hat{p}_{j'}] = 0\; .
\end{equation}
We can now make the canonical transformation\index{transformation!canonical} 
to the operators $\hat{a}_j^{\dagger}$ and $\hat{a}_j$:
\begin{eqnarray}
 \hat{a}_j e^{-i\nu_j t} &=& \frac{1}{\sqrt{2m_j\hbar\nu_j}} \left( m_j\nu_j
 \hat{q}_j + i\hat{p}_j \right)\; , \cr
 \hat{a}^{\dagger}_j e^{i\nu_j t} &=& \frac{1}{\sqrt{2m_j\hbar\nu_j}} \left( 
 m_j\nu_j \hat{q}_j - i\hat{p}_j \right)\; .
\end{eqnarray}
The Hamiltonian in terms of the creation and annihilation operators thus 
becomes
\begin{equation}
 H = \hbar \sum_j \nu_j \left( \hat{a}_j^{\dagger} \hat{a}_j + \frac{1}{2} 
 \right)\; . 
\end{equation}
With every operator $\hat{a}_j^{\dagger}$ corresponds a mode $a_j$. This is a
Hamiltonian for a {\em massless} quantum field.\index{field!massless quantum-} 
At this point I should briefly clarify my notation. Since the creation and 
annihilation operators are closely related to the modes they act upon, I make
the distinction between modes and operators by writing the operators with a 
hat. Although observables like the Hamiltonian and unitary transformations are 
also operators, they do not yield such a potential ambiguity, and I will not 
write them with hats. 

Using the canonical commutation relations given by Eq.~(\ref{cancomrel}) we
immediately see that 
\begin{equation}
 [\hat{a}_j,\hat{a}^{\dagger}_{j'}] = \delta_{jj'}\; , \qquad\mbox{and}\qquad
 [\hat{a}_j,\hat{a}_{j'}] = [\hat{a}_j^{\dagger},\hat{a}_{j'}^{\dagger}] =0\; .
\end{equation}
The electric and magnetic fields after second 
quantisation\index{quantisation!second} thus read
\begin{eqnarray}
 \hat{E}_x(z,t) &=& \sum_j {\mathcal{E}}_j \left( \hat{a}_j e^{-i\nu_j t} 
 + \hat{a}_j^{\dagger} e^{i\nu_j t} \right) \sin k_j z \cr
 \hat{B}_x(z,t) &=& -i\frac{\epsilon_0 c}{\mu_0} \sum_j {\mathcal{E}}_j \left( 
 \hat{a}_j e^{-i\nu_j t} -\hat{a}_j^{\dagger} e^{i\nu_j t} \right)\cos k_jz\; ,
\end{eqnarray}
with the field strength\index{field!strength}
\begin{equation}
 {\mathcal{E}}_j = \left( \frac{\hbar\nu_j}{\epsilon_0 V} \right)^{1/2}\; .
\end{equation}

Just as in the case of the linear harmonic oscillator,\index{harmonic 
oscillator} we can write the energy eigenstates of one mode of the 
electro-magnetic\index{field!electro-magnetic} field as $|n\rangle$:
\begin{equation}
 H|n\rangle = \hbar\nu \left(\hat{a}^{\dagger} \hat{a} +\frac{1}{2} \right)
 = E_n |n\rangle\; .
\end{equation}
By applying the annihilation operator\index{operator!annihilation} on the last 
two sides of this equation and using the commutation relations 
\index{commutation relation} we easily find \cite{scully97}:
\begin{equation}
 \hat{a}|n\rangle = \sqrt{n}|n-1\rangle\; , \quad \hat{a}^{\dagger}|n\rangle 
 = \sqrt{n+1}|n+1\rangle\; , \quad  \hat{a}^{\dagger}\hat{a}|n\rangle = 
 \hat{n} |n\rangle = n|n\rangle\; ,
\end{equation}
with $E_n = \hbar\nu(n+1/2)$. The eigenstates are orthonormal:
$\langle m|n\rangle = \delta_{mn}$.

Rather than interpreting the eigenstates\index{eigen!-states} $|n\rangle$ as the 
energy levels for a fixed system, in quantum optics the state $|n\rangle$ 
denotes a state of $n$ {\em quanta}.\index{quanta} The quanta corresponding 
to the electro-magnetic field are the light-quanta or {\em photons}.
\index{photon} The operators $\hat{a}^{\dagger}$ and 
$\hat{a}$ thus {\em create} and {\em destroy} photons. Generally, in quantum 
field theory a field or a wave function is quantised, and the excited modes 
is given a particle interpretation\index{particle interpretation} 
\cite{brown88,ryder96}.

At this point I would like to stress that a single photon {\em does not have a
wave function} \cite{newton49}. It is the excitation of the electro-magnetic 
field (see also
Ref.\ \cite{scully97}). In the rest of this thesis I let $|n\rangle_{a_j}$ 
denote the state of the field, giving the number of photons $n$ in mode $a_j$.

In quantum mechanics, with every physical system corresponds a Hilbert space. 
\index{Hilbert space}
An orthonormal basis for a single-mode system $a$ is given by $\{|n\rangle\}$,
which spans an infinite dimensional Hilbert space. When we have several modes 
$\{ a_j\}$ in our system, the total Hilbert space of the system is a tensor 
product of the Hilbert spaces of the separate modes with orthonormal basis
$\{ |\vec{n}\rangle\equiv|n_1,\ldots,n_N\rangle\}$. This total Hilbert space 
can be uniquely decomposed into subspaces with fixed photon number. A Hilbert
space with this property is called a Fock space $\fock$:\index{Fock space}
\begin{equation}
 {\fock} = {\hilbert}_0\oplus{\hilbert}_1\oplus{\hilbert}_2\oplus\cdots\; ,
\end{equation}
where ${\hilbert}_k$ denotes the subspace\index{subspace} spanned by the 
vectors $|n_1,\ldots, n_N\rangle$ with $\sum_i n_i=k$ for an $N$-mode system. 
The states $|n\rangle$ are also called {\em Fock states}.\index{state!Fock} 
The subspace ${\hilbert}_0$ is a one-dimensional subspace, better known as 
the {\em vacuum}.\index{vacuum}

As a last remark, a tensor product state\index{state!tensor product} having a 
total of $k$ photons may involve one or more modes in vacuum.\index{vacuum} 
For example $|n,0,m\rangle$, where
$|0\rangle$ denotes the vacuum. The state $|n,0,m\rangle$ is part of the basis 
spanning the subspace ${\hilbert}_{k=n+m}$. The second mode is also said to be 
in the vacuum state. This ambiguity is further explored in section 
{\ref{sec:vacuum}}.

\subsection{Coherent and squeezed states}\index{state!coherent}
\index{state!squeezed}

The creation\index{operator!creation} and 
annihilation\index{operator!annihilation} operators are very important in 
quantum optics. We can therefore ask what the eigenstates\index{eigen!-states} 
of, e.g., the annihilation operator are. Consider the eigenvalue equation 
\index{eigen!-value!equation} for the annihilation operator:
\begin{equation}
 \hat{a} |\alpha\rangle = \alpha |\alpha\rangle\; .
\end{equation}
The eigenstate can be expanded in terms of number states\index{state!number} 
\cite{scully97}:
\begin{equation}
 |\alpha\rangle = e^{-|\alpha|^2/2} \sum_n \frac{\alpha^n}{\sqrt{n!}}
 |n\rangle\; ,
\end{equation}
which can be written as
\begin{equation}
 |\alpha\rangle = e^{-|\alpha|^2/2} e^{\alpha\hat{a}^{\dagger}} |0\rangle\; .
\end{equation}
The corresponding {\em displacement} operator\index{operator!displacement} 
can then be written as
\begin{equation}
 |\alpha\rangle = D(\alpha) |0\rangle \equiv e^{-|\alpha|^2/2} 
 e^{\alpha\hat{a}^{\dagger}-\alpha^*\hat{a}} |0\rangle\; ,
\end{equation}
since the second term in the exponential does not change its behaviour when 
applied to the vacuum.\index{vacuum} The operator $D(\alpha)$ is unitary, with 
$D^{\dagger}(\alpha)=D(-\alpha)=D^{-1}(\alpha)$. When acting on a creation or 
annihilation operator, we have
\begin{eqnarray}\label{2coherent}
 D^{-1}(\alpha) \hat{a} D(\alpha) &=& \hat{a} + \alpha\; , \cr
 D^{-1}(\alpha) \hat{a}^{\dagger} D(\alpha) &=& \hat{a}^{\dagger}+\alpha^*\; .
\end{eqnarray}
The states $|\alpha\rangle$ are called {\em coherent 
states}.\index{state!coherent} On a single mode,
there are no two coherent states which are orthogonal:
\begin{equation}
 \langle\alpha|\alpha'\rangle = \exp\left( -\frac{1}{2}|\alpha|^2 + 
 \alpha'\alpha^* - \frac{1}{2}|\alpha'|^2 \right)\; .
\end{equation}
This is only zero when $\alpha=\alpha'=0$.

The creation and annihilation operators do not commute, and as a consequence,
(exponential) functions of these operators generally cannot be rewritten 
according to the rules of normal arithmetic. In particular, when $[A,B]\neq 0$ 
we have $e^{A+B}\neq e^A e^B$. From a computational point of view, it is often 
convenient to deal with the annihilation operators first, and then the creation
operators. Especially when the state acted upon is the vacuum, the annihilation
operators will yield zero, thus simplifying the task. A function of these 
operators which is written as
\begin{equation}
 f(\hat{a}_1^{\dagger},\hat{a}_1,\ldots,\hat{a}_N^{\dagger},\hat{a}_N) =
 \sum_{j} g_j(\hat{a}_1^{\dagger},\ldots,\hat{a}_N^{\dagger})\,
 h_j(,\hat{a}_1,\ldots,\hat{a}_N)
\end{equation}
is said to be in {\em normal ordered form}.\index{normal ordering} For every 
term in the sum, the annihilation operators\index{operator!annihilation} are 
placed on the right and the creation operators\index{operator!creation} on 
the left. When the positions of these operators are reversed (i.e., creation 
operators on the right), we speak of {\em anti-normal 
ordering}.\index{anti-normal ordering}

Let me consider a simple example. The second order term in the displacement
operator (with $\alpha$ real for simplicity) is proportional to 
$(\hat{a}^{\dagger}-\hat{a})^2$. In normal ordered form, this is equal to 
$(\hat{a}^{\dagger})^2 + \hat{a}^2 - 2\hat{a}^{\dagger}\hat{a} - 1$. Note the 
$-1$ in this expression. For higher order terms, the `non-arithmetic' addition 
becomes more complicated, until finally, we arrive at \cite{scully97}:
\begin{equation}
 D(\alpha) = e^{\alpha\hat{a}^{\dagger} - \alpha^*\hat{a}} =
 e^{-|\alpha|^2/2} e^{\alpha\hat{a}^{\dagger}} e^{-\alpha^*\hat{a}}\; .
\end{equation}
Later in this section I will return to the normal ordering in more general 
terms.

\bigskip

Just as the displacement operator\index{operator!displacement} creates 
coherent states, we can construct a squeezing 
operator\index{operator!squeezing} which creates so-called {\em squeezed} 
states \index{state!squeezed}
\cite{walls83,loudon87,scully97}:
\begin{equation}\label{singlemodesq}
 S(\xi)|0\rangle = \exp\left( \frac{1}{2}\xi^*\hat{a}^2 - 
 \frac{1}{2}\xi(\hat{a}^{\dagger})^2 \right)|0\rangle = |\xi\rangle\; .
\end{equation}
This operator is also unitary: $S^{\dagger}(\xi)=S(-\xi)=S^{-1}(\xi)$. It 
transforms the creation and annihilation operators according to 
($\xi=re^{i\theta}$)
\begin{eqnarray}\label{2squeez}
 S^{-1}(\xi)\, \hat{a}\, S(\xi) &=& \hat{a} \cosh r - \hat{a}^{\dagger} 
 e^{i\theta}\sinh r\; , \cr
 S^{-1}(\xi)\, \hat{a}^{\dagger}\, S(\xi) &=& \hat{a}^{\dagger} \cosh r - 
 \hat{a}\, e^{-i\theta} \sinh r\; .
\end{eqnarray}

The squeezing operator can also be written in normal ordered form 
\cite{fisher84}:
\begin{eqnarray}
 S(\xi) &=& \exp\left[\frac{e^{i\theta}}{2}\tanh r (\hat{a}^{\dagger})^2\right]
 \exp\left[ -\ln(\cosh r) \left( \hat{a}^{\dagger}\hat{a}+\frac{1}{2} \right) 
 \right]\cr && \qquad\times \exp\left(-\frac{e^{-i\theta}}{2} \tanh r 
 \hat{a}^2 \right)\; .
\end{eqnarray}
Rather than deriving this formula, I will now concentrate on the so-called
Baker-Campbell-Hausdorff formula.\index{Baker-Campbell-Hausdorff}

We have seen that for non-commuting operators $A$ and $B$ we have $e^A e^B\neq
e^{A+B}$. The natural question to ask is then: what {\em is} $e^A e^B$? The 
relationship between the two is given by a Baker-Campbell-Hausdorff formula.
There are several ways in which we can write this formula, and here I will 
give two (without proof; the interested reader is referred to, e.g., Gilmore
\cite{gilmore94}):
\begin{eqnarray}\label{2bch}
 e^A e^B &=& e^{A+B + \frac{1}{2}[A,B] + \frac{1}{12}[A,[A,B]] + 
 \frac{1}{12}[[A,B],B] + \cdots}\; ,\cr
 e^{-B} A e^B &=& A + [A,B] + \frac{1}{2!} [[A,B],B] + \frac{1}{3!} 
 [[[A,B],B],B] + \cdots \; .
\end{eqnarray}
When $A=\alpha\hat{a}^{\dagger}$ and $B=\alpha^*\hat{a}$, the first BCH 
formula\index{Baker-Campbell-Hausdorff} immediately gives the normal ordered 
form\index{normal ordering} for the displacement\index{operator!displacement} 
operator: $[A,B]=|\alpha|^2$, which is a constant. Therefore, repeated 
commutators are zero and the BCH formula terminates.

The sum over repeated commutators does {\em not} terminate in general, in 
particular when $A$ and $B$ form a Lie algebra\index{algebra} (possibly with 
a set of other operators $C$, $D$,\ldots). When $A$ and $B$ generate an 
$su(1,1)$\index{algebra!$su(1,1)$} or an $su(2)$\index{algebra!$su(2)$} 
algebra (with $C = \pm[A,B]/2$ the third generator\index{group!generator} of 
the respective 
algebras) the BCH formula consists of an infinite number of terms, which 
converge to exponential functions of the generators $A$, $B$ and $C$ 
\cite{truax85}. But even this convergence is not guaranteed. In some cases it 
is just not possible to write a function of operators in a concise normal 
ordered form. In appendix \ref{app:forms} I further discuss the squeezing
operator\index{operator!squeezing} in connection with the $su(1,1)$ algebra. 
Because of the particular non-compactness for the group $SU(1,1)$, 
\index{group!$SU(1,1)$} we recognise a one-to-one 
correspondence between squeezing and $SU(1,1)$.

More properties of the squeezing and displacement operators can be found in
Ref.\ \cite{scully97}. I will now turn my attention to so-called {\em 
multi-mode} squeezing.\index{squeezer!multi-mode}

Eq.~(\ref{singlemodesq}) is defined for a single mode $a$. However, we can 
apply this single-mode squeezing operator to several distinct modes
\begin{eqnarray}
 |\vec{\xi}\rangle &\equiv& S(\vec{\xi}) |0\rangle =
 S(\xi_1) \times \ldots\times S(\xi_N) |0\rangle\cr && \cr &=& 
 \exp\left[ \frac{1}{2}\sum_{j=1}^N \left( \xi_j^* \hat{a}_j^2 - \xi_j
 (\hat{a}_j^{\dagger})^2 \right)\right] |0\rangle\; .
\end{eqnarray}
After an $N$-mode basis transformation\index{transformation!basis} $U$ we 
obtain
\begin{equation}
 |\vec{\xi}'\rangle = U^{\dagger} S(\vec{\xi}) U |0\rangle
 = S'(\vec{\xi}) |0\rangle\; .
\end{equation}
The last operator $S'$ can in general be written as 
\begin{equation}\label{2gaussian}
 S'(\vec{\xi}) = \exp\left[ \frac{1}{2} \sum_{j,k=1}^N \left( A^*_{jk}
 \hat{a}_j \hat{a}_k + B_{jk}\hat{a}^{\dagger}_j\hat{a}_k - 
 A_{jk}\hat{a}^{\dagger}_j\hat{a}^{\dagger}_k\right)\right]\; ,
\end{equation}
with $A$ and $B$ complex symmetric matrices. This is multi-mode squeezing. 
It has been studied among others by Caves \cite{caves82}, Barnett and Knight 
\cite{barnett85}, and Caves and Schumaker \cite{caves85,schumaker85}. The 
normal-ordering of these operators has been studied by Yuen \cite{yuen76}, 
Fisher {\em et al}.\ \cite{fisher84} and Truax \cite{truax85}.

Multi-mode squeezing\index{squeezer!multi-mode} is of fundamental importance 
to this thesis. In the next chapter I study whether operators of the form of 
Eq.~(\ref{2gaussian}) can yield so-called {\em event-ready entanglement}. 
\index{entanglement!event-ready}In chapter \ref{chap5} I 
determine the general state of Eq.~(\ref{2gaussian}) when, in addition, 
conditional measurements\index{measurement!conditional} are included. Also 
parametric down-conversion,\index{down-converter} a 
technique which will appear frequently in this thesis, can be described by 
this evolution. It is now also clear why I present unitary evolutions 
\index{unitary evolution} in terms of generators\index{group!generator} rather 
than the Schr\"odinger equation.\index{Schr\"odinger!equation} 
Eq.~(\ref{2gaussian}) is not necessarily an interaction 
Hamiltonian,\index{interaction Hamiltonian} but we 
can still consider the evolution it yields.

\subsection{Optical components}\index{optical!components}

The multi-mode displacement\index{operator!displacement} and squeezing 
operators\index{operator!squeezing} are not just mathematical
inventions, they correspond to physical devices. A coherent displacement of 
the vacuum\index{vacuum} yields a state which can be generated by a laser.
\index{laser} Squeezed states\index{state!squeezed}
can be generated by, for instance, optical parametric oscillators 
\cite{walls94} or parametric down-conversion \cite{kwiat95}.

Another type of optical components is given by unitary evolutions, the 
generator of which leaves the photon number\index{photon!-number} invariant. 
The most important one is the {\em beam-splitter}.\index{beam-splitter} 
Physically, the beam-splitter consists of a semi-reflective mirror: when light 
falls on this mirror part will be reflected
and part will be transmitted. 

\begin{figure}[t]
  \begin{center}
  \begin{psfrags}
     \psfrag{ain}{$a_{\rm in}$}
     \psfrag{aout}{$b_{\rm out}$}
     \psfrag{bin}{$b_{\rm in}$}
     \psfrag{bout}{$a_{\rm out}$}
     \psfrag{BS}{BS}
     \epsfxsize=8in
     \epsfbox[-100 20 700 380]{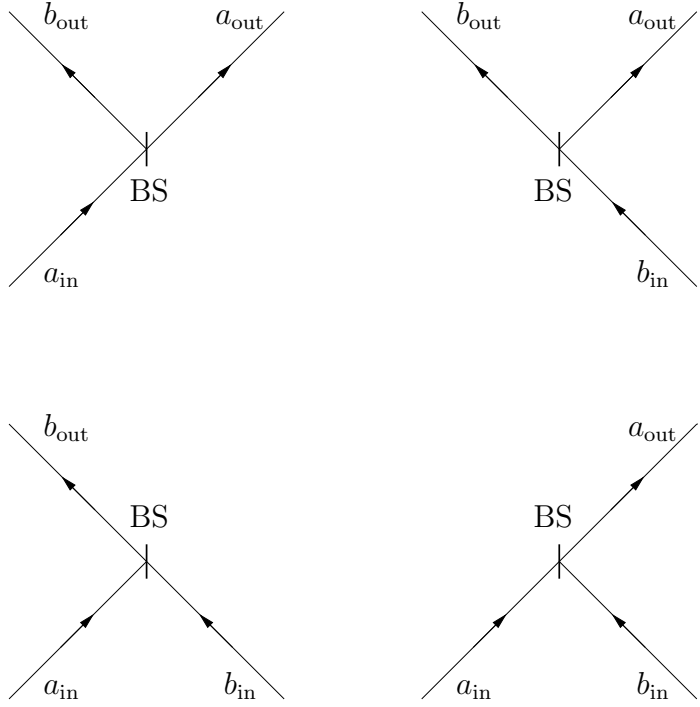}
  \end{psfrags}
  \end{center}
  \caption{The four global modes of the beam-splitter.}
  \label{fig:2.1}
\end{figure}

Let the two incoming modes be denoted by $\hat{a}_{\rm in}$ and $\hat{b}_{\rm 
in}$ respectively. The outgoing modes are denoted by $\hat{a}_{\rm out}$ and 
$\hat{b}_{\rm out}$. There are four {\em global} modes,\index{mode!global} 
depicted in figure \ref{fig:2.1}.

When a photon\index{photon} is incident on a beam-splitter, it has a certain 
probability of 
being reflected and a certain probability that it is transmitted. When we 
parametrise the probability amplitudes of these possibilities as $\cos\theta$
and $\sin\theta$, then in operator form the beam-splitter yields an evolution
\begin{eqnarray}\label{2bs}
 \hat{a}_{\rm out}^{\dagger} &=& \cos\theta\,\hat{a}_{\rm in}^{\dagger} + 
 \sin\theta\,\hat{b}_{\rm in}^{\dagger} \; , \cr
 \hat{b}_{\rm out}^{\dagger} &=& \sin\theta\,\hat{a}_{\rm in}^{\dagger} - 
 \cos\theta\,\hat{b}_{\rm in}^{\dagger} \; ,
\end{eqnarray}
and similar relations for the annihilation operators.
The reflection and transmission coefficients $R$ and $T$ of the beam-splitter 
are $R=\cos^2\theta$ and $T=1-R=\sin^2\theta$. The relative phase shift in the
second relation ensures that the transformation is unitary. This means that 
the beam-splitter is an asymmetric device.

Alternatively, we can write the beam-splitter evolution in terms of a unitary
operator generated by an Hermitian operator.\index{operator!Hermitian} 
Eq.~(\ref{2bs}) can be interpreted
as the beam-splitter version of Eqs.~(\ref{2coherent}) and (\ref{2squeez}).
The question is therefore what the corresponding unitary 
transformation\index{transformation!unitary} is 
(analogous to $D(\alpha)$ and $S(\xi)$). Using the second line in 
Eq.~(\ref{2bch}) we can easily verify that 
\begin{eqnarray}
 \hat{a}_{\rm out}^{\dagger} &=& e^{{\theta}(\hat{a}_{\rm in}^{\dagger}
 \hat{b}_{\rm in} - \hat{a}_{\rm in}\hat{b}_{\rm in}^{\dagger})} \,
 \hat{a}_{\rm in}^{\dagger}\, e^{-{\theta}(\hat{a}_{\rm in}^{\dagger}
 \hat{b}_{\rm in} - \hat{a}_{\rm in}\hat{b}_{\rm in}^{\dagger})} =
 \cos\theta\,\hat{a}_{\rm in}^{\dagger} + \sin\theta\,
 \hat{b}_{\rm in}^{\dagger}\; , \cr && \cr
 \hat{b}_{\rm out}^{\dagger} &=& e^{{\theta}(\hat{a}_{\rm in}^{\dagger}
 \hat{b}_{\rm in} - \hat{a}_{\rm in}\hat{b}_{\rm in}^{\dagger})} \,
 \hat{b}_{\rm in}^{\dagger}\, e^{-{\theta}(\hat{a}_{\rm in}^{\dagger}
 \hat{b}_{\rm in} - \hat{a}_{\rm in}\hat{b}_{\rm in}^{\dagger})} =
 \sin\theta\,\hat{a}_{\rm in}^{\dagger} - \cos\theta\,
 \hat{b}_{\rm in}^{\dagger}\; .
\end{eqnarray}
In general, the generator\index{group!generator} $H_{\rm BS}$ of the 
beam-splitter\index{beam-splitter} evolution 
$U=\exp(iH_{\rm BS})$ is given by
\begin{equation}\label{bsint}
 H_{\rm BS} = -i\lambda\hat{a}_{\rm in}^{\dagger}\hat{b}_{\rm in} +
 i\lambda^* \hat{a}_{\rm in}\hat{b}_{\rm in}^{\dagger}\; .
\end{equation}
Since the photon-number\index{photon!-number} is conserved in the 
beam-splitter, the operator $H_{\rm BS}$ commutes with the number operator: 
\index{operator!number}$[H_{\rm BS},\hat{n}]=0$. Furthermore,
$H_{\rm BS}$ is a generator of an $su(2)$ algebra.\index{algebra!$su(2)$}

The same mathematical description applies to the evolution due to a 
{\em polarisation rotation}.\index{polarisation rotator} Instead of having 
two different spatial modes 
$a_{\rm in}$ and $b_{\rm in}$, the two incoming modes have different 
polarisations. We write $\hat{a}_{\rm in}\rightarrow\hat{a}_x$ and 
$\hat{b}_{\rm in}\rightarrow\hat{a}_y$, for some rectilinear set of 
coordinates $x$ and $y$. The parameter $\theta$ is now the angle of rotation:
\begin{eqnarray}\label{2polrot}
 \hat{a}_{x'}^{\dagger} &=& \cos\theta\,\hat{a}_x^{\dagger} + 
 \sin\theta\,\hat{a}_y^{\dagger} \; , \cr
 \hat{a}_{y'}^{\dagger} &=& \sin\theta\,\hat{a}_x^{\dagger} - 
 \cos\theta\,\hat{a}_y^{\dagger} \; .
\end{eqnarray}
This evolution has the same generator as the beam-splitter, except that the
angle $\theta$ is real for the polarisation rotation, whereas for the 
beam-splitter we admit complex $\lambda$'s in Eq.\ (\ref{bsint}).

Another important optical component is the single-mode {\em phase shift}:
\index{phase shift}
\begin{equation}
 \hat{a}^{\dagger}_{\rm out} = e^{i\varphi}\hat{a}^{\dagger}_{\rm in}\; .
\end{equation}
It is easily verified that 
\begin{equation}
 e^{i\varphi\hat{a}_{\rm in}^{\dagger}\hat{a}_{\rm in}}\,
 \hat{a}_{\rm in}^{\dagger}\, e^{-i\varphi\hat{a}_{\rm in}^{\dagger}
 \hat{a}_{\rm in}} = e^{i\varphi} \hat{a}_{\rm in}^{\dagger}\; .
\end{equation}
The corresponding generator\index{group!generator} is given by 
$H_{\varphi}=\varphi\,\hat{a}_{\rm in}^{\dagger}\hat{a}_{\rm in}$. It also 
commutes with the number operator.\index{operator!number}

In general, when some unitary evolution leaves the photon 
number\index{photon!-number} invariant,
that evolution corresponds to some {\em passive} optical circuit. 
Alternatively, when a unitary evolution\index{unitary evolution} does not 
conserves the photon number, we speak of {\em active} optical devices or 
{\em photon sources}.\index{photon!source}

\section{Quantum information}

Quantum information\index{quantum!information} theory delivers the foundations 
for quantum computation,\index{quantum!computation}
and in order to perform a quantum computation we need to be able to, among
other things, prepare certain quantum states. In this thesis, I will not 
consider any quantum computation algorithms, but I {\em do} consider state 
preparation. In later chapters, I will need some concepts from quantum 
information theory, such as the {\em fidelity},\index{fidelity} to assess 
various aspects of a state preparation\index{state preparation} process. 
Furthermore, since this thesis revolves around 
states, I have to develop an understanding of what subtleties are involved 
when talking about quantum states.\index{state}

\subsection{The computational basis and alphabets}\index{computational basis}
\index{alphabet}

Suppose we have a single system with a corresponding Hilbert 
space\index{Hilbert space} of dimension $N$. Rather than giving an orthonormal 
basis\index{orthonormal basis} of a system as wave\index{wave!function} 
functions in configuration space\index{configuration space} (like, for 
instance the eigenstates of the harmonic oscillator\index{harmonic oscillator} 
in Eq.~(\ref{lho})), we can ignore the particular
spatial behaviour of these states and {\em enumerate} them from 0 to $N-1$. 
The corresponding basis $\{ |j\rangle\}$, with $j\in\{ 0,\ldots,N-1\}$, is 
then called the {\em computational} basis. When we have a two-level system 
($N=2$), the computational basis states are $|0\rangle$ and $|1\rangle$, and
we speak of a {\em qubit}.\index{qubit}

The advantage of the computational basis is that it is independent of the 
physical representation. {\em Any} quantum mechanical two-level system is a
qubit, for example an electron in a magnetic field, a polarised photon or a 
SQUID with clockwise or counter-clockwise current.

When we have two or more systems, the computational basis can be extended
accordingly. If we have $M$ systems, we can choose a computational basis
$\{ |j_1,\ldots,j_M \}$, where $j_i\in\{ 0,\ldots,N_i-1 \}$, with $N_i$ the
dimensionality of the $i^{\rm th}$ system. For instance, the computational 
basis\index{computational basis} for two qubits\index{qubit} is given by 
$\{ |0,0\rangle, |0,1\rangle, |1,0\rangle,|1,1\rangle\}$. In most of the rest 
of this thesis, I will concentrate on qubits.

As a final point in this section, I present the concept of an {\em alphabet}
\index{alphabet}
of states. It is a finite set of possibly non-orthogonal states. It may be 
over-complete or it may not span the total Hilbert space.\index{Hilbert space} 
Furthermore, an alphabet of states can generate a POVM,\index{POVM} 
corresponding to a generalised\index{measurement!generalised} 
measurement. I am now ready to discuss some information-theoretic aspects
of quantum states.

\subsection{Shannon entropy and quantum information}\index{entropy!Shannon}
\index{quantum!information}

Prior to the measurement\index{measurement} of a system (in, for example, the 
computational basis), we have a probability distribution
\index{distribution!probability} $\{ p_j\}$ with $\sum_j p_j =1$ 
over all possible outcomes. We do {\em not} know the measurement outcome 
\index{measurement!outcome}
beforehand. Can we quantify our ignorance of this measurement outcome?

Obviously, when one probability $p_k$ is 1 and the others are all 0, there
is no ignorance about the measurement outcome: we will find the system in the
state $|k\rangle$, corresponding to the probability $p_k=1$. On the other hand,
when all $p_j$'s are equal, our ignorance about the measurement outcome is 
maximal. We are looking for a function of the set of probabilities $\{ p_j\}$
which is zero when one of the $p_j$'s is zero, and maximal when all $p_j$'s
are equal. Such a function is given by
\begin{equation}\label{2shannon}
 S_{\rm Shannon} = - \sum_{j=0}^{N-1} p_j \log p_j\; .
\end{equation}
This is called the {\em Shannon entropy}\index{entropy!Shannon} of a 
probability distribution
\cite{shannon48,peres95}. It is immediately verified that $S_{\rm Shannon}=0$ 
if all $p_j$'s are 0, except $p_k=1$, and by differentiation $\partial 
S_{\rm Shannon}/\partial p_k$ we find that the only extremum (a maximum) 
occurs when all $p_j$'s are equal.

The Shannon entropy is a classical entropy.\index{entropy!classical} Quantum 
mechanically, we can also
define an entropy which gives a measure for our ignorance of a state. Suppose 
the state can be written as a mixture $\rho$:
\begin{equation}
 \rho = \sum_{j=0}^{N-1} p_j |j\rangle\langle j|\; ,
\end{equation}
with $\{|j\rangle\}$ some suitable basis. The Shannon entropy can be 
calculated according to Eq.~(\ref{2shannon}) using the probability 
distribution $\{ p_i\}$. Quantum mechanically, our ignorance of the state is 
given by the {\em Von Neumann entropy}:\index{entropy!Von Neumann}
\begin{equation}
 S_{\rm VonNeumann} = - {\rm Tr}\left( \rho\log\rho \right)\; ,
\end{equation}
When $\rho$ is pure, it is easy to verify that $S_{\rm VonNeumann}=0$ by using
the fact that basis transformations inside the trace leave $S_{\rm VonNeumann}$
invariant. Therefore $S_{\rm Shannon}\geq S_{\rm VonNeumann}$, the Von Neumann
entropy\index{entropy!Von Neumann} is a {\em lower bound} on our ignorance of 
the state.

As an example, consider a source which produces pure right-handedly polarised
photons. In the linear polarisation basis this state is given by
\begin{equation}
 |\circlearrowright\rangle = \frac{1}{\sqrt{2}} \left( |\leftrightarrow\rangle
 + i|\updownarrow\rangle \right)\; .
\end{equation}
A measurement in the basis $\{|\leftrightarrow\rangle,|\updownarrow\rangle\}$
would yield a probability distribution\index{distribution!probability} over 
the measurement outcomes $p_{\leftrightarrow}=p_{\updownarrow}=1/2$, so the 
Shannon entropy\index{entropy!Shannon} is maximal.
However, since we {\em could have} measured in the circular basis, the 
probability distribution would have been $\{ p_{\circlearrowright}=1,
p_{\circlearrowleft}=0\}$, with a corresponding $S_{\rm Shannon}=0$. In both
cases the Von Neumann entropy is 0, corresponding to the lower bound of the
Shannon entropy. 

Massar and Popescu \cite{massar99} proved that the minimal ignorance 
about a state after a measurement is given by the Von Neumann 
entropy. For more details on the Shannon entropy in quantum theory I refer the 
reader to Peres \cite{peres95}.

\subsection{Fidelity and the partition ensemble fallacy}\index{fidelity}
\index{PEF}

In the previous section, we used the knowledge of the probability distribution 
to quantify our ignorance of the measurement 
outcomes\index{measurement!outcome} prior to the 
measurement.\index{measurement}
However, in general this probability distribution is not known. When we measure
the polarisation of a photon we will find definite outcomes, not probabilities.
In this section, I ask the question how much information can be gained in a 
single-shot measurement\index{measurement!single-shot} when the state of the 
system is not known beforehand.
One measure of the information we can extract from a state is given by the {\em
fidelity} \cite{helstrom76,holevo82,fuchs96,fuchs96b,massar99}.

Suppose we want to measure an unknown state and use the knowledge gained by 
the measurement outcome to reconstruct that state. We then need a measure 
quantifying the accuracy of the reconstruction. Such a measure is given by the 
{\em fidelity}. Suppose further that the initial unknown state is given by 
$|\psi\rangle$. We now measure this state along 
$|\phi\rangle\langle\phi|$\footnote{In general, $|\phi\rangle\langle\phi|$ is 
part of a POVM,\index{POVM} yielding a generalised measurement.}, which is 
our {\em estimate}. We can define the measure of success of our estimate by 
\cite{massar99}
\begin{equation}
 F_{\phi\psi} = |\langle\phi|\psi\rangle|^2\; .
\end{equation}
This is not the only possible measure, but for our present purposes it is the 
simplest. When we have a probability distribution over a set of initial states
(i.e., we have a mixed state $\rho = \sum_i p_i |\psi_i\rangle\langle\psi_i|$),
and a POVM $E_k = \sum_j \mu_{jk} |\phi_j\rangle\langle\phi_j|$, we can 
average $F_{\phi\psi}$ over the two alphabets\index{alphabet} of states:
\begin{equation}
 F = \sum_j \langle\phi_j| \left( \sum_{i} p_i |\psi_i\rangle\langle\psi_i|
 p_{j|i} \right) |\phi_j\rangle\; ,
\end{equation}
where $p_{j|i}$ is the probability of estimating the state $|\phi_j\rangle$
when the prepared state is $|\psi_i\rangle$. $F$ is called the average 
fidelity of state reconstruction.\index{fidelity!of state reconstruction}

Consider a source which creates either randomly polarised photons, or linearly
polarised photons $|\leftrightarrow\rangle$ and $|\updownarrow\rangle$ with 
equal probabilities in a given coordinate frame. When we measure the 
polarisation in $\{|\leftrightarrow\rangle,|\updownarrow\rangle\}$, the 
outcomes will always be a horizontally or vertically polarised photon for both 
randomly and linearly polarised photons. We now reconstruct the photon 
state according to this measurement outcome\index{measurement!outcome} 
(creating a photon in the direction corresponding to the measurement outcome). 
If the photons are linearly polarised, the fidelity\index{fidelity} of the 
reconstruction is equal to 1, whereas
in the case of randomly polarised photons the fidelity is 2/3 \cite{massar95}.
In the last case we recognise the state very well ($F=1$), and in the former,
we recognise the state quite badly. Therefore, the {\em same} measurement with
the {\em same} outcomes (with the {\em same} relative frequencies) yield a 
{\em different} fidelity. As a consequence, this fidelity is a measure of the 
information we extract from the state \cite{massar99}.

Let's now ask a slightly different question: what is the probability that a
state $\rho$ is mistaken for another `estimated' state $|\phi\rangle$? This 
probability is given by the overlap between the two states:
\begin{equation}\label{fiddef}
 F = \text{Tr}[\rho|\phi\rangle\langle\phi|] \; .
\end{equation}
This equation gives the definition of fidelity in a different context. It 
corresponds to the lower bound for the probability of mistaking $\rho$ for 
$|\phi\rangle$ in any possible (single) measurement \cite{fuchs96}. When $\rho$
is an exact replica of $|\phi\rangle$ then $F=1$, and when $\rho$ is an 
imprecise copy of $|\phi\rangle$ then $F<1$. Finally, when $\rho$ is completely
orthogonal to $|\phi\rangle$ the fidelity is zero. 

\bigskip

In the above discussion I constructed a mixed state $\rho$ for randomly 
polarised photons. Consider a polarisation state\index{state!polarisation} 
\begin{equation}
 |\psi(\theta)\rangle = \cos\theta|\leftrightarrow\rangle+\sin\theta|\updownarrow\rangle\; .
\end{equation}
The mixed state\index{state!mixed} of randomly polarised photons is obtained 
by integrating over 
all $\theta$ which yield a different state $|\psi(\theta)\rangle$:
\begin{equation}
 \rho = \frac{1}{\pi} \int_0^{\pi} d\theta\, 
 |\psi(\theta)\rangle\langle\psi(\theta)| = \frac{1}{2}|\leftrightarrow\rangle
 \langle\leftrightarrow| + \frac{1}{2} 
 |\updownarrow\rangle\langle\updownarrow|\; . 
\end{equation}
But this mixture is equal to that which we would have obtained by randomly
choosing only horizontally and vertically polarised photons. In other words, a
mixed state does not contain information\index{information} about the 
preparation process! In
general, we can construct infinitely many physically different sources which
generate the same mixed outgoing state. Or equivalently, there exist 
infinitely many decompositions,\index{decomposition} or {\em partitions}, 
\index{partition} of any given mixed state.

Another example. Consider the state $\rho$ of the form of 
\begin{equation}
 \rho=\alpha|\phi_1\rangle\langle\phi_1|+\beta|\phi_2\rangle\langle\phi_2|\; .
\end{equation}
It is the sum of two pure states. Again, this is not a unique partition. 
Whereas in a chemical mixture\index{chemical mixture} of, say, nitrogen and 
oxygen there is a unique 
partition (into N$_2$ and O$_2$), a quantum mixture can be decomposed in many 
ways. For instance, $\rho$ can equally be written in terms of
\begin{equation}
 |\psi_1\rangle = \alpha|\phi_1\rangle + \beta|\phi_2\rangle 
 \quad\mbox{and}\quad
 |\psi_2\rangle = \alpha|\phi_1\rangle - \beta |\phi_2\rangle
\end{equation}
as
\begin{equation}\label{novacuum}
 \rho_{\rm out} = \frac{1}{2} |\psi_1\rangle\langle\psi_1| + \frac{1}{2} 
	|\psi_2\rangle\langle\psi_2|\; .
\end{equation}
This is just one of an infinite number of possible decompositions. 
Quantum mechanics dictates that all partitions are equivalent to each other 
\cite{peres95}. They are indistinguishable. To elevate one partition over 
another is to commit the `Partition Ensemble Fallacy.'\index{PEF}

Why is this so important? Suppose we have a mixture of the vacuum 
$|0\rangle\langle 0|$ and a single-photon state $|1\rangle\langle 1|$, yielding
$\rho=\alpha|0\rangle\langle 0|+\beta|1\rangle\langle 1|$. It is very tempting 
to interpret such a mixture as: `with probability $|\beta|^2$ there {\em is}
a photon, and with probability $|\alpha|^2$ there {\em is no} photon'. However,
quantum theory does not say anything about what {\em is} without referring to
measurement outcomes. If we were to measure an observable whose eigenstates 
are {\em not} number states (like, for instance, coherent states), the outcome 
would not involve any reference to photon numbers. Therefore, in the context 
of quantum mechanics, the above statement is meaningless. These considerations 
will become important in chapter \ref{teleportation}.

\subsection{Non-locality issues}\label{sec:vacuum}\index{non-locality}

Quantum theory is a local theory, in the sense that space-like and time-like
separated operators $A(x_{\mu})$ and $B(x_{\mu}')$ always commute: 
$[A(x_{\mu}),B(x_{\mu}')]\propto\delta^4(x_{\mu}-x_{\mu}')$ \cite{haag91}. 
However, when one 
seeks a {\em classical deterministic}\index{classical!determinism} underlying 
explanation for the correlations observed in quantum mechanics, one has to 
allow non-local influences. This was first noted by Bell\index{Bell, J.S.} 
\cite{bell64}, who formulated his now 
famous inequalities \cite{hilgevoord93,redhead87}.

Let me set up a simple version of Bell's argument. Alice and Bob, who are 
sufficiently far away from each other, both receive a photon with some unknown
polarisation. Alice randomly chooses a polarisation measurement out of two 
possible directions $\mathbf{a}$ and $\mathbf{a}'$. Similarly, Bob randomly 
chooses a polarisation measurement out of $\mathbf{b}$ and $\mathbf{b}'$.
Let's denote the two possible measurement outcomes\index{measurement!outcome} 
of a polarisation 
measurement by $\pm 1$. Then the eigenvalues $a$, $a'$, $b$ and $b'$ are all 
either $+1$ or $-1$. We repeat this procedure a large number of times.

We now define the expression \cite{redhead87}
\begin{equation}
 \gamma_n \equiv a_n b_n + a_n' b_n + a_n b_n' - a_n' b_n' = a_n (b_n + b_n')
 + a_n' (b_n - b_n')\; , 
\end{equation}
where the subscript $n$ indicates the $n^{\rm th}$ trial. The value of 
$\gamma_n$ is an integer between $-2$ and $+2$. The absolute value of the 
average of $\gamma_n$ over all the $N$ trials is given by
\begin{equation}
 \left|\frac{1}{N} \sum_{n=1}^N\gamma_n \right| = \left|\frac{1}{N}\sum_{n=1}^N
 \left( a_n b_n + a_n' b_n + a_n b_n' - a_n' b_n'\right)\right| \leq 2\; .
\end{equation}
When we define the correlation coefficients
\begin{equation}
 c(\mathbf{x},\mathbf{y}) \equiv \lim_{N\rightarrow\infty} \frac{1}{N}
 \sum_{n=1}^N x_n y_n\; ,
\end{equation}
the above inequality becomes
\begin{equation}
 \left| c(\mathbf{a},\mathbf{b}) + c(\mathbf{a}',\mathbf{b}) + c(\mathbf{a},
 \mathbf{b}') - c(\mathbf{a}',\mathbf{b}') \right| \leq 2\; .
\end{equation}
This is one form of the Bell inequality.\index{Bell!inequality} We can 
calculate these correlation 
coefficients for the case where the two photons are part of the singlet state
\index{state!singlet}
$(|\leftrightarrow,\updownarrow\rangle - |\updownarrow,\leftrightarrow\rangle)
/\sqrt{2}$, which yields $c(\mathbf{a},\mathbf{b})=-\cos(\theta_{ab}/2)$. For
suitably chosen angles $\theta_{ab}$, the Bell inequality is violated by 
quantum mechanics.

What does this mean? All I assumed in the above derivation of the inequality
was statistical independence of the measurement outcomes obtained by Alice
and Bob. The violation therefore implies that the measurements performed by 
Alice and Bob, though possibly in different galaxies, and thus well and truly
separated, can not be considered statistically 
independent!\index{statistical independence} This has led to 
wild speculations about superluminal signalling, but all quantum mechanics 
predicts are correlations which cannot be given a local 
realistic\index{local!realism} 
interpretation. If we want a classical picture, we therefore have to give up
either realism or locality. The choice is yours.

Suppose we have a bi-partite state\index{state!bi-partite} which violates a 
Bell inequality. Then that state is said to be {\em 
entangled}.\index{state!entangled} The contrary is not necessarily true: a
state which is entangled does not have to violate any Bell inequalities
\cite{popescu94} (see also appendix \ref{app:povm}). Several different Bell 
inequalities have been experimentally verified by many groups, the first of 
which was led by Aspect \cite{aspect81,aspect82}. Nowadays, experimental tests
of the violation of a Bell inequality is used mostly to indicate whether a 
state is entangled. Alternatively, tests of non-locality without Bell 
inequalities have been proposed by DiGiuseppe and Boschi 
\cite{digiuseppe97,boschi97}.

In the context of quantum optics, there are nonlocal 
effects\index{non-local correlations} in Fock space\index{Fock space} 
which are of some interest in this thesis. Hardy and Peres showed that a 
single photon can exhibit non-local properties, following the work by Tan,
Walls and Collett \cite{tan91,tan92,santos92}. Here, I will follow Peres' 
argument \cite{hardy93,hardy94,vaidman94,greenberger95,peres95c}.

Consider a pure one-particle state 
\begin{equation}\label{focksinglet}
 |\psi\rangle = \frac{1}{\sqrt{2}} \left( |0\rangle_a |1\rangle_b -
 |1\rangle_a |0\rangle_b \right)\; ,
\end{equation}
where $|0\rangle$ is the vacuum and $|1\rangle$ a single-photon 
state.\index{state!single-photon} The 
subscripts $a$ and $b$ denote the different modes, possibly spatially separated
over a large distance. Note that this state has the same mathematical structure
as a singlet state. Alice and Bob can demonstrate a violation of a 
Bell\index{Bell!inequality} 
inequality\footnote{More precisely, a Clauser-Horne 
inequality.\index{CHSH inequality}} using the 
strategy described above: Alice and Bob both randomly choose an observable
from a set of two non-commuting observables. The first observable for both
parties is obviously the one spanned by $\{ |0\rangle\langle 0|, 
|1\rangle\langle 1|\}$, i.e., whether there is a photon in mode $a$ or $b$ 
respectively. Let $P_a$ denote the projector\index{projector} 
$|1\rangle_a\langle 1|$ and $P_b$ the projector $|1\rangle_b\langle 1|$.

Another observable for Alice might include the projector $P_{a'}$ along the 
eigenvector $(|1\rangle_a + \sqrt{3}|0\rangle_a)/2$ and Bob can choose to 
measure along the projector $P_{b'}$ along the eigenvector $(|1\rangle_b 
- \sqrt{3}|0\rangle_b)/2$. Given Eq.~(\ref{focksinglet}) quantum theory yields
\cite{peres95c,greenberger96}:
\begin{eqnarray}
 & \langle P_{a'} \rangle = \langle P_{b'} \rangle = 0.5\; , & \cr &&\cr
 & \langle P_{a} P_{b} \rangle = 0\; , & \cr &&\cr
 & \langle P_a P_{b'} \rangle = \langle P_{a'} P_b \rangle = 0.375\; , & \cr
 &&\cr
 & \langle P_{a'} P_{b'} \rangle = 0.375\; , & 
\end{eqnarray}
which violates the inequality
\begin{equation}
 0\leq\langle P_{a'} + P_{b'} - P_{a'} P_{b'} - P_{a'} P_b - P_a P_{b'} + 
 P_{a} P_{b} \rangle\leq 1\; .
\end{equation}
This proves non-locality. However, the fact that $P_{a'}$ and $P_{b'}$ do not
conserve photon number means that active detection devices, i.e., detectors 
which can create photons, have to be used. This has provoked many comments 
\cite{santos92,vaidman95,greenberger95,greenberger96}, but treating them all 
would lead me too far from the main subject of this thesis. Let us therefore 
move on to the creation of maximal entanglement\index{entanglement!maximal} 
in quantum optics, the subject 
of the next chapter.

\newpage
\thispagestyle{empty}

\chapter{Creation of Maximal Entanglement}\label{chap3}

\noindent
$\qquad\qquad\qquad${\sl That's the wacky thing about these entangled}

\noindent
$\qquad\qquad\qquad${\sl photon pairs---They're sort of the Bill Clinton and}

\noindent
$\qquad\qquad\qquad${\sl Monica Lewinsky of the quantum world: they're}

\noindent
$\qquad\qquad\qquad${\sl heavily entangled until somebody `looks' at them.}

\medskip

{\hfill ---Jonathan P.\ Dowling}

\bigskip

\noindent
Entanglement\index{entanglement} is one of the key ingredients in quantum 
communication\index{quantum!communication} and 
information.\index{quantum!information} For instance, quantum protocols such 
as dense coding\index{dense coding}
\cite{bennett92}, quantum error correction\index{quantum!error correction} 
\cite{steane95,shor95} and quantum teleportation\index{quantum teleportation} 
\cite{bennett93} rely on the non-classical correlations provided 
by entanglement. Currently, substantial efforts are being made to use {\em 
optical} implementations for quantum communication. 

The advantages of this are obvious: light travels at high speed and it weakly 
interacts with the environment. However, exactly this weak interaction poses 
serious drawbacks. The fact that photons do not interact with each other makes 
it hard to manipulate them. For example, it has recently been shown that it is 
impossible to perform so-called complete Bell 
measurements\index{Bell!measurement} on two-mode 
polarisation states in linear quantum optics \cite{lutkenhaus99,vaidman99} 
(although theoretical schemes involving Kerr media\index{Kerr medium} 
\cite{scully99} and atomic coherence\index{atomic coherence} \cite{paris99} 
have been reported). Furthermore, maximally polarisation-entangled two-photon 
states\index{state!maximally entangled} have not been 
unconditionally produced. In this chapter I investigate the possibility of 
creating such 
states with linear optics and a specific class of non-linear elements.

Before that, however, I will have to introduce the terminology I will use in 
this chapter (and throughout this thesis). In the next section I will 
discuss various issues connected to entanglement, such as 
separability,\index{separability} maximal 
entanglement,\index{entanglement!maximal} multi-partite 
entanglement\index{entanglement!multi-partite} and 
purification.\index{purification} In section
{\ref{sec:pdc}} I will study parametric 
down-conversion,\index{down-converter} currently the most 
common entanglement source in quantum optics. Finally, I give limitations for 
the creation of maximal entanglement with a special class of optical circuits.
This chapter is based on Kok and Braunstein \cite{kok00a,kok00b}. 

\section{Separability and entanglement}

In this section I discuss the concept of entanglement. First, I define 
separable and entangled states, and then I introduce {\em event-ready}
entanglement.\index{entanglement!event-ready} Three-particle entanglement is 
briefly considered, and finally, 
I discuss the entanglement measure for pure states and entanglement {\em 
purification}.

\subsection{What is maximal entanglement?}

Two quantum systems in a pure state, labelled by $x_1$ and $x_2$ respectively, 
are called entangled when the state $\Psi(x_1,x_2)$ describing the total system
cannot be factorised into two separate states $\psi_1(x_1)$ and $\psi_2(x_2)$:
\begin{equation}\label{entanglement}
 \Psi(x_1,x_2) \neq \psi_1(x_1) \psi_2(x_2)\; .
\end{equation}
All possible states $\Psi(x_1,x_2)$ accessible to the combined state of the 
two quantum systems form a set $\mathcal{S}$. These states are generally 
entangled. Only in extreme cases is $\Psi(x_1,x_2)$ {\em 
separable},\index{separability} i.e., 
it can be written as a product of states describing the separate systems. The 
set of separable states form a subset of $\mathcal{S}$ with measure zero.

We arrive at another extremum when the states $\Psi(x_1,x_2)$ are {\em 
maximally} entangled. The set of maximally entangled 
states\index{state!maximally entangled}\index{entanglement!maximal} also 
forms a subset of $\mathcal{S}$ with measure zero. I will now give a 
definition of maximal entanglement for two finite-dimensional systems.
\begin{description}
 \item[Definition:] Two $N$-level systems are called {\em maximally entangled} 
 when their total state $|\Psi\rangle$ in the Schmidt 
 decomposition\index{Schmidt decomposition} can be
 written as
 \begin{equation}\label{maxent}
  |\Psi\rangle = \frac{1}{\sqrt{N}} \sum_{k=0}^{N-1} e^{i\theta_k} 
  |n_k,m_k\rangle\; ,
 \end{equation}
 with $\{ |n_k\rangle \}$ and $\{ |m_k\rangle \}$ two orthonormal
 bases\index{orthonormal basis} and 
 $\{\theta_k\}$ a set of arbitrary phases.
\end{description}

Suppose we have two (not necessarily identical) two-level systems, 1 and 2, 
whose states can be written in the orthonormal basis $\{ |0\rangle_k,
|1\rangle_k\}$ (where $k=1,2$). This is defined as the computational basis 
for these two systems. Physically, those systems could be for example polarised
photons or electrons in a magnetic field. Every possible 
state of the two systems together can be written on the basis of four 
orthonormal states $|0,0\rangle_{12}$, $|0,1\rangle_{12}$, $|1,0\rangle_{12}$ 
and $|1,1\rangle_{12}$. These basis states generate a four-dimensional Hilbert 
space.\index{Hilbert space} Another possible basis for this space is given 
by the so-called Bell states:\index{state!Bell}
\begin{eqnarray}\label{bellstates}
 |\Psi^{\pm}\rangle_{12} &=& ( |0,1\rangle_{12} \pm |1,0\rangle_{12} )
	/\sqrt{2}\; ,\cr
 |\Phi^{\pm}\rangle_{12} &=& ( |0,0\rangle_{12} \pm |1,1\rangle_{12} )
	/\sqrt{2}\; .
\end{eqnarray}
These states are also orthonormal. They are examples of {\em maximally} 
entangled states. The Bell states are not the only maximally entangled states
(as an alternative, we can include a relative phase $e^{i\theta}$ in one of 
the branches; see also the definition on this page), but they are a convenient
and common choice. All maximally entangled states can be 
transformed into each other by a local unitary transformation (see Appendix
\ref{app:trans}).\index{transformation!unitary}

If we want to conduct an experiment which makes use of maximal entanglement, 
in particular $|\Psi^-\rangle_{12}$, we would most straightforwardly like to 
have a source which produces these states at the push of a button. In practice,
this might be a bit much to ask. A second option might be to have a source 
which only produces $|\Psi^-\rangle_{12}$ randomly, but flashes a red light 
when it happens. Such a source would create so-called {\em event-ready}
\index{entanglement!event-ready} 
entanglement\footnote{The term first seems to appear in the context of 
detector efficiencies \cite{zukowski93} in 1993, and subsequently with the 
meaning used here by Pavi\v ci\' c \cite{pavicic96} in 1996.}: it produces 
$|\Psi^-\rangle_{12}$ only part of the time, but when it does, it tells you.

More formally, the outgoing state $|\psi_{\rm out|red~light~flashes}\rangle$ 
conditioned on the red light flashing is said to exhibit event-ready 
entanglement if it can be written as
\begin{equation}\label{ere}
 |\psi_{\rm out|red~light~flashes}\rangle_{12} \simeq |\Psi^-\rangle_{12} 
 + O(\xi)\; ,
\end{equation}
where $\xi\ll 1$. In what follows I shall omit the subscript
`$|$red light flashes' since it is clear that we can only speak of event-ready 
entanglement conditioned on the red light flashing.

Non-maximal entanglement has been created in the context of quantum 
optics by means of parametric down-conversion\index{down-converter} 
\cite{shih88}. Rather than a 
(near) maximally entangled state, as in Eq.\ (\ref{ere}), this process 
produces states with a large vacuum contribution.\index{vacuum contribution} 
Only a minor part consists 
of an entangled photon state. Every time parametric down-conversion is 
employed, there is only a small probability\footnote{This probability is kept 
small so that the occurrence of higher order double photon-pairs is 
negligible.} of creating an entangled 
photon-pair (notice my use of PEF; see chapter II). We will call this {\em 
randomly produced} entanglement.\index{entanglement!randomly produced}

\subsection{Tri-partite entanglement}\index{entanglement!tri-partite}

So far, I have only considered the entanglement of two systems. But quantum 
mechanics does not give a limit to the number of systems which can be 
entangled. For instance, we can define a maximally entangled state for three 
systems 1, 2 and 3:
\begin{equation}\label{smit}
 |{\rm GHZ}\rangle_{123} = \frac{1}{\sqrt{2}} \left( |0,0,0\rangle_{123} + 
 |1,1,1\rangle_{123}\right)\; .
\end{equation}
Such multi-partite entangled\index{entanglement!multi-partite} states are 
called Greenberger-Horne-Zeilinger- or GHZ-states\index{state!GHZ-} 
\cite{greenberger89}. They also represent a particular state of
three maximally entangled systems. Post-selected\index{post-selection} 
three-particle GHZ entanglement was observed experimentally
by Bouwmeester {\em et al}.\ in 1999 \cite{bouwmeester99b}. In chapter 
\ref{teleportation} I will extensively discuss the post-selected nature of
this and related experiments.

The state $|{\rm GHZ}\rangle$ can also be interpreted as a Schmidt 
decomposition\index{Schmidt decomposition} for three systems (i.e., it 
can be written as a single sum over orthonormal basis\index{orthonormal basis} 
states). Contrary to the bi-partite\index{entanglement!bi-partite} case, it 
is {\em not} true that a Schmidt 
decomposition exists for any three-partite state. The Schmidt decomposition 
for two systems follows from the existence of the unitary transformations 
\index{transformation!unitary}
$U_1$ and $U_2$ which diagonalise\index{diagonalisation} a matrix $T$ 
according to $\Lambda = U_1 T U_2$, giving \cite{acin00}
\begin{equation}
 |\Psi\rangle_{12} = \sum_{i,j} T_{ij} |i\rangle_1|j\rangle_2 \rightarrow
 \sum_{\mu} \lambda_{\mu} |\mu\rangle_1|\mu\rangle_2 \; ,
\end{equation}
with $\{ |i\rangle\}$, $\{ |j\rangle\}$ and $\{ |\mu\rangle\}$ orthonormal 
bases\index{orthonormal basis} and $\lambda_{\mu}$ the eigenvalues
\index{eigen!-values} of the diagonal matrix $\Lambda$. 
For the general three-system case this is no longer true:
\begin{equation}
 |\Psi\rangle_{123} = \sum_{i,j,k} T_{ijk} |i\rangle_1|j\rangle_2|k\rangle_3 
 \rightarrow \sum_{\mu,\nu} \lambda_{\mu\nu} |\mu\rangle_1|\nu\rangle_{23}\; ,
\end{equation}
by virtue of the Schmidt decomposition\index{Schmidt decomposition} (again 
with $\{|\mu\rangle\}$ and 
$\{|\nu\rangle\}$ orthonormal bases). This is converted to a single sum {\em 
if and only if} \cite{peres95b}
\begin{equation}
 |\nu\rangle_{23} = \sum_{j,k} (\Omega_{\mu})_{jk} |j\rangle_2 |k\rangle_3 = 
 |\zeta_{\mu}\rangle_2 |\xi_{\mu}\rangle_3 \; ,
\end{equation}
or $\Omega_{\mu}^{\dagger}\Omega_{\nu} = \Omega_{\mu}\Omega_{\nu}^{\dagger}=0$
with $(\mu\neq\nu)$.

For multi-partite entanglement,\index{entanglement!multi-partite} we can no 
longer completely define maximal 
entanglement\index{entanglement!maximal} in terms of the Schmidt 
decomposition. For example, it can easily be verified that the state
\begin{equation}\label{w}
 |W\rangle = \frac{1}{\sqrt{3}} \left( |0,0,1\rangle + |0,1,0\rangle + 
 |1,0,0\rangle \right)\; ,
\end{equation}
although representing three maximally entangled systems, cannot be written in 
terms of the Schmidt decomposition in Eq.~(\ref{smit}) (the number of linear 
independent terms exceeds the number of orthonormal basis states of the 
separate systems). More specifically, D\"ur {\em et al}.\ prove that ensembles
of three-partite entangled\index{entanglement!tri-partite} states of qubits 
can be transformed either to the 
state $|{\rm GHZ}\rangle$ or to the state $|W\rangle$ by 
stochastic\footnote{`Stochastic' meaning that the transformation is successful 
with non-zero probability.} local operations and classical communication 
(SLOCC) alone \cite{dur00}. That is, under SLOCC,\index{SLOCC} 
$|{\rm GHZ}\rangle$ 
and $|W\rangle$ generate two distinct invariant 
subspaces\index{subspace!invariant} of the total Hilbert\index{Hilbert space} 
space $({\hilbert}_2^{\otimes 3})^{\otimes\infty}$ spanned by 
the three qubits.\index{qubit} In conclusion, the definition of maximally 
entangled states 
in terms of the Schmidt decomposition given on page \pageref{maxent} only works
for bi-partite systems. However, in the rest of this chapter (and indeed, this 
thesis) I will concentrate on entanglement between two systems, and this
definition is sufficient.

\subsection{Purification}\index{purification}

In this section, I will define a measure of 
entanglement\index{entanglement!measure} $E$ for (pure) states 
which are non-maximally entangled \cite{bennett96a,bennett96b}\footnote{There 
are subtleties in defining measures of entanglement for mixed states; see Ref.\
\cite{bouwmeester00} and references therein.}. The natural measure of 
entanglement is defined by the Von Neumann entropy\index{entropy!Von Neumann} 
of the reduced density matrix\index{density matrix} of the subsystems. Suppose 
we have two systems 1 and 2 held by Alice and
Bob respectively in a (non-maximally) entangled state $|\Psi\rangle_{12}$. The 
reduced density matrices of the two subsystems are 
\begin{equation}
 \rho_1 \equiv {\rm Tr}_2 \left[|\Psi\rangle_{12}\langle\Psi|\right]
 \qquad\mbox{and}\qquad
 \rho_2 \equiv {\rm Tr}_1 \left[|\Psi\rangle_{12}\langle\Psi|\right]\; .
\end{equation}
In chapter \ref{chap2} we defined the Von Neumann 
entropy\index{entropy!Von Neumann} of a density matrix 
$\rho$ as
\begin{equation}
 S(\rho) = -{\rm Tr} \left[ \rho\log\rho\right]\; .
\end{equation}
The measure of entanglement\index{entanglement!measure} $E(\Psi)$ is now defined as
\begin{equation}
 E(\Psi) \equiv S(\rho_1) = S(\rho_2)\; .
\end{equation}
This measure has a number of pleasant properties: it is zero for separable 
states and maximal for maximally entangled states. Furthermore, $E$ remains 
the same whether we trace out system 1 or system 2.

Can we in some way increase the entanglement in non-maximally entangled 
(pure) states?\index{state!non-maximally entangled} The answer to this 
question is the domain of {\em entanglement purification}.\index{purification} 
I will now briefly discuss the general idea behind purification.

Suppose we distribute a set of singlet states\index{state!singlet} among Alice 
and Bob for quantum communication\index{quantum!communication} purposes. After 
the distribution, however, the actual states 
held by Alice and Bob will in general no longer be maximally entangled. This 
is because noise\index{noise} and decoherence\index{decoherence} in the 
distribution process degrade the 
entanglement. In order to obtain singlet states again we have to {\em purify} 
the ensemble\index{ensemble} of states shared between Alice and Bob with only 
local operations\index{local!operations} 
\cite{bennett96a}. Because operations of this kind cannot increase the amount 
of entanglement, the total entanglement shared by Alice and Bob remains the 
same or decreases. However, we still have the possibility of converting 
several copies of poorly entangled states into a few highly entangled states.
In general, when we have $N$ non-maximally entangled initial states 
$|\Psi\rangle$ with entanglement content $E(\Psi)$, we wish to obtain $M<N$ 
entangled states $|\Phi\rangle$ with $E(\Phi) > E(\Psi)$ using a restricted 
class of operations. The local operations can be divided into three groups 
\cite{bouwmeester00}:
\begin{enumerate}
 \item {\sl Local transformations\index{transformation!local} and 
	measurements\index{measurement}}. These can be modelled
	in general by local POVM's, i.e., POVM's acting on only one subsystem,
 \item {\sl classical communication\index{classical!communication}}. This 
	creates the opportunity to 
	classically correlate the actions of the two (distant) parties holding 
	the entanglement,
 \item {\sl post-selection\index{post-selection}}. Depending on the outcome 
	of local measurements 
	and classical communication, we can select a subset from our ensemble
	of states.
\end{enumerate} 
A procedure which manages to increase the entanglement in a subset of the $N$ 
initial states using only operations from these three classes is called an 
{\em entanglement purification} protocol.

As an example, I consider the original purification protocol\footnote{In 
chapter \ref{teleportation} we will encounter another purification protocol.} 
\cite{bennett96a}. We distribute (at least) two singlets, written in the 
computational basis $\{ |0\rangle_k,|1\rangle_k \}$ with $k=1,2$, between 
Alice and Bob:
\begin{equation}
 |\Psi^-\rangle_{12} = \frac{1}{\sqrt{2}} \left( |0,1\rangle_{12} - 
 |1,0\rangle_{12} \right)\; .
\end{equation}
Similarly, the other Bell states\index{state!Bell} are given by 
$|\Psi^+\rangle_{12}$ and $|\Phi^{\pm}\rangle_{12}$. A simple model of the 
noise\index{noise} due to the distribution 
implies that upon arrival the singlets\index{state!singlet} have become 
Werner\index{state!Werner} 
states\footnote{Decoherence tends to evolve pure states towards (maximally) 
mixed states.} \cite{werner89} $\rho$:
\begin{equation}\label{werner}
 \rho_{12} = \varepsilon |\Psi^-\rangle_{12}\langle\Psi^-| + (1-\varepsilon)
	\frac{{\unity}}{4}\; . 
\end{equation}
Note that there is a singlet contribution in the maximally 
mixed\index{state!maximally mixed} part $\unity$ 
as well. Two states with equal density matrices\index{density matrix} cannot 
be distinguished in 
{\em any} physical way, which means that the method of preparation of $\rho$
is irrelevant. In other words, we do not care {\em how} 
decoherence\index{decoherence} took place 
in the distribution process, the occurrence of $\unity$ defines a class of 
noise types which give rise to Eq.~(\ref{werner}). The noise gain is 
parametrised by $\varepsilon$.

In order to purify\index{purification} $\rho_{12}$ we need at least two 
such systems in a total 
state $\rho_{12}\otimes\rho_{12}$. As a first step, Bob operates with the 
Pauli matrix\index{Pauli matrices} $\sigma_{2y}$ on his part of the density 
matrix, yielding a transformation
\begin{equation}
 \rho_{12}' = (\sigma_{2y}\otimes\sigma_{2y})\;\rho\otimes\rho\;(\sigma_{2y}
 \otimes\sigma_{2y})^{\dagger}\; ,
\end{equation}
This transformation is equivalent to the symbol swapping $\Psi^{\pm} 
\leftrightarrow \Phi^{\mp}$ in the Bell states given in Eq.~(\ref{bellstates}).

Next, Bob applies the `controlled NOT'\index{controlled!NOT} to his two 
subsystems, after which both Alice and Bob measure the target in the 
computational basis.\index{computational basis} They compare their
measurement\index{measurement} outcome by means of classical 
communication,\index{classical!communication} and if the two
outcomes are the same (00 or 11), Bob again applies the $\sigma_y$ operator 
to his remaining state. If the measurement outcomes are not the same, the 
purification failed. This is the post-selection\index{post-selection} stage. 
Note the {\em probabilistic}\index{purification!probabilistic} character of 
the purification protocol. After a successful
purification run, Alice and Bob share a mixed state $\rho_{12}''$.

Before purification the fidelity\index{fidelity} of the distributed system was
\begin{equation}
 F = \langle\Psi^-|\rho_{12}|\Psi^-\rangle = \frac{1+3\varepsilon}{4}\; .
\end{equation} 
After a successful purification round the new fidelity $F'$ 
is \cite{bennett96a}
\begin{equation}
 F' = \langle\Psi^-|\rho_{12}''|\Psi^-\rangle = \frac{1 + 2 \varepsilon + 
 5\varepsilon^2}{4(1 + \varepsilon)}\; .
\end{equation} 
For purification to be meaningful we need $F'>F$, or $\varepsilon>\frac{1}{3}$.
In other words, if the decoherence is too strong, this type of purification 
cannot increase the entanglement in a subset of the distributed ensemble.
\index{ensemble}

Numerous other purification protocols have been proposed (see Ref.\ 
\cite{bouwmeester00} and references therein). In general, it provides a 
procedure to obtain (near) maximal entanglement.\index{entanglement!maximal} 
Also called distillation,\index{distillation} entanglement purification 
reduces an ensemble of poorly entangled states to a few highly entangled
states. 

By contrast, in this thesis I study the creation of event-ready entanglement,
\index{entanglement!event-ready} in particular with quantum optics. 
The difference with purification is that it is {\em 
dynamic}:\index{purification!dynamic} an event-ready entangler does not need 
to store entanglement
(something which is very difficult for photons), it tries a (large) number of
times and `flashes a red light' upon success. In the next section I will study
a more modest device, called a parametric down-converter. It produces 
entanglement randomly.

\section{Entanglement sources in quantum optics}\label{sec:pdc}
\index{entanglement!sources}

\begin{figure}[t]
\begin{center}
  \begin{psfrags}
     \psfrag{p}{pump}
     \psfrag{c}{crystal}
     \psfrag{l}{$|\updownarrow\rangle$}
     \psfrag{r}{$|\leftrightarrow\rangle$}
     \epsfxsize=8in
     \epsfbox[-70 20 470 120]{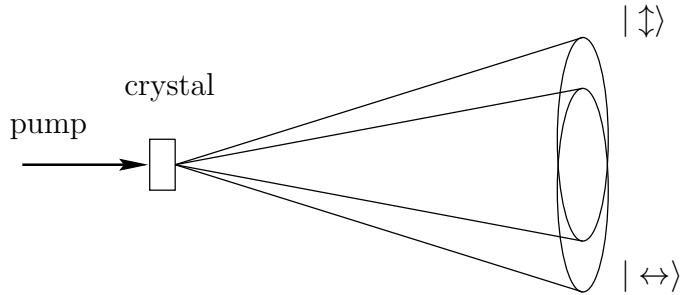}
  \end{psfrags}
  \end{center}
  \caption{A schematic representation of type II parametric 
	down-conversion. A high-intensity laser pumps a non-linear crystal. 
	With some probability a photon in the pump beam will be split into 
	two photons with orthogonal polarisation $|\updownarrow\rangle$ and 
	$|\leftrightarrow\rangle$ along the surface of the two respective 
	cones. Depending on the optical axis of the crystal, the two cones 
	are slightly tilted from each other. Selecting the spatial modes at 
	the intersection of the two cones yields the outgoing state
	$(1-\xi^2) |0\rangle + \xi |\Psi^-\rangle + O(\xi^2)$.}
\label{fig:3.1}
\end{figure}

In this section I will look at a particular class of devices capable of 
creating entanglement in quantum optics. These devices are commonly known as
{\em down-converters},\index{down-converter} since they convert a high-energy 
photon into two
lower-energy photons. Physically, in parametric down-conversion a crystal is 
pumped\index{pump} by a high-intensity laser, which we will treat classically 
(the parametric approximation).\index{parametric!approximation} The crystal is 
special in the sense that it has different refractive 
indices\index{refraction index} for horizontally and vertically polarised 
light.\index{polarised light} 
In the case of degenerate (type II)\index{down-converter!type II} parametric 
down-conversion a photon from 
the pump is split into two photons with half the energy of the pump photon. 
Furthermore, the process can be set up such that the two photons have 
orthogonal polarisations. The outgoing modes of the crystal constitute two 
intersecting cones with orthogonal polarisations $|\updownarrow\rangle$ and 
$|\leftrightarrow\rangle$ as depicted in Fig.\ \ref{fig:3.1}. 

Due to the conservation of momentum,\index{momentum conservation} the two 
produced photons are always in opposite modes with respect to the central 
axis (determined by the direction of the pump).\index{pump} In the two 
spatial modes where the different polarisation cones 
intersect we can no longer infer the polarisation of the photons, and as a 
consequence the two photons become entangled in their 
polarisation.\index{entanglement!polarisation} Parametric 
down-conversion\index{down-converter} as a device to create entangled photons 
was introduced by 
Shih and Alley in 1988 \cite{shih88} and is being continuously improved 
\cite{shih94,kwiat95,kwiat99,kim00,oberparleiter00}.

However, parametric down-converters do {\em not} produce pure Bell-states 
\index{state!Bell}
\cite{zukowski93,braunstein98,kok00a}. Because of the spontaneous nature of
the down-conversion process, only a in a small number of cases (i.e., a 
fraction of the trials) will a photon from the pump be split into two photons.
We thus have randomly produced entanglement.\index{entanglement!randomly
produced} Furthermore, there is an even 
smaller probability of creating more photon pairs, originating from several 
pump photons. The probability of this happening decreases with the number
of created pairs. I will derive the value of these probabilities in due course.

The outgoing state of the parametric down-converter I am interested in here is
\begin{equation}\label{3pdcout}
 |\psi_{\rm out}\rangle = (1-\xi^2) |0\rangle + \xi|\Psi^-\rangle +O(\xi^2)\; ,
\end{equation}
where $\xi\ll 1$ is a parameter indicating the strength of the down-conversion.
Note that $|0\rangle$ denotes the {\em vacuum}\index{state!vacuum} here, rather 
than a computational basis\index{computational basis} state. In the next 
section I will give a mathematical 
description of down-converters and subsequently I will determine the 
statistical properties of these devices.

\subsection{The physics of down-converters}

In this section I will describe the physical properties of parametric
down-con\-ver\-sion. Consider a down-converter with outgoing field modes $a_i$ 
and $b_j$. The indices denote the particular polarisation along the $x$- and 
$y$-axis of a given coordinate system. We are working in the interaction 
picture\index{interaction picture} of the Hamiltonian\index{Hamiltonian} which 
governs the dynamics of creating two entangled field modes $a$ and $b$ using 
weak parametric down-conversion. In the rotating wave 
approximation\index{rotating wave approximation} this Hamiltonian reads 
($\hbar =1$):
\begin{equation}\label{hamilton}
  H = i\kappa ( \hat{a}_x^{\dagger} \hat{b}_y^{\dagger} -
    \hat{a}_y^{\dagger} \hat{b}_x^{\dagger} ) + \text{H.c.}
\end{equation}
In this equation H.c.\ means Hermitian 
conjugate,\index{Hermitian conjugate} and $\kappa$ is the product
of the pump amplitude and the coupling constant between the electro-magnetic 
field\index{field!electro-magnetic} and the crystal. The operators 
$\hat{a}^{\dagger}_i$, $\hat{b}^{\dagger}_i$ and $\hat{a}_i$, $\hat{b}_i$ are 
creation\index{operator!creation} and annihilation 
operators\index{operator!annihilation} for polarisations $i\in\{ x,y \}$ 
respectively. They satisfy the following commutation 
relations:\index{commutation relation}
\begin{eqnarray}\label{commutation}
 [\hat{a}_i,\hat{a}_j^{\dagger}] = \delta_{ij}~ , &\qquad & 
 [\hat{a}_i,\hat{a}_j] = [\hat{a}_i^{\dagger},\hat{a}_j^{\dagger}] = 0, \cr
 [\hat{b}_i,\hat{b}_j^{\dagger}] = \delta_{ij}~ , &\qquad &
 [\hat{b}_i,\hat{b}_j] = [\hat{b}_i^{\dagger},\hat{b}_j^{\dagger}] = 0 \; ,
\end{eqnarray}
where $i,j\in\{ x,y \}$. The time evolution\index{time evolution} due to 
this Hamiltonian\index{Hamiltonian} is given by
\begin{equation}\label{unitary}
 U(t)\equiv\exp(-iHt) \; ,
\end{equation}
where $t$ is the time it takes for the pulse to travel through the crystal.
By applying this unitary transformation\index{transformation!unitary} to the 
vacuum\index{state!vacuum} $|0\rangle$ the state 
$|\Psi_{\text{src}}\rangle$ is obtained:
\begin{equation}\label{evolution}
  |\Psi_{\text{src}}\rangle = U(t) |0\rangle = \exp(-iHt) |0\rangle \; .
\end{equation}

We are interested in the properties of $|\Psi_{\text{src}}\rangle$. Define the 
$L_+$ and the $L_-$ operator to be
\begin{equation}
 L_+ = \hat{a}_x^{\dagger} \hat{b}_y^{\dagger} - \hat{a}_y^{\dagger} 
 \hat{b}_x^{\dagger}= L_-^{\dagger}\; .
\end{equation}
This will render Eqs.\ (\ref{hamilton}) and (\ref{unitary}) into:
\begin{equation}\label{unitl+}
 {H} = i\kappa L_+ - i\kappa^* L_- \qquad\text{and}\qquad
 U(t) = \exp[\kappa t L_+ - \kappa^* t L_-] \; .
\end{equation}

Applying $L_+$ to the vacuum will yield a singlet state\index{state!singlet} 
(up to a normalisation\index{normalisation} factor) in modes $a$ and $b$:
\begin{eqnarray}
  L_+ |0\rangle &=& |\leftrightarrow,\updownarrow\rangle_{ab} - 
		|\updownarrow,\leftrightarrow\rangle_{ab} \cr
  &=& |1,0;0,1\rangle_{a_x a_y b_x b_y}-|0,1;1,0\rangle_{a_x a_y b_x b_y} \; ,
\end{eqnarray}
we henceforth use the latter notation where $|i,j;k,l\rangle_{a_x a_y b_x b_y}$
is shorthand for $|i\rangle_{a_x}\!\otimes|j\rangle_{a_y}\!\otimes|k
\rangle_{b_x}\!\otimes|l\rangle_{b_y}$, a tensor product\index{tensor product} 
of photon number\index{state!number} states. Applying this operator $n$ times gives a state $|\Phi^n\rangle$ (where 
we have included a normalisation factor $N_n$, so that 
$\langle\Phi^n|\Phi^n\rangle=1$):
\begin{equation}
  |\Phi^n\rangle \equiv N_n L_+^n |0\rangle = \sum_{m=0}^{n} (-1)^m
  \sqrt{\frac{n!}{(n+1)!}} |m_x,(n-m)_y;(n-m)_x,m_y \rangle_{ab} \; ,
\end{equation}
where the normalisation constant $N_n$ is given by
\begin{equation}\label{normalization}
  N_n^2 = \frac{1}{n!(n+1)!} \; .
\end{equation}
We interpret $|\Phi^n\rangle$ as the state of $n$ entangled photon-pairs on
\index{state!of $n$ photon-pairs}
two spatial modes $a$ and $b$.

We want the unitary operator $U(t)$ in Eq.\ (\ref{unitl+}) to be in a 
normal ordered\index{normal ordering} form, because then the annihilation 
operators\index{operator!annihilation} will `act' on 
the vacuum first, in which case Eq.\ (\ref{evolution}) simplifies. In order to 
obtain the normal ordered form of $U(t)$ we examine the properties of $L_+$ 
and $L_-$. Given the commutation relations\index{commutation relation} 
(\ref{commutation}), it is straightforward to show that:
\begin{eqnarray}
  [L_-,L_+] &=& \hat{a}_x^{\dagger} \hat{a}_x + \hat{a}_y^{\dagger} \hat{a}_y 
 + \hat{b}_x^{\dagger} \hat{b}_x + \hat{b}_y^{\dagger} \hat{b}_y + 2
 \equiv  2 L_0 \quad\text{and}\cr
  [L_0,L_{\pm}] &=& \pm L_{\pm} \; .
\end{eqnarray}
An algebra which satisfies these commutation 
relations\index{commutation relation} (together with the 
properties $L_- = L_+^{\dagger}$ and $L_0 = L_0^{\dagger}$) is an $su(1,1)$ 
\index{algebra!$su(1,1)$}
algebra\footnote{The interaction Hamiltonian from Eq.\ (\ref{hamilton}) thus 
generates unitary evolutions which are closely related to the group elements 
of $SU(1,1)$.}. The normal ordering\index{normal ordering} for this algebra is 
known \cite{truax85} 
(with $\hat\tau = \tau/|\tau|$) (see also appendix \ref{app:forms} for more 
details):
\begin{equation}\label{BCHSU11}
 e^{\tau L_+ - \tau^* L_-} = e^{\hat\tau\tanh|\tau| L_+} 
 e^{-2\ln(\cosh|\tau|) L_0} e^{-\hat\tau^*\tanh|\tau|L_-}\; .
\end{equation}
The scaled time $\tau$ is defined as $\tau\equiv\kappa t$. Without loss 
of generality we can take $\tau$ to be real. Since the `lowering' operator 
\index{operator!lowering}
$L_-$ is placed on the right, it will yield zero when applied to the 
vacuum\index{state!vacuum} and
the exponential reduces to the identity. Similarly, the exponential containing 
$L_0$ will yield a $c$-number, contributing only an overall phase.

Parametric down-conversion\index{down-converter} is an example of so-called 
{\em multi-mode squeezed vacuum}.\index{squeezer!multi-mode} The 
photon-statistics\index{photon!-statistics} of two-mode squeezed 
states\index{state!squeezed} have been studied 
in Refs.\ \cite{caves91,artoni91,schrade93}. Here, I study the particular case
of the down-conversion process used to create randomly produced maximal 
entanglement.\index{entanglement!randomly produced} 

Are the pairs formed in parametric down-conversion independent of each other?
If they are, the number of pairs should give a Poisson 
distribution.\index{distribution!Poisson} I will
now calculate whether this is the case.

Suppose $P_{\text{PDC}}(n)$ is the probability of creating $n$ photon-pairs 
with parametric down-conversion and let 
\begin{equation}\label{substitutions}
  r \equiv \tanh\tau\qquad\text{and}\qquad
  q \equiv 2\ln(\cosh\tau) \; , 
\end{equation}
then the probability of finding $n$ entangled photon-pairs is:
\begin{eqnarray}\label{EPRdist}
  P_{\text{PDC}}(n) &\equiv& |\langle \Phi^n |\Psi_{\text{src}}\rangle|^2 \cr 
  && \cr
  &=& |\langle 0 | \left( L_-^n N_n \right) \left( e^{rL_+}e^{-qL_0} e^{-rL_-} 
      \right) |0\rangle|^2 \cr
  &=& e^{-2q} |\langle 0 | L_-^n N_n \left[ \sum_{l=0}^{\infty} \frac{r^l}{l!} 
  L_+^l\right] |0\rangle|^2 \cr 
  &=& (n+1) r^{2n} e^{-2q} \; .
\end{eqnarray}
It should be noted that this is a normalised probability distribution in 
\index{distribution!probability}
the limit of $r,q\rightarrow 0$. 

Given Eqs.\ (\ref{substitutions}) $P_{\text{PDC}}(n)$ deviates from the 
Poisson distribution, and the pairs are therefore not independent. For weak 
sources,\index{weak!sources} however, one might expect that 
$P_{\text{PDC}}(n)$ approaches the 
Poisson distribution sufficiently closely. This hypothesis can be tested by 
studying the distinguishability of the two 
distributions.\index{distribution!distinguishability}

\subsection{Statistical properties of down-converters}

Here, I study the distinguishability between the pair 
distribution\index{distribution!distinguishability} calculated 
in the previous section and the Poisson distribution. The Poisson distribution 
\index{distribution!Poisson}
for independently created objects is given by
\begin{equation}\label{poisson}
 P_{\text{poisson}}(n) = \frac{p^n e^{-p}}{n!} \; .
\end{equation}
Furthermore, rewrite the pair distribution in Eq.\ (\ref{EPRdist}) as
\begin{equation}\label{EPRdist2}
  P_{\text{PDC}}(n) = (n+1) \left(\frac{p}{2}\right)^n e^{-p}\; , \qquad
  \text{for}~p\ll 1 \; ,
\end{equation}
using $q\approx r^2$ and $p \equiv 2r^2 = 2\tanh^2\tau$ for small scaled times.
Here $p/2$ is the probability of creating one entangled photon-pair. Are these 
probability distributions distinguishable? Naively one would say that for 
sufficiently weak down-conversion\index{down-converter} (i.e., when $p\ll 1$) 
these distributions 
largely coincide, so that instead of the complicated pair-distribution 
(\ref{EPRdist2}) we can use the Poisson distribution, which is much easier 
from a mathematical point of view. The distributions are distinguishable when 
the `difference' between them is larger than the size of an average statistical
fluctuation\index{statistical!fluctuation} of the difference. This 
fluctuation depends on the number of samplings. 

Consider two nearby discrete probability distributions $\{ p_j\}$ and $\{ p_j 
+ dp_j\}$. A natural difference between these distributions is given by 
the so-called (infinitesimal) {\em statistical 
distance}\index{statistical!distance|see{distance}} $ds$ 
\cite{wootters81,braunstein94,hilgevoord91} (see also appendix 
\ref{app:distance}):
\begin{equation}
 ds^2 = \sum_j \frac{dp_j^2}{p_j} \; .
\end{equation}
When the typical statistical fluctuation after $N$ samplings is 
$1/\sqrt{N}$, the two probability distributions are distinguishable if:
\begin{equation}\label{stdstN}
 ds \gtrsim \frac{1}{\sqrt{N}} \quad\Leftrightarrow\quad 
	N ds^2 \gtrsim 1 \; .
\end{equation}
The statistical distance between (\ref{poisson}) and (\ref{EPRdist2}), and 
therefore the distinguishability criterion is: 
\begin{equation}\label{32p2}
 ds^2 \propto \frac{p^2}{8} \quad\rightarrow\quad N \gtrsim 
	\frac{8}{p^2} \; .
\end{equation}
On the other hand, the average number of trials in the teleportation 
\index{quantum teleportation}
experiment required to get one photon-pair from both down-converters is:
\begin{equation}\label{speed}
 N = \frac{1}{p^2} \; .
\end{equation}
The minimum number of trials in the experiment thus almost immediately renders 
the two probability distributions\index{distribution!probability} 
distinguishable, and we therefore cannot approximate the actual probability 
distribution with the Poisson distribution.\index{distribution!Poisson}

Since the Poisson distribution in Eq.\ (\ref{poisson}) is derived by requiring 
statistical independence\index{statistical!independence} of $n$ pairs and the 
pair distribution is distinguishable from the Poisson distribution, the 
photon-pairs cannot be considered to be independently produced, {\em even in 
the weak limit}.\index{weak!limit}

This concludes my study of parametric down-conversion\index{down-converter} 
here. I will now return to the problem of maximal entanglement creation in 
quantum optics.

\section{The creation of maximal entanglement}\index{entanglement!maximal}\label{sec:max}

Now that I have investigated the properties of one of the most common
entanglement sources, i.e., down-conversion, I am ready to consider the 
creation of maximal, or event-ready, entanglement. The optical circuits 
discussed here consist of so-called {\em passive}\index{component!passive} 
and {\em active} components.\index{component!active}
The passive components leave the photon-number\index{photon!-number} 
invariant, i.e., they correspond to unitary operators which commute with the 
number operator.\index{operator!number} Examples of such components are 
beam-splitters,\index{beam-splitter} phase-shifters\index{phase shift} and 
polarisation rotators.\index{polarisation rotator}

Active components correspond to unitary evolutions\index{unitary evolution} 
which do {\em not} commute with the photon-number operator, like the 
parametric down-converter. Other examples of active components are single-mode 
squeezers\index{squeezer!single-mode} and pumped $\chi^{(3)}$ 
media\index{x3@$\chi^{(3)}$ medium} \cite{dariano00}. They can generally be 
characterised by an interaction Hamiltonian\index{interaction Hamiltonian} 
which is a polynomial function of creation\index{operator!creation} and 
annihilation\index{operator!annihilation} 
operators. Later in this chapter I will restrict the discussion to interaction
Hamiltonians which are quadratic\index{interaction Hamiltonian!quadratic} in 
these operators.

In this section, I will first show that the creation of maximal entanglement 
with only passive components from a pure separable state\index{state!separable}
is impossible. Then, a general condition for an optical setup is derived, 
which should be satisfied in order to yield event-ready 
entanglement.\index{entanglement!event-ready} I subsequently examine this 
condition for a specific class of optical circuits.\index{optical!circuit}

\subsection{Passive optical components}

So far, I have hardly paid attention to passive optical 
components.\index{component!passive} In this
section I will show that they cannot transform a completely separable state
of two photons into a maximally entangled state.\index{state!maximally 
entangled}

Suppose we have a linear interferometer\index{interferometer!linear} which 
consists only of passive components. Such an interferometer is described by 
a unitary matrix $U$ \cite{reck94}, which transforms the creation operators 
\index{operator!creation}of the electro-magnetic 
field\index{field!electro-magnetic} according to
\begin{equation}\label{unitary2}
 \hat{a}_i \rightarrow \sum_j u_{ij} \hat{b}_j \qquad\text{and}\qquad
 \hat{a}_i^{\dagger} \rightarrow \sum_j^{\phantom{n}} u_{ij}^* 
 \hat{b}_j^{\dagger}\; ,
\end{equation}           
where the $u_{ij}$ are the components of $U$ and $i,j$ enumerate both the
modes and polarisations. There is no mixing between the 
creation\index{operator!creation} and annihilation 
operators,\index{operator!annihilation} because photons do not interact with 
each other. I will now show that we cannot create maximal (event-ready) 
entanglement\index{entanglement!event-ready} with such
linear interferometers\index{interferometer!linear} when the input state is a 
separable state.\index{state!separable}

Without loss of generality I consider the separable state $|x,y\rangle = 
\hat{a}_x^{\dagger} \hat{b}_y^{\dagger} |0\rangle$. In order to create maximal 
entanglement, the creation operators should be transformed according to
\begin{equation}
 \hat{a}_x^{\dagger} \hat{b}_y^{\dagger} ~\rightarrow~ (
 \hat{c}_x^{\dagger} \hat{d}_y^{\dagger} - \hat{c}_y^{\dagger}
 \hat{d}_x^{\dagger})/\sqrt{2} \; .
\end{equation}
Relabel the modes $a_x$, $a_y$, $b_x$ and $b_y$ as $a_1$ to $a_4$ respectively.
Without loss of generality (and leaving the 
normalisation\index{normalisation} aside for the 
moment) we can then write
\begin{equation}\label{trans}
 \hat{a}_1^{\dagger} \hat{a}_2^{\dagger} ~\rightarrow~ \hat{b}_1^{\dagger} 
 \hat{b}_2^{\dagger} - \hat{b}_3^{\dagger} \hat{b}_4^{\dagger} \; .
\end{equation}
Substituting Eq.\ (\ref{unitary2}) into Eq.\ (\ref{trans}) generates ten
equations for eight variables $u^*_{ij}$:
\begin{eqnarray}
 u^*_{11} u^*_{21} = u^*_{12} u^*_{22} = u^*_{13} u^*_{23} = 
 u^*_{14} u^*_{24} &=& 0 \cr && \cr
 (u^*_{11} u^*_{23} + u^*_{13} u^*_{21}) = 
 (u^*_{11} u^*_{24} + u^*_{14} u^*_{21}) &=& 0 \cr
 (u^*_{12} u^*_{23} + u^*_{13} u^*_{22}) = 
 (u^*_{12} u^*_{24} + u^*_{14} u^*_{22}) &=& 0 \cr && \cr
 (u^*_{11} u^*_{22} + u^*_{12} u^*_{21}) &=& 1 \cr
 (u^*_{13} u^*_{24} + u^*_{14} u^*_{23}) &=& -1\; .
\end{eqnarray}
It can be easily verified that there are no solutions for the $u^*_{ij}$
which satisfy these ten equations simultaneously: since $u^*_{11} u^*_{21}=0$,
choose $u^*_{11}=0$ and $u^*_{21}\neq 0$. From the fourth line above follows
that $u^*_{12}\neq 0$. Hence $u^*_{22}=0$. The second line (second equation) 
then determines $u^*_{14}=0$, which (third line, second equation) implies that 
$u^*_{12} u^*_{24}=0$. We already set $u^*_{12}\neq 0$, so we obtain 
$u^*_{24}=0$. We now derive a contradiction between $u^*_{14}=0$, $u^*_{24}=0$
and the last line of the set of equations above. This means that there is 
no passive interferometer which transforms pure separable states into 
maximal (event-ready) entangled states.

\subsection{General optical circuits}\index{optical!circuit}

\begin{figure}[t]
\begin{center}
  \begin{psfrags}
     \psfrag{a}{a)}
     \psfrag{b}{b)}
     \psfrag{psi}{$|\Psi^-\rangle$}
     \psfrag{u0}{$U^{(0)}$}
     \psfrag{u1}{$U^{(1)}$}
     \psfrag{un}{$U^{(n)}$}
     \psfrag{u}{$U$}
     \psfrag{in}{$|\psi_{\rm in}\rangle$}
     \psfrag{to}{$\Longrightarrow$}
     \psfrag{d}{\ldots}
     \psfrag{b5}{$a_5$}
     \psfrag{bn}{$a_N$}
     \epsfxsize=8in
     \epsfbox[-150 40 860 300]{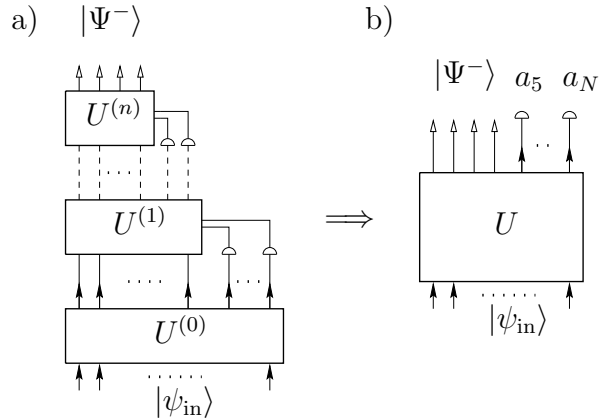}
  \end{psfrags}
  \end{center}
  \caption{If an optical circuit with feed-forward detection (a) 
	produces a specific state, the same output can be obtained by an optical
	circuit where detection of the {\em auxiliary} modes takes place at 
	the end (b). The efficiency of the latter, however, will generally be 
	smaller.}
\label{fig:3.5}
\end{figure}

In order to make $|\Psi^-\rangle$, I will assume that we have several 
resources at our disposal. The class of elements will consist of 
beam-splitters,\index{beam-splitter} phase-shifters,\index{phase shift} 
photo-detectors\index{photo-detector} and non-linear components such 
as down-converters,\index{down-converter} squeezers,\index{squeezer} etc. 
These elements are then arranged to give a specific optical circuit (see 
Fig.\ {\ref{fig:3.5}}). Part of this setup might be so-called {\em 
feed-forward} detection.\index{detection!feed-forward} In this scheme the 
outcome of the detection of a number of modes dynamically chooses the internal 
configuration of the subsequent optical circuit\index{optical!circuit} based 
on the interim detection results (see also Ref.\ \cite{lutkenhaus99}). 
Conditioned on these detections we want to obtain a freely propagating 
$|\Psi^-\rangle$ Bell state\index{state!Bell} in the remaining undetected 
modes. 

I now introduce two simplifications for such an optical circuit. First, I
will show that we can discard feed-forward detection. Secondly, we only have 
to consider the detection of modes with at most one photon.

\begin{description}
 \item[Theorem 1:] In order to show that it is {\em possible} to produce a 
 specific outgoing state, any optical circuit with feed-forward
 detection\index{detection!feed-forward} can be replaced by a {\em fixed}
 optical circuit\index{optical!circuit} where detection only takes place 
 at the end.
 \item[Proof:] Suppose a feed-forward optical circuit (like the one depicted 
 in Fig.\ \ref{fig:3.5}a) giving $|\Psi^-\rangle$ exists. That means that
 the circuit creates $|\Psi^-\rangle$ conditioned on one of potentially many
 patterns of detector responses. It is sufficient to consider a single
 successful pattern. We can then take every interferometer
 \index{interferometer!linear} to be fixed and postpone all detections of the
 auxiliary modes\index{mode!auxiliary} to the very end (Fig.\
 \ref{fig:3.5}b). Note that this procedure selects generally only {\em one}
 setup in which entanglement\index{entanglement} is produced, whereas a 
 feed-forward optical circuit potentially allows more setups. It therefore
 might reduce the efficiency\index{efficiency} of the process. However, since 
 we are only interested in the {\em
 possibility} of creating $|\Psi^-\rangle$, the efficiency is irrelevant.
 \hfill $\square$
 \item[Theorem 2:] Suppose an optical circuit produces a specific outgoing
 state conditioned on $n_1$ detected photons in mode 1, $n_2$ detected photons 
 in mode 2, etc.\ (with $n_i=0,1,2,\ldots$). The same output can be obtained
 by a circuit where in every detected mode {\em at most} one photon is found.
 \item[Proof:] If there are more photons in a mode, we can replace the 
 corresponding detector by a so-called detector {\em 
 cascade}\index{detector!cascade} \cite{kok99}.
 This device splits the mode into many modes which are all detected (see also
 chapter \ref{chap4}). For a 
 sufficiently large cascade there is always a non-vanishing probability to
 have at most one photon in each outgoing mode. In that case, the same state
 is created while at most one photon enters each detector. Note that this
 again yields a lower efficiency.\index{efficiency} \hfill $\square$
\end{description}

Applying these results to the creation of $|\Psi^-\rangle$, it is sufficient 
to consider a single {\em fixed} interferometer\index{interferometer!linear} 
acting on an incoming state, followed at the end by detection of the so-called 
{\em auxiliary} modes.\index{mode!auxiliary} 
$|\Psi^-\rangle$ is signalled by at least one fixed detection pattern with at 
most one photon in each detector.

How do I proceed in trying to make the $|\Psi^-\rangle$ Bell 
state?\index{state!Bell} Let the time independent interaction 
Hamiltonian\index{interaction Hamiltonian} ${{H}}_I$ incorporate both the
interferometer $U$ and the creation of $|\psi_{\rm in}\rangle$ (see Fig.\ 
\ref{fig:3.5}b). The outgoing state prior to the detection can be 
formally written as
\begin{equation}\label{evolution2}
 |\psi_{\rm out}\rangle = U |\psi_{\rm in}\rangle \equiv \exp\left( -it 
 {{H}}_I \right) |0\rangle\; ,
\end{equation}
with $|0\rangle$ the vacuum.\index{state!vacuum} This defines an effective 
Hamiltonian ${{H}}_I$ which is generally not unique.

\subsection{The Bargmann representation}\index{representation!Bargmann}

At this point it is useful to change the description. Since the creation
\index{operator!creation} and annihilation 
operators\index{operator!annihilation} satisfy the same commutation relations 
\index{commutation relation}as c-numbers
and their derivatives, we can make the substitution $\hat{a}^{\dagger}_i
\rightarrow\alpha_i$ and $\hat{a}_i\rightarrow\partial_i$, where $\partial_i
\equiv\partial/\partial\alpha_i$. Furthermore, we define $\vec\alpha = 
(\alpha_1,\ldots,\alpha_N)$. Quantum states are then represented by functions 
of c-numbers and their derivatives. This is called the Bargmann representation 
\cite{bargmann61}.

Furthermore, suppose we can normal order\index{normal ordering} the operator 
$\exp(-it{{H}}_I)$ in Eq.\ (\ref{evolution2}). This would yield a function of 
only the creation operators, acting on the vacuum. In the Bargmann 
representation we then obtain a function of complex numbers without their 
derivatives. In particular, an optical circuit\index{optical!circuit} 
consisting of $N$ distinct modes (for notational convenience I treat distinct 
polarisations like, for instance, $x$ and $y$ as separate modes), can be 
written as a function $\psi_{\rm out} (\vec{\alpha})$ after the unitary 
evolution\index{unitary evolution} $U$ and 
normal ordering. The normal ordering of the evolution operator in conjunction 
with the vacuum input state is crucial, since it allows a significant 
simplification of the problem.

I now treat the (ideal) detection of the auxiliary modes in the Bargmann 
representation. Suppose the outgoing state after the detection of $M$ 
photons emerges in modes $a_1$, $a_2$, $a_3$ and $a_4$. After a suitable 
reordering of the detected modes the state which is responsible for the 
detector coincidence indicating success can be written as $|1_5,\ldots\!,
1_{M+4}, 0_{M+5},\ldots\rangle$ (possibly on a countably infinite number of 
modes). We then obtain the post-selected\index{post-selection} state 
$|\psi_{\rm post}\rangle$
\begin{eqnarray}\label{detcs}
 |\psi_{\rm post}\rangle_{1..4} &\propto& \langle 1_5,\ldots\!,1_{M+4}, 
 0_{M+5}, \ldots| \psi_{\rm out}\rangle\cr 
 &=& \langle 0|\, \hat{a}_5 \cdots \hat{a}_{M+4} |\psi_{\rm out}\rangle\; .
\end{eqnarray}
In the Bargmann representation\index{representation!Bargmann} the right-hand 
side of Eq.\ (\ref{detcs}) is
\begin{equation}
 \left. \partial_5 \cdots \partial_{M+4}\; \psi_{\rm out} (\vec\alpha) 
 \right|_{\vec\alpha' = 0} \; ,
\end{equation}
where I have written $\vec\alpha' = (\alpha_5,\ldots\!,\alpha_{M+4},\ldots)$.

Writing out the entanglement\index{entanglement} explicitly in the four modes 
(treating the polarisation implicitly), I arrived at the following condition 
for the creation of two photons in the antisymmetric Bell 
state:\index{state!Bell!anti-symmetric}
\begin{equation}\label{con}
 \left. \partial_5 \cdots \partial_{M+4}\; \psi_{\rm out} (\vec\alpha) 
 \right|_{\vec\alpha'=0} \propto \alpha_1 \alpha_2 - \alpha_3 \alpha_4 + 
 O(\xi)\; .
\end{equation}
The term $O(\xi)$ will allow for a small pollution ($\xi\ll 1$) in the 
outgoing state. I will show that for certain special classes of interaction 
Hamiltonians\index{interaction Hamiltonian} this condition is very hard 
(if not impossible) to satisfy. This renders the experimental realisation of 
two maximally polarisation entangled photons\index{entanglement!polarisation} 
at least highly impractical.

\subsection{Physical limitations on event-ready entanglement}
 
\begin{figure}[t]
\begin{center}
  \begin{psfrags}
     \psfrag{5}{$a_5$}
     \psfrag{N}{$a_N$}
     \psfrag{u}{$U'$}
     \psfrag{s}{\small s}
     \psfrag{in}{$|0\rangle$}
     \psfrag{out}{$|\Psi^-\rangle$}
     \psfrag{un}{$\underbrace{\phantom{xxxxxxxxxxxx.}}$}
     \epsfxsize=8in
     \epsfbox[-200 40 600 180]{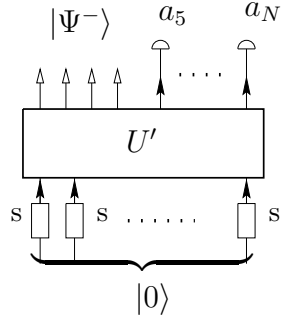}
  \end{psfrags}
  \end{center}
  \caption{The unitary interferometer $U'$ with conditional photo-detection 
	and single-mode squeezers which should transform $|0\rangle$ into 
	$|\Psi^-\rangle$.}
\label{fig:3.6}
\end{figure}

I am now ready to shape $\psi_{\rm out}$ in more detail. Consider optical 
circuits\index{optical!circuit} including mode-mixing,\index{mode!-mixing} 
squeezers\index{squeezer} and down-converters.\index{down-converter} The 
corresponding interaction Hamiltonians ${{H}}_I$ are {\em 
quadratic}\index{interaction Hamiltonian!quadratic} in the creation 
operators.\index{operator!creation} There are no linear terms, so there are 
no coherent displacements.\index{coherent displacement} More formally
\begin{equation}\label{source}
 {{H}}_I = \sum_{i,j=1}^N \hat{a}^{\dagger}_i A^{(1)}_{ij} 
 \hat{a}^{\dagger}_j +  \sum_{i,j=1}^N \hat{a}^{\dagger}_i A^{(2)}_{ij} 
 \hat{a}_j +  {\rm H.c.}\; .
\end{equation}
With $A^{(1)}$ and $A^{(2)}$ complex matrices (see also Appendix 
\ref{app:forms} for more details about the dependence of $H$ on $A$).
According to Braunstein \cite{braunstein99}, such an active 
interferometer\index{interferometer!active} is 
equivalent to a passive interferometer\index{interferometer!active} $V$, 
followed by a set of single-mode squeezers\index{squeezer!single-mode} and 
another passive interferometer $U'$. The photon source\index{photon!source} 
described by Eq.\ (\ref{source}) can be viewed as an active bilinear 
component\index{interaction Hamiltonian!bilinear} of an 
interferometer.\index{interferometer} For vacuum input and after normal 
ordering\index{normal ordering} \cite{truax85}, the 
optical setup then gives rise to
\begin{equation}\label{bilinear}
 \psi_{\rm out}(\vec\alpha) = \exp\left[(\vec\alpha,B\vec\alpha)\right]\; ,
\end{equation} 
with $(\vec\alpha,B\vec\alpha)=\sum_{ij}^N \alpha_i B_{ij} \alpha_j$.
Such an optical setup would correspond to a collection of single-mode squeezers
\index{squeezer!single-mode} acting on the vacuum,\index{state!vacuum} 
followed by a passive optical interferometer\index{interferometer!passive} 
$U'$. Here, $B$ is a complex symmetric matrix determined by the interaction 
Hamiltonian\index{interaction Hamiltonian} ${{H}}_I$ and the interferometer 
$U'$. We take $B$ to be proportional to a common coupling 
constant\index{coupling constant} $\xi$. The outgoing auxiliary 
modes\index{mode!auxiliary} $a_5$ to $a_N$ are detected (see Fig.\ 
\ref{fig:3.6}). I will now investigate whether the production of 
$|\Psi^-\rangle$ conditioned on a given number of detected photons is possible.

In the case of a bilinear interaction 
Hamiltonian\index{interaction Hamiltonian!bilinear} (see Eq.\ (\ref{source})), 
photons are always created in pairs. In addition, we seek to create {\em two} 
maximally entangled photons.\index{entanglement!maximal} An odd number of 
detected photons can never give 
$|\Psi^-\rangle$ and the number of detected photons should therefore be even. 
The lowest even number is zero. In this case no photons are detected and 
$\psi_{\rm out}$ in Eq.\ (\ref{bilinear}) is proportional to $1+O(\xi)$, which 
corresponds to the vacuum state.

The next case involves two detected photons. To have entanglement in modes 
$\alpha_1$ to $\alpha_4$ after detecting {\em two} photons requires
\begin{equation}\label{2con}
 \left. \partial_5 \partial_6 \; e^{(\vec\alpha, B \vec\alpha)}
 \right|_{\vec\alpha'=0} \propto \alpha_1\alpha_2-\alpha_3\alpha_4 + O(\xi)\; .
\end{equation}

The left-hand side of Eq.\ (\ref{2con}) is equal to 
\begin{equation}
 \Bigl( B_{56} + \sum_{i,j=1}^4 \alpha_i B_{i5} B_{j6} \alpha_j \Bigr) \left.
 e^{(\vec\alpha,B \vec\alpha)} \right|_{\vec\alpha'=0} \; .
\end{equation}
To satisfy Eq.\ (\ref{2con}), the vacuum contribution\index{vacuum 
contribution} $B_{56}$ would have to 
be negligible. I now investigate whether the second term can give us 
entanglement. The right hand side of Eq.\ (\ref{2con}) can be rewritten 
according to $\alpha_1\alpha_2 - \alpha_3 \alpha_4 = \sum_{i,j=1}^4 \alpha_i 
E_{ij} \alpha_j$, where $E_{ij}$ are the elements of a symmetric matrix $E$:
\begin{equation}
 E = 
 \begin{pmatrix}
  0 & 1 &  0 &  0 \cr
  1 & 0 &  0 &  0 \cr
  0 & 0 &  0 & -1 \cr
  0 & 0 & -1 &  0
 \end{pmatrix}\; ,
\end{equation}
from which it is immediate that seen that $\det E = 1$. 

Let $M_{ij} = B_{i5}B_{j6}$. Since only the symmetric part of $M$ contributes, 
consider $\widetilde{M}_{ij} = (M_{ij} + M_{ji})/2$. The condition for two 
detected photons now yields
\begin{equation}
 \sum_{i,j=1}^4 \alpha_i \widetilde{M}_{ij} \alpha_j = \sum_{i,j=1}^4
 \alpha_i E_{ij} \alpha_j + O(\xi)\; ,
\end{equation}
If this equality is to hold, we need $\det E = \det\widetilde{M} + O(\xi) = 1$.
However, it can be shown that $\det\widetilde M = 0$. $\widetilde M$ can 
therefore never have the same form as $E$ for small $\xi$, so it is {\em not} 
possible to create maximal polarisation 
entanglement\index{entanglement!maximal} conditioned upon two 
detected photons.

Finally, consider the outgoing state conditioned on four detected photons.
Define $X_i \equiv \sum_{j} B_{ij} \alpha_j$. The left-hand side of Eq.\ 
(\ref{con}) for four detected photons then gives
\begin{multline}\nonumber
 \left( B_{56} B_{78} + B_{57} B_{68} + B_{58} B_{67} + B_{56} X_7 X_8 +
 B_{57} X_6 X_8 + B_{58} X_6 X_7 \right. \cr \left.\left. + B_{67} X_5 X_8 + 
 B_{68} X_5 X_7 + B_{78} X_5 X_6 + X_5 X_6 X_7 X_8 \right) 
 e^{(\vec\alpha,B\vec\alpha)}\right|_{\vec\alpha'=0} \; .
\end{multline}
I have not been able either to prove or disprove that $|\Psi^-\rangle$ can be 
made this way. The number of terms which contribute to the bilinear part in 
$\alpha$ rapidly increases for more detected photons. 

Suppose we {\em could} create maximal entanglement conditioned upon 
four detected photons, how efficient would this process be? For four detected 
photons yielding $|\Psi^-\rangle$ we need at least three photon-pairs. These 
are created with a probability of the order of $|\xi|^6$. Currently, 
$|\xi|^2$, the probability per mode, has a value of $10^{-4}$ 
\cite{weinfurter98}. For experiments operating at a repetition rate of 100 MHz 
using ideal detectors, the procedure conditioned on four detected photons will 
amount to approximately one maximally 
entangled pair every few hours. For realistic detectors this is much less. 

So far, there have been no experiments which exceeded the detection of more 
than two {\em auxiliary} photons (not including the actual detection of the 
maximally entangled state). This, and the estimation of the above efficiency 
appears to place strong practical limitations on the creation of maximal 
entanglement. 

\subsection{Six detected photons}

Recently, Knill, Laflamme and Milburn have discovered a method which allows us 
to create event-ready entanglement\index{entanglement!event-ready} conditioned 
on {\em six} detected photons \cite{knill00}. This method involves the 
construction of the C-SIGN\index{controlled!SIGN} operator
\begin{equation}
 U_{\rm C-SIGN} = 
 \begin{pmatrix}
  1 & 0 & 0 & 0 \cr
  0 & 1 & 0 & 0 \cr
  0 & 0 & 1 & 0 \cr
  0 & 0 & 0 & -1
 \end{pmatrix}\; ,
\end{equation}
on the basis $\{ |0,0\rangle,|0,1\rangle,|1,0\rangle,|1,1\rangle \}$. In 
quantum optics, these two qubits\index{qubit} are defined on four distinct 
modes $a_1$, $a_2$, $a_3$ and $a_4$.

The C-NOT\index{controlled!NOT} $U_{\rm C-NOT}$ is then defined using the 
Hadamard transform\index{Hadamard} $H$ on modes $a_3$ and $a_4$ as 
\begin{eqnarray}
 U_{\rm C-NOT} &=& H^{\dagger}_{a_3,a_4} U_{\rm C-SIGN} H_{a_3,a_4} \cr &&\cr
 &=& \frac{1}{2}
 \begin{pmatrix}
  1 & 1 & 0 & 0 \cr
  1 & -1 & 0 & 0 \cr
  0 & 0 & 1 & 1 \cr
  0 & 0 & 1 & -1
 \end{pmatrix}
 \begin{pmatrix}
  1 & 0 & 0 & 0 \cr
  0 & 1 & 0 & 0 \cr
  0 & 0 & 1 & 0 \cr
  0 & 0 & 0 & -1
 \end{pmatrix}
 \begin{pmatrix}
  1 & 1 & 0 & 0 \cr
  1 & -1 & 0 & 0 \cr
  0 & 0 & 1 & 1 \cr
  0 & 0 & 1 & -1
 \end{pmatrix} \cr && \cr &=& 
 \begin{pmatrix}
  1 & 0 & 0 & 0 \cr
  0 & 1 & 0 & 0 \cr
  0 & 0 & 0 & 1 \cr
  0 & 0 & 1 & 0
 \end{pmatrix}\; .
\end{eqnarray}
Applying the C-NOT and the Hadamard transformation on a separable state yields 
a maximally entangled state.\index{state!maximally entangled}

How do we construct the C-SIGN\index{controlled!SIGN} operator? Following 
Knill {\em et al}.\ this amounts to the construction of the operator
\begin{equation}
 A : \alpha_0|0\rangle + \alpha_1|1\rangle + \alpha_2|2\rangle~\longrightarrow~
 \alpha_0|0\rangle + \alpha_1|1\rangle - \alpha_2|2\rangle\; . 
\end{equation}
First, we apply a beam-splitter\index{beam-splitter} $B_{\theta}$:
\begin{equation}  
 B_{\theta} = 
 \begin{pmatrix}
  \cos\theta & \sin\theta \cr
  -\sin\theta & \cos\theta
 \end{pmatrix}\; ,
\end{equation}
with $\theta=\pi/4$ to modes $a_1$ and $a_3$, which 
transforms $|1,1\rangle$ into $(|2,0\rangle - |0,2\rangle)/\sqrt{2}$. 
Subsequently, we apply the operator $A$ to modes $a_1$ and $a_3$, and finally,
we apply a beam-splitter $B_{-\pi/4}$ again to these modes. 
 
\begin{figure}[t]
\begin{center}
  \begin{psfrags}
     \psfrag{a1}{$a_1$}
     \psfrag{a2}{$a_2$}
     \psfrag{a3}{$a_3$}
     \psfrag{a4}{$a_4$}
     \psfrag{1}{\small 1}
     \psfrag{2}{\small 2}
     \psfrag{01}{$|1,0\rangle$}
     \psfrag{een}{$|1\rangle$}
     \psfrag{nul}{$|0\rangle$}
     \psfrag{H}{$H$}
     \psfrag{UA}{$U_A$}
     \psfrag{B+}{$B_{\pi/4}$}
     \psfrag{B-}{$B_{-\pi/4}$}
     \epsfxsize=8in
     \epsfbox[-20 40 780 200]{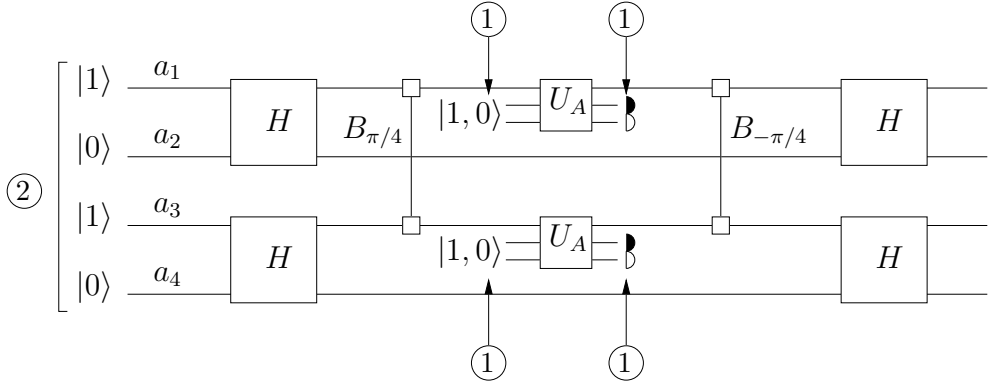}
  \end{psfrags}
  \end{center}
  \caption{Circuit for event-ready entanglement conditioned on six detected
	photons. Here, $U_A$ is given by Eq.~(\ref{ua}), $H$ is the Hadamard
	transform and $B_{\theta}$ is a beam-splitter. The encircled numbers
	denote the number of detected photons needed to create the 
	corresponding states.}
\label{fig:3.7}
\end{figure}

The operator A is defined as a unitary transformation $U_A$ on three modes,
one main mode and two auxiliary modes\index{mode!auxiliary} which are detected:
\begin{equation}\label{ua}
 U_A = 
 \begin{pmatrix}
  1-\sqrt{2} & \frac{1}{\sqrt[4]{2}} & \sqrt{\frac{3}{\sqrt{2}}-2} \cr
  \frac{1}{\sqrt[4]{2}} & \frac{1}{2} & \frac{1}{2}- \frac{1}{\sqrt{2}} \cr
  \sqrt{\frac{3}{\sqrt{2}}-2} & \frac{1}{2}- \frac{1}{\sqrt{2}} & \sqrt{2} - 
  \frac{1}{2}
 \end{pmatrix}\; .
\end{equation}
This transformation can be explicitly constructed using the techniques 
developed by Reck {\em et al}.\ \cite{reck94}. The input state on the two 
auxiliary modes $b_1$ and $b_2$ is $|1,0\rangle$, and the operator is 
conditioned on a state $|1,0\rangle$ in the outgoing auxiliary modes. This 
post-selection\index{post-selection} means that $A$ is a probabilistic 
operator\index{operator!probabilistic} with a probability of success of 1/4.

Event-ready entanglement\index{entanglement!event-ready} can now be created 
using the setup shown in figure
\ref{fig:3.7}. The incoming state is given by $|1,0,1,0\rangle$, which can be 
made conditioned on two detected photons. The input states on the auxiliary
modes can be created conditioned on one detected photon and the outgoing 
auxiliary modes also involve one detected photon. Therefore the total number
of detected photons is six.

\section{Summary}

I have demonstrated strong limitations on the possibility of creating maximal 
entanglement\index{entanglement!maximal} with quantum optics. To this end, I 
introduced two simplifications to the hypothetical optical 
circuit:\index{optical!circuit} I replaced feed-forward 
detection\index{detection!feed-forward} by a fixed set of detectors at the 
end, and secondly, every detector needs to detect at most one photon. 
Conditioned on two detected photons, multi-mode squeezed 
vacuum\index{vacuum!multi-mode squeezed} fails to create maximal entanglement. 
\index{entanglement!maximal}

What happens when we have a combination of 
squeezing\index{squeezer} and coherent displacements?\index{coherent 
displacement} In that case the approach taken here fails due to the 
more complex normal ordering of the interaction 
Hamiltonian.\index{interaction Hamiltonian} Also, I have 
only considered ideal detections, but how do realistic 
detectors\index{detector!real} affect the 
outgoing state? This is the subject of the next chapter.

\newpage
\thispagestyle{empty}

\chapter{Auxiliary Resources: Detection Devices}\label{chap4}

Wouldn't it be nice to have a machine which creates the quantum states of your 
choice at the push of a button? Unfortunately, these machines do not yet 
exist\footnote{Quantum computers will be able to make such states for qubits.}.
There are currently machines which create certain specific states, like for
instance lasers (creating coherent states) and down-converters (creating
squeezed states), but notwithstanding their importance for scientific and 
technological applications, these devices create only a limited class of 
quantum states. 

When we want to create more exotic quantum states, we need to extend our 
resources: in addition to the devices mentioned above we may use passive 
transformations (like, for instance, beam-splitters and phase-shifters in 
quantum optics) and {\em measurements}.\index{measurement} With this new set 
of tools we can build more sophisticated state preparation devices, or 
`circuits'. As I have shown in the previous chapter, depending on the 
particular physical implementation of these circuits we can create more exotic 
quantum states. State preparation has been studied among others by Vogel {\em 
et al}.\ \cite{vogel93}, Harel {\em et al}.\ \cite{harel96}, Dakna {\em et 
al}.\ \cite{dakna99} and Rubin \cite{rubin00}.

Having extended our resources to state preparation circuitry, the next issue 
is the {\em quality} of the state preparation. Suppose we want to create a 
particular state. In practice, we can never obtain this state perfectly, 
due to uncontrollable effects like decoherence\index{decoherence} and 
measurement errors. Nevertheless, we want our maximise the quality of the 
state preparation\index{state preparation} process.

Formulating this more precisely, we want to prepare a {\em single} (pure) 
state $|\phi\rangle$ by means of some process, and we want the resulting state 
$\rho$ to be as `close' to $|\phi\rangle$ as possible. In chapter \ref{chap2} 
we have seen that a measure of resemblance between states is given by the 
fidelity\index{fidelity} $F$:
\begin{equation}
 F = {\rm Tr} [\rho |\phi\rangle\langle\phi|]\; .
\end{equation}
The quality of a state preparation process can therefore be measured by the
fidelity. When $F=1$, the process gives exactly $|\phi\rangle$ and when $F=0$,
the prepared state is orthogonal to $|\phi\rangle$. In practice, the fidelity 
will not reach these extreme measures, but will lie between 0 and 1.

In short, we have a state preparation circuit which creates states with some 
fidelity. Generally, the preparation process is {\em conditioned} on 
measurements \cite{klyshko77}. For example, if we want to prepare a 
single-photon state $|1\rangle$ in quantum optics we can use the following 
process: a parametric down-converter\index{down-converter} creates a state 
$|\psi\rangle_{ab}$ on two spatial modes $a$ and $b$ (see chapter \ref{chap3}):
\begin{equation}
 |\psi\rangle_{ab} \propto |0\rangle_a |0\rangle_b +\xi|1\rangle_a |1\rangle_b 
 + O(\xi^2)\; ,
\end{equation}
where $|0\rangle$ denotes the vacuum state and we assume $\xi\ll 1$. The 
higher order terms (included in $O(\xi^2)$) consist of states with more than 
two photons. We now place a photo-detector\index{detector!photo-} in mode $a$, 
which `clicks' when it 
sees one or more photons (typically, standard detectors can see single photons,
but fail to distinguish between one and two photons). Conditioned on such a 
click, mode $b$ will be in a state
\begin{equation}
 \rho \propto |1\rangle_b\langle 1| + O(|\xi|^2)\; .
\end{equation}
The fidelity\index{fidelity} of this process is high: $F=\langle 
1|\rho|1\rangle\simeq 1$,
and this is therefore typically a very good single-photon state 
preparation\index{state preparation!single photon} 
process (although the situation changes drastically when multiple 
down-converters are considered \cite{braunstein98,kok00a}). Due to the large 
vacuum contribution,\index{vacuum contribution} however, the probability of 
the detector giving a `click' will be small (of order $O(|\xi|^2)$). When the 
detector does not click, that particular trial is dismissed, hence the {\em 
conditional} character of the detection.\index{detection!conditional} 

In this example the outcome of the detection is used to either accept or 
reject a particular run of the state preparation device. However, in general 
the outcome of the detector can be used to determine a more complicated 
operation on the remainder of the state preparation process. This is detection
plus {\em feed-forward},\index{detection!feed-forward} since the outcome is 
used further on in the 
process. An example of this is quantum teleportation,\index{quantum 
teleportation} where the outcome of 
the Bell measurement\index{Bell!measurement} determines the unitary 
transformation needed to retrieve the original input state.

\medskip

When the measurements in the state preparation process are prone to errors,
the state we want to create may not be the state we actually create. This 
means that errors in the detection devices can lead to reduced fidelities. In
this chapter I study the effect of detection errors on state preparation.
To this end I introduce the concept of the {\em confidence} of preparation. 
Using this measure I evaluate different types of detection devices. This 
chapter is based on Kok and Braunstein \cite{kok99}.

\section{Confidence}\label{confidence}\index{confidence}

Consider a preparation device which prepares a state conditioned on a single 
measurement. For simplicity, I employ two subsystems. One subsystem will be 
measured, leaving a quantum state in the other. It is clear that prior to
the measurement the two systems have to be entangled.\index{state!entangled} 
Otherwise conditioning
on the measurement does not have any effect on the state of the second system.

\begin{figure}[t]
  \begin{center}
  \begin{psfrags}
     \psfrag{a}{$a_k$}
     \psfrag{rho}{$\rho_{a_k}$}
     \psfrag{psi}{$|\psi\rangle$}
     \epsfxsize=8in
     \epsfbox[-230 50 670 142]{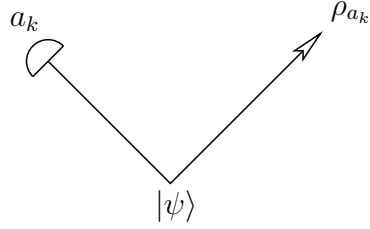}
  \end{psfrags}
  \end{center}
  \caption{A schematic representation of state preparation conditioned on a 
	measurement. One branch of the entanglement $|\psi\rangle$ is detected,
	yielding an eigenvalue $a_k$. The other branch is now in a state 
	$\rho_{a_k}$.}
  \label{fig:4.0}
\end{figure}

We can write the total state $|\psi\rangle_{12}$ prior to the measurement in
the Schmidt decomposition:\index{Schmidt decomposition}
\begin{equation}\label{4schmidt}
 |\psi\rangle_{12} = \sum_k c_k |a_k\rangle_1 |b_k\rangle_2\; ,
\end{equation}
with $\{ |a_k\rangle\}$ and $\{ |b_k\rangle\}$ orthonormal sets of states 
for system 1 and 2 respectively. These states correspond to 
eigenstates\index{eigen!-states} of observables\index{observable} $A$ and $B$ 
with sets of eigenvalues\index{eigen!-values} $\{ a_k\}$ and $\{ b_k\}$ 
respectively. We now measure the observable $A$ in system 1, yielding an 
outcome $a_k$ (see Fig.\ \ref{fig:4.0}).

We can model this measurement using so-called projection operator valued
measures, or POVM's\index{POVM}\index{projection operator valued 
measure|see{POVM}} for short. For {\em ideal} measurements, 
\index{measurement!ideal} we can describe the measurement of mode 1 as a 
projection\index{projection} $P_k = |a_k\rangle\langle a_k|$ 
operating on the state $|\psi\rangle_{12}$. When we trace out the first system 
the (normalised) state of the second system will be
\begin{equation}
 \rho_{a_k} = \frac{{\rm Tr}_1 [(P_k\otimes{\unity})|\psi\rangle_{12}
 \langle\psi|]}{{\rm Tr}_1 [(P_k\otimes{\unity})|\psi\rangle_{12}\langle\psi|]}
 = |b_k\rangle\langle b_k|\; .
\end{equation}

For non-ideal measurements\index{measurement!non-ideal} we do not use a 
projection operator, but rather 
a {\em projection operator valued measure}. In general, a POVM $E_{\nu}$
can be written as
\begin{equation}\label{povm1}
 E_{\nu} = \sum_{\mu} d_{\mu\nu} {\mathcal{P}}_{\mu} \geq 0\; ,
\end{equation}
where the ${\mathcal{P}}_{\mu}$'s form a set\footnote{This set is possibly 
over-complete, hence the difference in notation from $P_k$.} of projection 
operators $\{ |\mu\rangle\langle\mu|\}_{\mu}$. We also require a completeness 
relation
\begin{equation}
 \sum_{\nu} E_{\nu} = \unity\; .
\end{equation}
a more general definition of POVM's is given by (see appendix \ref{app:povm}):
\begin{eqnarray}\label{4povm}
 E_{\nu} &=& \sum_{\mu} d_{\mu\nu} {\mathcal{P}}_{\mu} = \sum_{\mu} d_{\mu\nu} 
 |\mu\rangle\langle\mu| = \sum_{\mu} d_{\mu\nu} |\mu\rangle\langle\nu
 |\nu\rangle\langle\mu| \cr 
 &=& \sum_{\mu} \left( u^*_{\mu\nu}|\mu\rangle\langle\nu| \right) 
 \left( u_{\mu\nu}|\nu\rangle\langle\mu| \right) = \sum_{\mu} 
 {\mathcal{A}}_{\mu\nu}^{\dagger} {\mathcal{A}}_{\mu\nu}\; .
\end{eqnarray}
The operator ${\mathcal{A}}_{\mu\nu}$ is generally not unique. These POVM's 
are used to model non-ideal measurements.

As mentioned before, a measurement outcome $a_k$ in mode 1 gives rise to 
an outgoing state $\rho_{a_k}$ in mode 2. We cannot describe a non-ideal
measurement\index{measurement!non-ideal} with the projection\index{projection} 
$P_k = |a_k\rangle\langle a_k|$. Instead, we have a POVM\index{POVM} $E_k$ 
(corresponding to the outcome $a_k$), which reduces to $P_k$ in the case of 
an ideal measurement.\index{measurement!ideal} Let $\rho_{12} = 
|\psi\rangle_{12}\langle\psi|$, the entangled state\index{state!entangled} 
prior to the measurement. The outgoing state in mode $b$ will then be 
\begin{equation}\label{4rhoout}
 \rho_{a_k} = \frac{{\rm Tr}_1[(E_k\otimes{\unity})\rho_{12}]}{{\rm 
 Tr}[(E_k\otimes{\unity})\rho_{12}]}\; ,
\end{equation}
where the total trace over both systems in the denominator gives the proper
normalisation.

If we had an ideal detector (corresponding to $E_k = |a_k\rangle\langle a_k|$),
the outgoing state would be $\rho_{a_k} = |b_k\rangle\langle b_k|$. However, 
with the general POVM $E_k$, this will not be the case. The resulting state 
will be different. In order to quantify the reliability of a state preparation 
process I introduce the {\em confidence} of a process.
\begin{description}
 \item[Definition:] The {\em confidence}\index{confidence} in the preparation 
	of a particular state is given by the fidelity\index{fidelity} of the 
	preparation process.
\end{description}
That means that using Eqs.\ (\ref{4schmidt}) and (\ref{4rhoout}) the 
confidence $C$ is given by 
\begin{equation}\label{conf1}
 C = \frac{{\rm Tr}[(E_k\otimes|b_k\rangle\langle b_k|)\rho_{12}]}{{\rm 
 Tr}[(E_k\otimes{\unity})\rho_{12}]} = \frac{|c_k|^2 \langle a_k| E_k 
 |a_k\rangle}{\sum_l |c_l|^2 \langle a_l| E_k |a_l\rangle}\; ,
\end{equation}
where the $|c_l|^2$ are the diagonal elements of the density matrix. The 
confidence $C$ can be interpreted as the probability of obtaining outcome 
$a_k$ from the `branch' containing $|a_k\rangle$ in Eq.\ (\ref{4schmidt}) 
divided by the unconditional probability of obtaining outcome $a_k$. We will 
also call this the `confidence of state 
preparation'.\index{confidence!of state preparation}

This interpretation suggests that there does not need to be a 
second system to give the idea of confidence meaning. Suppose, for instance, 
that we have an `electron factory' which produces electrons with random spin. 
A Stern-Gerlach apparatus\index{Stern-Gerlach apparatus} in the path of 
such an electron will make a spin 
measurement along a certain direction {\bf r}. Suppose we find that the 
electron has spin `up' along {\bf r}. Before this measurement the electron was 
in a state of random spin ($\rho_{\rm in}=\frac{1}{2}{|\uparrow\rangle\langle
\uparrow|} +\frac{1}{2}{|\downarrow\rangle\langle\downarrow|}$), and after the 
measurement the electron is in the `spin up' state ($\rho_{\rm out}=
|\uparrow\rangle\langle\uparrow|$). The state of the electron has {\em 
collapsed} into the `spin up' state. I will now investigate how we can define 
the confidence of the detection of a single system.

Formally, we can model state collapse\index{state!collapse} by means of the 
super-operator\index{operator!super-} $\hat{\mathcal{F}}_{a_k}$, where $a_k$ 
is again the outcome of the measurement of observable\index{observable} 
$A$ (`spin up' in the above example). In general, a 
super-operator yields a (non-normalised) mapping $\rho\rightarrow
\hat{\mathcal{F}}_{\mu}(\rho)$. In the POVM representation used above (see 
Eq.\ (\ref{4povm})) we can write this as
\begin{equation}\label{family}
 \hat{\mathcal{F}}_{\mu} : \rho ~\longrightarrow~ \sum_{\nu} 
 {\mathcal{A}}_{\mu\nu}\,\rho\, {\mathcal{A}}_{\mu\nu}^{\dagger}\; .
\end{equation}
When the eigenstate corresponding to $a_k$ is given by $|a_k\rangle$, we can
define the confidence of this measurement\index{confidence!of measurement} as 
\begin{equation}
 C_{\rm m} = \frac{\langle a_k|\hat{\mathcal{F}}_{a_k}(\rho)|a_k\rangle}{{\rm 
 Tr}[\hat{\mathcal{F}}_{a_k}(\rho)]} = \frac{{\rm Tr}[\hat{\mathcal{F}}_{a_k}
 (\rho)|a_k\rangle\langle a_k|]}{{\rm Tr}[\hat{\mathcal{F}}_{a_k}(\rho)]}\; ,
\end{equation}
with ${\rm Tr}[\hat{\mathcal{F}}_{a_k}(\rho)]$ the proper normalisation. 
However, this expression depends strongly on the details of the family of 
operators ${\mathcal{A}}_{\mu\nu}$. This is a more complicated generalisation 
than the POVM's\index{POVM} $E_k$. The confidence of state preparation, 
\index{confidence!of state preparation}on the other hand, 
is a function of the POVM $E_k$. Furthermore, $C_{\rm m}$ will in general {\em 
not} be equal to the confidence of state preparation derived in Eq.\ 
(\ref{conf1}).

In conclusion, there are two distinct versions of the confidence: the 
confidence of {\em measurement} and the confidence of {\em state preparation}.
Later in this chapter I will use the concept of the confidence to make a
quantitative comparison between different detection devices. This suggests 
that we need to calculate the confidence of measurement with all its 
difficulties. One way to circumvent this problem is to calculate the the 
confidence of state preparation using a fixed state. Instead of concentrating 
on the state preparation\index{state preparation} process we now choose a 
standard input state and calculate the confidence for different types of 
measurement devices. One such choice might be the maximally entangled 
state\index{state!maximally entangled}
\begin{equation}
 |\Psi\rangle_{12} = \frac{1}{\sqrt{N}} \sum_{k=0}^{N-1} |a_k,a_k\rangle\; .
\end{equation}
When $N\rightarrow\infty$, this is perhaps not the ideal choice and another 
state may be preferred. For any choice, the confidence offers a quantitative 
measure of performance for different types of measurement devices. 

\section{Optical detection devices}\index{detection!devices!optical}

Having set the stage for state preparation conditioned on measurement outcomes,
I will now restrict the remainder of this chapter to optical implementations.
Let's consider the measurement of optical Fock states\index{state!Fock} using 
photo-detectors.\index{detector!photo-} In order to classify different types 
of detectors I use the following terminology: a detector is said to have a 
{\em single-photon sensitivity}\index{detector!single-photon!sensitivity} when 
it is sensitive enough to detect a single-photon wave-packet. When a detector 
can distinguish between $n$- and $(n+1)$-photon wave-packets, it is said to 
have a {\em single-photon resolution}.\index{detector!single-photon!resolution}
\label{sps}

Real detectors\index{detector!real} have a variety of characteristics. 
Most common detectors do not 
have single-photon resolution, although they can distinguish between a few and 
many photons. When small photon numbers are detected, however, these are  
single-photon sensitivity detectors to a good approximation. There are also 
single-photon resolution detectors \cite{kim99,takeuchi99}. Currently, these
detectors require demanding operating conditions.

\begin{figure}[t]
  \begin{center}
  \begin{psfrags}
     \psfrag{N}{$N$}
     \psfrag{N2}{$N$}
     \psfrag{p}{\!\!$N$-port}
     \psfrag{<}{$\Biggl\{$}
     \psfrag{>}{$\Biggr\}$}
     \epsfxsize=8in
     \epsfbox[-270 40 830 90]{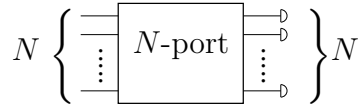}
  \end{psfrags}
  \end{center}
  \caption{An $N$-port with unit-efficiency, non-resolving detectors. The $N$ 
	incoming modes are unitarily transformed into $N$ output modes. The 
	$N$-ports considered here consist of mirrors and beam-splitters and do 
	not mix creation operators with annihilation operators.}
  \label{fig:4.1}
\end{figure}

When we need single-photon resolution\index{resolution!single-photon} but do 
not have the resources to employ single-photon resolution 
detectors,\index{detector!single-photon!resolution} we can use a so-called 
{\em detector cascade}\index{detector!cascade} \cite{song90}. 
In a detector cascade an incoming mode (populated by 
a number of photons) is split into $N$ output modes with equal amplitude which 
are all detected with single-photon sensitivity detectors. The idea is to 
choose the number of output modes large enough, so that the probability that 
two photons enter the same detector becomes small. In general, an optical 
setup which transforms $N$ incoming modes into $N$ outgoing modes is called an 
$N$-port (see Fig.\ \ref{fig:4.1}) \cite{reck94}. A detector cascade is a 
symmetric $N$-port\index{symmetric $N$-port} with detectors at the outgoing 
modes and vacuum states in 
all input modes except the first mode. In the next section I will study the
statistics of symmetric $N$-ports, but first I need to elaborate on the types
of errors which occur in detectors.

There are two sources of errors for a detector: it might fail to detect a 
photon, or it might give a signal although there wasn't actually a photon 
present. The former may be characterised as a `detector 
loss'\index{detector!loss} and the latter as a `dark count'.\index{dark counts}
Here, the emphasis will be on detector losses. Later I will give a model for 
incorporating dark counts in a realistic detector\index{detector!real} model. 
In some experiments (like the Innsbruck teleportation experiment 
\cite{bouwmeester97}) the detectors operate within short gated time intervals. 
This greatly reduces the effect of dark counts.

Detector losses are not so easily dismissed. Every photon entering a detector 
has a certain probability of triggering it. This probability is called the
{\em efficiency}\index{detector!efficiency} of the detector. For the purposes 
of brevity, when a detector is perfectly efficient, we will call it a {\em 
unit-efficiency}\index{efficiency!unit-} detector. When it has some lower 
efficiency, we speak of a {\em finite-efficiency}\index{efficiency!finite-} 
detector. Here, I study detector cascading with unit-efficiency detectors, as 
well
as cascading with finite-efficiency detectors \cite{yuen80}. I am interested 
in the case where cascading distinguishes between photon-number states 
$|k\rangle$ and $|k'\rangle$ with $k\simeq k'$. 

\section{{\it N}-ports}\label{sec:nports}

In this section I treat the properties of detector cascades, or symmetric 
$N$-ports with single-photon sensitivity 
detectors\index{detector!single-photon!sensitivity} in the outgoing modes. 
Symmetric $N$-ports yield a (unitary) transformation $U$ of the spatial field
modes $a_k$, with $j,k=1,\ldots,N$:
\begin{equation}\label{trans1}
 \hat{b}_k \rightarrow \sum_{j=1}^N U_{jk} \hat{a}_j
 \qquad\mbox{and}\qquad
 \hat{b}^{\dagger}_k \rightarrow \sum_{j=1}^N U_{jk}^*\hat{a}^{\dagger}_j\; ,
\end{equation} 
where the incoming modes of the $N$-port are denoted by $a_j$ and the outgoing 
modes by $b_j$. Here, $\hat{a}^{\dagger}_j$ and $\hat{a}_j$ are the respective 
creation and annihilation operators of mode $a_j$. Similarly for mode $b_k$.
The unitary matrix $U$ can be chosen to be
\begin{equation}\label{trans2}
  U_{jk} = \frac{1}{\sqrt{N}}\exp[2\pi i (j-1)(k-1)/N]
\end{equation} 
without loss of generality up to an overall phase-factor. Paul {\em et al}.\ 
have studied such devices in the context of tomography\index{tomography} and 
homodyne detection \cite{smithey93,dariano94,paul96}. 

Here, I study $N$-ports in the context of optical state preparation, where 
only one copy of a state is given, instead of an ensemble. I will use
the concept of the {\em confidence},\index{confidence} introduced in section 
{\ref{confidence}}.

\subsection{Statistics of {\it N}-ports}

\index{cascade|see{detector cascade}}
Suppose we have a detector cascade,\index{detector!cascade} consisting of a 
symmetric $N$-port\index{symmetric $N$-port} with single-photon sensitivity 
detectors\index{detector!single-photon!sensitivity} in the outgoing modes. 
According to 
Eqs.\ (\ref{trans1}) and (\ref{trans2}) incoming photons will be redistributed 
over the outgoing modes. In this section I study the photon statistics
\index{photon!-statistics} of this 
device. In particular, I study the case where $k$ photons enter a single input 
mode of the $N$-port, with vacuum in all other input modes. This device (i.e., 
the detector cascade) will act as a sub-ideal single-photon resolution 
detector since there is a probability that some of the photons end up in the 
same outgoing mode, thus triggering the same detector.

To quantify the single-photon resolution\index{resolution!single-photon} of 
the cascade we use the confidence given by Eq.\ (\ref{conf1}). Suppose we have 
two spatially separated entangled modes\index{mode!entangled} of the 
electro-magnetic field\index{field!electro-magnetic} $a$ and $b$ with number 
states\index{state!number} $|m\rangle$
in $a$ and some other orthogonal states $|\phi_m\rangle$ in $b$:
\begin{equation}
 |\Psi\rangle = \sum_m \gamma_m |m\rangle_a |\phi_m\rangle_b\; ,
\end{equation}
where the second mode is used only to give the confidence an operational 
meaning. The POVM\index{POVM} governing the detection can be written as 
$E_k = \sum_m 
p_N (k|m) |m\rangle\langle m|$, since we assume that the photons are not 
lost in the $N$-port. In this expression $p_N(k|m)$ is the probability that 
$m$ incoming photons cause a $k$-fold detector 
coincidence\index{detector!coincidence} in the $N$-port 
cascade. The confidence can then be written as
\begin{equation}\label{conf2}
 C = \frac{|\gamma_k|^2 \langle k|E_k|k\rangle}{\sum_l|\gamma_m|^2 \langle m|
 E_k|m\rangle} =\frac{|\gamma_k|^2 p_N (k|k)}{\sum_m|\gamma_m|^2 p_N (k|m)}\; .
\end{equation}
In order to find the confidence,\index{confidence} I therefore first have to 
calculate the probability distribution\index{distribution!probability} $p_N$. 
This will allow us to compare single-photon resolution 
detectors\index{detector!single-photon!resolution} with various arrangements 
($N$-ports) of single-photon sensitivity 
detectors.\index{detector!single-photon!sensitivity}

Suppose $k$ photons enter the first input mode and all other input modes are
in the vacuum state.\index{state!vacuum} The density matrix of the pure input 
state $\rho_0 = |k\rangle\langle k|$ will be transformed according to 
$\rho = U_N \rho_0 U^{\dagger}_N$ with $U_N$ the unitary transformation 
associated with the symmetric $N$-port.\index{symmetric $N$-port} Let 
$\vec{n}$ be the $N$-tuple of the photon number in 
every outgoing mode: $\vec{n}= (n_1,n_2,\ldots, n_N)$. The probability of 
finding $n_1$ photons in mode 1 and $n_2$ photons in mode 2, et cetera, is 
given by $p_{\vec{n}} = \langle\vec{n}|\rho|\vec{n}\rangle$. Using the 
$N$-port transformation this probability yields 
\begin{equation}
 p_{\vec{n}} = \langle\vec{n}|U_N\rho_0 U^{\dagger}_N|\vec{n}\rangle = 
 |\langle\vec{n}|U_N|\vec{k}\rangle|^2\; ,
\end{equation}
where $\vec{k}=(k,0,\ldots, 0)$, since only the first input mode inhabits
photons and the rest are vacuum. From Refs.\ \cite{dodonov94} and 
\cite{dodonov96} we find that this can be rewritten as 
\begin{equation}\label{prob}
 p_{\vec{n}} = \frac{\left[ H^R_{\vec{k}\vec{n}}(0)\right]^2}{n_1!
  \cdots n_N! k!}\; .
\end{equation}
Here, $H^R_{\vec{k}\vec{n}}(\vec{x})$ is a so-called multi-dimensional 
Hermite polynomial 
(MDHP)\index{MDHP}\index{Hermite polynomial!multi-dimensional|see{MDHP}} 
\cite{dodonov84} (this is a non-trivial result; see 
appendix \ref{app:mdhp} for a comprehensive treatment of multi-dimensional 
Hermite polynomials) and the matrix $R$ is defined as
\begin{equation}\label{matrix}
 R \equiv
 \begin{pmatrix}
  0 & -U^{\dagger} \cr -U^{\dagger} & 0
 \end{pmatrix}\; .
\end{equation}
For our present purposes it is convenient to characterise the $N$-port by 
its transformation of the field modes given by Eqs.\ (\ref{trans1}) and 
(\ref{trans2}). I therefore concentrate on $U$ rather than $U_N$.

Since there is a one-to-one correspondence between the $N$-port ($U$) and the 
matrix $R$, knowledge of $U$ is sufficient to calculate the confidence 
of a given event using the $N$-port. The MDHP for $N$ input modes with $k$ 
photons in the first mode and zero in the others (giving an $N$-tuple 
$\vec{k}$) and $N$ output modes $\vec{n}$ is given by
\begin{equation}
 H_{\vec{k}\vec{n}}^R (\vec{x}) = (-1)^{2k}\; e^{\frac{1}{2}
 \vec{x}\,R\,\vec{x}^T}\; \nabla_{\vec{k}\vec{n}}^{2k}\; e^{-\frac{1}{2}\vec{x}
 \,R\,\vec{x}^T} \; ,
\end{equation}
where $\vec{x}\,R\,\vec{x}^T = \sum_{ij} x_i R_{ij} x_j$, $\vec{x} = (x_1,
\ldots,x_{2N})$ and
\begin{equation}\nonumber
 \nabla_{\vec{k}\vec{n}}^{2k} \equiv \frac{\partial^{2k}}{\partial x^k_1 
  \partial x^{n_1}_{N+1} \cdots \partial x^{n_N}_{2N}}\; .
\end{equation}
The number of photons in the input mode is equal to the total number of 
photons in the output modes. The dimension of $\vec{x}$ obeys $\dim\vec{x} = 
\dim \vec{k} + \dim\vec{n} = 2N$. For example, for a two-photon input state we 
have
\begin{equation}\label{2photon}
 e^{\frac{1}{2}\vec{x}\,R\,\vec{x}^T} \frac{\partial^4}{\partial x_1^2\partial 
 x_l \partial x_k} \left. 
 e^{-\frac{1}{2}\vec{x}\,R\,\vec{x}^T} \right|_{\vec{x}=0} = 2R_{1l}R_{1k}\; .
\end{equation}

There are many different ways in which $k$ incoming photons can trigger a 
$k$-fold detector coincidence.\index{detector!coincidence} These different 
ways correspond to different 
photon distributions in the outgoing (detected) modes, and are labelled by 
$\vec{n}_r$. The probability that all $k$ photons enter a different 
detector is found by determining the $p_{\vec{n}_r}$s where every $n_i$ in 
$\vec{n}_r$ is at most one. The sum over all these $p_{\vec{n}_r}$'s is equal 
to the probability $p_N (k|k)$ of a $k$-fold coincidence in an $N$-port 
conditioned on $k$ incoming photons:
\begin{equation}\label{pkn}
 p_N(k|k) = \sum_{\vec{n}_r} p_{\vec{n}_r} = \frac{k!}{N^k}\binom{N}{k}\; .
\end{equation}

Finally, in order to find the probability of a $k$-fold detector coincidence 
conditioned on $m$ photons in the input state (with $m\geq k$) we need to sum 
all probabilities in Eq.\ (\ref{prob}) with $k$ non-zero entries in the 
$N$-tuple $\vec{n}$:
\begin{equation}\label{pkm1}
 p_N (k|m) = \sum_{\vec{n}\in{\mathcal{S}}_k }\frac{\left[ H^R_{\vec{m}
 \vec{n}}(0)\right]^2}{n_1!\cdots n_N! m!}\; ,
\end{equation}
where ${\mathcal{S}}_k$ is the set of all $\vec{n}$ with exactly $k$ non-zero
entries. 

\subsection{Realistic {\it N}-ports}{\label{realisticnports}}
\index{realistic $N$-ports}

\begin{figure}[t]
  \begin{center}
  \begin{psfrags}
     \psfrag{2n}{$2N$}	
     \psfrag{n1}{$N$}	
     \psfrag{n2}{$N$}	
     \psfrag{p}{$2N$-port}	
     \psfrag{<}{\large $\Biggl\{$}	
     \psfrag{>}{\large $\Biggr\}$}
     \psfrag{2}{$\overbrace{\phantom{xxxxx}}$}	
     \epsfxsize=8in
     \epsfbox[-220 40 800 150]{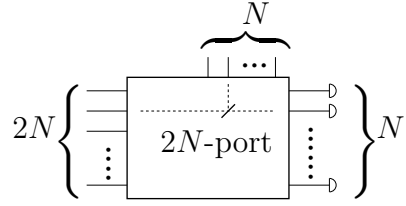}
  \end{psfrags}
  \end{center}
  \caption{A $2N$-port with $N$ modes which are detected with ideal
	detectors and $N$ undetected modes. These modes are associated with 
	the detector losses.}
  \label{fig:4.2}
\end{figure}

I now consider a symmetric $N$-port\index{symmetric $N$-port} cascade with 
finite-efficiency\index{efficiency!finite-} single-photon 
sensitivity\index{detector!single-photon!sensitivity}
detectors. Every one of the $N$ detectors has a certain loss, which means that
some photons do not trigger the detector they enter. We can model this 
situation by putting a beam-splitter with intensity transmission coefficient 
$\eta^2$ in front of the ideal detectors \cite{yuen80}. The reflected photons 
are sent into the environment and can be associated with the loss. The 
transmitted photons are detected (see Fig.\ \ref{fig:4.2}). 

Before I continue with the description of realistic detector cascades,
\index{detector!cascade!realistic} let me return to the question of the 
dark counts.\index{dark counts} In the model for detectors with
finite efficiency I assumed a beam-splitter with intensity transmission 
coefficient $\eta^2$ and vacuum in the second input mode. We can now model dark
counts by replacing this vacuum state with a thermal input state $\rho_{\rm 
th}$\index{state!thermal}
\cite{scully97}:
\begin{equation}\label{thermalstate}
 \rho_{\rm th} = \sum_{n=0}^{\infty} \int_0^{\infty} d\nu f(\nu) 
 \left[ 1-\exp\left(\frac{-\hbar\nu}{k_B T_{\rm eff}}\right) \right] 
 \exp\left(\frac{-n\hbar\nu}{k_B T_{\rm eff}}\right) |n\rangle\langle n|\; ,
\end{equation}
with $\nu$ the frequency, $f(\nu)$ the appropriate frequency distribution 
($\int_0^{\infty} d\nu f(\nu) = 1$), $k_B$ Boltzmann's constant and $T_{\rm 
eff}$ the effective temperature `seen' by the detector. However, for the 
remainder of this chapter I will assume that there are no dark counts 
($T_{\rm eff}=0$).

Let us now consider cascades\index{detector!cascade} with finite-efficiency 
detectors.\index{detector!finite-efficiency} The 
implementation of the beam-splitters responsible for the detector 
losses\index{detector!loss} 
transform our $N$-port into a $2N$-port and the unitary transformation $U$ 
of the field modes in this $N$-port now becomes a $2N\times 2N$ unitary matrix 
$U~\rightarrow~ U\otimes {\unity}_2$ (where ${\unity}_2$ is the two dimensional
unit matrix). Applying a transformation $V_{\eta}$ to implement the 
beam-splitters with transmission coefficient $\eta^2$ will give a new unitary 
transformation governing the behaviour of the $2N$-port. Although nothing 
holds us from considering detectors with different efficiencies, for 
simplicity I will assume that all detectors have the same 
efficiency\index{efficiency} $\eta^2$.
In terms of the original unitary matrix $U$ from Eq.\ (\ref{trans2}) the new 
unitary matrix $\widetilde{U}$ becomes 
\begin{equation}\nonumber
 U ~\rightarrow~ \widetilde{U} =
 \begin{pmatrix}
  \eta\; U & \sqrt{1-\eta^2}\; U \cr
  -\sqrt{1-\eta^2}\; U & \eta\; U
 \end{pmatrix}\; .
\end{equation} 
This changes the matrix $R$ of the MDHP\index{MDHP} accordingly:
\begin{equation}
  R ~\rightarrow~ \widetilde{R} =
 \begin{pmatrix}
  0 & -\widetilde{U}^{\dagger} \cr
  -\widetilde{U}^{\dagger} & 0
 \end{pmatrix}
\end{equation}
and $\widetilde{R}$ is now a $4N \times 4N$ matrix dependent on $\eta$. The 
probability of finding a $k$-fold detector 
coincidence\index{detector!coincidence} in an $N$-port cascade
with finite-efficient detectors then becomes
\begin{equation}\label{pkm}
 p_{N}(k|m) = \sum_{\vec{n}\in{\mathcal{S}}_k}\frac{\left[ H_{\vec{m}\vec{n}}^{
 \widetilde{R}}(0)\right]^2 }{n_1! \cdots n_{2N}!m!} \; ,
\end{equation}
where ${\mathcal{S}}_k$ is the set of all $\vec{n}$ with exactly $k$ non-zero
entries in the detected modes (note that I still call it an $N$-port although 
technically it is a $2N$-port). The confidence\index{confidence} of 
having a total of $k$ photons
in a $k$-fold detector coincidence is again given by Eq.\ (\ref{conf2}).
The variables of the MDHP will be a $2N$-tuple $\vec{k} = (k, 0, \ldots 
0)$. The output photon number $2N$-tuple can now be written as $\vec{n} = 
(n^d_1, n^d_2,\ldots n^d_N, n_1^u, \ldots n^u_N)$, where the superscripts $d$ 
and $u$ again denote the detected and undetected modes respectively. 
Furthermore we have $\sum_{i=1}^N n^d_i \equiv N_d$ and $\sum_{i=1}^N n^u_i 
\equiv N_u$.

Using Eq.\ (\ref{pkn}) and observing that every detected photon carries a 
factor $\eta^2$ it is quite straightforward to obtain the probability that $k$ 
photons give a $k$-fold coincidence in an efficient $N$-port 
cascade:\index{detector!cascade}
\begin{equation}
 p_{N}(k|k) = \frac{\eta^{2k} N!}{N^k (N-k)!}\; .
\end{equation}

\subsection{The single-photon resolution of {\it N}-ports}
\index{resolution!single-photon}

Having determined the probability 
distribution\index{probability!distribution|see{distribution}} 
$p_N$, I can now calculate the confidence\index{confidence} of detector 
cascading. First of all, in order to obtain a high 
confidence in the outcome of a detector cascade, the possible number of photons
should be much smaller than the number of modes in the cascade: $N\gg k$. In 
practice there is a limit to the number of detectors we can build a cascade 
with, so I only look at the lowest order: distinguishing between one and two 
photons.

I will calculate the confidence of having outgoing state $|\phi_1\rangle$ 
conditioned a single detector giving a `click' in the detector cascade when 
the input state is given by
\begin{equation}\label{confinput}
 |\Psi\rangle_{12} = \alpha |0\rangle_1 |\phi_0\rangle_2 + 
 \beta|1\rangle_1|\phi_1\rangle_2 + \gamma|2\rangle_1|\phi_2\rangle_2\; .
\end{equation}
This state corresponds, for example, to the output of a down-converter 
\index{down-converter} when we
ignore higher-order terms. The confidence is then
\begin{equation}
 C_N (1,|\Psi\rangle_{12}) = \frac{|\beta|^2 p_{N}(1|1)}{|\alpha|^2 p_N(1|0) + 
 |\beta|^2p_{N}(1|1) + |\gamma|^2 p_{N}(1|2)}\; .
\end{equation}
Eqs.\ (\ref{pkm}) and (\ref{2photon}) allow us to calculate the probabilities
of a zero-, one- and two-fold detector coincidence conditioned on one or two
incoming photons:
\begin{eqnarray}
 p_N (0|0) &=& 1 \\
 p_N (1|0) &=& 0 \\ && \cr
 p_N (0|1) &=& 1-\eta^2 \\ 
 p_N (1|1) &=& \eta^2 \\ && \cr
 p_N (0|2) &=& (1-\eta^2)^2 \\
 p_N (1|2) &=& \frac{\eta^4}{N} + 2\eta^2 (1-\eta^2) \label{fout} \\ 
 p_N (2|2) &=& \frac{N-1}{N} \eta^4\; ,
\end{eqnarray}
For example, using these probabilities, together with Eq.\ (\ref{confinput}), 
gives us an expression for the confidence\index{confidence} that a single 
detector hit was triggered by one photon ($\delta = |\gamma|^2 / |\beta|^2$):
\begin{equation}\label{2con2}
 C = \frac{N}{N + \delta[\eta^2 + 2N(1-\eta^2)]}\; ,
\end{equation}
where, for simplicity, we omitted the functional dependence of $C$ on the 
incoming state, the size of the cascade and the order of the detector 
coincidence.\index{detector!coincidence}

A close look at Eq. (\ref{fout}) shows us that $p_N (1|2)$ includes a term
which is independent of the number of modes in the $N$-port 
cascade.\index{detector!cascade} This term
takes on a maximum value of $1/2$ for $\eta^2=\frac{1}{2}$. However, the 
confidence is a monotonously increasing function of $\eta^2$. As expected, 
for small $\delta$'s the confidence $C_N (1,|\Psi\rangle)$ approaches 1.
Detector cascading thus turns a collection of single-photon sensitivity
detectors\index{detector!single-photon!sensitivity} into a device with {\em 
some} single-photon resolution.\index{detector!single-photon!resolution} In 
the next section I will give a quantitative estimation of this resolution.

\section{Comparing detection devices}

Let's return again to the schematic state preparation\index{state preparation} 
process depicted in 
figure \ref{fig:4.0}. There we had two modes, one of which was detected, giving
the prepared outgoing state in the other. I argued that different detection 
devices yield different output states, and the comparison of these states with 
the ideal case (where we used an ideal detector) led to the introduction of the
confidence of a state preparation process. Here, I will use the confidence
to make a comparison of different {\em detection devices}, rather than output
states. This can be done by choosing a fixed entangled input 
state.\index{state!entangled} The 
confidence then quantifies the performance of these detection devices.

Consider the state preparation process in the setting of quantum optics. We
have two spatial modes of the electro-magnetic 
field,\index{field!electro-magnetic} one of which is detected.
In this thesis I am mostly interested in states containing a few photons, and
the detection devices I consider therefore include single-photon sensitivity
detectors, single-photon resolution detectors and detector cascades. As an 
example, I set the task of distinguishing between one and two photons. Since
single-photon sensitivity detectors are not capable of doing this, I will
compare the performance of detector cascading with that of a single-photon
resolution detector. Let the state prior to the detection be given by 
\begin{equation}
 |\Psi\rangle = \frac{1}{\sqrt{3}} \left( |0\rangle |\phi_0\rangle + 
 |1\rangle |\phi_1\rangle + |2\rangle |\phi_2\rangle \right)\; .
\end{equation}
This state is maximally entangled\index{state!maximally entangled} and will 
serve as our `benchmark' state. It
corresponds to the choice $\delta=1$ in the previous section. Suppose the 
outgoing state conditioned on a `one-photon' indication in the detection 
device is $\rho$. The confidence is then again given by $C = \langle\phi_1|
\rho|\phi_1\rangle$.

First, consider the single-photon resolution 
detector\index{detector!single-photon!resolution} described in Refs.\ 
\cite{kim99,takeuchi99}. This detector can distinguish between one and two
photons very well, but it does suffer from detector losses\index{detector!loss}
(the efficiency\index{efficiency} was
determined at 88\%). That means that a two-photon state can be identified as 
a single-photon state when one photon is lost. The confidence\index{confidence}
of this detector is therefore not perfect.

In order to model the finite efficiency\index{efficiency!finite-} of the 
single-photon resolution 
detector we employ the beam-splitter model from section 
{\ref{realisticnports}}. We write the input state as
\begin{equation}
 |\Psi\rangle = \frac{1}{\sqrt{3}} \left( |0\rangle|\phi_0\rangle + 
 \hat{a}^{\dagger} |0\rangle |\phi_1\rangle + \frac{(\hat{a}^{\dagger})^2}
 {\sqrt{2}} |0\rangle |\phi_2\rangle \right)\; .
\end{equation}
When we make the substitution $\hat{a}^{\dagger}\rightarrow \eta 
\hat{b}^{\dagger} + \sqrt{1-\eta^2}\hat{c}^{\dagger}$ we obtain a state $\rho$.
The outgoing density matrix conditioned on a single photon in mode $b$ is then
\begin{equation}
 \rho_{\rm out} = \frac{{\rm Tr}_{bc}[(|1\rangle_b\langle 1|\otimes{\unity}_c)
 \rho]}{{\rm Tr}[(|1\rangle_b\langle 1|\otimes{\unity}_c)\rho]} = 
 \frac{\eta^2}{4-3\eta^2} |\phi_1\rangle\langle\phi_1| +  \frac{4(1-\eta^2)}
 {4-3\eta^2} |\phi_2\rangle\langle\phi_2|\; .
\end{equation}
With $\eta^2=0.88$ the confidence of the single-photon resolution detector is 
easily calculated to be $C=0.65$.

\begin{figure}[t]
  \begin{center}
  \begin{psfrags}
     \psfrag{0.2}{\scriptsize 0.2}	
     \psfrag{0.4}{\scriptsize 0.4}	
     \psfrag{0.6}{\scriptsize 0.6}	
     \psfrag{0.8}{\scriptsize 0.8}	
     \psfrag{1}{\scriptsize 1}	
     \psfrag{0}{\scriptsize 0}	
     \psfrag{e}{$\!\!\eta^2$}
     \psfrag{c}{$\!\!\!\!C$}
     \psfrag{a}{\small $N=1$}
     \psfrag{b}{\small $N=4$}
     \psfrag{f}{\small $N=16$}
     \psfrag{g}{\small $N=\infty$}
     \psfrag{h}{$\!\!\!\!\!\!\!\!\!\!\!\!\! |\Psi\rangle = 
	(|0\rangle|\phi_0\rangle + |1\rangle|\phi_1\rangle + 
	|2\rangle|\phi_2\rangle)/\sqrt{3}$}
     \epsfxsize=8in
     \epsfbox[-100 90 600 250]{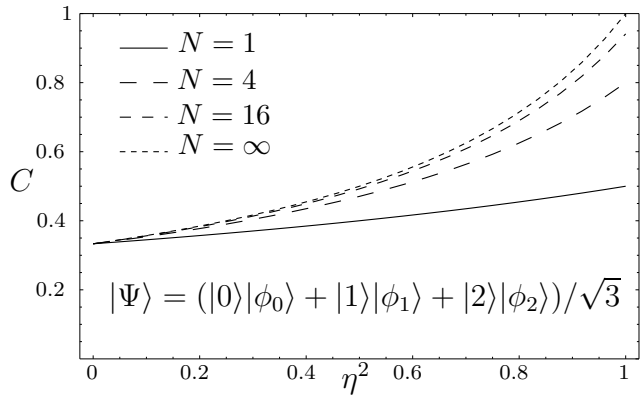}
  \end{psfrags}
  \end{center}
  \caption{The single-photon confidence $C$ [Eq.\ (\ref{2con2})] as 
	a function of the detector efficiency $\eta^2$. The solid line 
	corresponds to a single-detector cascade (no cascading: $N=1$), the 
	dashed lines correspond to $N=4$, $N=16$ and $N=\infty$ in ascending 
	order. We consider a maximally entangled input state $|\Psi\rangle = 
	(|0\rangle|\phi_0\rangle + |1\rangle|\phi_1\rangle + 
	|2\rangle|\phi_2\rangle)/\sqrt{3}$ to serve as a benchmark.}
  \label{fig:4.3}
\end{figure}

Now we consider a detector cascade\index{detector!cascade} with single-photon 
sensitivity detectors.\index{detector!single-photon!sensitivity} 
In Fig.\ \ref{fig:4.3} the confidence of a single-photon detection with 
$N$-port cascades is depicted. When the cascade consists of four detectors 
($N=4$) it can be easily calculated from Eq.\ (\ref{2con2}) that the detectors 
need an efficiency of 0.84 to achieve a 0.65 confidence. In the case of 
infinite cascading ($N=\infty$) the single-photon confidence of 0.65 is met 
only if the efficiency is roughly 0.73. This puts a severe practical limit on 
the efficiency of the single-photon sensitivity detectors in the cascade. 
\index{detector!cascade}

Detector cascading would be practically useful if a reasonably small 
number of finite-efficiency detectors\index{detector!finite-efficiency} yields 
a high confidence.\index{confidence} In particular 
when cascading is viewed as an economical alternative to a detector with 
single-photon resolution\index{detector!single-photon!resolution} the number 
of detectors in the cascade should be 
small. Additionally, cascading should yield a confidence similar to 
single-photon resolution detectors. Unfortunately, as a practical application, 
detector cascading only appears to yield a modest boost in resolution, unless 
the detectors with single-photon 
sensitivity\index{detector!single-photon!sensitivity} have a very high 
efficiency. Real 
single-photon resolution detectors are therefore superior to detector cascading
with currently available detectors, notwithstanding the demanding operating 
conditions. 

\section{Summary}

In this chapter I have studied the use of detection devices in entanglement
based state preparation.\index{state preparation!entanglement-based} In 
particular I considered optical devices such as
single-photon sensitivity detectors, single-photon resolution detectors and
detector cascades. 

Detector cascading has generally been regarded as a good way to enhance 
single-photon resolution and consequently the fidelity of a state preparation
process \cite{song90}. However, an extensive theory for the use of these
detection devices has not been available so far. The statistics of $N$-ports
have been considered in the context of tomography\index{tomography} 
\cite{paul96}, which relies 
on the availability of a large number of copies of a quantum state. In
state preparation, however, we perform measurements on single systems, and we 
therefore need precise bounds on the distinguishability of these measurements.

To this end, I introduced the confidence of preparation, which can also be 
used to quantify the performance of a detection device. Thus, I compared a
single-photon resolution detector with a cascade of single-photon sensitivity
detectors and found that cascading {\em does not give a practical advantage} 
over detectors with single-photon resolution.

\newpage
\thispagestyle{empty}

\chapter{Mathematical Description of Optical Circuits}\label{chap5}

In the previous chapters, I have discussed state preparation in quantum 
optics with realistic detectors. I now ask what the general outgoing state 
of an optical circuit is. 

Suppose we have an optical circuit,\index{optical!circuit|(} that is, a 
collection of connected optical 
components. It is usually important to know what the outgoing state of this 
circuit is. In this chapter, I give a description of the outgoing state for a
special class of optical circuits. First, in section {\ref{circuit}}, I define
this class of optical circuits and show that they can be described by 
so-called multi-dimensional Hermite polynomials.\index{MDHP|(} In section 
\ref{example}, I 
give an example of this description. Section \ref{hp} discusses the Hermite 
polynomials, and finally, in section \ref{sec:det}, I briefly consider the 
effect of imperfect detectors\index{detector!imperfect} on the outgoing state. 
This chapter is based on Kok and Braunstein \cite{kok00g}.

\section{The Optical Circuit}\label{circuit}

What do we mean by an optical circuit? We can think of a black box with 
incoming and outgoing modes of the electro-magnetic field.
\index{field!electro-magnetic} The black
box transforms a state of the incoming modes into a (different) state of the
outgoing modes. The black box is what we call an {\em optical circuit}. We 
can now take a more detailed look inside the black box. We will consider three 
types of components.

First, the modes might be mixed by beam-splitters,\index{beam-splitter} or 
they may pick up a relative phase shift\index{phase shift} or polarisation 
rotation.\index{polarisation rotator} These operations all belong  
to a class of optical components which preserve the photon number. We 
call them {\em passive} optical 
components.\index{optical!passive --- components} 

Secondly, we may find optical components such as lasers,\index{laser} 
down-converters\index{down-converter} or (optical) parametric amplifiers
\index{parametric!amplifier} in the black box. These components can be 
viewed as photon sources, since they do not leave the photon number invariant. 
We will call these components {\em active} optical 
components.\index{optical!active --- components}  

And finally, the box will generally include measurement devices, the outcomes
of which may modify optical components on the remaining modes depending on the 
detection outcomes. This is called {\em feed-forward} 
detection.\index{detection!feed-forward} We can 
immediately simplify optical circuits using feed-forward detection, by 
considering the family of fixed circuits
corresponding to the set of measurement outcomes (see also Ref.~\cite{kok00b}).
In addition, we can postpone the measurement to the end, where all the optical
components have `acted' on the modes. 

These three component types have their own characteristic mathematical 
description. A passive component yields a unitary evolution $U_i$, which can 
be written as 
\begin{equation}
 U_i = \exp\left( -i\kappa \sum_{jk} c_{jk} \hat{a}_j \hat{a}_k^{\dagger}
 - {\rm H.c.}\right)\; ,
\end{equation}
where H.c.\ denotes the Hermitian conjugate. This unitary evolution commutes 
with the total number operator $\hat{n}=\sum_j \hat{a}_j^{\dagger}\hat{a}_j$.

Active components\index{optical!active --- components}  also correspond to 
unitary transformations, which can be written as $\exp(-itH_I^{(j)})$. Here 
$H_I^{(j)}$ is the interaction Hamiltonian\index{interaction Hamiltonian} 
associated with the $j^{\rm th}$ active component in a sequence. 
This Hamiltonian does not necessarily commute with the total number operator. 
\index{operator!number}
To make a typographical distinction between passive and active components, we 
denote the $i^{\rm th}$ passive component by $U_i$, and the $j^{\rm th}$ 
active component by its evolution in terms of the interaction Hamiltonian.
The mathematical description of the (ideal) measurement will correspond to 
taking the inner product of the outgoing state prior to the measurement with 
the eigenstate\index{eigen!-states} corresponding to the measurement. 

\subsection{The state prior to detection}

Now that we have the components of an optical circuit of $N$ modes, we have to 
combine them into an actual circuit. Mathematically, this corresponds to 
applying the unitary evolutions of
the successive components to the input state. Let $|\psi_{\rm in}\rangle$ be 
the input state and $|\psi_{\rm prior}\rangle$ the output state {\em prior} to 
the measurement. We then have (with $K>0$ some integer)
\begin{equation}\label{stateprior}
 |\psi_{\rm prior}\rangle = U_K e^{-it H_I^{(K)}} \ldots
 U_1 e^{-it H_I^{(1)}} U_0 |\psi_{\rm in}\rangle\; ,
\end{equation}
where it should be noted that $U_i$ might be the identity 
operator\index{operator!identity} $\unity$
or a product of unitary transformations corresponding to passive 
components:\index{optical!passive --- components}  
\begin{equation}
 U_i = \prod_k U_{i,k}\; .
\end{equation}
When the (multi-mode) eigenstate\index{eigen!-states!multi-mode} corresponding 
to the measurement outcome for 
a limited set of modes labelled $1,\ldots,M$ with $M<N$ is given by 
$|\gamma\rangle=|n_1,n_2,\ldots,n_M\rangle$ with $M$ the number of detected 
modes out of a total of $N$ modes, and $n_i$ the number of photons found in 
mode $i$, the state leaving the optical circuit in the undetected modes is 
given by 
\begin{equation}
 |\psi_{\rm out}\rangle_{M+1,\ldots,N} ~=~ _{1,\ldots,M}\!\langle\gamma|
 \psi_{\rm prior}\rangle_{1,\ldots,N}\; .
\end{equation}

In this chapter, I study the outgoing states $|\psi_{\rm out}\rangle$ for a 
special class of optical circuits. First, I assume that the input state is 
the vacuum\index{vacuum} on all modes. Thus, I effectively study optical 
circuits as state preparation devices.\index{state preparation!device} 
Secondly, our class of optical circuits include all 
possible passive components, but {\em only} active components with quadratic
interaction Hamiltonians:
\begin{equation}
 H_I^{(j)} = \sum_{kl} \hat{a}_k^{\dagger} R^{(j)}_{kl} \hat{a}_l^{\dagger} +
 \sum_{kl} \hat{a}_k R^{(j)*}_{kl} \hat{a}_l \; ,	
\end{equation}
where $R^{(j)}$ is some complex symmetric matrix. This matrix determines the 
behaviour of the $j^{\rm th}$ active component, which can be any combination 
of down-converters\index{down-converter} and squeezers.\index{squeezer} 
Finally, we consider ideal photo-detection,\index{photo-detector} 
where the eigenstate corresponding to the measurement outcome can be written as
$|\gamma\rangle = |n_1,\ldots,n_M\rangle$.

The class of optical circuits I consider here is not the most general class, 
but it still includes important experiments like quantum 
teleportation\index{quantum teleportation} \cite{bouwmeester97}, entanglement 
swapping\index{entanglement!swapping} \cite{pan98} and the demonstration
of GHZ correlations\index{GHZ correlations} \cite{bouwmeester99b}. In section 
{\ref{example}}, I show how teleportation can be modelled using the methods 
presented here.

The state $|\psi\rangle$ prior to the photo-detection can be written 
in terms of the components of the optical circuit as
\begin{equation}\label{incoming}
 |\psi\rangle = U_K e^{-it {\mathcal{H}}_I^{(K)}} \ldots U_1 
 e^{-it H_I^{(1)}} |0\rangle\; .
\end{equation}
The creation\index{operator!creation} and annihilation 
operators\index{operator!annihilation} $\hat{a}_i^{\dagger}$ and $\hat{a}_i$ 
for mode $i$ satisfy the standard canonical commutation 
relations\index{commutation relation}
\begin{equation}\label{comm}
 [\hat{a}_i,\hat{a}_j^{\dagger}] = \delta_{ij} \quad\mbox{and}\quad 
 [\hat{a}_i,\hat{a}_j] = [\hat{a}_i^{\dagger},\hat{a}_j^{\dagger}] = 0\; ,
\end{equation}
with $i,j=1\ldots N$. 

For any unitary evolution $U$, we have the relation
\begin{equation}
 U e^{R} U^{\dagger} = \sum_{l=0}^{\infty} \frac{U R^l U^{\dagger}}{l!} =
 \sum_{l=0}^{\infty} \frac{(U R U^{\dagger})^l}{l!} = e^{U R U^{\dagger}}\; .
\end{equation}
Furthermore, if $U$ is due to a collection of only passive components, such an 
evolution leaves the vacuum invariant:\index{vacuum!invariance} $U|0\rangle = 
|0\rangle$. Using these 
two properties it can be shown that Eq.\ (\ref{incoming}) can be written as 
\begin{equation}\label{in2}
 |\psi\rangle = \exp\left[ -\frac{1}{2} \sum_{i,j=1}^N \hat{a}^{\dagger}_i 
 A^{\dagger}_{ij} \hat{a}^{\dagger}_j + \frac{1}{2} \sum_{i,j=1}^N \hat{a}_i 
 A_{ij} \hat{a}_j + \frac{1}{2} \sum_{i,j=1}^N \hat{a}_i^{\dagger} B_{ij}
 \hat{a}_j \right] |0\rangle\; ,
\end{equation}
where $A$ is some complex symmetric matrix and $B=B^{\dagger}$. I will now 
simplify this expression by normal-ordering\index{normal ordering|(} this 
evolution.

Define $(\vec{a},A\vec{a}) \equiv \sum_{ij} \hat{a}_i A_{ij} \hat{a}_j$. As 
shown by Braunstein \cite{braunstein99}, we can rewrite Eq.\ (\ref{in2}) using 
two passive unitary transformations $U$ and $V$ as:
\begin{equation}\label{lambda}
 |\psi\rangle = U e^{-\frac{1}{2}(\vec{a}^{\dagger},\Lambda^{\dagger}
 \vec{a}^{\dagger}) + \frac{1}{2}(\vec{a},\Lambda\vec{a})} V^T |0\rangle\; ,
\end{equation}
where $\Lambda$ is a diagonal matrix with real non-negative eigenvalues 
$\lambda_i$.
This means that, starting from vacuum, the class of optical circuits I
consider here is equivalent to a set of single-mode 
squeezers,\index{squeezer!single-mode} followed by a 
passive unitary transformation $U$ and photo-detection. Since $\Lambda$ is 
diagonal, we can write Eq.~(\ref{lambda}) as 
\begin{equation}
 |\psi\rangle = U \left( \prod_{i=1}^N \exp\left[-\frac{\lambda^*_i}{2} 
 (\hat{a}_i^{\dagger})^2 + \frac{\lambda_i}{2} \hat{a}_i^2\right]\right) 
 |0\rangle\; .
\end{equation}
We can now determine the normal ordering of every factor $\exp[-
\frac{\lambda^*_i}{2} (\hat{a}_i^{\dagger})^2 + \frac{\lambda_i}{2} 
\hat{a}_i^2]$ separately. Note that the operators $\hat{a}_i^2$, 
$(\hat{a}_i^{\dagger})^2$ and $2\hat{a}_i^{\dagger}\hat{a}_i + 1$ generate an 
su(1,1) algebra.\index{algebra!$su(1,1)$} According to Refs.\ 
\cite{yuen76,fisher84,truax85}, this may be normal-ordered as
\begin{equation}
 e^{-\frac{\lambda^*_i}{2} (\hat{a}_i^{\dagger})^2 + \frac{\lambda_i}{2} 
 \hat{a}_i^2} =
 e^{-\hat{\lambda_i}^* \tanh|\frac{\lambda_i}{2}| (\hat{a}_i^{\dagger})^2}
 e^{-2 \ln (\cosh|\frac{\lambda_i}{2}|) \hat{a}_i^{\dagger} \hat{a}_i}
 e^{\hat{\lambda_i} \tanh|\frac{\lambda_i}{2}| \hat{a}_i^2}\; ,
\end{equation}
where $\hat{\lambda_i} = \lambda_i / |\lambda_i|$. In general, when $L_{\pm}$
and $L_0$ are generators of an $su(1,1)$ algebra (i.e., when $A$ is unitary) we
find \cite{truax85}
\begin{equation}
 e^{-\frac{1}{2}(\tau L_+ \tau^* L_-)} = e^{-\hat{\tau}\tanh|\tau| L_+}
 e^{-2 \ln (\cosh|\tau|) L_0} e^{\hat{\tau}\tanh|\tau| L_-}\; ,
\end{equation}
with $\tau$ a complex coupling constant and $\hat{\tau}$ its orientation in 
the complex plane. When we now apply this operator to the vacuum,\index{vacuum}
the annihilation operators\index{operator!annihilation} will vanish, leaving 
only the exponential function of the creation 
operators.\index{operator!creation} We thus have 
\begin{equation}\label{in3}
 |\psi\rangle = U e^{-\frac{1}{2}(\vec{a}^\dagger,\Lambda^*\vec{a}^{\dagger})} 
 V^T |0\rangle = e^{-\frac{1}{2}(\vec{a}^\dagger, B\vec{a}^{\dagger})}
 |0\rangle\; ,
\end{equation}
with $B \equiv U \Lambda^* U^{\dagger}$, again by virtue of the invariance
property of the vacuum.\index{vacuum!invariance} This is the state of the 
interferometer\index{interferometer} prior to photo-detection.\index{detection}
It corresponds to {\em multi-mode squeezed vacuum}.
\index{vacuum!multi-mode squeezed}

\subsection{Photo-detection and Bargmann representation}
\index{representation!Bargmann}

The photo-detection itself can be modelled by successive application of 
annihilation operators. Every annihilation operator $\hat{a}_i$ removes a 
photon in 
mode $i$ from the state $|\psi\rangle$. Suppose the optical circuit employs 
$N$ distinct modes. We will now detect $M$ modes, finding $n_1+\ldots +n_M = 
N_{\rm tot}$ photons (with $M<N$). These modes can be relabelled 1 to $M$. 
The vector $\vec{n}$ denotes the particular detector `signature': $\vec{n} = 
(n_1,\ldots,n_M)$ means that $n_1$ photons are detected in mode 1, $n_2$ in 
mode 2, and so on. The freely propagating outgoing state 
$|\psi_{\vec{n}}\rangle$ can then be described as
\begin{equation}\label{det}
 |\psi_{\vec{n}}\rangle = ~_{1..M}\langle n_1,\ldots, n_M | \psi\rangle_{1..N} 
 = c_{\vec{n}} \langle 0| \hat{a}_1^{n_1}\cdots \hat{a}_M^{n_M}|\psi\rangle\; .
\end{equation}
Here, $c_{\vec{n}} = (n_1!\cdots n_M!)^{-\frac{1}{2}}$.

At this point it is convenient to introduce the $N$-mode Bargmann
representation\index{representation!Bargmann} we encountered in chapter 
\ref{chap3} \cite{bargmann61}. The creation 
\index{operator!creation} and annihilation 
operators\index{operator!annihilation} obey 
the commutation relations given in Eq.\ (\ref{comm}). We can replace these 
operators with c-numbers and their derivatives according to
\begin{equation}
 \hat{a}_i^{\dagger} \rightarrow \alpha_i \quad\mbox{and}\quad 
 \hat{a}_i \rightarrow \partial_i \equiv \frac{\partial}{\partial\alpha_i}\; .
\end{equation}
The commutation relations\index{commutation relation} then read
\begin{equation}
 [\partial_i,\alpha_j] = \delta_{ij} \quad\mbox{and}\quad
 [\partial_i,\partial_j] = [\alpha_i,\alpha_j] = 0\; .
\end{equation}
Note that the actual values of $\alpha_i$ are irrelevant (the creation
\index{operator!creation} and annihilation \index{operator!annihilation} 
operators do not have numerical values either); what matters here 
is the functional relationship between $\alpha_i$ and $\partial_{\alpha_i}$.

The state created by the optical circuit in this representation (prior to the 
detections, analogous to Eq.\ (\ref{in3})) in the Bargmann representation 
\index{representation!Bargmann} is
\begin{equation}
 \psi(\vec\alpha) = \exp\left[ -\frac{1}{2}(\vec\alpha,B\vec\alpha)\right]
 = \exp\left[ -\frac{1}{2}\sum_{ij}\alpha_i B_{ij}\alpha_j \right]\; .
\end{equation}
Returning to Eq.\ (\ref{det}), we can write the freely propagating state 
after detection of the auxiliary modes\index{mode!auxiliary} in the Bargmann 
representation as
\begin{equation}
 \psi_{\vec{n}}(\vec\alpha) \propto c_{\vec{n}}\; \partial_1^{n_1} \cdots 
 \partial_M^{n_M} \left. e^{-\frac{1}{2}(\vec\alpha,B\vec\alpha)} 
 \right|_{\vec\alpha'=0}\; ,
\end{equation}
up to some normalisation factor, where $\vec\alpha' = (\alpha_1,\ldots,
\alpha_M)$. By setting $\vec\alpha'=0$ we ensure that {\em no more} than 
$n_i$ photons are present in mode $i$. It plays the role of the vacuum
\index{vacuum} bra in Eq.~(\ref{det}).

\subsection{The outgoing state in terms of Hermite polynomials}

Now that we have an expression for the freely propagating state emerging
from our optical setup after detection, we seek to simplify it. We can 
multiply $\psi_{\vec{n}}(\vec\alpha)$ by the identity operator $\unity$, 
written as
\begin{equation}
 \unity = (-1)^{2N_{\rm tot}} \exp\left[-\frac{1}{2}(\vec\alpha,B\vec\alpha)
 \right] \; \exp\left[\frac{1}{2}(\vec\alpha,B\vec\alpha)\right]\; ,
\end{equation}
where $N_{\rm tot}$ is the total number of detected photons. We then find the
following expression for the unnormalised freely propagating state created by 
our optical circuit:
\begin{equation}\label{5out}
 \psi_{\vec{n}}(\vec\alpha) \propto \left. c_{\vec{n}} (-1)^{N_{\rm tot}} 
 H_{\vec{n}}^B (\vec\alpha)\; e^{-\frac{1}{2}(\vec\alpha,B\vec\alpha)}
 \right|_{\vec\alpha'=0}\; .
\end{equation}

Now I introduce the so-called {\em multi-dimensional} Hermite polynomial,
or MDHP for short:
\begin{equation}\label{mdhp}
 H_{\vec{n}}^B (\vec\alpha) = (-1)^{N_{\rm tot}} e^{\frac{1}{2}(\vec\alpha,
 B\vec\alpha)}\; \frac{\partial^{n_1}}{\partial\alpha_1^{n_1}} \cdots 
 \frac{\partial^{n_M}}{\partial\alpha_M^{n_M}}\; e^{-\frac{1}{2}
 (\vec\alpha,B\vec\alpha)}\; .
\end{equation}
The use of multi-dimensional Hermite polynomials and Hermite polynomials of 
two variables have previously been used to describe $N$-dimensional
first-order systems \cite{dodonov84,klauderer93} and photon statistics
\index{photon!-statistics}
\cite{vourdas87,dodonov94,kok99} (see also chapter \ref{chap4}). Here, I have 
shown that the lowest order of 
the {\em outgoing} state of optical circuits with quadratic components (as 
described by Eq.\ (\ref{in2})) and conditional 
photo-detection\index{detection!conditional} can be expressed
directly in terms of an MDHP.

In physical systems, the coupling constants (the $\lambda_i$'s) are usually
very small (i.e., $\lambda_i\ll 1$ or possibly $\lambda_i \lesssim 1$). This 
means that for all practical purposes only the first order term in 
Eq.~(\ref{5out}) is important (i.e., for small $\lambda_i$'s we can approximate
the exponential by 1). Consequently, studying the multi-dimensional Hermite 
polynomials yields knowledge about the typical states we can produce using 
Gaussian sources\index{Gaussian source} 
without coherent displacements. In section \ref{hp} I take a closer look at 
these polynomials, but first I consider the description of quantum 
teleportation in this representation.

\section{Example: Quantum Teleportation}\label{example}
\index{quantum teleportation}

As an example of how to determine the outgoing state of an optical circuit,
consider the teleportation experiment by Bouwmeester {\em et al}.\ 
\cite{bouwmeester97}. The optical circuit corresponding to this experiment
consists of eight incoming modes, all in the vacuum state. Physically, there 
are four spatial modes $a$, $b$, $c$ and $d$, all with two polarisation 
components $x$ and $y$. Two down-converters\index{down-converter} create 
entangled polarisation 
states; they belong to the class of active Gaussian components
\index{optical!active --- components} without 
coherent displacements. Mode $a$ undergoes a polarisation rotation over an 
angle $\theta$ and modes $b$ and $c$ are mixed in a 50:50 beam-splitter.
\index{beam-splitter}
Finally, modes $b$ and $c$ emerging from the beam-splitter are detected with 
polarisation insensitive detectors and mode $a$ is detected using a 
polarisation sensitive detector. The state which is to be teleported is 
therefore given by
\begin{equation}
 |\Psi\rangle = \cos\theta |x\rangle - \sin\theta |y\rangle\; .
\end{equation}

The state prior to the detection and normal ordering (corresponding to 
Eq.\ (\ref{stateprior})) is given by ($\tau$ is a coupling constant) 
\begin{equation}
 |\psi_{\rm prior}\rangle = U_{\rm BS} U_{\theta} 
 e^{\tau (\vec{u}^{\dagger},L\vec{u}^{\dagger})/2 + \tau^* (\vec{u},L\vec{u})/2
 + \tau (\vec{v}^{\dagger},L\vec{v}^{\dagger})/2 + \tau^* (\vec{v},L\vec{v})/2}
 |0\rangle\; ,
\end{equation}
with 
\begin{equation}
 L = \frac{1}{\sqrt{2}} \left( 
 \begin{array}{rrrr}
  0 & 0 & 0 & 1 \\
  0 & 0 & 1 & 0 \\
  0 & 1 & 0 & 0 \\
  1 & 0 & 0 & 0 
 \end{array}
 \right)
\end{equation}
and $\vec{u}^{\dagger}=(\hat{a}^{\dagger}_x,\hat{a}^{\dagger}_y,
\hat{b}^{\dagger}_x,\hat{b}^{\dagger}_y)$, $\vec{v}^{\dagger}=
(\hat{c}^{\dagger}_x,\hat{c}^{\dagger}_y,\hat{d}^{\dagger}_x,
\hat{d}^{\dagger}_y)$. This can be written as
\begin{equation}\label{priorstate}
 |\psi_{\rm prior}\rangle = \exp\left[ \frac{\tau}{2} 
 (\vec{a}^{\dagger},A\vec{a}^{\dagger}) + \frac{\tau^*}{2} (\vec{a},A\vec{a}) 
 \right] |0\rangle\; ,
\end{equation}
with $\vec{a}\equiv(\hat{a}_x,\ldots,\hat{d}_y)$ and $A$ the (symmetric) matrix
\begin{equation}
 A = \frac{1}{\sqrt{2}} \left(
 \begin{array}{rrrrrrrr}
  0 & 0 & -\sin\theta & \cos\theta & -\sin\theta & \cos\theta & 0 & 0 \\ 
  & 0 & \cos\theta & \sin\theta & \cos\theta & \sin\theta & 0 & 0 \\ 
  & & 0 & 0 & 0 & 0 & 0 & -1 \\ 
  & & & 0 & 0 & 0 & -1 & 0 \\ 
  & & & & 0 & 0 & 0 & 1 \\ 
  & & & & & 0 & 1 & 0 \\ 
  & & & & & & 0 & 0 \\ 
  & & & & & & & 0 
 \end{array}
 \right)\; .
\end{equation}

We now have to find the normal ordering of Eq.~(\ref{priorstate}). Since $A$
is unitary, the polynomial $(\vec{a}^{\dagger},A\vec{a}^{\dagger})$ is a 
generator of an $su(1,1)$ algebra.\index{algebra!generators of $su(1,1)$} 
According to Truax \cite{truax85}, the 
normal ordering\index{normal ordering|)} of the exponential thus yields a state
\begin{equation}\label{telprior}
 |\psi_{\rm prior}\rangle = \exp\left[ \frac{\xi}{2}
 (\vec{a}^{\dagger},A\vec{a}^{\dagger}) \right] |0\rangle\; ,
\end{equation}
with $\xi = (\tau\tanh|\tau|)/|\tau|$. The lowest order contribution after 
three detected photons is due to the term $\xi^2(\vec{a}^{\dagger},
A\vec{a}^{\dagger})^2/8$. However, first I write Eq.~(\ref{telprior}) in the 
Bargmann representation:\index{representation!Bargmann}
\begin{equation}
 \psi_{\rm prior}(\vec{\alpha}) = \exp\left[ \frac{\xi}{2}
 (\vec{\alpha},A\vec{\alpha}) \right]\; ,
\end{equation}
where $\vec{\alpha}=(\alpha_{a_x},\ldots,\alpha_{d_y})$ and $\vec{\alpha}'=
(\alpha_{a_x},\ldots,\alpha_{c_y})$. The polarisation independent 
photo-detection (the Bell measurement)\index{Bell!measurement} is then modelled by the differentiation $(\partial_{b_x}
\partial_{c_y} - \partial_{b_y}\partial_{c_x})$. Given a detector hit in mode
$a_x$, the polarisation sensitive detection of mode $a$ is modelled by 
$\partial_{a_x}$:
\begin{equation}
 \psi_{\rm out}(\vec{\alpha}) = \left. \partial_{a_x}\left( \partial_{b_x}
 \partial_{c_y} - \partial_{b_y}\partial_{c_x} \right) \exp\left[ \frac{\xi}{2}
 (\vec{\alpha},A\vec{\alpha}) \right] \right|_{\vec{\alpha}'=0}\; .
\end{equation}
The outgoing state in the Bargmann representation is thus given by
\begin{equation}
 \psi_{\rm out}(\vec{\alpha}) = \left(\cos\theta\, \alpha_{d_x} + \sin\theta\,
 \alpha_{d_y}\right) e^{\frac{\xi}{2} (\vec{\alpha},A\vec{\alpha})}\; ,
\end{equation}
which is the state teleported from mode $a$ to mode $d$ in the Bargmann 
representation. This procedure 
essentially amounts to evaluating the multi-dimensional Hermite polynomial
$H_{\vec{n}}^A(\vec{\alpha})$. Note that the polarisation independent 
Bell-detection of modes $b$ and $c$ yield a {\em superposition} of the MDHP's.
\index{MDHP!superposition of}

\section{The Hermite Polynomials}\label{hp}
\index{Hermite polynomial}

The one-dimensional Hermite polynomials are of course well known from the
description of the linear harmonic oscillator\index{harmonic oscillator} in 
quantum mechanics. These polynomials may be obtained from a generating 
function\index{generating function} $G$ (see appendix \ref{app:mdhp}). 
Furthermore, there exist two recursion relations\index{recursion relation} and 
an orthogonality relation\index{orthogonality!relation} between 
them. The theory of multi-dimensional Hermite polynomials with real variables 
has been developed by Appell and Kemp\'e de F\'eriet \cite{appell26} and in 
the Bateman project \cite{erdely53}. Mizrahi derived an expression for real 
MDHP's from an $n$-dimensional generalisation of the Rodriguez formula
\index{Rodriguez formula} 
\cite{mizrahi75}. I will now give the generating function for the complex 
MDHP's given by Eq.\ (\ref{mdhp}) and consecutively derive the recursion 
relations and the orthogonality relation (see also Ref.\ \cite{klauderer93}). 

\subsection{Generating functions and recursion relations}
\index{generating function}\index{recursion relation}

Define the generating function $G_B(\vec{\alpha},\vec{\beta})$ to be
\begin{equation}\label{genfunc}
 G_B(\vec{\alpha},\vec{\beta}) = e^{(\vec{\alpha},B\vec{\beta}) -
 \frac{1}{2} (\vec{\beta},B\vec{\beta})} = \sum_{\vec{n}} 
 \frac{\beta_1^{n_1}}{n_1!} \cdots \frac{\beta_M^{n_M}}{n_M!}
 H^B_{\vec{n}} (\vec{\alpha})\; .
\end{equation}
$G_B(\vec{\alpha},\vec{\beta})$ gives rise to the MDHP in 
Eq.\ (\ref{mdhp}), which determines this particular choice. Note that the
inner product $(\vec{\alpha},B\vec{\beta})$ does not involve any complex 
conjugation. If complex conjugation was involved, we would have obtained 
different polynomials (which we could also have called multi-dimensional 
Hermite polynomials, but they would not bear the same relationship to optical
circuits). 

In the rest of the chapter I use the following notation: by $\vec{n}-e_j$ I
mean that the $j^{\rm th}$ entry of the vector $\vec{n}=(n_1,\ldots,n_M)$ is
lowered by one, thus becoming $n_j-1$. By differentiation of both sides of 
the generating function in Eq.\ (\ref{genfunc}) we can thus show that the 
first recursion relation becomes
\begin{equation}
 \frac{\partial}{\partial\alpha_i} H^B_{\vec{n}} (\vec{\alpha}) = 
 \sum_{j=1}^M B_{ij} n_j H^B_{\vec{n}-e_j} (\vec{\alpha})\; .
\end{equation}

The second recursion relation is given by
\begin{equation}
 H_{\vec{n}+e_i}^B (\vec{\alpha}) - \sum_{j=1}^M B_{ij}\alpha_j H_{\vec{n}}^B
 (\vec{\alpha})+\sum_{j=1}^M B_{ij} n_j H_{\vec{n}-e_j}^B (\vec{\alpha}) =0\; ,
\end{equation}
which can be proved by mathematical induction using
\begin{equation}
 \sum_{k=1}^M B_{ik} n_k H_{\vec{n}-e_k+e_i}^B (\vec{\alpha}) - B_{ii}
 H_{\vec{n}}^B (\vec{\alpha}) = \sum_{k=1}^M B_{ik} m_k H_{\vec{m}+e_i}^B
 (\vec{\alpha})\; .
\end{equation}
Here, I have set $\vec{m}=\vec{n}-e_k$.

\subsection{Orthogonality relation}\index{orthogonality!relation}

The orthogonality relation is somewhat more involved. Ultimately, we want 
to use this relation to determine the normalisation constant of the states
given by Eq.\ (\ref{5out}). To find this normalisation\index{normalisation} we 
have to evaluate the integral
\begin{equation}\nonumber
 \int_{\mathbb{C}^N} d\vec\alpha\, \psi_{\vec{n}}^* (\vec\alpha) \psi_{\vec{m}}
 (\vec\alpha)\; .
\end{equation}
The state $\psi_{\vec{n}}$ includes $|_{\vec\alpha'=0}$, which translates into
a delta-function\index{delta-function} $\delta(\vec\alpha')$ in the integrand. 
The relevant integral thus becomes
\begin{equation}\nonumber
 \int_{\mathbb{C}^N} d\vec\alpha\, e^{-{\rm Re}(\vec\alpha,B\vec\alpha)}
 \left[ H_{\vec{n}}^B (\vec\alpha) \right]^* H_{\vec{m}}^B (\vec\alpha)\,
 \delta(\vec\alpha')\; .
\end{equation}
From the orthonormality of different quantum states we know that this 
integral must be proportional to $\delta_{\vec{n},\vec{m}}$. 

Since in the Bargmann representation\index{representation!Bargmann} we are 
only concerned with the {\em functional} relationship between $\alpha_i$ 
and $\partial_{\alpha_i}$ and not the actual values, we can choose $\alpha_i$ 
to be real. To stress this, we write $\alpha_i\rightarrow x_i$. The 
orthogonality relation\index{orthogonality!relation} is thus derived from 
\begin{equation}
 \int_{\mathbb{R}^N} d\vec{x}\, \psi^*_{\vec{n}}(\vec{x})
 \psi_{\vec{m}}(\vec{x}) = \int_{\mathbb{R}^N} d\vec{x}\,
 e^{-(\vec{x},{\rm Re}(B)\vec{x})} H_{\vec{n}}^{B^*} 
 (\vec{x}) H_{\vec{m}}^B (\vec{x})\, \delta(\vec{x}') \; .
\end{equation}
where $\delta(\vec{x}')$ is the real version of $\delta(\vec\alpha')$. 
Following Klauderer \cite{klauderer93} we find that
\begin{multline}
 \int d\vec{x}\, e^{-(\vec{x},{\rm Re}(B)\vec{x})}
 H_{\vec{n}}^{B^*} (\vec{x}) H_{\vec{m}}^B (\vec{x}) = \cr 
 (-1)^{N_{\rm tot}}\, \int d\vec{x}\, e^{-\frac{1}{2}(\vec{x},B\vec{x})}
 \partial_{\vec{x}}^{\vec{n}}\, 
 \left[ e^{-\frac{1}{2}(\vec{x},B^*\vec{x})}\right] H_{\vec{m}}^B
 (\vec{x})\; ,
\end{multline}
where $\partial_{\vec{x}}^{\vec{n}}$ is the differential 
operator\index{operator!differential}
$\partial_{x_1}^{n_1}\cdots\partial_{x_M}^{n_M}$ acting solely on the 
exponential function. We now integrate the right-hand side by parts, yielding
\begin{multline}
 (-1)^{N_{\rm tot}}\, \int d\vec{x}\, e^{-\frac{1}{2}(\vec{x},B\vec{x})}
 \partial_{\vec{x}}^{\vec{n}}\, 
 e^{-\frac{1}{2}(\vec{x},B^*\vec{x})} H_{\vec{m}}^B (\vec{x})=\cr
 \left.(-1)^{N_{\rm tot}}\, \int d'\vec{x}\, e^{-\frac{1}{2}(\vec{x},B\vec{x})}
 \partial_{\vec{x}}^{\vec{n}-e_i}\,
 e^{-\frac{1}{2}(\vec{x},B^*\vec{x})} H_{\vec{m}}^B (\vec{x})
 \right|^{+\infty}_{x_i = -\infty} \cr - (-1)^{N_{\rm tot}}\, \int 
 d\vec{x}\, e^{-\frac{1}{2}(\vec{x},B\vec{x})}
 \partial_{\vec{x}}^{\vec{n}-e_i}\, e^{-\frac{1}{2}
 (\vec{x},B^*\vec{x})} \partial_{x_i} H_{\vec{m}}^B (\vec{x})\; ,
\end{multline}
with $d'\vec{x}=dx_1\cdots dx_{i-1} dx_{i+1}\cdots dx_{N}$.
The left-hand term is equal to zero when ${\rm Re}(B)$ is positive definite,
i.e., when $(\vec{x},{\rm Re}(B)\vec{x})>0$ for all non-zero $\vec{x}$. 
Repeating this procedure $n_i$ times yields
\begin{multline}
 \int d\vec{x}\, e^{-(\vec{x},{\rm Re}(B)\vec{x})}
 H_{\vec{n}}^{B^*} (\vec{x}) H_{\vec{m}}^B (\vec{x}) = \cr
 (-1)^{N_{\rm tot}+n_i}\, \int d\vec{x}\, e^{-\frac{1}{2}(\vec{x},B\vec{x})}
 \partial_{\vec{x}}^{\vec{n}-n_ie_i}\,
 e^{-\frac{1}{2}(\vec{x},B^*\vec{x})} \partial_{x_i}^{n_i} 
 H_{\vec{m}}^B (\vec{x})\; .
\end{multline}
When there is at least one $n_i>m_i$, differentiating the MDHP $n_i$ times to
$x_i$ will yield zero. Thus we have 
\begin{equation}
 \int d\vec{x}\, e^{-(\vec{x},{\rm Re}(B)\vec{x})} 
 H_{\vec{n}}^{B^*} (\vec{x}) H_{\vec{m}}^B (\vec{x}) = 0 \quad\mbox{for}~
 \vec{n}\neq\vec{m}
\end{equation}
when ${\rm Re}(B)$ is positive definite and $n_i \neq m_i$ for any $i$. The
case where $\vec{n}$ equals $\vec{m}$ is given by
\begin{equation}
 \int d\vec{x}\, e^{-\frac{1}{2}(\vec{x},{\rm Re}(B)\vec{x})} 
 H_{\vec{n}}^{B^*} (\vec{x}) H_{\vec{m}}^B (\vec{x}) = \delta_{\vec{n}\vec{m}}
 {\mathcal{N}}\; ,
\end{equation}
where $\delta_{\vec{n}\vec{m}}$ denotes the product of $\delta_{n_i m_i}$ with
$1\leq i \leq N$. Here, $\mathcal{N}$ is equal to
\begin{equation}
 {\mathcal{N}} \equiv 2^{N_{\rm tot}}\, B_{11}^{n_1} \cdots B_{NN}^{n_N}\, 
 n_1!\cdots n_N! \, \left|\pi^{-1} B\right|^{-\frac{1}{2}}\; .
\end{equation}
For the proof of this identity I refer to Ref.\ \cite{klauderer93}. 

\section{Imperfect Detectors}\label{sec:det}\index{detector!imperfect}

So far, I only considered the use of ideal photo-detection. That is, I
assumed that the detectors tell us exactly and with unit 
efficiency\index{efficiency!unit-} how many 
photons were present in the detected mode. However, in reality such detectors
do not exist. In particular we have to incorporate 
losses\index{detector!loss} (non-perfect efficiency) and dark 
counts\index{dark counts} (see chapter \ref{chap4}). Furthermore, we have to 
take into account the fact that most detectors do not have a single-photon 
resolution\index{resolution!single-photon} (i.e., they
cannot distinguish a single photon from two photons) \cite{kok99}. 

This model is not suitable when we want to include dark counts. These
unwanted light sources provide thermal light,\index{state!thermal} which is 
not of the form of Eq.~(\ref{in2}) but given by Eq.~(\ref{thermalstate}). In 
single-shot experiments,\index{experiment!single-shot} however, dark counts 
can be neglected when the detectors operate only within a narrow time interval.

We can model the efficiency of a detector\index{detector!efficiency} by placing
a beam-splitter\index{beam-splitter} with transmission amplitude $\eta$ in 
front of a perfect detector \cite{kok99}. The part of the
signal which is reflected by the beam-splitter (and which will therefore never 
reach the detector) is the loss due to the imperfect detector. Since 
beam-splitters are part of the set of optical devices\index{optical!device} we 
allow, we can make this generalisation without any problem. We now trace out 
all the reflected modes (they are truly `lost'), and end up with a 
mixture\index{state!mixed} in the remaining undetected modes.

Next, we can model the lack of single-photon 
resolution\index{resolution!single-photon} by using the 
relative probabilities $p(n|k)$ and $p(m|k)$ of the actual number $n$ or $m$ 
of detected photons conditioned on the indication of $k$ photons in the 
detector (as described in Ref.\ \cite{kok99} and chapter \ref{chap4}). We can 
determine the pure states
according to $n$ and $m$ detected photons, and add them with relative weights
$p(n|k)$ and $p(m|k)$. This method is trivially generalised for more than two 
possible detected photon numbers.

Finally, we should note that my description of this class of optical circuits
\index{optical!circuit|)}
(in terms of multi-dimensional Hermite polynomials) is essentially a one-way
function. Given a certain setup, it is relatively straightforward to determine
the outgoing state of the circuit. The other way around, however, is very 
difficult. As exemplified by our efforts in Ref.\ \cite{kok00b} and chapter 
\ref{chap3}, it is almost
impossible to obtain the matrix $B$ associated with an optical circuit which 
produces a particular predetermined state from a Gaussian 
source.\index{Gaussian source} 

\section{Summary}

In this chapter, I have derived the general form of squeezed multi-mode vacuum 
\index{vacuum!squeezed multi-mode} states conditioned on photo-detection of 
some of the modes. To lowest order, the outgoing states in the Bargmann 
representation\index{representation!Bargmann} are proportional to 
multi-dimensional Hermite polynomials.\index{MDHP|)} As an example, I showed 
how teleportation can be described this way. 

\newpage
\thispagestyle{empty}

\part*{Some Applications}\addcontentsline{toc}{part}{$\quad\,\,$Some 
	Applications}

\chapter{Teleportation and Entanglement Swapping}\label{teleportation}

In this chapter I study the experimental realisations of quantum 
teleportation\index{quantum teleportation} by Bouwmeester {\em et al}.\ 
\cite{bouwmeester97}, entanglement swapping\index{entanglement!swapping} by 
Pan {\em et al}.\ \cite{pan98} and the observation of three-photon 
GHZ-entanglement\index{entanglement!GHZ-} by Bouwmeester {\em et al}.\ 
\cite{bouwmeester99b}. I will 
show that these experiments heavily relied on {\em post-selection}. 

In section \ref{postselection} I briefly discuss the issues concerned with 
post-selection. Then, in section \ref{sec:teleportation} the quantum 
teleportation 
experiment performed in Innsbruck will be studied. This section is based on 
Ref.\ \cite{kok00a}. Section \ref{es} is based on Refs.\ \cite{kok00c,kok00d}, 
and discusses entanglement swapping and entanglement 
purification.\index{purification} Finally, I 
briefly consider the experimental observation of three-photon GHZ-entanglement 
in section \ref{sec:ghz}. This chapter is based on Kok and Braunstein
\cite{kok00a,kok00c,kok00d}.

\section{Post-selection in quantum optics}\label{postselection}

In this section, I will discuss the concept of {\em 
post-selection}.\index{post-selection} 
Suppose we measure an observable\index{observable} $A$ with respect to an 
ensemble\index{ensemble} of systems in a state $|\phi\rangle$. In general we 
have a set of different measurement\index{measurement} outcomes $\{ a_k\}$, 
where the $a_k$'s denote the eigenvalues\index{eigen!-values} of $A$. We speak 
of post-selection when a {\em subset} of the set of outcomes $\{ a_k\}$ is 
discarded. The remaining {\em post-selected} set of outcomes may be used for 
subsequent data-analysis.

For example, if we had a tri-partite\index{entanglement!tri-partite} optical 
system in the state
\begin{equation}\label{ps}
 |\phi\rangle \propto |0,1,1\rangle + |1,0,1\rangle + |1,1,0\rangle +
 |1,1,1\rangle\; ,
\end{equation}
we could place three\index{detector!photo-} photo-detectors\footnote{Here we 
use three measurements of
the same observable, i.e., photon number. There is, however, no reason why we
can't measure three different observables in the respective modes.} in the 
three outgoing modes. For ideal detectors, there are four possible measurement 
outcomes: one photon in any two of the three detectors, or one photon in all 
three detectors. When we repeat this `experiment' a large number of times, we 
might discard all the measurements which do not yield a three-fold detector 
coincidence.\index{detector!coincidence} This would correspond to 
post-selecting our data set on a 
three-fold coincidence. (Admittedly, this is not a very interesting experiment.
However, later we will see that using post-selection we can even partially 
perform a Bell measurement.)\index{Bell!measurement}

Note that there is a fundamental difference between a {\em conditional 
measurement}\index{measurement!conditional} and post-selection. In chapter 
\ref{chap4} I discussed 
entanglement-based state 
preparation,\index{state preparation!entanglement-based} in which one 
subsystem was measured, the
outcome of which was used to accept or reject the state of the remaining
system. The crucial property of such a conditional 
measurement\index{measurement!conditional} is that at the
end of the procedure, there is a physically propagating state remaining. 
Post-selection\index{post-selection} offers a completely different type of 
control to the 
experimenter. Since all the subsystems are measured there is no physically 
propagating state left over, but a subset of the data can be selected for 
further analysis. 

The question is now whether in our `experiment' above we have demonstrated 
the {\em existence} of the state $|1,1,1\rangle$. The answer has to be `no':
immediately before the measurement\index{measurement} the state had the form 
of Eq.~(\ref{ps}),
whereas afterwards, there was no state left at all.

Post-selection can be very powerful, though. In section \ref{sec:ghz} we will
see that non-local correlations\index{non-local correlations} can be inferred 
from post-selected data which
was obtained in an experiment designed to create a three-photon GHZ-state.
\index{state!GHZ-}
In addition, it does {\em not necessarily} mean that a post-selected state 
cannot be used further in, say, a quantum computer.\index{quantum!computer} 
As long as the 
post-selection can be made in the end, the relevant branch (or branches) in 
the superposition\index{superposition} undergo(es) the quantum computation. 
With this in mind, we can now consider the experimental demonstration of 
quantum teleportation\index{quantum teleportation} and 
entanglement swapping.\index{entanglement!swapping}

\section{Quantum teleportation}\label{sec:teleportation}

We speak of quantum teleportation when a (possibly unknown) quantum state 
$|\phi\rangle$ held by Alice is sent to Bob without actually traversing the 
intermediate space. The protocol uses an entangled state\index{state!entangled}
of two systems which 
is shared between Alice and Bob. In the next section I will present the 
teleportation protocol for discrete variables. In the subsequent sections 
I confine the discussion to the teleportation experiment performed by 
Bouwmeester {\em et al}., and study the difficulties which arise using its
particular experimental setup.

\subsection{The discrete teleportation protocol}

Quantum teleportation was first introduced by Bennett {et al}.\ in 1993
\cite{bennett93}. In this protocol, a quantum state held by Alice is sent to 
Bob by means of what the authors called `dual classical and 
Einstein-Podolsky-Rosen channels'.\index{EPR-channel} How does this work?

Suppose we have a set of three two-level systems, or qubits, the states of 
which can be written in the computational basis\index{computational basis} 
$\{|0\rangle_k,|1\rangle_k\}$, 
where $k=1,2,3$ denotes the system. Let Alice and Bob share a maximally 
entangled state\index{state!maximally entangled} (the Einstein-Podolsky-Rosen 
channel \cite{einstein35}), for instance one of the Bell 
states\index{state!Bell} $|\Psi^-\rangle$ in systems 2 and 3:
\begin{equation}
 |\Psi^-\rangle_{23} = \frac{1}{\sqrt{2}} \left( |0,1\rangle_{23} - 
 |1,0\rangle_{23} \right)\; .
\end{equation}
System 1 is in an unknown state $|\phi\rangle_1$, which can be written as
\begin{equation}
 |\phi\rangle_1 = \alpha |0\rangle + \beta |1\rangle\; .
\end{equation}
The other three Bell states\index{state!Bell} are given by
\begin{eqnarray}
 |\Psi^+\rangle &=& \frac{1}{\sqrt{2}}\left( |0,1\rangle+|1,0\rangle\right)\cr
 |\Phi^-\rangle &=& \frac{1}{\sqrt{2}}\left( |0,0\rangle-|1,1\rangle\right)\cr
 |\Phi^+\rangle &=& \frac{1}{\sqrt{2}}\left( |0,0\rangle+|1,1\rangle\right)\; .
\end{eqnarray}
We can write the total state of the three systems as
\begin{multline}\label{teltotal}
 |\phi\rangle_1 |\Psi^-\rangle_{23} = \frac{1}{\sqrt{2}} \left( \alpha
 |0,0,1\rangle_{123} - \alpha |0,1,0\rangle_{123} \right. \cr +
 \left. \beta
 |1,0,1\rangle_{123} - \beta |1,1,0\rangle_{123} \right)\; .
\end{multline}
The computational basis\index{computational basis} states of two 
qubits\index{qubit} can also be written in the Bell basis:\index{Bell!basis}
\begin{eqnarray}
 |0,0\rangle &=& \frac{1}{\sqrt{2}}\left(|\Phi^+\rangle + |\Phi^-\rangle
 \right)\; ,\cr
 |0,1\rangle &=& \frac{1}{\sqrt{2}}\left(|\Psi^+\rangle + |\Psi^-\rangle
 \right)\; ,\cr
 |1,0\rangle &=& \frac{1}{\sqrt{2}}\left(|\Psi^+\rangle - |\Psi^-\rangle
 \right)\; ,\cr
 |1,1\rangle &=& \frac{1}{\sqrt{2}}\left(|\Phi^+\rangle - |\Phi^-\rangle
 \right)\; .
\end{eqnarray}
When we make this substitution for qubits 1 and 2, Eq.~(\ref{teltotal})
becomes
\begin{eqnarray}
 |\phi\rangle_1 |\Psi^-\rangle_{23} &=& \frac{1}{2} \left[ 
 |\Phi^+\rangle_{12} \left( \alpha|1\rangle_3 - \beta|0\rangle_3 \right) + 
 |\Phi^-\rangle_{12} \left( \alpha|1\rangle_3 + \beta |0\rangle_3 \right) 
 \right. \cr &&\quad \left. -
 |\Psi^+\rangle_{12} \left( \alpha|0\rangle_3 - \beta|1\rangle_3 \right) - 
 |\Psi^-\rangle_{12} \left(  \alpha + \beta |1\rangle_3 \right) \right]\; .
\end{eqnarray}
Alice is in possession of qubits 1 and 2, while Bob holds qubit 3. When Alice 
now performs a Bell measurement,\index{Bell!measurement} Bob's qubit is 
transformed into the unknown state {\em up to one of four unitary 
transformations}.\index{transformation!unitary} Alice's measurement 
outcome determines which one of these transformations should be inverted on 
Bob's qubit to return it to the original state $|\phi\rangle$. This completes 
the teleportation protocol.

Quantum teleportation\index{quantum teleportation} is not restricted to qubits.
For example, suppose we have an $N$-level system\footnote{Sometimes called a 
`qu$N$it'\index{qunit@qu$N$it} or `qudit'.\index{qudit}} in the 
state
\begin{equation}
 |\phi\rangle_1 = \sum_j \alpha_j |j\rangle_1\; ,
\end{equation}
and a maximally entangled state\index{state!maximally entangled} shared 
between Alice and Bob:
\begin{equation}
 |\Psi\rangle_{23} = \frac{1}{\sqrt{N}} \sum_{j} |j,j\rangle_{23}\; .
\end{equation}
We measure system 1 and 2, held by Alice, in the basis 
$\{ |\psi_{nm}\rangle_{12}\}$, with
\begin{equation}
 |\psi_{nm}\rangle_{12} = \frac{1}{\sqrt{N}} \sum_k e^{2\pi ijn/N} 
 |j,j\oplus m\rangle_{12}\; .
\end{equation}
In this notation we have $j\oplus m = j+m\mod N$. Conditioned on the 
measurement outcome $(n,m)$ corresponding to $|\psi_{nm}\rangle_{12}$, Bob's
system 3 is transformed into $|\phi\rangle_3$ after a transformation
\cite{bennett93}:
\begin{equation}
 U_{nm} = \sum_k e^{2\pi ikn/N} |k\rangle_3 \langle k\oplus m|\; .
\end{equation}
Thus the state of system 1 is transferred to system 3. System 1 can itself be 
mixed\index{state!mixed} or part of an entangled state.\index{state!entangled} 
Note that, since Bob needs the measurement outcome, Alice has to send a 
classical message\index{classical!message} of $2\log_2 N$
classical bits.\index{classical!bit} Sending this classical message is, like 
all classical communication, bounded by the speed of light. Therefore, quantum 
teleportation\index{quantum teleportation}
does not yield an information transfer\index{information!transfer} faster than 
light.

After the invention of discrete quantum 
teleportation,\index{quantum teleportation!discrete} Vaidman and Braunstein 
and Kimble introduced teleportation for states of dynamical variables with 
continuous spectra \cite{vaidman94,braunstein98b}. In 1997, teleportation was 
experimentally realised by Bouwmeester {\em et al}.\ in Innsbruck 
\cite{bouwmeester97} and Boschi {\em et al}.\ in Rome \cite{boschi98}, 
followed by Furusawa {\em et al}.\ in Pasadena \cite{furusawa98} in 1998. 
This last experiment involved the teleportation of continuous 
variables.\index{quantum teleportation!continuous} Quantum teleportation was 
also reported using nuclear magnetic resonance\index{NMR} by Nielsen
{\em et al}.\ in 1998 \cite{nielsen98}. In 2000, Kim {\em et al}.\ performed 
quantum teleportation of polarised single-photon states using complete Bell 
detection\index{Bell!measurement!complete} \cite{kim00b}.

In this chapter, however, I will focus 
mainly on the teleportation experiment of Bouwmeester {\em et al}.

\subsection{The `Innsbruck Experiment'}\label{experiment}

\begin{figure}[t]
  \begin{center}
  \begin{psfrags}
     \psfrag{Alice}{Alice}
     \psfrag{Bob}{Bob}
     \psfrag{a}{$a$}
     \psfrag{b}{$b$}
     \psfrag{c}{$c$}
     \psfrag{d}{$d$}
     \psfrag{uv-pulse}{{\sc uv}-pulse}
     \psfrag{bs}{BS}
     \psfrag{mirrors}{mirrors}
     \psfrag{crystal}{crystal}
     \psfrag{45}{$\pm$45}
     \epsfxsize=8in
     \epsfbox[ -130 20 670 260]{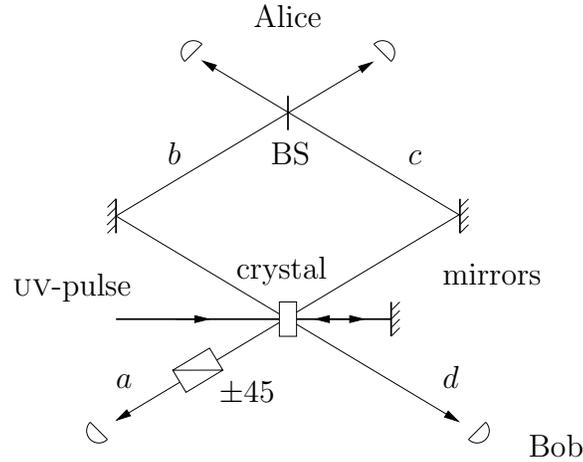}
  \end{psfrags}
  \end{center}
  \caption{Schematic representation of the teleportation experiment 
	con\-duc\-ted 
	in Innsbruck. A {\sc uv}-pulse is sent into a non-linear crystal, thus 
	creating an entangled photon-pair. The {\sc uv}-pulse is reflected by a
	mirror and returned into the crystal again. This reflected pulse 
	creates the second photon-pair. Photons $b$ and $c$ are sent into a 
	beam-splitter and are detected. This is the Bell measurement. Photon 
	$a$ is detected to prepare the input state and photon $d$ is the 
	teleported output state Bob receives. In order to rule out the 
	possibility that there are no photons in mode $d$, Bob detects this 
	mode as well.}
  \label{fig:innsbruck}
\end{figure}

In this section, I study the experimental realisation of quantum teleportation 
of a single polarised photon as performed in Innsbruck, henceforth called the 
`Innsbruck experiment'\index{Innsbruck experiment} (Bouwmeester {\em et al}.\ 
\cite{bouwmeester97}). In the Innsbruck experiment, parametric down-conversion
\index{down-converter} is used to create two 
entangled photon-pairs. One pair constitutes the entangled state shared 
between Alice and Bob, while the other is used by Victor to create an `unknown'
single-photon polarisation state\index{state!polarisation} 
$|\phi\rangle$: Victor detects mode $a$, shown
in figure \ref{fig:innsbruck} to prepare the single-photon input state in mode 
$b$. This mode is sent to Alice. A coincidence in the 
detection\index{detector!coincidence} of the two outgoing modes of the 
beam-splitter\index{beam-splitter} (Alice's --- incomplete --- Bell 
measurement)\index{Bell!measurement} tells us that Alice's two photons are in 
a $|\Psi^-\rangle$ Bell state\index{state!Bell} 
\cite{weinfurter94,braunstein95,braunstein96}. The remaining photon 
(held by Bob) is now in the same unknown state as the photon prepared by 
Victor because in this case the unitary 
transformation\index{transformation!unitary} Bob has to apply 
coincides with the identity, i.e., doing nothing. Bob verifies this by 
detecting his state along the same polarisation axis which was used by Victor. 
A four-fold coincidence in the detectors of Victor's state 
preparation,\index{state preparation} 
Alice's Bell measurement and Bob's outgoing state indicate that quantum 
teleportation\index{quantum teleportation} of a single-photon state is 
complete.

There is however a complication which gave rise to a different interpretation
of the experiment \cite{braunstein98,bouwmeester98,bouwmeester99,kok00a}. 
Analysis shows that the state before detection by Bob (but conditioned on the 
other three detector `hits') is a mixture of the vacuum\index{state!vacuum} 
and the original state \cite{bouwmeester97,braunstein98} (to lowest order). 
This vacuum contribution\index{vacuum contribution} occurs when the 
down-converter\index{down-converter} responsible for creating 
the input state $|\phi\rangle$ yields two photon-pairs,\index{photon-pair} 
while the other gives 
nothing. The detectors used in the experiment cannot distinguish between one 
or several photons coming in, so Victor's detection of mode $a$ in figure 
\ref{fig:innsbruck} will not reveal the presence of more than one photon. A 
three-fold coincidence in the detectors\index{detector!coincidence} of Victor 
and Alice alone is still possible, but Bob has not received a photon and 
quantum teleportation\index{quantum teleportation} has not 
been achieved. Bob therefore needs to detect his state in order to identify 
successful quantum teleportation. Were Victor to use a detector which can 
distinguish between one or several photons this problem would disappear. 
However, currently such detectors require an operating environment of roughly 
6K \cite{kwiat94,kwiat98,kim99,takeuchi99}.

I evaluate the suggestions to `improve' the experiment in order to yield 
non-post-selected\index{post-selection} operation, as made by Braunstein and 
Kimble \cite{braunstein98} (I will discuss the reply by Bouwmeester {\em et 
al}.\ \cite{bouwmeester98,bouwmeester99} in section {\ref{sec:fid+eff}}). 
These suggestions include the employment of a detector 
cascade\index{detector!cascade} (as proposed in 
chapter \ref{chap4}) in the state preparation mode, and enhancement of the 
down-converter\index{down-converter} responsible for the 
entanglement\index{entanglement} channel 
(see chapter \ref{chap3}) relative to the one responsible for the initial 
state preparation. Subsequently, I hope to clarify some of the differences 
in the interpretation of the Innsbruck experiment\index{Innsbruck experiment} 
\cite{kok00a}. 

As pointed out by Braunstein and Kimble \cite{braunstein98}, to lowest order 
the teleported state in the Innsbruck experiment is a mixture of the vacuum 
\index{state!vacuum} and a single-photon state.\index{state!single-photon} 
However, we cannot interpret this {\em state} as a 
low-efficiency\index{efficiency} teleported state, where sometimes a photon 
emerges from the apparatus and sometimes not. This reasoning is based on the 
so-called 
`Partition Ensemble Fallacy',\index{Partition Ensemble Fallacy|see{PEF}} or 
PEF\footnote{This term was coined by 
Samuel L.\ Braunstein and first appeared in Kok and Braunstein \cite{kok00a}.} 
for short. It will be studied more extensively in section {\ref{sec:fid+eff}}.
{\sc PEF}\index{PEF} relies on a particular partition of the outgoing density 
matrix, \index{density matrix} and this is not consistent with quantum 
mechanics\index{quantum mechanics} \cite{peres95}.
Circumventing {\sc pef} leads to the notion of {\em post-selected} 
teleportation, in which the teleported state is detected. The post-selected 
teleportation indeed has a high fidelity\index{fidelity} and a low efficiency. 
Although 
generally {\sc PEF} is harmless (it might even be considered a useful tool in 
understanding aspects of quantum theory), to my knowledge, this is the first 
instance where it leads to a {\em quantitatively } different evaluation of an 
experiment.

It will turn out that the suggested improvements require near perfect 
efficiency photo-detectors\index{detector!photo-} or a considerable increase 
in the time
needed to run the experiment. The remaining practical alternative in order to 
obtain {\em non-post-selected} quantum teleportation (i.e., teleportation {\em 
without} the need for detecting the teleported photon) is to employ a
single-photon resolution detector\index{detector!single-photon!resolution} in 
the state-preparation mode (a technology currently 
requiring approximately 6K operating conditions) 
\cite{kim99,takeuchi99} (see also chapter \ref{chap4}).

\subsection{The generalised experiment}\label{generalised}

\begin{figure}[t]
  \begin{center}
  \begin{psfrags}
     \psfrag{Alice}{Alice}
     \psfrag{Bob}{Bob}
     \psfrag{Bell-detection}{Bell-detection}
     \psfrag{state-preparation}{state-preparation}
     \psfrag{source1}{Source 1}
     \psfrag{source2}{Source 2}
     \psfrag{t}{$\theta$}
     \psfrag{a}{$a$}
     \psfrag{b}{$b$}
     \psfrag{c}{$c$}
     \psfrag{d}{$d$}
     \psfrag{u}{$u$}
     \psfrag{v}{$v$}
     \epsfxsize=8in
     \epsfbox[ -130 30 700 240]{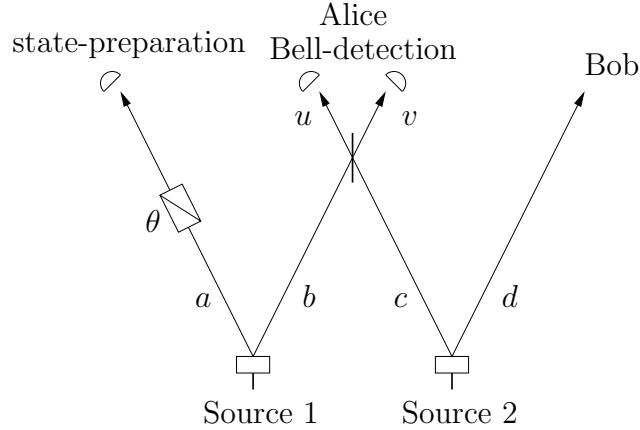}
  \end{psfrags}
  \end{center}
  \caption{Schematic `unfolded' representation of the teleportation 
	experiment with two independent down-converters (Source 1 and Source 2)
	and a polarisation rotation $\theta$ in mode $a$. The state-preparation
	detector is actually a detector cascade and Bob does not detect the 
	mode he receives.}
  \label{model}
\end{figure}

In the rest of this section I consider a generalised scheme for the Innsbruck 
experiment which enables us to establish the requirements to obtain 
non-post-selected quantum teleportation (based on a three-fold coincidence of 
Victor and Alice's detectors). The generalisation consists of a detector 
cascade (Chapter \ref{chap4} and Ref.\ \cite{song90} for Victor's 
state preparation\index{state preparation} detection and parametric 
down-converters\index{down-converter} with different 
specifications, rather than two identical down-converters. Furthermore, an 
arbitrary polarisation rotation in the state-preparation mode allows us to 
consider any superposition\index{superposition} of $x$- and $y$-polarisation. 
I calculate the output state and give an expression for the teleportation 
fidelity\index{fidelity} in terms 
of the detector efficiencies\index{detector!efficiency} and down-conversion 
rates. To this end, I 
consider a simplified `unfolded' schematic representation of the experiment, 
shown in figure \ref{model}. 

\subsection{Detectors}\label{detectors}

As explained in chapter \ref{chap4}, there are two sources of errors for a 
detector: losses\index{detector!loss} and dark counts.\index{dark counts} 
Dark counts are negligible in the teleportation\index{quantum teleportation} 
experiment because the {\sc uv}-pump is fired during very short 
time intervals and the probability of finding a dark count in such a small 
interval is negligible. Consequently, the model for real, finite-efficiency 
detectors\index{detector!finite-efficiency} I presented in chapter \ref{chap4} 
only takes into account detector 
losses. Furthermore, the detectors cannot distinguish between one or several 
photons. In my terminology: {\it finite-efficiency single-photon sensitivity} 
detectors\index{detector!single-photon!sensitivity} (see page \pageref{sps}).

\begin{figure}[t]
  \begin{center}
  \begin{psfrags}
     \psfrag{eta}{$\eta$}
     \psfrag{in}{$|0\rangle$}
     \psfrag{a}{$a$}
     \psfrag{b}{$b$}
     \psfrag{c}{$c$}
     \psfrag{d}{$d$}
     \epsfxsize=8in
     \epsfbox[-180 40 520 140]{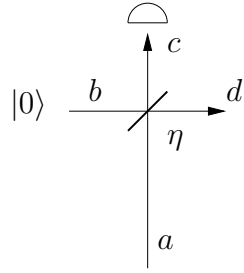}
  \end{psfrags}
  \end{center}
  \caption{A model of an inefficient detector. The beam-splitter with 
	transmission amplitude $\eta$ will 
	reflect part of the incoming mode $a$ to mode $d$, which is thrown 
	away. The transmitted part $c$ will be sent into a ideal detector. 
	Mode $b$ is vacuum.}
  \label{fig:detector}
\end{figure}

To simulate a realistic detector I make use of projection operator valued 
measures, or POVM's\index{POVM} for short \cite{kraus83} (see also appendix 
\ref{app:povm}). Consider a beam-splitter\index{beam-splitter} 
in the mode which is to be detected so that part of the signal is reflected 
(see figure \ref{fig:detector}). The second incoming mode of the beam-splitter 
is the vacuum\index{state!vacuum} (I neglect higher photon number states 
\index{state!number} because they hardly 
contribute at room temperature). The transmitted signal $c$ is sent into an 
ideal detector. We identify mode $d$ with the detector loss.

Suppose in mode $a$ there are $n$ $x$-polarised and $m$ $y$-polarised photons.
Furthermore, let these photons all be reflected by the 
beam-splitter.\index{beam-splitter} The projector\index{projector} for 
finding these photons in the $d$-mode is given by:
\begin{equation}\label{povmnm}
  {E}_d = |n,m\rangle_{d_x d_y}\langle n,m| = \frac{1}{n!m!}
  (\hat{d}_x^{\dagger})^n(\hat{d}^{\dagger}_y)^m |0,0\rangle_{d_x d_y}
	\langle 0,0| \hat{d}^n_x \hat{d}^m_y \; .
\end{equation}
The beam-splitter equations are taken to be ($\widetilde\eta\equiv
\sqrt{1-\eta^2}$):
\begin{equation}\label{bs}
  \hat{c} = \eta \hat{a} + \widetilde\eta \hat{b} \qquad\text{and}\qquad
  \hat{d} = \widetilde\eta \hat{a} - \eta \hat{b} \; .
\end{equation}
Substituting these equations in (\ref{povmnm}), summing over all $n$ and 
$m$ and using the binomial expansion yields
\begin{eqnarray}
  {E}_{ab} &=& \sum_{n,m} \binom{n}{k}^2 \binom{m}{l}^2 
	\frac{(-1)^{2(k+l)}}{n!m!}
	(\widetilde\eta \hat{a}^{\dagger}_x)^{n-k} (\eta\hat{b}^{\dagger}_x)^k 
	(\widetilde\eta \hat{a}^{\dagger}_y)^{m-l} (\eta\hat{b}^{\dagger}_y)^l~
	|0\rangle_{ab} \cr && \times \langle 0| (\widetilde\eta\hat{a}_x)^{n-k}
	(\eta\hat{b}_x)^k (\widetilde\eta\hat{a}_y)^{m-l}(\eta\hat{b}_y)^l \; .
\end{eqnarray}
Since the $b$-mode is the vacuum,\index{state!vacuum} the only contributing 
term is $k=l=0$. So the POVM\index{POVM} ${E}_a^{(0)}$ of finding {\em no} 
detector counts in mode $a$ is
\begin{equation}
 {E}_a^{(0)} = \sum_{n,m} \frac{ \widetilde{\eta}^n 
  (\hat{a}_x^{\dagger})^n \widetilde{\eta}^m (\hat{a}_y^{\dagger})^m}{n!m!} 
	~|0\rangle_{a_x a_y}\langle 0|~
  \widetilde{\eta}^n \hat{a}_x^n \widetilde{\eta}^m \hat{a}_y^m
  = \sum_{n,m} \widetilde\eta^{2(n+m)} |n,m\rangle_{a_x a_y} \langle n,m| \; .
\end{equation}
The required POVM for finding a detector count is 
\begin{equation}
  {E}_a^{(1)} = {\unity} - {E}_a^{(0)} = \sum_{n,m} 
  [1-\widetilde\eta^{2(n+m)}] |n,m\rangle_{a_x a_y} \langle n,m| \; ,
\end{equation}
where $\unity$ is the identity operator, $\eta^2$ is the detector 
efficiency\index{detector!efficiency} and $\widetilde\eta^2\equiv 1-\eta^2$ 
the detector loss.\index{detector!loss} When we let ${E}_a^{(1)}$ 
act on the total state and trace out mode $a$, we have modelled the inefficient
detection of this mode. In the case of continuous detection we need a more 
elaborate model (see for example Ref.\ \cite{wiseman93}).

In order for Victor to distinguish between one or more photons in the state
preparation mode $a$, I consider a detector cascade\index{detector!cascade} 
(Victor doesn't have a 
detector which can distinguish between one or several photons coming in). When 
there is a detector coincidence in the cascade, more than one photon was 
present in mode $a$, and the event should be dismissed. In the case of ideal 
detectors, this will improve the fidelity\index{fidelity} of the 
teleportation\index{quantum teleportation} up to an 
arbitrary level (we assume there are no beam-splitter losses). Since we employ 
the cascade in the $a$-mode (which was used by Victor to project mode $b$ onto 
a superposition\index{superposition} in the polarisation basis) we need to 
perform a {\em polarisation sensitive} 
detection.\index{detector!polarisation-sensitive}

In order to model this I separate the incoming state $|n,m\rangle_{a_x a_y}$ 
of mode $a$ into two spatially separated modes $|n\rangle_{a_x}$ and 
$|m\rangle_{a_y}$ by means of a polarisation 
beam-splitter.\index{beam-splitter} The modes $a_x$ 
and $a_y$ will now be detected. The POVM's\index{POVM} corresponding to 
inefficient detectors are derived along the same lines as in the previous 
section and read:
\begin{eqnarray}\label{cascadepovm}
 {E}^{(0)}_{a_j} &=& \sum_n\widetilde{\eta}^{2n}|n\rangle_{a_j}\langle 
	n| \qquad\text{and}\cr
 {E}^{(1)}_{a_j} &=& \sum_n[1-\widetilde{\eta}^{2n}]|n\rangle_{a_j}
	\langle n| \; .
\end{eqnarray}
with $j\in\{ x,y \}$. we choose to detect the $x$-polarised mode. This means
that we only have to make sure that there are no photons in the $y$-mode.
The output state will include a product of the two POVM's: one for
finding a photon in mode $a_x$, and one for finding {\em no} photons in mode
$a_y$: $E^{(1)}_{a_x}E^{(0)}_{a_y}$.

To make a cascade with two detectors in $a_x$ and one in $a_y$ employ another 
50:50 beam-splitter in mode $a_x$ and repeat the above procedure of detecting 
the outgoing modes $c$ and $d$ (\ref{cascadepovm}). Since we can detect a 
photon in either one of the modes, we have to include the sum of the 
corresponding POVM's, yielding a transformation $E^{(1)}_{c_x} 
E^{(0)}_{d_x} + E^{(0)}_{c_x}E^{(1)}_{d_x}$. This is easily expandable to 
larger cascades by using more beam-splitters and summing over all possible 
detector hits.

\subsection{Output state}

In this section I incorporate the finite-efficiency 
detectors\index{detector!finite-efficiency} and the detector
cascade in the calculation of the undetected teleported output state. This 
calculation includes the creation of two photon-pairs\index{photon-pair} 
(lowest order) and three photon pairs (higher order corrections due to four or 
more photon-pairs in the experiment are highly negligible). A formula for the 
vacuum contribution\index{vacuum contribution} to 
the teleportation fidelity is given for double-pair production (lowest order). 

Let the two down-converters\index{down-converter} in the generalised 
experimental setup yield evolutions $U_{\text{src1}}$ and $U_{\text{src2}}$ on 
modes $a$, $b$ and $c$, $d$ respectively (see figures \ref{fig:innsbruck} and 
\ref{model}) according to Eq.~(\ref{evolution}). The beam-splitter which 
transforms modes $b$ and $c$ into $u$ and $v$ (see figure \ref{model}) is 
incorporated by a suitable unitary transformation\index{transformation!unitary}
$U_{\text{BS}}$, as is the polarisation rotation $U_{\theta}$ over an angle 
$\theta$ in mode $a$. The $N$-cascade\index{detector!cascade} will be 
modelled by $N-1$ beam-splitters\index{beam-splitter} in the $x$-polarisation 
branch of the cascade, and can therefore be expressed in terms of a unitary 
transformation $U_{a_1\ldots a_N}$ on the Hilbert space\index{Hilbert space} 
corresponding to modes $a_1$ to $a_N$ (i.e., replace mode $a$ with modes $a_1$ 
to $a_N$):
\begin{equation}\label{subout}
 |\Psi_{\theta}\rangle\langle\Psi_{\theta}| = 
  U_{a_1\ldots a_N} U_{\theta} U_{\text{BS}}U_{\text{scr1}}U_{\text{scr2}}
    |0\rangle\langle 0| U^{\dagger}_{\text{scr1}} U^{\dagger}_{\text{scr2}} 
    U^{\dagger}_{\text{BS}} U^{\dagger}_{\theta} U^{\dagger}_{a_1\ldots a_N}\; 
    .
\end{equation}
Detecting modes $a_1\ldots a_N$, $u$ and $v$ with real (inefficient) detectors 
\index{detector!real} means taking the partial trace\index{partial trace} over 
the detected modes, including the POVM's\index{POVM} derived in section 
{\ref{detectors}}:
\begin{equation}\label{out}
 \rho_{\text{out}} = \text{Tr}_{a_1\ldots a_N u v} \left[ 
 {E}_{N\text{-cas}} {E}^{(1)}_u {E}^{(1)}_v 
 ~|\widetilde{\Psi}_{\theta}\rangle_{a_1\ldots a_N uvd}\langle
 \widetilde{\Psi}_{\theta}| \right] \; ,
\end{equation}
with ${E}_{N\text{-cas}}$ the superposition\index{superposition} of {\sc 
POVM}'s for a polarisation sensitive detector cascade\index{detector!cascade} 
having $n$ detectors with finite efficiency. In the case $N=2$ this expression 
reduces to the 2-cascade {\sc 
POVM}-superposition derived in the previous section. Eq.\ (\ref{out}) is an 
analytic expression of the undetected outgoing state in the generalisation of 
the Innsbruck experiment.\index{Innsbruck experiment}

The evolutions $U_{\text{src1}}$ and $U_{\text{src2}}$ are exponentials of 
creation operators.\index{operator!creation} In the computer simulation (using 
{\sc Mathematica},\index{Mathematica} see appendix \ref{app:math}) I
truncated these exponentials at first and second order. The terms that remain 
correspond to double and triple pair production in the experimental setup. To 
preserve the order of the creation operators we put them as arguments in a 
function $f$. I defined the following algebraic rules for $f$ (see appendix 
\ref{app:math}):
\begin{verbatim}
  f[x__, y__ + w__, z__]    :=   f[x, y, z] + f[x, w, z]
  f[x__, n_ a__, y__]       := n f[x, a, y] 
  f[x__, n_ adagger__, y__] := n f[x, adagger, y]
\end{verbatim}
where {\tt x,y,z} and {\tt w} are arbitrary expressions including creation and 
annihilation operators\index{operator!annihilation} ({\tt adagger} and {\tt a})
and {\tt n} some expression {\em not} depending on creation or annihilation 
operators. The last entry of $f$ is always a photon number 
state\index{state!number} (including the initial vacuum 
state).\index{state!vacuum}

Since we now have functions of creation and annihilation operators, it is 
quite straightforward to define (lists of) substitution rules for a
beam-splitter (see also Eq.\ (\ref{bs})), polarisation rotation, POVM's 
and the trace operation. I then use these substitution rules
to `build' a model of the generalised experimental setup.

\subsection{Results}

The probability of creating one entangled 
photon-pair\index{photon-pair!entangled} using the weak 
parametric down-conversion\index{down-converter} source 1 or 2 is $p_1$ or 
$p_2$ respectively (see figure \ref{model}). I calculated the output state 
both for an $N$-cascade\index{detector!cascade} up 
to order $p^2$ (i.e.\ $p_1^2$ or $p_1 p_2$) and for a 1-cascade up to the order
$p^3$ ($p_1^3$, $p_1^2 p_2$ or $p_1 p_2^2$). The results are given below. For 
brevity, we take:
\begin{eqnarray}
 |\Psi_\theta\rangle &=& \cos\theta |0,1\rangle + e^{i\varphi}\sin\theta 
 |1,0\rangle \qquad\text{and}\cr
 |\Psi^{\perp}_\theta\rangle &=& e^{i\varphi}\sin\theta |0,1\rangle - 
 \cos\theta|1,0\rangle
\end{eqnarray}
as the ideally prepared state and the state orthogonal to it. Suppose 
$\eta^2_u$ and $\eta^2_v$ are the efficiencies of the 
detectors\index{detector!efficiency} in mode $u$ 
and $v$ respectively, and $\eta^2_c$ the efficiency of the detectors in the 
cascade (for simplicity I assume that the detectors in the cascade have the 
same efficiency). Define $g_{uvc} = \eta^2_u \eta^2_v \eta^2_c$. The detectors 
in modes $u$ and $v$ are polarisation insensitive, whereas the cascade 
consists of polarisation sensitive detectors. Bearing this in mind, we have up 
to order $p^2$ for an $N$-cascade in mode $a_x$ and finding no detector click 
in the $a_y$-mode:
\begin{equation}\label{ord2cas2}
 \rho_{\text{out}} \propto \frac{p_1}{8}g_{uvc} \biggl\{ \frac{p_1}{N} 
  [1+(5N-3)(1-\eta^2_c)]|0\rangle\langle 0|+ p_2 
  |\Psi_\theta\rangle\langle\Psi_\theta|\biggr\} + O(p^3) \; ,
\end{equation}
where the vacuum contribution\index{vacuum contribution} formula was 
calculated and found to be correct 
for $N\leq 4$ (and $N\neq 0$). 

In order to have non-post-selected quantum 
teleportation,\index{quantum teleportation} the fidelity\index{fidelity} $F$ 
must be larger than 3/4 \cite{massar95,massar99,fuchs97}. Since I only 
estimated 
the two lowest order contributions (to $p^2$ and $p^3$), the fidelity is 
also correct up to $p^2$ and $p^3$, and I write $F^{(2)}$ and $F^{(3)}$ 
respectively. Using Eqs.\ (\ref{fidelity}) and (\ref{ord2cas2}) we have:
\begin{equation}\label{2oinns}
 F^{(2)} = \frac{Np_2}{p_1 [1+(5N-3)(1-\eta_c^2)]+Np_2} \geq\frac{3}{4} \; ,
\end{equation}
\begin{equation}\label{2oeff}
 \Longleftrightarrow\qquad \eta_c^2 \geq \frac{(15N-6)p_1 - Np_2}{(15N-9)p_1}
 \; .
\end{equation}
This means that in the limit of infinite detector cascading 
($N\rightarrow\infty$) and $p_1 = p_2$ the efficiency of the detectors must be 
better than $\frac{14}{15}$ or 93.3\% to achieve non-post-selected quantum 
teleportation. When we 
have detectors with efficiencies of $98\%$, we need at least four detectors in 
the cascade to get unequivocal quantum teleportation. The necessity of a 
lower bound on the efficiency of the detectors used in the cascade might seem
surprising, but this can be explained as follows. Suppose the detector
efficiencies become smaller than a certain value $x$. Then upon a two-photon 
state entering the detector, finding only one click becomes more likely than 
finding a coincidence,\index{detector!coincidence} and `wrong' events end up 
contributing to the output 
state. Eq.\ (\ref{2oeff}) places a severe limitation on the practical use of 
detector cascades in this situation.

In the experiment in Innsbruck,\index{Innsbruck experiment} no detector 
cascade was employed and also the
$a_y$-mode was left undetected. The state entering Bob's detector therefore 
was (up to order $p^2$):
\begin{equation}\label{actual}
  \rho_{\text{out}} \propto \frac{p^2}{8} g_{uvc} \left[ 
  (3-\eta^2_c)|0\rangle\langle 0| + |\Psi_\theta\rangle
	\langle\Psi_\theta| \right] + O(p^3) \; .
\end{equation}
Remember that $p_1 = p_2$ since the experiment involves one source which
is pumped twice. The detector efficiency\index{detector!efficiency} $\eta_c^2$ 
in the Innsbruck experiment\index{Innsbruck experiment} was 10\% 
\cite{weinfurter98}, and the fidelity\index{fidelity} without detecting the
outgoing mode therefore would have been $F^{(2)}\simeq 26 \% $ (conditioned 
only on successful Bell detection\index{Bell!measurement} and state 
preparation).\index{state preparation} This clearly 
exemplifies the need for Bob's detection. Braunstein and Kimble 
\cite{braunstein98} predicted a theoretical maximum of 50\% for the 
teleportation\index{quantum teleportation} fidelity, which was conditioned upon (perfect) detection of both 
the $a_x$- and the $a_y$-mode.

Rather than improving the detector efficiencies and using a detector 
cascade,\index{detector!cascade} 
Eq.\ (\ref{2oinns}) can be satisfied by adjusting the probabilities $p_1$ and 
$p_2$ of creating entangled photon-pairs\index{photon-pair!entangled} 
\cite{braunstein98}. From Eq.\ (\ref{2oinns}) we have
\begin{equation}
 p_1 \leq \frac{N}{3[1+(5N-3)(1-\eta_c^2)]} p_2 \; .
\end{equation}
Experimentally, $p_1$ can be diminished by employing a 
beam-splitter\index{beam-splitter} with a 
suitable reflection coefficient rather than a mirror to reverse the pump beam 
(see figure \ref{fig:innsbruck}). Bearing in mind that $\kappa$ is proportional
to the pump amplitude, the equation $p_i = 2\tanh^2(\kappa_i t)$ [see the 
discussion following Eq.\ (\ref{EPRdist2}) with $i=1,2$] gives a relation 
between the pump amplitude and the probability of creating a photon-pair. In 
particular when $p_2 = x p_1$:
\begin{equation}
  \frac{\tanh(\kappa_2 t)}{\tanh(\kappa_1 t)} = \sqrt{x} \; .
\end{equation}
Decreasing the production rate of one photon-pair source will increase the 
time needed to run the experiment. In particular, we have from Eq.\ 
(\ref{actual}) that
\begin{equation}
 p_2 \geq 3(3-\eta_c^2) p_1 \; .
\end{equation}
With $\eta^2_c=10\%$, we obtain $p_2 \geq 8.7 p_1$. Using Eq.\ (\ref{speed}) I
estimated that diminishing the probability $p_1$ by a factor 8.7 will increase 
the running time by that same factor (i.e., running the experiment about nine 
days, rather than twenty four hours).

The third-order contribution to the outgoing density 
matrix\index{density matrix} without cascading and without detecting the 
$a_y$-mode is
\begin{multline}\label{3oinns}
 \rho_{\text{out}} \propto \frac{p_1}{8} g_{uvc} (4 - \eta_u^2 - \eta_v^2)
 \frac{1}{16} \left[ 6p_1^2 (6 - 4\eta_c^2 + \eta_c^4)~|0\rangle\langle 0| 
  \right. \cr 
  + 2 p_1 p_2 (2 - \eta_c^2) \left( |\Psi_\theta\rangle\langle\Psi_\theta| + 
 |\Psi^{\perp}_\theta\rangle\langle\Psi^{\perp}_\theta|  \right) + \cr
 \left. 8 p_1 p_2 (3 - \eta_c^2)\rho_1 + 12 p_2^2 \rho_2 \right] 
\end{multline}
with (we assume from now on that the phase factor $e^{i\varphi}$ in 
$|\Psi_\theta\rangle$ is real)
\begin{equation}
 \rho_1 = \frac{1}{2} \left( |1,0\rangle\langle 1,0| + |0,1\rangle
	\langle 0,1| \right) \; ,
\end{equation}
\begin{eqnarray}
 \rho_2 &=& \frac{1}{6} \left[ (2 + \cos 2\theta) |0,2\rangle\langle 0,2| +
	(2 - \cos 2\theta) |2,0\rangle\langle 2,0| \right. \cr &&
        ~+ 2 |1,1\rangle\langle 1,1| + \frac{1}{2}\sqrt{2}\sin 2\theta 
	\left( |2,0\rangle\langle 1,1| \right. \cr && \left.\left. ~+ 
	|1,1\rangle\langle 2,0| + |0,2\rangle\langle 1,1| + |1,1\rangle
	\langle 0,2| \right) \right] \; .
\end{eqnarray}
I have explicitly extracted the state which is to be teleported
($|\Psi_\theta\rangle\langle\Psi_\theta|$) from the density 
matrix\index{density matrix} contribution $\rho_1$ (this is not {\em 
necessarily} the decomposition\index{decomposition} with the largest 
$|\Psi_\theta\rangle\langle\Psi_\theta|$ contribution). As expected, this term 
is less important in the third order than it is in the second\footnote{The 
density matrix consists of several distinct parts: a vacuum 
contribution,\index{vacuum contribution} a 
contribution due to one photon in mode $d$, two photons, and so on. Suppose 
there are $n$ photon-pairs created in the whole system, and $m$ photon-pairs 
out of $n$ are produced by the second source (modes $c$ and $d$). The outgoing 
mode must then contain $m$ photons. Reversing this argument, when we find $m$ 
photons in the outgoing mode the probability of creating this particular 
contribution must be proportional to $p_1^{n-m} p_2^m$. Expanding the $n$-th 
order output state into parts of definite photon number we can write
\begin{equation}\nonumber
  \rho_{\text{out}}^{(n)} = \sum_{m=0}^{n-1} p_1^{n-m} p_2^m \rho_m^{(n)} \; ,
\end{equation}
where $\rho_m^{(n)}$ is the (unnormalised) $n$-th order contribution containing
all terms with $m$ photons. 

An immediate corollary of this argument is that all the cross-terms between
different photon number states in the density matrix must vanish. The 
cross-terms {\em are} present in Eq.\ (\ref{subout}), and I must therefore 
show that the partial trace\index{partial trace} in Eq.\ (\ref{out}) makes 
them vanish. Suppose 
there are $n$ photons in the total system. A cross-term in the density matrix
will have the form
\begin{equation}\nonumber
 | j, k, l, m \rangle_{auvd}\langle j', k', l', m' | \; ,
\end{equation}
with $m\neq m'$. We also know that $j+k+l+m=j'+k'+l'+m'=n$,
so that at least one of the other modes must have the cross-term property as 
well. Suppose $k$ is not equal to $k'$. Since we have 
Tr$[|k\rangle\langle k'|] = \delta_{k,k'}$, the cross-terms must vanish.}.

The teleportation\index{quantum teleportation} fidelity\index{fidelity} 
including the third-order contribution (\ref{3oinns}) can be derived along the 
same lines as (\ref{2oinns}). Assuming that all detectors have the same 
efficiency\index{efficiency} $\eta^2$ and $p_1=p_2=p$, the 
teleportation fidelity up to third order is
\begin{equation}
 F^{(3)} = \frac{4+p(2 - \eta^2)^2}{4 (4 - \eta^2) + 
	   p (80 - 76 \eta^2 + 34 \eta^4 - 3 \eta^6)} \; .
\end{equation}
With $p=10^{-4}$ and a detector efficiency of $\eta^2 = 0.1$, this 
fidelity differs from (\ref{2oinns}) with only a few parts in ten thousand:
\begin{equation}\label{ratio}
 \frac{F^{(2)}-F^{(3)}}{F^{(2)}} ~\propto~ p ~\sim~ 10^{-4} \; .
\end{equation} 

On the other hand, let me compare two {\em gedanken} 
experiments\index{gedanken@{\em gedanken} experiment} in which the 
cascades\index{detector!cascade} have different detector 
efficiencies\index{detector!efficiency} (but all the detectors in one cascade 
still have the same efficiency). The ratio between the teleportation 
fidelity\index{fidelity} with 
detector efficiencies $\eta^2_-$ and $\eta^2_+$ (with $\eta^2_-$ and $\eta^2_+$
the lower and higher detector efficiencies respectively) up to lowest order is
\begin{equation}
 \frac{F^{(2)}_{95\%}-F^{(2)}_{10\%}}{F^{(2)}_{95\%}} ~\propto~ \frac{\Delta
  \eta^2}{2-\eta^2_-} ~\sim~ 0.1 \; ,
\end{equation}
where $\Delta\eta^2$ is the difference between these efficiencies. This shows 
that detector efficiencies have a considerably larger influence on the 
teleportation fidelity than the higher-order pair production, as expected.

To summarise my results, I have found that detector cascading is only useful
for this realisation of quantum teleportation\index{quantum teleportation} 
when the detectors in the cascade have near unit efficiency, in accordance with
the results of chapter \ref{chap4}. In particular, there is a lower bound to 
the efficiency below which an increase in the number of detectors in the 
cascade actually {\em decreases} the ability to distinguish between one or 
several photons entering the cascade. Finally, enhancement of the photon-pair 
source responsible for the entanglement\index{entanglement} channel relative 
to the one responsible for the state preparation\index{state preparation} 
increases the time needed to run the experiment by roughly an order of 
magnitude. 

\subsection{Fidelity versus efficiency}\label{sec:fid+eff}

In the context of the Innsbruck experiment,\index{Innsbruck experiment} the 
fidelity is used to distinguish between quantum teleportation and 
teleportation which could have been achieved `classically'. Here, classical 
teleportation\index{classical!teleportation} is the disembodied 
transport of some quantum state from Alice to Bob by means of a classical 
communication channel alone. There is {\em no} shared entanglement between 
Alice and Bob. Since classical communication can be duplicated, such a scheme 
can lead to many copies of the transported output state (so-called {\em 
clones}).\index{clone} Classical teleportation with perfect fidelity (i.e., 
$F=1$) would then lead to the possibility of {\em perfect} cloning, thus 
violating the no-cloning theorem\index{no-cloning theorem} 
\cite{wootters82,dieks82}. 
This means that the maximum fidelity for classical teleportation has an upper 
bound which is less than one. 

Quantum teleportation, on the other hand, can achieve perfect fidelity (and 
circumvents the no-cloning theorem by disrupting the original). To demonstrate 
{\em quantum} teleportation therefore means that the teleported state should 
have a higher fidelity than possible for a state obtained by any scheme 
involving classical communication\index{classical!communication} {\em 
alone}\footnote{The fidelity captures this one particular feature of quantum 
teleportation very well and is already extensively studied.}.

For classical teleportation of randomly sampled polarisations, the maximum 
attainable fidelity is $F=2/3$ \cite{fuchs96,fuchs97}. When only linear 
polarisations are to be teleported, the maximum attainable 
fidelity\index{fidelity} is $F=3/4$ \cite{massar95,massar99}. These are the 
values which the fidelity of true quantum 
teleportation\index{quantum teleportation} should exceed.

In the case of the Innsbruck experiment,\index{Innsbruck experiment} 
$|\phi\rangle$ denotes the `unknown' linear polarisation state of the photon 
issued by Victor. I can write the undetected outgoing state (to lowest order 
and conditioned on a successful Bell state 
measurement)\index{Bell!measurement} as
\begin{equation}\label{rhoout}
 \rho_{\text{out}} \propto |\alpha|^2 |0\rangle\langle 0| + |\beta|^2 
   |\phi\rangle\langle\phi| \; ,
\end{equation}
where $|0\rangle$ is the vacuum state.\index{state!vacuum} The overlap between 
$|\phi\rangle$ and $\rho_{\text{out}}$ is given by Eq.\ (\ref{fiddef}). In the 
Innsbruck experiment the fidelity $F$ is then given by
\begin{equation}\label{fidelity}
 F \equiv \text{Tr}[\rho_{\text{out}}|\phi\rangle\langle\phi|] = 
 \frac{|\beta|^2}{|\alpha|^2 + |\beta|^2}\; .
\end{equation}
This should be larger than 3/4 in order to demonstrate quantum teleportation. 
The vacuum contribution in Eq.\ (\ref{rhoout}) arises from the fact that 
Victor cannot distinguish between one or several photons entering his detector,
i.e., Victor's inability to properly prepare a single-photon state.
\index{state!single-photon}

As pointed out by Braunstein and Kimble \cite{braunstein98}, the fidelity of 
the Innsbruck experiment remains well below the lower bound of 3/4 due to the 
vacuum contribution. Replying to this, Bouwmeester {\em et al}.\ 
\cite{bouwmeester98,bouwmeester99} argued that `when a photon appears, it has 
all the properties required by the teleportation protocol'. The vacuum 
contribution in Eq.\ (\ref{rhoout}) should therefore only affect the 
efficiency\index{efficiency} of the experiment, with a consequently high 
fidelity. However, 
this is a potentially ambiguous statement. If by `appear' we mean `appearing 
in a photo-detector',\index{detector!photo-} I agree that a high fidelity (and 
low efficiency) can 
be inferred. However, this yields a so-called {\em post-selected} 
fidelity,\index{post-selection}\index{fidelity!post-selected} 
where the detection destroys the teleported state. The fidelity prior to (or 
without) Bob's detection is called the {\em non}-post-selected fidelity. The 
question is now whether we can say that a photon appears when {\em no} 
detection is made, thus yielding a high {\em non}-post-selected fidelity. 

This turns out not to be the case. Making an {\em ontological} distinction 
between a photon and {\em no} photon in a mixed state (without a detection) is 
based on what we call the `Partition Ensemble Fallacy,'\index{PEF} introduced 
in chapter \ref{chap2}. In the absence of Bob's detection, the density matrix 
of the teleported state (i.e., the {\it non}-post-selected state) may be 
decomposed 
into an infinite number of partitions. These partitions do not necessarily 
include the vacuum state at all. It would therefore be incorrect to say that 
teleportation did or did not occur except through some operational means 
(e.g., a detection performed by Bob).

Bob's detection thus leads to a high post-selected fidelity. However, the
vacuum term in Eq.\ (\ref{rhoout}) contributes to the {\em non}-post-selected
fidelity, decreasing it well below the lower bound of 3/4. Due to this vacuum 
contribution, the Innsbruck experiment did {\em not} demonstrate {\em 
non}-post-selected quantum teleportation. Nonetheless, teleportation was 
demonstrated using post-selected data obtained by detecting the teleported 
state. By selecting events where a photon was observed in the teleported 
state, a post-selected 
fidelity\index{post-selection}\index{fidelity!post-selected} higher than 
$3/4$ could be inferred (estimated 
at roughly 80\% \cite{bouwmeester99})\footnote{We recall that this entire 
discussion is restricted to the subset of events where successful Bell-state 
and state-preparation have occurred.}.

\section{Entanglement swapping and purification}\label{es}\index{purification}

\subsection{Teleportation of entanglement: swapping}

In the previous section, I discussed quantum 
teleportation\index{quantum teleportation} \cite{bennett93}, in
which a quantum state is sent from Alice to Bob using (maximal) 
entanglement.\index{entanglement!maximal} 
If this quantum state is itself part of an entangled state, i.e., if Alice's 
system is entangled with Charlie's system, this entanglement is `transferred' 
from Alice to Bob. In other words, Bob's system becomes entangled with 
Charlie's system, even though these two systems might never have physically 
interacted. This is called {\em entanglement 
swapping}\index{entanglement!swapping} \cite{zukowski93}.

For example, suppose we have a system of two independent polarisation entangled
photon-pairs\index{photon-pair!entangled} in modes $a,b$ and $c,d$ 
respectively. If we restrict ourselves to the Bell states,\index{state!Bell} 
we have
\begin{equation}\label{3ent}
 |\Psi\rangle_{abcd} = |\Psi^-\rangle_{ab} \otimes |\Psi^-\rangle_{cd}\; .
\end{equation}
However, on a different basis this state can be written as:
\begin{eqnarray}\label{entswap}
 |\Psi\rangle_{abcd} &=& 
 \frac{1}{2}|\Psi^-\rangle_{ad} \otimes |\Psi^-\rangle_{bc} + 
 \frac{1}{2}|\Psi^+\rangle_{ad} \otimes |\Psi^+\rangle_{bc} \cr && \qquad + 
 \frac{1}{2}|\Phi^-\rangle_{ad} \otimes |\Phi^-\rangle_{bc} + 
 \frac{1}{2}|\Phi^+\rangle_{ad} \otimes |\Phi^+\rangle_{bc}\; .
\end{eqnarray}
This can be easily checked by writing out the Bell states. The non-cancelling
terms can be rewritten as Eq.\ (\ref{3ent}). 

If we make a Bell measurement\index{Bell!measurement} on modes $b$ and $c$, we 
can see from Eq.\
(\ref{entswap}) that the undetected remaining modes $a$ and $d$ become 
entangled. For instance, when we find modes $b$ and $c$ in a $|\Phi^+\rangle$ 
Bell state, the remaining modes $a$ and $d$ must be in the $|\Phi^+\rangle$ 
state as well. In appendix \ref{app:trans} I show that a suitably chosen
unitary transformation\index{transformation!unitary} of Bob's branch can 
return the state to 
$|\Psi^-\rangle_{ad}$, just as in the teleportation of a single state.

\begin{figure}[t]
\begin{center}
  \begin{psfrags}
     \psfrag{the}{\small\sc The Physical State}
     \psfrag{u}{$D_u$}
     \psfrag{v}{$D_v$}
     \psfrag{pbs}{\sc pbs}
     \psfrag{a}{$a$}
     \psfrag{b}{$b$}
     \psfrag{c}{$c$}
     \psfrag{d}{$d$}
     \psfrag{BS}{\sc bs}
     \psfrag{DC 1}{\sc pdc1}
     \psfrag{DC 2}{\sc pdc2}
     \epsfxsize=8in
     \epsfbox[-130 20 570 275]{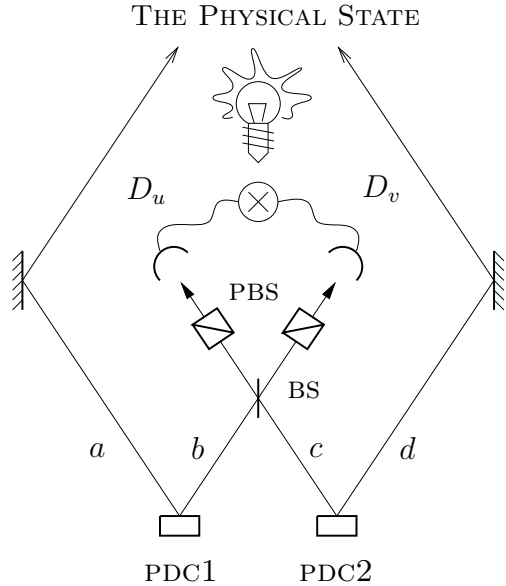}
  \end{psfrags}
  \end{center}
\caption{A schematic representation of the entanglement swapping setup.
 Two parametric down-converters ({\sc pdc}) create states which exhibit
 polarisation entanglement. One branch of each source is sent into a beam
 splitter ({\sc bs}), after which the polarisation beam splitters ({\sc pbs})
 select particular polarisation settings. A coincidence in detectors $D_u$
 and $D_v$ ideally identify the $|\Psi^-\rangle$ Bell state. However, since
 there is a possibility that one down-converter produces two photon-pairs
 while the other produces nothing, the detectors $D_u$ and $D_v$ no longer
 constitute a Bell-detection, and the freely propagating {\sc physical state}
 is no longer a pure Bell state.}
\label{fig:3.2}
\end{figure}

Entanglement swapping was performed in Innsbruck by Pan {\em et 
al}.\ in 1998 \cite{pan98}. In this experiment two parametric down-converters
\index{down-converter} were employed to create polarisation 
\index{entanglement!polarisation}entanglement\footnote{In the experiment,
the two down-converters were implemented by a single BBO 
crystal\index{BBO crystal} pumped twice 
in opposite directions. The experiment thus closely resembled the quantum
teleportation experiment six months earlier \cite{bouwmeester97} (see also 
chapter \ref{chap2}).}. The schematics of the experimental setup are depicted 
in Fig.~\ref{fig:3.2}. One branch of each down-converter is sent into a 50:50 
beam-splitter.\index{beam-splitter} The outgoing modes of the beam-splitter 
are detected. A detector\index{detector!coincidence} coincidence indicates 
that the state $|\Psi^-\rangle$ was present, and thus acts as an (incomplete) 
Bell measurement.\index{Bell!measurement} I have included two polarisation 
beam-splitters in the outgoing modes of the beam-splitter. These were not 
present in the actual experiment, but they play an important r\^ole in the 
subsequent discussion \cite{zukowski99}.

Since entanglement swapping\index{entanglement!swapping} is formally the 
teleportation\index{quantum teleportation} of one branch of 
an entangled state, it should not come as a surprise that the entanglement
swapping experiment performed by Pan {\em et al}.\ suffers from the same 
complication as the quantum teleportation experiment performed by Bouwmeester
{\em et al}.\ \cite{bouwmeester97}: apart from both 
down-converters\index{down-converter} creating
a single pair, there is also the possibility that one of the down-converters
creates two pairs, while nothing happens in the other. 

Now I use the extra information about the detected photons due to the 
polarisation beam-splitters in figure \ref{fig:3.2}. Conditioned on the 
polarisation $(j,k)$, with $j,k\in\{ x,y\}$, of the detected photons we obtain
the outgoing states
\begin{eqnarray}\label{entswapout}
 |\Upsilon_{(x,x)}\rangle_{14} &=& \frac{1}{\sqrt{2}} \left(
 |0,y^2\rangle - |y^2,0\rangle \right)\; , \cr
 |\Upsilon_{(x,y)}\rangle_{14} &=& \frac{1}{2} \left( |xy,0\rangle -
 |x,y\rangle + |y,x\rangle - |0,xy\rangle \right)\; , \cr
 |\Upsilon_{(y,x)}\rangle_{14} &=& \frac{1}{2} \left( |xy,0\rangle + 
 |x,y\rangle - |y,x\rangle - |0,xy\rangle \right)\; , \cr
 |\Upsilon_{(y,y)}\rangle_{14} &=& \frac{1}{\sqrt{2}} \left(
 |0,x^2\rangle - |x^2,0\rangle \right)\; .
\end{eqnarray}
The outgoing state of the entanglement swapping\index{entanglement!swapping} 
experiment (without the polarisation beam-splitter)\index{beam-splitter} is a 
random mixture of these four states.

Let me define the following states:
\begin{eqnarray}\label{entswapout2}
 |\Phi_{xy}\rangle &\equiv& \frac{1}{\sqrt{2}} \left( |xy,0\rangle - 
 |0,xy\rangle \right)\; , \cr
 |\Phi_{x^2}\rangle &\equiv& \frac{1}{\sqrt{2}} \left( |x^2,0\rangle - 
 |0,x^2\rangle \right)\; , \cr
 |\Phi_{y^2}\rangle &\equiv& \frac{1}{\sqrt{2}} \left( |y^2,0\rangle - 
 |0,y^2\rangle \right)\; , \cr
 |\Psi^-\rangle &\equiv& \frac{1}{\sqrt{2}} \left( |x,y\rangle - 
 |y,x\rangle \right)\; .
\end{eqnarray}
After some involved, but essentially straightforward algebra it can be shown 
that the outgoing state of the entanglement swapping experiment performed by 
Pan {\em et al}.\ can also be written as the mixed state\index{state!mixed} 
$\rho$:
\begin{equation}
 \rho = \frac{1}{4} \left( |\Phi_{xy}\rangle_{ad}\langle\Phi_{xy}| + 
 |\Phi_{x^2}\rangle_{ad}\langle\Phi_{x^2}| + |\Phi_{y^2}\rangle_{ad}
 \langle\Phi_{y^2}| + |\Psi^-\rangle_{ad}\langle\Psi^-| \right)
\end{equation}
to lowest order. {\em Conditioned} on detected photons in the outgoing modes 
a high entanglement swapping fidelity can be inferred ($F\sim 1$). However, 
the fidelity\index{fidelity} for non-post-selected\index{post-selection} 
entanglement swapping is $F=\frac{1}{4}$. This argument is completely 
analogous to the argument presented in section
\ref{sec:teleportation}{\ref{sec:fid+eff}}. 

\subsection{Entanglement swapping as purification}\index{purification}

The non-post-selected fidelity $F=\frac{1}{4}$ of having a maximally 
entangled state\index{state!maximally entangled} $|\Psi^-\rangle$ as the 
output of the entanglement swapping experiment is much higher than the that of 
the down-converter\index{down-converter} output state,
where $F\sim 10^{-4}$. This suggests that entanglement swapping can be viewed
as a purification protocol (see also chapter \ref{chap3}). Indeed, this has 
been suggested by Bose {\em et al}.\ \cite{bose99}. This protocol was 
subsequently extended by Shi {\em et al}.\ \cite{shi00}.

The Bose protocol works as follows: let $x$ and $y$ denote photons with 
polarisations in the $x$- and $y$-direction of a Cartesian coordinate system.
We consider (an ensemble of) non-maximally entangled states for two systems 1 
and 2
\begin{equation}\label{bose1}
 |\Phi(\theta)\rangle_{12} = \cos\theta |x,x\rangle_{12} + \sin\theta 
 |y,y\rangle_{12}\; ,
\end{equation}
and similarly for systems 3 and 4:
\begin{equation}\label{bose2}
 |\Phi(\theta)\rangle_{34} = \cos\theta |x,x\rangle_{34} + \sin\theta 
 |y,y\rangle_{34}\; .
\end{equation}
The purification\index{purification} protocol now employs entanglement 
swapping\index{entanglement!swapping} from systems 1, 2 and 3, 4 to the two 
systems 1 and 4. These two systems were previously unentangled. Making a Bell 
measurement\index{Bell!measurement} of system 2 and 3 entangles the
remaining systems 1 and 4.

There are four different outcomes of the Bell measurement, which give rise 
to four different entangled states in systems 1 and 4. These are \cite{bose99}
\begin{eqnarray}
 |\Phi^+\rangle_{23} :&& |\Psi_{\rm out}\rangle_{14} = \frac{1}{N} \left(
 \cos^2\theta |x,x\rangle_{14} + \sin^2\theta |y,y\rangle_{14}\right)\; , \cr
 |\Phi^-\rangle_{23} :&& |\Psi_{\rm out}\rangle_{14} = \frac{1}{N} \left(
 \cos^2\theta |x,x\rangle_{14} - \sin^2\theta |y,y\rangle_{14}\right)\; , \cr
 |\Psi^+\rangle_{23} :&& |\Psi_{\rm out}\rangle_{14} = \frac{1}{\sqrt{2}} 
	\left( |x,y\rangle_{14} + |y,x\rangle_{14} \right)\; , \cr
 |\Psi^-\rangle_{23} :&& |\Psi_{\rm out}\rangle_{14} = \frac{1}{\sqrt{2}} 
	\left( |x,y\rangle_{14} - |y,x\rangle_{14} \right)\; .
\end{eqnarray}
The normalisation factor $N$ is given by $N=\sqrt{\cos^4\theta+\sin^4\theta}$.
It is easily seen that the measurement outcomes $|\Phi^+\rangle_{23}$ and
$|\Phi^-\rangle_{23}$ actually {\em degrade} the entanglement compared to
the entanglement of the systems 1 and 2 or 3 and 4. In the case of measurement
outcomes $|\Psi^+\rangle_{23}$ and $|\Psi^-\rangle_{23}$, however, the 
resulting (pure) states are {\em maximally} entangled. With probability 
$\sqrt{2}\cos\theta\sin\theta$ we will obtain a maximally entangled state,
\index{state!maximally entangled} and with probability 
$\sqrt{\cos^4\theta+\sin^4\theta}$ we degrade the entanglement.

When we compare this protocol with the swapping experiment by Pan {\em et al}.,
we note that there is a crucial difference: The outgoing state of the 
experiment is not confined to the Hilbert space\index{Hilbert space} spanned 
by the basis $\{ |x,x\rangle, |x,y\rangle, |y,x\rangle, |y,y\rangle\}$, 
contrary to the protocol by Bose {\em et al}. In addition, we have to include 
states like the vacuum\index{state!vacuum} ($|0\rangle$) and two-photon states 
($|x^2\rangle$, $|y^2\rangle$ and $|xy\rangle$).

However, if the swapping protocol used by Pan {\em et al}.\ can increase the
entanglement content upon repetition of the procedure in this larger Hilbert 
space (or, more precisely, this truncated Fock space),\index{Fock space} we 
can still call it a purification protocol. I will now investigate this.

\begin{figure}[t]
\begin{center}
  \begin{psfrags}
     \psfrag{1}{1}
     \psfrag{2}{2}
     \psfrag{3}{3}
     \psfrag{N}{$N$}
     \psfrag{ldots}{$\ldots$}
     \psfrag{Bell detections}{photo-detections}
     \epsfxsize=8in
     \epsfbox[-100 40 700 175]{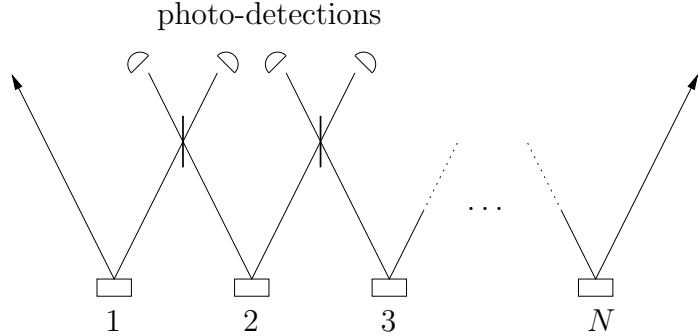}
  \end{psfrags}
  \end{center}
\caption{A series of parametric down-converters 1 to $N$, of which the 
	outgoing modes are connected by beam-splitters to form a string. The
	photo-detections are essentially {\em polarisation sensitive} 
	photo-detectors (an incomplete Bell measurement would require the
	loss of the polarisation information). This can be interpreted as 
	repeated entanglement swapping. However, is it also a repeated 
	purification protocol?}
\label{fig:string}
\end{figure}

First, I have to determine precisely what I mean by a repetition of the 
swapping procedure. It means that the {\em outgoing states} of two distinct
entanglement swapping setups are again used as an entanglement 
source\index{entanglement!sources} for a swapping experiment. Repeating this 
$N$ times, we can depict this as a string of 
down-converters\index{down-converter} connected by 
beam-splitters\index{beam-splitter} (see figure \ref{fig:string}).
Such a string can consist of an even or an odd number of 
down-converters.\index{down-converter} When $N$ is odd, we have a string of 
`entanglement swappers' the outgoing states of which are again used in 
entanglement swapping.\index{entanglement!swapping}

Instead of Bell detections,\index{Bell!measurement} we consider polarisation 
sensitive photo-detection.\index{detector!photo-}
This allows us to condition the outgoing state on $x$- and $y$-polarised 
photons in the detectors. In the case $N=1$ this led to the outgoing state
\begin{equation}
 |\Upsilon_{(x,y)}\rangle = \frac{1}{\sqrt{2}} \left( |xy,0\rangle -
	|0,xy\rangle \right)\; .
\end{equation}
In the case of $N$ down-converters, we can keep track of the single- and 
double-pair production in the following table:
\begin{center}
\begin{tabular}{r|l|l|l|c|l}
 {\sc pdc} & 1 & 2 & 3 &  & $N$ \\
 \hline $\#$ pairs & 1 & 1 & 1 & \ldots & 1 \\
		 & 0 & 2 & 0 & \ldots & 0/2 (odd/even) \\
		 & 2 & 0 & 2 & \ldots & 2/0 (odd/even)
\end{tabular}
\end{center}
In the top row the parametric down-converters are enumerated (in accordance 
with figure \ref{fig:string}). The entries in the lower rows identify the 
number of photon-pairs created by the associated down-converter. These rows 
identify the only three possibilities in which the detectors (from left to 
right) signal the detection of polarised photons in the direction $x$, $y$,
$x$, $y\ldots$

There are several things to be noted. First of all, depending on the parity of 
$N$, the bottom three rows correspond to different orders of 
pair-creation.\index{pair-creation} The lowest order corresponds to no created 
pairs (vacuum),\index{state!vacuum} the next order is one 
created pair, and so on. Since parametric down-conversion has such a small 
probability of creating an entangled photon-pair,\index{photon-pair!entangled} 
the lowest order is always by far the leading order.

Secondly, we can easily verify that the possibility of every 
down-converter\index{down-converter} creating exactly {\em one} photon-pair 
never occurs alone. We can always construct a different photon-pair 
configuration of the same order which triggers the detectors in the same way, 
thus preventing the creation of maximal 
entanglement\index{entanglement!maximal} (this is, of course, not the proof I 
was looking for in section \ref{sec:max} of chapter \ref{chap3}). 
At the same time, it is easily 
verified that (for ideal detections) the possibility that one down-converter 
creates {\em three} pairs is dismissed, since it would mean that some 
detectors see at least two photons. Furthermore, this procedure can be 
immediately repeated for other polarisation choices in the detectors. 

Finally, we should note that the entanglement content alternates between almost
nothing ($F\sim 10^{-4}$) and about one quarter ($F=\frac{1}{4}$). This 
behaviour occurs because for an odd number of down-converters, the detectors 
in the setup can be triggered by $(N-1)/2$ down-converters creating a
double-pair. The total number of pairs is then $N-1$, which is the lowest 
order. 

When we have an even number of down-converters we have three possibilities 
given in the table above. When every down-converter creates one photon-pair,
the outgoing state will be (up to lowest order) the anti-symmetric Bell state.
\index{state!Bell} In this case, all down-converter modes are connected by 
means of the beam-splitter\index{beam-splitter} operations and entanglement 
swapping\index{entanglement!swapping} is successful. On the other 
hand when the down-converters create two and zero photon pairs in alternation,
the left-hand outgoing mode is independent from the right-hand outgoing mode.
There is only a classical correlation\index{classical!correlations} between 
them: if on the left there are two photons, then we have the vacuum 
state\index{state!vacuum} on the right and vice versa.

The outgoing state is 
\begin{equation}
 |\Upsilon\rangle = \frac{1}{2} \left( |0,xy\rangle - |x,y\rangle + |y,x\rangle
 - |xy,0\rangle \right)\; ,
\end{equation}
conditioned on a detector sequence $x$, $y$, $x$, $y$\ldots This is independent
of the number of down-converters, as long as it is even. As a consequence, 
we cannot interpret entanglement swapping as performed by Pan {\em et al}.\ as 
a purification\index{purification} protocol.

\subsection{Entanglement content of output states}

I will now return to the states given by Eq.~(\ref{entswapout}). These states 
can also be obtained by running the states $|x,x\rangle$, $|x,y\rangle$, 
$|y,x\rangle$ and $|y,y\rangle$ through a 50:50 beam-splitter. This raises the
question what the entanglement content of the states of Eq.~(\ref{entswapout}) 
is. After all, the states $|j,k\rangle$ (with $j,k\in\{ x,y\}$) are {\em 
separable}.\index{separability}

The non-locality\index{non-locality} of single photons has been studied by 
Hardy \cite{hardy94,hardy95} and Peres \cite{peres95c} (see also chapter 
\ref{chap2}). Let $a$ and $b$ denote 
different spatial modes, $|0\rangle$ is the vacuum and $|1\rangle$ is a
single-photon state.\index{state!single-photon} The essential idea is that 
the state 
\begin{equation}
 |\Psi\rangle = \frac{1}{\sqrt{2}} \left( |1\rangle_a |0\rangle_b -
 |0\rangle_a |1\rangle_b \right)
\end{equation}
can be used to construct the violation of the Clauser-Horne-Shimony-Holt, or
CHSH inequality\index{CHSH inequality} \cite{clauser69,peres95} (this is a 
variant of a Bell\index{Bell!inequality} inequality \cite{bell64}). 
Consequently, a single-photon state can exhibit non-local 
properties. This point was debated by Vaidman \cite{vaidman95} and Greenberger,
Horne and Zeilinger \cite{greenberger95,greenberger96}.

Here, I will discuss the entanglement content of the two-photon states 
which are obtained by mixing two single-photon states at a 50:50 beam-splitter.
\index{beam-splitter} To this end I use the so-called Peres-Horodecki partial 
transpose criterion\index{partial transpose!criterion} for density 
matrices\index{density matrix} \cite{horodecki96,peres96} (see appendix 
\ref{app:povm}).
Using the partial transpose criterion I will examine the entanglement 
content of the state $(|x^2,0\rangle - |0,x^2\rangle)/\sqrt{2}$ and the 
density matrix $\rho$, which is a mixture of the states given in 
Eq.~(\ref{entswapout}).

We can write the density matrix of the state $(|x^2,0\rangle - |0,x^2\rangle)
/\sqrt{2}$ as 
\begin{equation}
 \rho = \frac{1}{2} \left( |0,x^2\rangle\langle 0,x^2| -
 |0,x^2\rangle\langle x^2,0| - |x^2,0\rangle\langle 0,x^2| + 
 |x^2,0\rangle\langle x^2,0| \right)\; .
\end{equation}
We obtain the partial transpose by exchanging the second entries of the
bras and kets, yielding
\begin{equation}
 \rho' = \frac{1}{2} \left( |0,x^2\rangle\langle 0,x^2| -
 |0,0\rangle\langle x^2,x^2| - |x^2,x^2\rangle\langle 0,0| + 
 |x^2,0\rangle\langle x^2,0| \right)\; .
\end{equation}
In matrix representation\index{matrix!representation} on the basis 
$\{ |0,0\rangle,|0,x^2\rangle,
|x^2,0\rangle,|x^2,x^2\rangle, \}$ this becomes\footnote{This is a matrix on 
a truncated Fock space\index{Fock space} corresponding to the given basis.}
\begin{equation}
 \rho' = \frac{1}{2} 
 \begin{pmatrix}
   0 & 0 & 0 & -1 \cr
   0 & 1 & 0 &  0 \cr
   0 & 0 & 1 &  0 \cr
  -1 & 0 & 0 &  0 
 \end{pmatrix}
\end{equation}
The eigenvalues\index{eigen!-values} of this matrix are $1/2$ (with multiplicity 
3) and $-1/2$. As proved in appendix \ref{app:povm}, the negative eigenvalues 
imply that the state $(|0,x^2\rangle - |x^2,0\rangle)/\sqrt{2}$ is entangled.

In the experiment\index{entanglement!swapping} performed by Pan {\em et al}., 
no polarisation beam-splitters were used, and the outgoing state before 
post-selection\index{post-selection} was
a mixture of the states given in Eq.~(\ref{entswapout2}). The density matrix 
can be written as
\begin{eqnarray}
 \rho &=& \frac{1}{8} \biggl[ |0,x^2\rangle\langle 0,x^2| -
 |0,x^2\rangle\langle x^2,0| - |x^2,0\rangle\langle 0,x^2| \biggr. \cr 
 &&\quad + |x^2,0\rangle\langle x^2,0| + |0,y^2\rangle\langle 0,y^2| - 
 |0,y^2\rangle\langle y^2,0| - |y^2,0\rangle\langle 0,y^2| \cr 
 &&\quad + |y^2,0\rangle\langle y^2,0| + |xy,0\rangle\langle xy,0| - 
 |xy,0\rangle\langle 0,xy| - |0,xy\rangle\langle xy,0| \cr 
 &&\quad + |0,xy\rangle\langle 0,xy| + |x,y\rangle\langle x,y| - 
 |x,y\rangle\langle y,x| \cr &&\quad\biggl. - |y,x\rangle\langle x,y| + 
 |y,x\rangle\langle y,x| \biggr]\; .
\end{eqnarray}
The partial transpose\index{partial transpose} then becomes
\begin{eqnarray}
 \rho' &=& \frac{1}{8} \biggl[ |0,x^2\rangle\langle 0,x^2| -
 |0,0\rangle\langle x^2,x^2| - |x^2,x^2\rangle\langle 0,0| \biggr. \cr 
 &&\quad \left. + |x^2,0\rangle\langle x^2,0| + |0,y^2\rangle\langle 0,y^2| - 
 |0,0\rangle\langle y^2,y^2| - |y^2,y^2\rangle\langle 0,0| \right. \cr
 &&\quad \left. + |y^2,0\rangle\langle y^2,0| + |xy,0\rangle\langle xy,0| - 
 |xy,xy\rangle\langle 0,0| - |0,0\rangle\langle xy,xy| \right. \cr
 &&\quad + |0,xy\rangle\langle 0,xy| + |x,y\rangle\langle x,y| - 
 |x,x\rangle\langle y,y| \cr &&\quad\biggl. - |y,y\rangle\langle x,x| + 
 |y,x\rangle\langle y,x| \biggr]\; .
\end{eqnarray}
In matrix representation\index{matrix!representation} on the basis 
\begin{eqnarray}\nonumber
 && \{ |0,0\rangle,|0,x^2\rangle,|x^2,0\rangle,|x^2,x^2\rangle, |0,y^2\rangle, 
 |y^2,0\rangle, |y^2,y^2\rangle, \cr && \qquad\qquad|xy,0\rangle, |0,xy\rangle,
 |xy,xy\rangle, |x,y\rangle, |y,x\rangle, |x,x\rangle, |y,y\rangle \}\; ,
\end{eqnarray}
the partial transpose $\rho'$ becomes\footnote{Again on a truncated Fock
space.}
\begin{equation}
 \rho' = \frac{1}{8}
 \left(
 \begin{array}{rrrrrrrrrrrrrr}
   0 &  0 &  0 & -1 &  0 &  0 & -1 &  0 & 0 & -1 & 0 & 0 & 0 & 0 \cr
   0 &  1 &  0 &  0 &  0 &  0 &  0 &  0 & 0 &  0 & 0 & 0 & 0 & 0 \cr
   0 &  0 &  1 &  0 &  0 &  0 &  0 &  0 & 0 &  0 & 0 & 0 & 0 & 0 \cr
  -1 &  0 &  0 &  0 &  0 &  0 &  0 &  0 & 0 &  0 & 0 & 0 & 0 & 0 \cr
   0 &  0 &  0 &  0 &  1 &  0 &  0 &  0 & 0 &  0 & 0 & 0 & 0 & 0 \cr
   0 &  0 &  0 &  0 &  0 &  1 &  0 &  0 & 0 &  0 & 0 & 0 & 0 & 0 \cr
  -1 &  0 &  0 &  0 &  0 &  0 &  0 &  0 & 0 &  0 & 0 & 0 & 0 & 0 \cr
   0 &  0 &  0 &  0 &  0 &  0 &  0 &  1 & 0 &  0 & 0 & 0 & 0 & 0 \cr
   0 &  0 &  0 &  0 &  0 &  0 &  0 &  0 & 1 &  0 & 0 & 0 & 0 & 0 \cr
  -1 &  0 &  0 &  0 &  0 &  0 &  0 &  0 & 0 &  0 & 0 & 0 & 0 & 0 \cr
   0 &  0 &  0 &  0 &  0 &  0 &  0 &  0 & 0 &  0 & 1 & 0 & 0 & 0 \cr
   0 &  0 &  0 &  0 &  0 &  0 &  0 &  0 & 0 &  0 & 0 & 1 & 0 & 0 \cr
   0 &  0 &  0 &  0 &  0 &  0 &  0 &  0 & 0 &  0 & 0 & 0 & 0 & -1 \cr
   0 &  0 &  0 &  0 &  0 &  0 &  0 &  0 & 0 &  0 & 0 & 0 & -1 & 0 
 \end{array}
 \right)\; .
\end{equation}
The eigenvalues\index{eigen!-values} of this matrix are given by $0$ 
(multiplicity 2), $1/8$ (multiplicity 9), $-1/8$ (multiplicity 1), 
$\sqrt{3}/8$ (multiplicity 1) and 
$-\sqrt{3}/8$ (multiplicity 1). Since $\rho'$ has negative eigenvalues, $\rho$
is an entangled state\index{state!entangled} \cite{peres96,horodecki96}. This 
means that entanglement swapping\index{entanglement!swapping} as it was 
originally proposed really does work, although maximally
entangled (Bell) states\index{state!Bell} can only be seen in a 
post-selected\index{post-selection} manner.

This teaches us something interesting about entanglement. It is generally 
believed that when two systems are entangled, they have somehow interacted in
the past. However, in the beam-splitter\index{beam-splitter} the two photons 
do {\em not} interact
with each other, and yet the outgoing state (given a separable input state)
is entangled. This shows that entanglement does not necessarily originates 
from an interaction.

\section{Three-particle entanglement}\label{sec:ghz}
\index{entanglement!three-particle}

\begin{figure}[t]
\begin{center}
  \begin{psfrags}
     \psfrag{T}{$T$}
     \psfrag{D1}{$D_1$}
     \psfrag{D2}{$D_2$}
     \psfrag{D3}{$D_3$}
     \psfrag{PBS1}{PBS$_1$}
     \psfrag{PBS2}{PBS$_2$}
     \psfrag{BS}{BS}
     \psfrag{crystal}{crystal}
     \psfrag{lambda}{$\lambda/2$}
     \psfrag{a}{$a$}
     \psfrag{b}{$b$}
     \psfrag{pump}{pump}
     \epsfxsize=8in
     \epsfbox[-125 40 625 280]{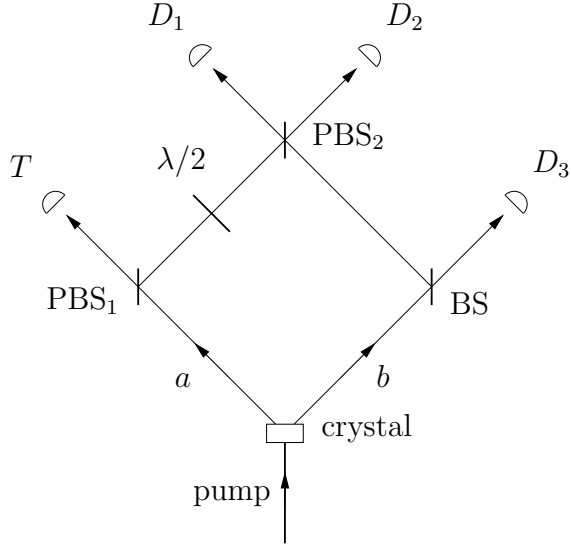}
  \end{psfrags}
  \end{center}
\caption{Schematic representation of the experimental setup which was used 
	to demonstrate the existence of three-photon GHZ-states in a 
	post-selected manner. A BBO crystal is pumped to create {\em two} 
	photon pairs. The subsequent interferometer is arranged such that
	conditioned on a detection event in detector $T$, the detectors $D_1$,
	$D_2$ and $D_3$ signal the detection of a GHZ-state. Furthermore,
	the interferometer includes polarisation beam-splitters (PBS$_1$ and 
	PBS$_2$), a beam-splitter (BS) and a $\lambda/2$ phase plate
	which transforms $|y\rangle$ into $(|x\rangle+|y\rangle)/\sqrt{2}$.}
\label{fig:ghz}
\end{figure}

Three-particle maximally entangled states have also been produced in a 
post-selected\index{post-selection} manner \cite{bouwmeester99b}. In figure 
\ref{fig:ghz} a schematic representation of this experiment is shown 
\cite{zeilinger97}. As in the teleportation\index{quantum teleportation} and 
entanglement swapping\index{entanglement!swapping} experiments, a non-linear 
crystal is pumped with a short-pulsed high-intensity laser. However, 
this time the setup is chosen such that the 
down-converter\index{down-converter} directly creates 
two photon-pairs in modes $a$ and $b$. The lowest order of the created state  
which can trigger the four detectors ($T$, $D_1$, $D_2$ and $D_3$) is given by
\begin{equation}
 |\Psi\rangle_{ab} \simeq \frac{1}{\sqrt{3}} \left( |x^2,y^2\rangle - 
 |xy,xy\rangle + |y^2,x^2\rangle \right) + O(\xi)\; ,
\end{equation}
with $\xi\ll 1$.

It is clear that only the branch $|xy,xy\rangle_{ab}$ can trigger all detectors
due to the polarisation beam-splitter\index{beam-splitter} PBS$_1$ (up to 
lowest order). This 
polarisation beam-splitter transmits $x$-polarised photons (which trigger $T$) 
and the retardation plate transforms the reflected photon $|y\rangle$ into 
$(|x\rangle+|y\rangle)/\sqrt{2}$. The part of the outgoing state (i.e., before 
detection) due to the input $|xy,xy\rangle_{ab}$ and conditioned on a photon 
in detector $T$ is thus
\begin{eqnarray}
 |\Phi\rangle &\propto& |xy,x,0\rangle + |x,x,y\rangle + |xy,0,x\rangle +
 |x,0,xy\rangle \cr &&\quad + |y,xy,0\rangle + |0,xy,y\rangle + |y,y,x\rangle +
 |0,y,xy\rangle\; .
\end{eqnarray}
From this state it is easily seen that three-photon 
entanglement\index{entanglement!three-photon} can only be observed in a 
post-selected\index{post-selection} manner by discarding the branches which 
include the vacuum\index{state!vacuum} $|0\rangle$. A three-fold coincidence 
in the detectors\index{detector!coincidence} $D_1$, $D_2$ 
and $D_3$ can thus only come from the branches $|x,x,y\rangle$ and 
$|y,y,x\rangle$.

As I argued in section \ref{postselection}, this experiment does {\em not} 
demonstrate the existence of the GHZ-state\index{state!GHZ-} $|x,x,y\rangle +
|y,y,x\rangle$. However, using post-selection of the data-set on a four-fold 
detector coincidence, non-local correlations\index{non-local correlations} 
could be inferred. This was demonstrated by Pan {\em et al}.\ in 2000 
\cite{pan00}. In the first experiment \cite{bouwmeester99b}
two tests were made: first it was shown that {\em only} the branches 
$|x,x,y\rangle$ and $|y,y,x\rangle$ contributed to a four-fold coincidence. 
Secondly it had to be shown that the two branches were in a coherent 
superposition,\index{superposition} since the post-selected events could be 
due to a statistical mixture\index{statistical!mixture} of the two branches. 
This was done by rotating the polarisation over 
$\pm 45^{\circ}$ before detection. The observed visibility in these 
experiments was 75\% \cite{bouwmeester99b}.

\section{Summary}

In this chapter I studied the optical implementations of quantum teleportation,
\index{quantum teleportation} entanglement 
swapping\index{entanglement!swapping} and the creation of three-photon 
Green\-berger-Horne-Zei\-linger entanglement. All these experiments succeeded 
in a {\em post-selected} manner and demonstrated their respective non-local 
features. 

The undetected outgoing state of the teleportation experiment is a mixture 
of the teleported state and the vacuum. As we have shown, this vacuum 
contribution\index{vacuum contribution} degrades the non-post-selected 
\index{post-selection} fidelity,\index{fidelity} rather than the 
efficiency\index{efficiency} of the experiment. The outgoing state of the 
entanglement swapping experiment is more complicated, since it is not simply a 
mixture of the vacuum and the swapped state. Here, the photo-detection 
post-selects particular branches from a superposition. 

The same happens in the creation of the three-photon GHZ-state. Furthermore,
since no physically propagating state left the apparatus after detection, we 
cannot say that the state $|x,x,y\rangle + |y,y,x\rangle$ was created. 
However, non-local correlations using post-selected data can be observed.

\newpage
\thispagestyle{empty}

\chapter{Quantum Lithography}\label{lithography}
\index{quantum lithography|(}

Optical lithography\index{lithography!optical} is a widely used printing 
method. In this process light is 
used to etch a substrate. The (un)exposed areas on the substrate then define
the pattern. In particular, the micro-chip industry uses 
lithography\index{lithography} to produce
smaller and smaller processors. However, classical optical lithography can 
only achieve a resolution\index{resolution} comparable to the wavelength of 
the used light 
\cite{brueck98,mack96,mansuripur00}. It therefore limits the scale of the 
patterns. To create smaller patterns we need to venture beyond this classical 
boundary \cite{yablonovich99}. Here, I investigate how we can beat this
boundary. This chapter is based on a collaboration with Agedi N.\ Boto, 
Daniel S.\ Abrams, Colin P.\ Williams and Jonathan P.\ Dowling at the Jet
Propulsion Laboratory, Pasadena \cite{boto00,kok00e,kok00f}. Recently, similar 
work was done by Bj\"ork, S\'anchez Soto and S\"oderholm \cite{bjork00}.

In Ref.\ \cite{boto00} we introduced a procedure called {\em quantum} 
lithography which offers an increase in resolution without an upper bound. 
This enables us to use quantum lithography to write closely spaced lines in 
one dimension. However, for practical purposes (like, e.g., optical surface 
etching)\index{surface etching} we need the ability to create more complicated 
patterns in both one 
and two dimensions. Here, we study how quantum lithography allows us to 
create these patterns.

This chapter is organised as follows: first I derive the classical resolution
limit in section \ref{resolution}. Section \ref{intro} reiterates 
the method introduced in Ref.\ \cite{boto00}. Then, in section \ref{1D}, I 
consider a generalised version of this procedure and show how we can tailor 
arbitrary one-dimensional patterns. Section \ref{2D} shows how a 
further generalisation of this procedure leads to arbitrary patterns in two 
dimensions. Finally, section \ref{phys} addresses the issues concerning the 
physical implementation of quantum lithography.

\section{Classical resolution limit}\label{resolution}

Classically, we can only resolve details of finite size. In this section 
we give a derivation of this classical resolution limit\index{resolution!limit}
using the so-called Rayleigh criterion\index{Rayleigh criterion} 
\cite{rayleigh1879}. Suppose two plane waves\index{plane wave} characterised 
by $\vec{k}_1$ and $\vec{k}_2$ hit a surface under an angle $\theta$ from the 
normal vector. The wave vectors are given by
\begin{equation}\label{planewave}
 \vec{k}_1 = k(\cos\theta,\sin\theta) \quad\mbox{and}\quad
 \vec{k}_2 = k(\cos\theta,-\sin\theta)\; ,
\end{equation}
where we used $|\vec{k}_1|=|\vec{k}_2|=k$. The wave number\index{wave!number} 
$k$ is related to the wavelength\index{wavelength} of the light according to 
$k=2\pi/\lambda$.

In order to find the interference pattern\index{pattern!interference} in the 
intensity, we sum the two plane waves\index{plane wave} at position $\vec{r}$ 
at the amplitude level:
\begin{equation}
 I(\vec{r}) \propto \left|e^{i\vec{k}_1\cdot\vec{r}}+e^{i\vec{k}_2\cdot\vec{r}}
 \right|^2 = 4\cos^2\left[ \frac{1}{2}(\vec{k}_1 - \vec{k}_2)\cdot\vec{r} 
 \right]\; .
\end{equation}
When we calculate the inner product $(\vec{k}_1 - \vec{k}_2)\cdot\vec{r}/2$ 
from Eq.\ (\ref{planewave}) we obtain the expression 
\begin{equation}\label{cos2}
 I(x) \propto \cos^2(kx\sin\theta)
\end{equation}
for the intensity along the substrate in direction $x$.

\begin{figure}[t]
\begin{center}
  \begin{psfrags}
     \psfrag{a}{a)}
     \psfrag{p}{$\theta$}
     \psfrag{x}{$x$}
     \psfrag{y}{$y$}
     \psfrag{k1}{$\vec{k}_1$}
     \psfrag{k2}{$\vec{k}_2$}
     \psfrag{cos}{$\!\!\!\!\!\!$b) $\cos^2\varphi$}
     \psfrag{phi}{$\varphi$}
     \psfrag{1.5708}[c]{$\qquad\pi/2$}
     \psfrag{Pi}[c]{$~\pi$}
     \psfrag{2 Pi}[c]{$\quad2\pi$}
     \epsfxsize=8in
     \epsfbox[-60 60 840 185]{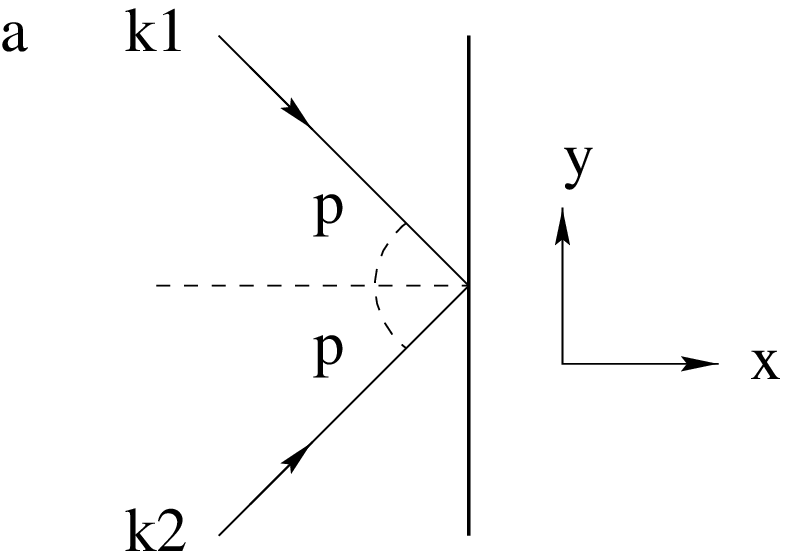}
     \epsfxsize=8in
     \epsfbox[-350 90 650 130]{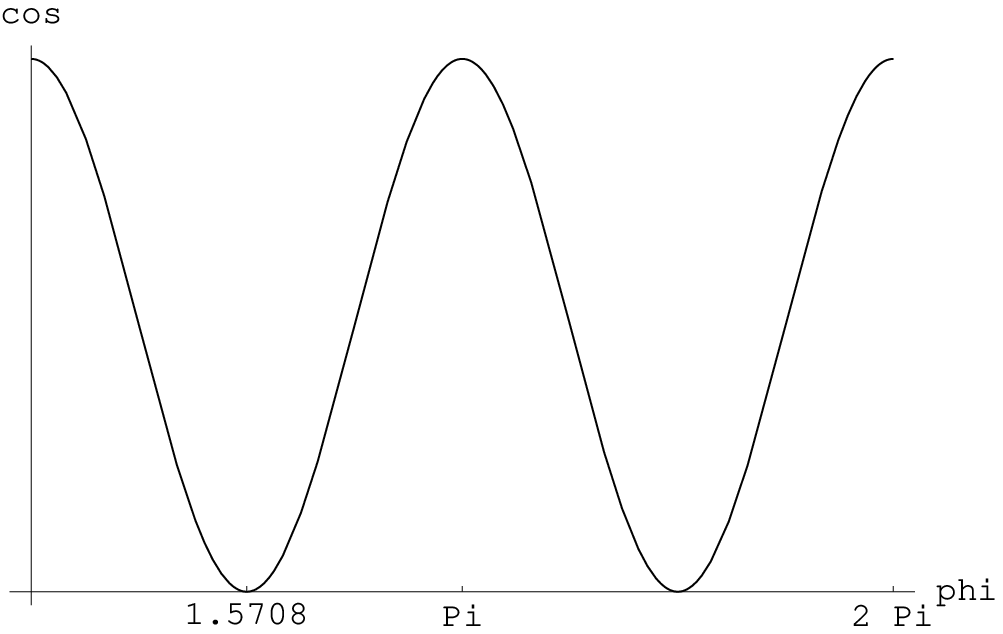}
  \end{psfrags}
  \end{center}
  \caption{a) Schematic representation of two light beams $\vec{k}_1$ and 
	$\vec{k}_2$ incident on a surface, yielding an interference pattern.
	b) The interference pattern for $\varphi=ky\sin\theta$.}
\label{fig:6.1}
\end{figure}

The Rayleigh criterion\index{Rayleigh criterion} states that the minimal 
resolvable feature size 
$\Delta x$ corresponds to the distance between an intensity maximum and an 
adjacent minimum (see figure \ref{fig:6.1}). From Eq.\ (\ref{cos2}) we obtain
\begin{equation}
 k\Delta x\sin\theta = \frac{\pi}{2}\; .
\end{equation} 
This means that the maximum resolution\index{resolution!maximum} is given by
\begin{equation}
 \Delta x = \frac{\pi}{2k\sin\theta} = \frac{\pi}{2\left(\frac{2\pi}{\lambda}
 \sin\theta\right)} = \frac{\lambda}{4\sin\theta}\; ,
\end{equation}
where $\lambda$ is the wavelength\index{wavelength} of the light. The maximum 
resolution is
therefore proportional to the wavelength and inversely proportional to the 
sine of the angle between the incoming plane waves and the normal. The 
resolution is thus maximal ($\Delta x$ is minimal) when $\sin\theta=1$, or 
$\theta=\pi/2$. This is the grazing limit.\index{grazing limit} The classical 
diffraction limit\index{diffraction limit} is 
therefore $\Delta x = \lambda/4$. 

\section{Introduction to Quantum Lithography}\label{intro}

\begin{figure}[t]
  \begin{center}
  \begin{psfrags}
     \psfrag{a}{$a$}
     \psfrag{b}{$b$}
     \psfrag{t}{$\!\!\!2\theta$}
     \psfrag{f}{$\varphi$}
     \psfrag{p}{substrate}
     \epsfxsize=8in
     \epsfbox[-250 40 750 90]{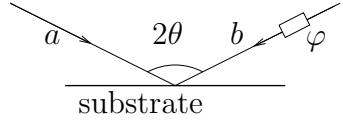}
  \end{psfrags}
  \end{center}
  \caption{Two light beams $a$ and $b$ cross each other at the
	surface of a photosensitive substrate. The angle between them is 
	$2\theta$ and they have a relative phase difference $\varphi$. We
	consider the limit case of $2\theta\rightarrow\pi$.}
  \label{fig:6.2}
\end{figure}

In this section we briefly reiterate our method of Ref.\ \cite{boto00}. 
Suppose we have two intersecting light beams $a$ and $b$. 
We place some suitable substrate at the position where the two beams meet, 
such that the interference pattern\index{pattern!interference} is recorded. 
For simplicity, we consider the grazing limit\index{grazing limit} in which 
the angle $\theta$ off axis for the two beams is 
$\pi/2$ (see figure \ref{fig:6.2}). Classically, the interference pattern on 
the substrate has a resolution\index{resolution} of the order of $\lambda/4$, 
where $\lambda$ is the wavelength
of the light. However, by using entangled photon-number states 
\index{state!number} (i.e., inherently {\em non-classical} states) we 
can increase the resolution well into the sub-wavelength regime 
\cite{jacobson95,rarity90,trifonov00,fonseca99}.

How does quantum lithography work? Let the two counter-propagating light beams 
$a$ and $b$ be in the combined entangled number state
\begin{equation}\label{n00n}
 |\psi_N\rangle_{ab} = \left( |N,0\rangle_{ab} + e^{iN\varphi} |0,N\rangle_{ab}
 \right) / \sqrt{2}\; ,
\end{equation}
where $\varphi=kx/2$, with $k=2\pi/\lambda$.
We define the mode operator $\hat{e} = (\hat{a}+\hat{b})/\sqrt{2}$ and its 
adjoint $\hat{e}^{\dagger} = (\hat{a}^{\dagger}+\hat{b}^{\dagger})/\sqrt{2}$. 
The deposition rate\index{deposition rate|(} $\Delta$ on a substrate sensitive
to $N$ photons with wavelength $\lambda$ (a so-called $N$-photon resist) 
\index{photon resist}
is then given by
\begin{equation}\label{delta}
 \Delta_N = \langle\psi_N|\hat{\delta}_N|\psi_N\rangle\qquad\text{with}\qquad
 \hat{\delta}_N=\frac{(\hat{e}^{\dagger})^N \hat{e}^N}{N!}\; ,
\end{equation}
i.e., we look at the higher moments of the electric field operator 
\index{operator!electric field!higher moments}
\cite{goppert31,javanainen90,perina98}.
The deposition rate $\Delta$ is measured in units of intensity. Leaving the 
substrate exposed for a time $t$ to the light source will result in an 
exposure pattern $P(\varphi)=\Delta_N t$. After a straightforward calculation 
we see that 
\begin{equation}\label{deprate}
 \Delta_N \propto (1 + \cos N\varphi)\; .
\end{equation}
We interpret this as follows. A path-differential phase-shift
\index{phase shift} $\varphi$ in 
light beam $b$ results in a displacement $x$ of the interference pattern on 
the substrate. Using two classical waves, a phase-shift of $2\pi$ will return 
the pattern to its original position. However, according to 
Eq.~(\ref{deprate}), one cycle is completed after a shift of $2\pi/N$. This 
means that a shift of $2\pi$ will displace the pattern $N$ times. In other 
words, we have $N$ times more maxima in the interference pattern. These need 
to be closely spaced, yielding an effective Rayleigh 
resolution\index{resolution!Rayleigh} of $\Delta x
= \lambda/4N$, a factor of $N$ below the classical interferometric result of 
$\Delta x=\lambda/4$ \cite{brueck98}.

Physically, we can interpret this result as follows: instead of having a state
of $N$ single photons, Eq.~(\ref{n00n}) describes an $N$-photon 
state.\index{state!nphoton@$N$-photon} Since 
the momentum of this state is $N$ times as large as the momentum for a 
single photon, the corresponding De Broglie wavelength
\index{wavelength!De Broglie} is $N$ times smaller. 
The interference of this $N$-photon state with itself on a substrate 
thus gives a periodic pattern with a characteristic resolution dimension of 
$\Delta x = \lambda/4N$.

\section{General Patterns in 1D}\label{1D}

So far, we have described a method to print a simple pattern of evenly
spaced lines of sub-wavelength resolution.\index{resolution!sub-wavelength} 
However, for any practical
application we need the ability to produce more complicated patterns. To this 
end, we introduce the state
\begin{equation}\label{nm}
 |\psi_{Nm}\rangle_{ab} = \left( e^{im\varphi}|N-m,m\rangle_{ab} 
 + e^{i(N-m)\varphi} e^{i\theta_m} |m,N-m\rangle_{ab} \right) / \sqrt{2}\; .
\end{equation}
This is a generalised version of Eq.\ (\ref{n00n}). In particular, Eq.\
(\ref{nm}) reduces to Eq.\ (\ref{n00n}) when $m=0$ and $\theta_m=0$. Note
that we included a relative phase $e^{i\theta_m}$, which will turn out to be 
crucial in the creation of arbitrary one-dimensional patterns.
\index{pattern!one-dimensional}

We can calculate the deposition rate again according to the procedure in section 
\ref{intro}. As we shall see later, in general, we can have superpositions
\index{superposition}
of the states given by Eq.\ (\ref{nm}). We therefore have to take into account 
the possibility of different values of $m$, yielding a quantity
\begin{equation}
 \Delta_{Nm}^{Nm'} = \langle\psi_{Nm}|\hat{\delta}_N|\psi_{Nm'}\rangle\; .
\end{equation}
Note that this deposition rate depends not only on the parameter $\varphi$,
but also on the relative phases $\theta_m$ and $\theta_{m'}$. The deposition
rate then becomes
\begin{eqnarray}\label{nmmn}
 \Delta_{Nm}^{Nm'} &\propto& \sqrt{\binom{N}{m}\binom{N}{m'}} \left[ 
 e^{i(m'-m)\varphi} + e^{i(N-m-m')\varphi} e^{i\theta_{m'}} + \right. \cr
 && \qquad\left. e^{-i(N-m-m')\varphi} e^{-i\theta_m} + e^{-i(m'-m)\varphi} 
 e^{i(\theta_{m'}-\theta_m)} \right] \; .
\end{eqnarray}
Obviously, $\langle\psi_{Nm}|\hat{\delta}_l|\psi_{N'm'}\rangle=0$ when 
$l\not\in \{N,N'\}$. 
For $m=m'$, the deposition rate takes on the form
\begin{equation}\label{genn00n}
 \Delta_{Nm} \propto \binom{N}{m} \left\{ 1 + \cos[(N-2m)\varphi+\theta_m]
 \right\}\; ,
\end{equation}
which, in the case of $m=0$ and $\theta_m=0$, coincides with Eq.\ 
(\ref{deprate}). When $\theta_m$ is suitably chosen, we see that we also have 
access to deposition rates $(1-\cos N\varphi)$ and $(1\pm\sin N\varphi)$.
Apart from this extra phase freedom, Eq.~(\ref{genn00n}) does not look like 
an improvement over Eq.\ (\ref{deprate}), since $N-2m\leq N$, which means that 
the resolution\index{resolution} decreases. However, we will show later how 
these states {\em can} be used to produce non-trivial patterns.

First, we look at a few special cases of $\theta_m$ and $\theta_{m'}$. When 
we write $\Delta_{Nm}^{Nm'}=\Delta_{Nm}^{Nm'}(\theta_m,\theta_{m'})$ we have
\begin{eqnarray}
 \Delta_{nm}^{Nm'} (0,0) &\propto& \cos\Bigl(\frac{N-2m}{2}\,
 \varphi\Bigr) \cos\Bigl(\frac{N-2m'}{2}\,\varphi\Bigr)\; , \\
 \Delta_{Nm}^{Nm'} (0,\pi) &\propto& \cos\Bigl(\frac{N-2m}{2}\,
 \varphi\Bigr) \sin\Bigl(\frac{N-2m'}{2}\,\varphi\Bigr)\; , \\
 \Delta_{Nm}^{Nm'} (\pi,0) &\propto& \sin\Bigl(\frac{N-2m}{2}\,
 \varphi\Bigr) \cos\Bigl(\frac{N-2m'}{2}\,\varphi\Bigr)\; , \\
 \Delta_{Nm}^{Nm'} (\pi,\pi) &\propto& \sin\Bigl(\frac{N-2m}{2}\,
 \varphi\Bigr) \sin\Bigl(\frac{N-2m'}{2}\,\varphi\Bigr)\; .
\end{eqnarray}
These relations give the dependence of the matrix elements $\Delta_{Nm}^{Nm'}$
on $\theta_m$ and $\theta_{m'}$ in a more intuitive way than Eq.\ (\ref{nmmn})
does. Finally, when $\theta_m = \theta_{m'}=\theta$ we obtain
\begin{equation}
 \Delta_{Nm}^{Nm'} \propto \cos\left[\frac{(N-2m)\varphi+\theta}{2}\right]
 \cos\left[\frac{(N-2m')\varphi-\theta}{2}\right]\; .
\end{equation}

So far we have only considered generalised deposition rates given by Eq.\ 
(\ref{nm}), with special values of their parameters. We will now turn our 
attention to the problem of creating more arbitrary patterns.

Note that there are two main, though fundamentally different, ways we can 
superpose the states given by Eq.\ (\ref{nm}). We can superpose states with 
different photon numbers $n$ and a fixed distribution $m$ over the two modes:
\begin{equation}\label{phot}
 |\Psi_m\rangle = \sum_{n=0}^N \alpha_n |\psi_{nm}\rangle\; .
\end{equation}
Alternatively, we can superpose states with a fixed photon number $N$, but with
different distributions $m$:
\begin{equation}\label{dist}
 |\Psi_N\rangle = \sum_{m=0}^{\lfloor N/2\rfloor} \alpha_m|\psi_{Nm}\rangle\; ,
\end{equation}
where $\lfloor N/2\rfloor$ denotes the largest integer $l$ with $l\leq N/2$.

These two different superpositions\index{superposition} can be used to tailor 
patterns which are
more complicated than just closely spaced lines. We will now study these two 
different methods.

\subsection{The Pseudo-Fourier Method}\index{pseudo-Fourier method}

The first method, corresponding to the superposition\index{superposition} 
given by Eq.\ 
(\ref{phot}), we will call the pseudo-Fourier method (this choice of name will 
become clear shortly). When we calculate the deposition rate $\Delta_m$ 
according to the state $|\Psi_m\rangle$ we immediately see that branches with 
different photon numbers $n$ and $n'$ do not exhibit interference:
\begin{equation}
 \Delta_m = \sum_{n=0}^N |\alpha_n|^2 \langle\psi_{nm}|\hat{\delta}_n|
 \psi_{nm}\rangle = \sum_{n=0}^N |\alpha_n|^2 \Delta_{nm} \; .
\end{equation}
Using Eq.\ (\ref{genn00n}) the exposure pattern\index{pattern!exposure} 
$P(\varphi)=\Delta_m t$ becomes
\begin{equation}\label{fourier}
 P(\varphi) = t \sum_{n=0}^{N} c_n \left\{1+\cos[(n-2m)\varphi+\theta_n] 
 \right\}\; , 
\end{equation}
where $t$ is the exposure time\index{exposure!time} and the $c_n$ are positive.
Since $m<n$ and $m$ is fixed, we have $m=0$. I will now prove that this is 
a Fourier series\index{Fourier!series} up to a constant. 

A general Fourier expansion\index{Fourier!expansion} of $p(\varphi)$ can be 
written as
\begin{equation}
 P(\varphi) = \sum_{n=0}^N (a_n \cos n\varphi + b_n \sin n\varphi)\; .
\end{equation}
Writing Eq.\ (\ref{fourier}) as 
\begin{equation}
 P(\varphi) = t\sum_{n=0}^{N} c_n + t \sum_{n=0}^{N} c_n \cos(n\varphi+
 \theta_n)\; ,
\end{equation}
where $t\sum_{n=0}^{N} c_n$ is a constant. If we ignore this constant (its
contribution to the deposition rate will give a general uniform background 
exposure\index{exposure!background} of the substrate, since it is independent 
of $\varphi$) we see that we need
\begin{equation}
 c_n \cos(n\varphi+\theta_n) = a_n \cos n\varphi + b_n \sin n\varphi
\end{equation}
with $c_n$ positive, $\theta_n\in[0,2\pi)$ and $a_n$, $b_n$ real. Expanding 
the left-hand side and equating terms in $\cos n\varphi$ and $\sin n\varphi$
we find
\begin{equation}
 a_n = c_n \cos\theta_n \qquad\mbox{and}\qquad b_n = c_n \sin\theta_n\; .
\end{equation}
This is essentially a co-ordinate change from 
Cartesian\index{Cartesian co-ordinates} to polar 
co-ordinates.\index{polar co-ordinates}
Thus, Eq.\ (\ref{fourier}) is equivalent to a Fourier series up to an additive 
constant. Since in the limit of $N\rightarrow\infty$ a Fourier series can 
converge to any well-behaved pattern $P(\varphi)$, this procedure allows us to 
approximate arbitrary patterns\index{pattern!arbitrary} in one dimension (up 
to a constant). It is now clear why we call this procedure the pseudo-Fourier 
method.

However, there is a drawback with this procedure. The deposition rate
$\Delta$ is a positive definite quantity, which means that once the substrate 
is exposed at a particular Fourier component,\index{Fourier!component} there 
is no way this can be undone. Technically, Eq.\ (\ref{fourier}) can be 
written as
\begin{equation}
 P(\varphi) = Q \cdot t + t \sum_{n=0}^{N} (a_n\cos n\varphi + 
 b_n\sin n\varphi)\; ,
\end{equation}
where $Q$ is the uniform background `penalty exposure\index{exposure!penalty} 
rate' $Q = \sum_{n=0}^{N} c_n$ we mentioned earlier. The second term on the 
right-hand side is a true Fourier series.\index{Fourier!series} Thus in the 
pseudo-Fourier method\index{pseudo-Fourier method} there is always a minimum 
exposure of the substrate. Ultimately, this penalty can be traced to the 
absence of interference between the terms with different photon number in Eq.\ 
(\ref{phot}). Next, we will investigate whether our second method of tailoring 
patterns can remove this penalty exposure.

\subsection{The Superposition Method}\index{superposition!method}

We will now study our second method of tailoring patterns, which we call the 
`superposition method' (lacking a better name). Here we keep the total number
of photons $N$ constant, and change how the photons are distributed between
the two beams in each branch [see Eq.\ (\ref{dist})]. A distinct advantage of 
this method is that it {\em does} exhibit interference between the different 
branches in the superposition,\index{superposition} which eliminates the 
uniform background penalty exposure. 

\begin{figure}[t]
  \begin{center}
  \begin{psfrags}
     \psfrag{D}{$\Delta_{20\, 9}^{20\, 5}$}
     \psfrag{phi}{$\varphi$}
     \psfrag{1.5708}[c]{$\qquad\pi/2$}
     \psfrag{Pi}{$~\pi$}
     \psfrag{4.71239}[c]{$\qquad 3\pi/2$}
     \psfrag{2 Pi}{$~2\pi$}
     \epsfxsize=8in
     \epsfbox[-130 70 620 250]{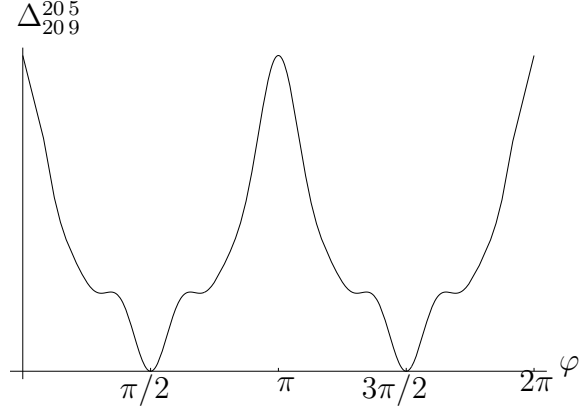}
  \end{psfrags}
  \end{center}
  \caption{A simple superposition of two states containing 20 photons
	with distributions $m=9$ and $m'=5$ ($\theta_m = \theta_{m'}=0$). The
	deposition rate at $\varphi=\pi/2$ and $\varphi=3\pi/2$ is zero, 
	which means that there is no general uniform background exposure using 
	the superposition method.}
  \label{fig:6.3}
\end{figure}

Take for instance a superposition of two distinct terms
\begin{equation}
 |\Psi_N\rangle = \alpha_m |\psi_{Nm}\rangle +\alpha_{m'}|\psi_{Nm'}\rangle\; ,
\end{equation}
with $|\alpha_m|^2 + |\alpha_{m'}|^2 =1$ and $|\psi_{nm}\rangle$ given by Eq.\ 
(\ref{nmmn}). After some algebraic manipulation the deposition rate can be
written as
\begin{eqnarray}
 \Delta_N &\propto& |\alpha_m|^2 
 \binom{N}{m} \left\{1+\cos[(N-2m)\varphi + \theta_m] \right\}\cr && 
 + |\alpha_{m'}|^2 \binom{N}{m'} \left\{1+\cos[(N-2m')\varphi + \theta_{m'}]
 \right\}\cr && 
  +8 r_m^{m'}\sqrt{\binom{N}{m}\binom{N}{m'}}
 \cos\left(\frac{\theta_{m'}}{2}-\frac{\theta_m}{2} + \xi_m^{m'} \right)\cr &&
 ~\times \cos\frac{1}{2}\left[(N-2m)\varphi + \theta_m \right] 
 \cos\frac{1}{2}\left[(N-2m')\varphi+ \theta_{m'} \right]\; ,
\end{eqnarray}
where the deposition rate $\Delta$ is now a function of 
$\alpha_m$ and $\alpha_{m'}$, where we have chosen the real numbers $r_m^{m'}$ 
and $\xi_m^{m'}$ to satisfy $\alpha_m^*\alpha_{m'} \equiv r_m^{m'}
\exp(i\xi_m^{m'})$. For the special values $N=20$, $m=9$, $m'=5$ and 
$\theta_m=\theta_{m'}=0$ we obtain the pattern shown in figure \ref{fig:6.3}. 
Clearly, there is no uniform background penalty 
exposure\index{exposure!penalty} here. 

For more than two branches in the superposition\index{superposition} this 
becomes a  complicated function, which is not nearly as well understood as a 
Fourier series.\index{Fourier!series} The 
general expression for the deposition rate can be written as
\begin{eqnarray}
 \Delta_N &\propto& \sum_{m=0}^{\lfloor N/2\rfloor}
 \sum_{m'=0}^{\lfloor N/2\rfloor} r_m^{m'}
 \sqrt{\binom{N}{m}\binom{N}{m'}}
 \cos\left(\frac{\theta_{m'}}{2}-\frac{\theta_m}{2} +\xi_m^{m'}\right)\cr 
 && \quad~\times \cos\frac{1}{2}\left[(N-2m)\varphi + \theta_m\right]
 \cos\frac{1}{2}\left[(N-2m')\varphi + \theta_{m'} \right]\; ,
\end{eqnarray}
where we have chosen $r_m^{m'}$ and $\xi_m^{m'}$ real to satisfy 
$\alpha_m^*\alpha_{m'} \equiv r_m^{m'}\exp(i\xi_m^{m'})$. Note that 
$\xi_m^m=0$. 

If we want to tailor a pattern $F(\varphi)$, it might be the case that this 
type of superposition will also converge to the required pattern. We will now 
compare the superposition method with the pseudo-Fourier 
method.\index{pseudo-Fourier method}

\subsection{Comparing the two methods}\label{arb}

So far, we discussed two methods of creating non-trivial patterns in one 
dimension. The pseudo-Fourier method is simple but yields a uniform background 
penalty exposure. The superposition method is far more complicated, but seems 
to get around the background exposure. Before we make a comparison between the 
two methods we will discuss the creation of `arbitrary' 
patterns.\index{pattern!arbitrary}

It is well known that any sufficiently well-behaved periodic function can be 
written as an infinite Fourier series (we ignore such subtleties which arise 
when two functions differ only at a finite number of points, etc.). However, 
when we create patterns with the pseudo-Fourier lithography method we do not 
have access to every component of the Fourier expansion, since this would 
involve an infinite number of photons ($n\rightarrow\infty$). This means that 
we can only employ truncated Fourier series,\index{Fourier!series!truncated} 
and these can merely approximate arbitrary patterns.

The Fourier expansion\index{Fourier!expansion} has the nice property that when 
a series is truncated
at $N$, the remaining terms still give the best Fourier expansion of the 
function up to $N$. In other words, the coefficients of a truncated Fourier 
series are equal to the first $N$ coefficients of a full Fourier series. If 
the full Fourier series is denoted by $F$ and the truncated series by $F_N$,
we can define the normed-distance quantity $D_N$:
\begin{equation}
 D_N \equiv \int_0^{2\pi} |F(\varphi)-F_N(\varphi)|^2 d\varphi\; ,
\end{equation}
which can be interpreted as a distance\index{distance} between $F$ and $F_N$. 
If quantum 
lithography yields a pattern $P_N(\varphi)=\Delta_N t$, we can introduce the 
following definition: quantum lithography can approximate arbitrary patterns 
\index{pattern!arbitrary} if 
\begin{equation}
 \int_0^{2\pi} |F(\varphi)-P_N(\varphi)|^2 d\varphi \leq \varepsilon D_N\; ,
\end{equation} 
with $\varepsilon$ some proportionality constant. This definition gives the 
concept of approximating patterns a solid basis.

We compare the pseudo-Fourier\index{pseudo-Fourier method} and the 
superposition method\index{superposition!method} for one special case. We 
choose the test function
\begin{eqnarray}\label{test}
 F(\varphi) = \left\{
 \begin{matrix}
  h ~\mbox{if}~ -\frac{\pi}{2} < \varphi < \frac{\pi}{2}\; , \cr
  \, 0 ~\mbox{otherwise}\; .\quad\qquad
 \end{matrix}
 \right.
\end{eqnarray}
With up to ten photons, we ask how well the pseudo-Fourier and the 
superposition method approximate this pattern.

In the case of the pseudo-Fourier method the solution is immediate. The Fourier
expansion\index{Fourier!expansion} of the `trench' 
function\index{trench function} given by Eq.\ (\ref{test}) is well known:
\begin{equation}\label{fourmin}
 F(\varphi) = \sum_{q=0}^{\infty} \frac{(-1)^q}{2q+1} \cos[(2q+1)\varphi]\; .
\end{equation}
Using up to $n=10$ photons we include terms up to $q=4$, since $2q+1\leq 10$.
The Fourier method thus yields a pattern $P(\varphi)$ (the two patterns 
$P(\varphi)$ and $F(\varphi)$ are generally not the same) which can be written 
as 
\begin{equation}
 P(\varphi) = \sum_{q=0}^4 \frac{c_q t}{2q+1}
 \left( 1+\cos\left[(2q+1)\varphi + \pi\kappa_q\right] \right)\; ,
\end{equation}
where $c_q$ is a constant depending on the proportionality constant of 
$\Delta_{2q+1}$, the rate of production of $|\psi_{nm}\rangle$ and the coupling
between the light field and the substrate. The term $\kappa_q$ is defined to
accommodate for the minus signs in Eq.\ (\ref{fourmin}): it is zero when $q$ 
is even and one when $q$ is odd. Note the uniform background penalty exposure 
\index{exposure!penalty}
rate $\sum_{q=0}^4 c_q/(2q+1)$. The result of this method is shown in 
figure \ref{fig:6.3}.

\begin{figure}[t]
  \begin{center}
  \begin{psfrags}
     \psfrag{D}{$p(\varphi) = \Delta(\varphi)t$}
     \psfrag{phi}{$\varphi$}
     \psfrag{b}{$p(\varphi)\sum_{n=1}^{10} \alpha_n \Delta_n t$}
     \psfrag{p}{$\Bigg\updownarrow$ penalty}
     \psfrag{pi2}{}     
     \psfrag{3 pi2}{}
     \psfrag{2 Pi}{$2 \pi$}
     \psfrag{Pi}{$\pi$}
     \psfrag{1.5708}[c]{$\qquad\pi/2$}
     \psfrag{4.71239}[c]{$\qquad 3\pi/2$}
     \epsfxsize=8in
     \epsfbox[-110 90 600 240]{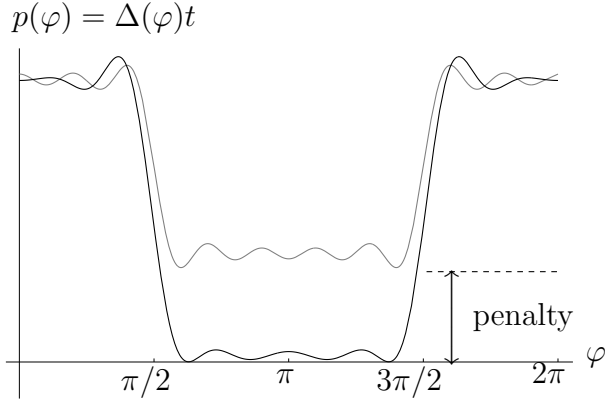}
  \end{psfrags}
  \end{center}
  \caption{The deposition rate on the substrate resulting from a 
	superposition of states with $n=10$ and different $m$ (black curve)
	and resulting from a superposition of states with different $n$ and
	$m=0$ (grey curve). The coefficients of the superposition yielding 
	the black curve are optimised using a genetic algorithm \cite{price97},
	while the grey curve is a truncated Fourier series. Notice the 
	`penalty' (displaced from zero) deposition rate of the Fourier series 
	between $\pi/2$ and $3\pi/2$.}
  \label{fig:6.4}
\end{figure}

Alternatively, the superposition method\index{superposition!method} employs a 
state
\begin{equation}
 |\Psi_{N}\rangle=\sum_{m=0}^{\lfloor N/2\rfloor}\alpha_m|\psi_{Nm}\rangle\; .
\end{equation}
The procedure of finding the best fit with the test function is more 
complicated. We have to minimise the absolute difference between the 
deposition rate $\Delta_{N}(\vec\alpha)$ times the exposure 
time\index{exposure!time} $t$ and the 
test function $F(\varphi)$. We have chosen $\vec\alpha=(\alpha_0,\ldots,
\alpha_{n/2})$. Mathematically, we have to evaluate the $\vec\alpha$ and $t$ 
which minimise $d_N$:
\begin{equation}\nonumber
 d_N = \int_0^{2\pi} |F(\varphi)-\Delta_N (\vec\alpha) t|^2 d\varphi\; ,
\end{equation}
with
\begin{equation}
 \Delta_N (\vec\alpha) = \langle\Psi_N |\hat{\delta}_N |\Psi_N \rangle\; .
\end{equation}
We have to fit both $t$ and $\vec\alpha$. Using a genetic optimalisation 
algorithm\index{algorithm!genetic} \cite{price97} (with $h=1$, a normalised 
height of the test function; see appendix \ref{app:genetic}) 
we found that the deposition rate is actually very close to zero in the 
interval $\pi/2 \leq\varphi\leq 3\pi/2$, unlike the pseudo-Fourier method,
\index{pseudo-Fourier method} 
where we have to pay a uniform background penalty.\index{exposure!penalty} 
This result implies that 
in this case a superposition of different photon distributions $m$, given a 
fixed total number of photons $N$, works better than a superposition of 
different photon number states (see figure \ref{fig:6.3}). In particular, {\em 
the fixed photon number method allows for the substrate to remain virtually 
unexposed in certain areas}.

We stress that this is merely a comparison for a specific example, namely that
of the trench target function $F(\varphi)$. We conjecture that the 
superposition method can approximate other arbitrary 
patterns\index{pattern!arbitrary} equally well, 
but we have not yet found a proof. Besides the ability to fit an arbitrary 
pattern, another criterion of comparison between the pseudo-Fourier method and 
the superposition method,\index{superposition!method} is the time needed to 
create the $N$-photon entangled states.\index{state!entangled!$N$-photon} 

Until now, we have only considered sub-wavelength 
resolution\index{resolution!sub-wavelength} in one direction,
namely parallel to the direction of the beams. However, for practical
applications we would like sub-wavelength resolution in both directions on
the substrate. This is the subject of the next section.

\section{General Patterns in 2D}\label{2D}

In this section we study how to create two-dimensional 
patterns\index{pattern!two-dimensional}
on a suitable substrate using the quantum lithography techniques developed
in the previous sections. As we have seen, the phase shift $\varphi$, in the
setup given by figure \ref{fig:6.1}, acts as a parametrisation for the 
deposition 
rate in one dimension. Let's call this the $x$-direction.

\begin{figure}[t]
  \begin{center}
  \begin{psfrags}
     \psfrag{a}{$a$}
     \psfrag{b}{$b$}
     \psfrag{c}{$c$}
     \psfrag{d}{$d$}
     \psfrag{t}{$\theta$}
     \psfrag{f}{$\varphi$}
     \psfrag{s}{substrate}
     \epsfxsize=8in
     \epsfbox[-225 40 775 230]{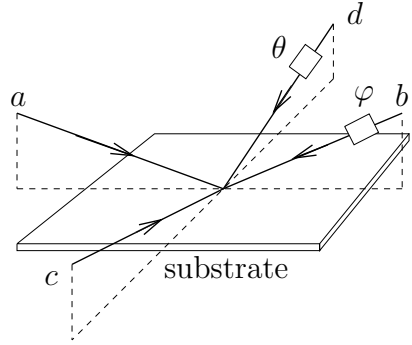}
  \end{psfrags}
  \end{center}
  \caption{Four light beams $a$, $b$, $c$ and $d$ cross each other at 
	the surface of a photosensitive substrate. The angles between $a$ and
	$b$ and $c$ and $d$ are again taken in the grazing limit of $\pi$. The 
	relative phase difference between $a$ and $b$ is $\varphi$ and the 
	relative phase difference between $c$ and $d$ is $\theta$.}
  \label{fig:6.5}
\end{figure}

We can now do the same for the $y$-direction, employing two counter-propagating
beams ($c$ and $d$) in the $y$-direction (see figure \ref{fig:6.4}). The same 
conditions apply: we consider the limit where the spatial angle $\theta$ off 
axis approaches $\pi/2$, thus grazing along the substrate's surface. 

Consider the region where the four beams $a$, $b$, $c$ and $d$ overlap. For 
real lithography we have to take into account the mode 
shapes,\index{mode!shape} but when we 
confine ourselves to an area with side lengths $\lambda$ (where $\lambda$ is 
the wavelength\index{wavelength} of the used light) this problem does not arise.

The class of states on modes $a$ to $d$ that we consider here are of the form
\begin{eqnarray}\label{nmk}
 |\psi^k_{Nm}\rangle &=& \frac{1}{2} 
 \biggl[ e^{im\varphi}|N-m,m;0,0\rangle 
 +\; e^{i(N-m)\varphi} e^{i\zeta_m}|m,N-m;0,0\rangle\biggr. \cr &&  
 \quad \biggl. +\; e^{ik\theta}|0,0;N-k,k\rangle
 +\; e^{i(N-k)\chi} e^{i\bar\zeta_k}|0,0;k,N-k\rangle \biggr] ,
\end{eqnarray}
where $\zeta_m$ and $\bar\zeta_k$ are two relative phases. 
This is by no means the only class of states, but we will restrict our 
discussion to this one for now. Observe that this is a 
superposition\index{superposition} on the 
amplitude level, which allows destructive interference in the deposition rate 
in order to create dark spots on the substrate. Alternatively, we could have 
used the one-dimensional method [with states given by Eq.\ (\ref{nm})] in the 
$x$- and $y$-direction, but this cannot give interference effects between the 
modes $a,b$ and $c,d$.

The phase-shifts $\varphi$ and $\chi$ in the light beams $b$ and $d$ (see
figure \ref{fig:6.4}) result in respective displacements $x$ and $y$ of the 
interference 
pattern on the substrate. A phase-shift of $2\pi$ in a given direction will 
displace the pattern, say, $N$ times. This means that the maxima are closer
together, yielding an effective resolution equal to $\Delta x = \Delta y =
\lambda/4N$. This happens in both the $x$- and the $y$-direction. 

We proceed again as in section \ref{intro} by evaluating the $N^{\rm th}$ order
moment $\hat{\delta}_N$ of the electric field 
operator\index{operator!electric field!higher moments} [see Eq.~(\ref{delta})].
On a substrate sensitive to  $N$ photons this gives the deposition rate 
$\Delta_{Nmk}^{Nm'k'} = \langle \psi^k_{Nm}|\delta_N|\psi^{k'}_{Nm'}\rangle$ 
[with $|\psi^k_{Nm}\rangle$ given by Eq.\ (\ref{nmk})]:
\begin{eqnarray}\label{2ddeprate}
 \Delta_{Nmk}^{Nm'k'} &\propto&
    \binom{N}{m}\binom{N}{m'} 
 \left( e^{-im\varphi} e^{im'\varphi} + 
  e^{-im\varphi} e^{i(N-m')\varphi} e^{i\zeta_{m'}} + e^{-i(N-m)\varphi}
  e^{im'\varphi} e^{-i\zeta_m} \right.\cr &&\qquad\qquad\qquad\qquad\left. + 
  e^{-i(N-m)\varphi} e^{i(N-m')\varphi}
  e^{-i(\zeta_m-\zeta_{m'})} \right) \cr
 && + \binom{N}{m}\binom{N}{k'} 
 \left( e^{-im\varphi} e^{ik'\chi} + 
  e^{-im\varphi} e^{i(N-k')\chi} e^{i\bar\zeta_{k'}} + e^{-i(N-m)\varphi}
  e^{ik'\chi} e^{-i\zeta_m} \right.\cr &&\qquad\qquad\qquad\qquad\left. + 
  e^{-i(N-m)\varphi} e^{i(N-k')\chi}
  e^{-i(\zeta_m-\bar\zeta_{k'})} \right) \cr
 && + \binom{N}{k}\binom{N}{m'} 
 \left( e^{-ik\chi} e^{im'\varphi} + 
  e^{-ik\chi} e^{i(N-m')\varphi} e^{i\zeta_{m'}} + e^{-i(N-k)\chi}
  e^{im'\varphi} e^{-i\bar\zeta_k} \right.\cr &&\qquad\qquad\qquad\qquad\left. 
  + e^{-i(N-k)\chi} e^{i(N-m')\varphi}
  e^{-i(\bar\zeta_k-\zeta_{m'})} \right) \cr
 && + \binom{N}{k}\binom{N}{k'} 
 \left( e^{-ik\chi} e^{ik'\chi} + 
  e^{-ik\chi} e^{i(N-k')\chi} e^{i\bar\zeta_{k'}} + e^{-i(N-k)\chi}
  e^{ik'\chi} e^{-i\bar\zeta_k} \right.\cr &&\qquad\qquad\qquad\qquad\left. + 
  e^{-i(N-k)\chi} e^{i(N-k')\chi}
  e^{-i(\bar\zeta_k-\bar\zeta_{k'})} \right)\; .
\end{eqnarray}
For the special choice of $m'=m$ and $k'=k$ we have 
\begin{eqnarray}
 \Delta_{Nm}^{k} &\propto&
 \binom{N}{m}^2 \left( 1 + \cos[(N-2m)\varphi+\zeta_m] \right) 
 + \binom{N}{k}^2 \left( 1 + \cos[(N-2k)\chi+\bar\zeta_k] \right)\cr && 
 + 4 \binom{N}{m} \binom{N}{k} \cos\frac{1}{2}\left[ N(\varphi-\chi)
 + (\zeta_m-\bar\zeta_k) \right] \cr && \qquad\times
 \cos\frac{1}{2}\left[ (N-2m)\varphi-\zeta_m\right] 
 \cos\frac{1}{2}\left[ (N-2k)\chi-\bar\zeta_k\right] \; .
\end{eqnarray}

We can again generalise this method and use superpositions of the states 
given in Eq.\ (\ref{nmk}). Note that there are now three numbers $N$, $m$ and 
$k$ which can be varied. Furthermore, as we have seen in the one-dimensional 
case, superpositions of different $n$ do not give interference terms in the 
deposition rate. 

Suppose we want to approximate a pattern $F(\varphi,\chi)$, with $\{\varphi,
\chi\}\in [0,2\pi)$. This pattern can always be written in a Fourier expansion:
\index{Fourier!expansion}
\begin{eqnarray}\label{2Dfourier}
 F(\varphi,\chi) &=& \sum_{p,q=0}^{\infty} a_{pq} \cos p\varphi\cos q\chi +
 b_{pq} \cos p\varphi\sin q\chi \times\cr
 && \qquad c_{pq} \sin p\varphi\cos q\chi + d_{pq} \sin p\varphi\sin 
 q\chi \; .
\end{eqnarray}
with $a_p$, $b_p$, $c_q$ and $d_q$ real. In the previous section 
we showed that quantum lithography could approximate the Fourier 
series\index{Fourier!series} of a one-dimensional 
pattern\index{pattern!one-dimensional} up to a constant displacement. This 
relied on absence 
of interference between the terms with different photon numbers. The question 
is now whether we can do the same for patterns in {\em two} dimensions. 
Or alternatively, can general superpositions of the state 
$|\psi_{Nm}^k\rangle$ approximate the pattern $F(\varphi,\chi)$?

From Eq.\ (\ref{2ddeprate}) it is not obvious that we can obtain the four 
trigonometric terms given by the Fourier expansion of Eq.\ (\ref{2Dfourier}):
\begin{eqnarray}
 \Delta &~\propto~& \cos p\varphi\, \cos q\chi\; , \\
 \Delta &~\propto~& \cos p\varphi\, \sin q\chi\; , \\
 \Delta &~\propto~& \sin p\varphi\, \cos q\chi\; , \\
 \Delta &~\propto~& \sin p\varphi\, \sin q\chi\; .
\end{eqnarray}
We can therefore not claim that two-dimensional quantum lithography can 
approximate arbitrary patterns\index{pattern!arbitrary} in the sense of 
one-dimensional lithography.
Only simple patterns like the one given in figure \ref{fig:6.5} can be inferred 
from Eq.\
(\ref{2ddeprate}). In order to find the best fit to an arbitrary pattern 
one has to use a minimisation procedure. 

For example, we calculate the total deposition rate due to the quantum state 
$|\Psi_N\rangle$, where
\begin{equation}
 |\Psi_N\rangle = \sum_{m=0}^{\lfloor N/2\rfloor} \sum_{k=0}^{\lfloor N/2
 \rfloor} \alpha_{mk} |\psi_{Nm}^k\rangle\; .
\end{equation}
Here, $\alpha_{mk}$ are complex coefficients. We now proceed by choosing a 
particular intensity pattern\index{pattern!intensity} $F(\varphi,\chi)$ and 
optimising the 
coefficients $\alpha_{mk}$ for a chosen number of photons. The deposition rate 
due to the state $|\Psi_N\rangle$ is now 
\begin{equation}
 \Delta_N (\vec\alpha) = \sum_{m,m'=0}^{\lfloor N/2\rfloor} 
 \sum_{k,k'=0}^{\lfloor N/2\rfloor} \alpha^*_{mk} \alpha_{m'k'} 
 \Delta_{Nmk}^{Nm'k'}\; ,
\end{equation}
with $\vec\alpha=(\alpha_{0,0},\alpha_{0,1}\ldots,\alpha_{N/2,N/2})$.
We again have to evaluate the $\vec\alpha$ and $t$ which minimise
\begin{equation}
 \int_{0}^{2\pi} \int_{0}^{2\pi} \left| F(\varphi,\chi)
 - \Delta_n (\vec\alpha) t \right|^2 d\varphi\, d\chi\; .
\end{equation}
The values of $\vec\alpha$ and $t$ can again be found using a genetic 
algorithm.\index{algorithm!genetic}

\begin{figure}[t]
  \begin{center}
  \begin{psfrags}
     \psfrag{X}{$\!\! y$}
     \psfrag{Y}[c]{$x$}
     \psfrag{Z}[t]{$\Delta$}
     \psfrag{0}{}
     \psfrag{2}{}
     \psfrag{4}{}
     \psfrag{6}{}
     \psfrag{2.5}{}
     \psfrag{5}{}
     \psfrag{7.5}{}
     \psfrag{10}{}
     \epsfxsize=8in
     \epsfbox[-150 40 680 290]{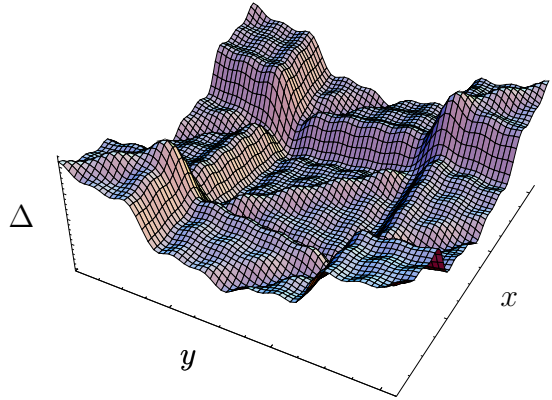}
  \end{psfrags}
  \end{center}
  \caption{A simulation of a two-dimensional intensity pattern on an 
	area $\lambda^2$, where $\lambda$ denotes the wavelength of the used 
	light. Here, I modelled a square area with sharp edges. The pattern 
	was generated by a Fourier series of up to ten photons (see also
	figure \ref{fig:6.4} for the one-dimensional case).}
  \label{fig:6.6}
\end{figure}

\section{Physical implementation}\label{phys}

With current experimental capabilities, the physical implementation of quantum 
lithography is very challenging. In particular, there are two major issues to
be dealt with before quantum lithography can become a mature technology. First
of all, we not only need the ability to create the entangled photon states 
\index{state!entangled!photon}
given by Eqs.\ (\ref{nm}) and (\ref{nmk}), but we should also be able to 
create coherent superpositions\index{superposition} of these states. One 
possibility might be to use optical components like parametric 
down-converters.\index{down-converter} Contrary to the 
results of Ref.\ \cite{kok00b}, we are not concerned with the usually large 
vacuum contribution\index{vacuum contribution} of these processes, since the 
vacuum will not contribute 
to the spatial profile of the deposition\index{deposition rate|)} [see Eqs.\ 
(\ref{n00n}) and (\ref{delta})]. 

Secondly, we need substrates which are sensitive to the higher moments of the 
electric field operator.\index{operator!electric field!higher moments} When we 
want to use the pseudo-Fourier method,\index{pseudo-Fourier method} up to 
$N$ photons for quantum lithography in one dimension, the substrate needs to be
reasonably sensitive to all the higher moments up to $N$, the maximum photon 
number. Alternatively, we can use the superposition 
method\index{superposition!method} for $N$ photons 
when the substrate is sensitive to predominantly one higher moment 
corresponding to $N$ photons. Generally, the method of lithography determines 
the requirements of the substrate.

There are also some considerations about the approximation of patterns. For
example, we might not {\em need} arbitrary patterns.\index{pattern!arbitrary} 
It might be the case that it is sufficient to have a set of patterns which can 
then be used to generate 
any desired circuit. This is analogous to having a universal set of logical 
gates, permitting any conceivable logical expression. In that case we only 
need to determine this elementary (universal) set of 
patterns.\index{pattern!universal set}

Furthermore, we have to study whether the uniform background penalty 
exposure\index{exposure!penalty} 
really presents a practical problem. One might argue that a sufficient 
difference between the maximum deposition rate and the uniform background 
penalty exposure is enough to accommodate lithography. This depends on 
the details of the substrate's reaction to the electro-magnetic 
field.\index{field!electro-magnetic}

Before quantum lithography can be physically implemented and used in the 
production of nano circuits, these issues have to be addressed satisfactorily.

\section{Summary}

In this chapter I have generalised the theory of quantum lithography as first 
outlined in Ref.\ \cite{boto00}. In particular, I have shown how we can create
arbitrary patterns in one dimension, albeit with a uniform background
penalty exposure. We can also create some patterns in two dimensions, but 
we have no proof that this method can be extended to give arbitrary patterns.

For lithography in one dimension we distinguish two methods: the 
pseudo-Fourier method' and the superposition method. The pseudo-Fourier method 
is conceptually easier since it depends on Fourier\index{Fourier!analysis} 
analysis, but it also involves a finite amount of unwanted exposure of 
the substrate. More specifically, the deposition rate equals the pattern in
its Fourier basis plus a term yielding unwanted background exposure.
The superposition method gets around this problem and seems to give better 
results, but lacks the intuitive clarity of the Fourier method. Furthermore,
we do not have a proof that this method can approximate arbitrary patterns.

Quantum lithography in two dimensions is more involved. Starting with a 
superposition of states, given by Eq.\ (\ref{nmk}), we found that we can indeed
create two-dimensional patterns with sub-wavelength resolution, but we do not
have a proof that we can create {\em arbitrary} 
patterns.\index{pattern!arbitrary} Nevertheless, we 
might be able to create a certain set of elementary basis patterns.

There are several issues to be addressed in the future. First, we 
need to study the specific restrictions on the substrate and how we can 
physically realize them. Secondly, we need to create the various entangled 
states\index{state!entangled} involved in the quantum lithography protocol.

Finally, G.S.\ Agarwal and R.\ Boyd have called to our attention that quantum 
lithography works also if the weak parametric down-converter source, described
in Ref.\ \cite{boto00} is replaced by a high-flux optical parametric 
amplifier \cite{agarwal00}. The visibility saturates at 20\% in the limit of 
large gain, but this is quite sufficient for some lithography purposes, as well
as for 3D optical holography\index{holography} used for data storage.
\index{quantum lithography|)}

\part*{Appendices}\addcontentsline{toc}{part}{$\quad\,\,$Appendices}

\appendix

\chapter{Complex vector spaces}\label{app:vectorspaces}

In this appendix I review some properties of complex vector 
spaces,\index{vector space!complex} since
quantum mechanics is defined in terms of a complex vector space. 

\section{Vector spaces}

A vector
space $\mathcal{V}$ consists of a set of vectors $\{\mathbf{v}_i\}$ on which 
two operations are defined:
\begin{description} 
 \item[Addition:]\index{addition!vector} for every 
	$\mathbf{x},\mathbf{y}\in{\mathcal{V}}$ the vector
	$\mathbf{x}+\mathbf{y}$ is an element of $\mathcal{V}$. 
 \item[Scalar multiplication:]\index{scalar!multiplication} for every 
	$\alpha\in\mathbb{C}$ and 
	$\mathbf{x}\in{\mathcal{V}}$ there is a unique element 
	$\alpha\mathbf{x}$ in $\mathcal{V}$.
\end{description}
Furthermore, for every (complex) vector space the following conditions hold:
\begin{enumerate}
 \item For all $\mathbf{x},\mathbf{y}\in{\mathcal{V}}$, $\mathbf{x}+\mathbf{y}
	= \mathbf{y}+\mathbf{x}$ (commutativity of 
	addition);\index{addition!commutativity of}
 \item for all $\mathbf{x},\mathbf{y},\mathbf{z}\in{\mathcal{V}}$, 
	$(\mathbf{x}+\mathbf{y})+\mathbf{z}=\mathbf{x}+(\mathbf{y}+\mathbf{z})$
	(associativity of addition);\index{addition!associativity of}
 \item there exists an element 0 in $\mathcal{V}$ such that $\mathbf{x}+0 
	=\mathbf{x}$ for every $\mathbf{x}\in{\mathcal{V}}$;
 \item for each element $\mathbf{x}$ in ${\mathcal{V}}$ there exists an element
	$\mathbf{y}$ in ${\mathcal{V}}$ such that $\mathbf{x}+\mathbf{y}=0$;
 \item for each element $\mathbf{x}$ in ${\mathcal{V}}$, 
	$1\mathbf{x}=\mathbf{x}$;
 \item for each pair $\alpha$ and $\beta$ in $\mathbb{C}$ and each 
	$\mathbf{x}\in{\mathcal{V}}$ we have $(\alpha\beta)\mathbf{x}
	=\alpha(\beta\mathbf{x})$;
 \item for each $\alpha\in\mathbb{C}$ and each pair $\mathbf{x},\mathbf{y}\in
	{\mathcal{V}}$ we have $\alpha(\mathbf{x}+\mathbf{y})=\alpha\mathbf{x}
	+\alpha\mathbf{y}$;
 \item for each pair $\alpha$ and $\beta$ in $\mathbb{C}$ and each 
	$\mathbf{x}\in{\mathcal{V}}$ we have $(\alpha+\beta)\mathbf{x} = 
	\alpha\mathbf{x}+\beta\mathbf{x}$.
\end{enumerate}

On a vector space we can also define an inner product\index{inner product} 
(sometimes called the \index{scalar!product}
scalar product, not to be confused with scalar multiplication). When two 
vectors in $\mathcal{V}$ are denoted by $\mathbf{x}$ and $\mathbf{y}$, their 
inner product is a (complex) number written as $(\mathbf{x},\mathbf{y})$. For
all $\mathbf{x},\mathbf{y},\mathbf{z}\in{\mathcal{V}}$ and 
$\alpha\in\mathbb{C}$ the inner product obeys the following rules
\begin{enumerate}
 \item $(\mathbf{x}+\mathbf{z},\mathbf{y})=(\mathbf{x},\mathbf{y})+
	(\mathbf{z},\mathbf{y})$;
 \item $(\alpha\mathbf{x},\mathbf{y})=\alpha(\mathbf{x},\mathbf{y})$;
 \item $(\mathbf{x},\mathbf{y})^*=(\mathbf{y},\mathbf{x})$, where $^*$ denotes
	complex conjugation;\index{complex conjugation}
 \item $(\mathbf{x},\mathbf{x})>0$ if $\mathbf{x}\neq 0$.
\end{enumerate}
A complex vector space\index{vector space!complex} with an inner product is 
called a {\em Hilbert space}.\index{Hilbert space}
Note that we only defined algebraic rules for the inner product, the actual 
form of $(\mathbf{x},\mathbf{y})$ depends on the 
representation.\index{representation}

In Dirac's bracket notation, the elements of the Hilbert space $\hilbert$ are 
written as so-called {\em kets}:\index{kets} 
$|\psi\rangle,|\phi\rangle\in{\hilbert}$. The adjoints of these kets are 
called {\em bras}:\index{bras} $\langle\psi|,\langle\phi|$. The 
inner product is given by $\langle\psi|\phi\rangle$. It necessarily obeys all 
the conditions given above. Two vectors $|\psi\rangle$ and $|\phi\rangle$ in
Hilbert space are {\em orthogonal}\index{orthogonality} if and only if their 
inner product vanishes: $\langle\psi|\phi\rangle=0$.

In a Hilbert space $\hilbert$ of dimension $d$ we can construct a set of $d$ 
orthogonal vectors, called an orthogonal basis of $\hilbert$. When the vectors
in the basis have unit length, i.e., if for every basis vector $|\psi_i\rangle$
we have $\langle\psi_i|\psi_i\rangle=1$, then the basis is {\em orthonormal}.
\index{orthonormal basis}

Next, we define {\em linear operators}\index{operator!linear} on $\hilbert$. 
Consider a 
transformation $A:{\hilbert}\rightarrow{\hilbert}$. $A$ is called a linear
operator on $\hilbert$ if for every $|\psi\rangle,|\phi\rangle\in{\hilbert}$ 
and $\alpha\in\mathbb{C}$
\begin{enumerate}
 \item $A(|\psi\rangle + |\phi\rangle) = A|\psi\rangle + A|\phi\rangle$;
 \item $A(\alpha|\psi\rangle) = \alpha (A|\psi\rangle)$.
\end{enumerate}
A linear operator transforms one vector in Hilbert space to another: 
$A|\psi\rangle=|\psi'\rangle$. In matrix notation $A$ corresponds to a matrix,
while kets correspond to column vectors and bras to row vectors.

Linear operators do not necessarily {\em commute}. That is, when we have two
linear operators $A$ and $B$ on $\hilbert$, their commutation 
relation\index{commutation relation} $[A,B]
= AB-BA$ is not necessarily zero. Non-zero commutation relations play an 
important r\^ole in quantum mechanics. For example, non-commuting operators 
lie at the heart of quantum cryptography.\index{quantum!cryptography}

Suppose a linear operator $A$ obeys the following relation:
\begin{equation}\label{eigeneq}
 A |\psi\rangle = \alpha|\psi\rangle\; ,
\end{equation}
where $|\psi\rangle$ is a vector in $\hilbert$ and $\alpha$ a complex number. 
This is called an eigenvalue equation\index{eigen!-value!equation} for $A$, 
where $\alpha$ is the {\em eigenvalue}\index{eigen!-values} and $|\psi\rangle$ 
the corresponding {\em eigenvector}.\index{eigen!-vectors} When the 
dimension of $\hilbert$ is $d$, every linear operator on $\hilbert$ has $d$ 
eigenvalue equations. An eigenvalue $\alpha$ might be $d_{\alpha}$-fold 
degenerate, in which case $\alpha$ generates a $d_{\alpha}$-dimensional {\em 
eigenspace}:\index{eigen!-space} there are $d_{\alpha}$ orthogonal vectors 
$|\psi_{\alpha}\rangle$ 
which obey Eq.~(\ref{eigeneq}), thus forming a basis for a 
$d_{\alpha}$-dimensional subspace of $\hilbert$.

There also exists a property called the {\em trace}\index{trace} of an 
operator $A$. We write
\begin{equation}\label{trace}
 {\rm Tr} A = \sum_i \langle\psi_i|A|\psi_i\rangle\; ,
\end{equation}
where $\{|\psi_i\rangle\}$ can be {\em any} complete orthonormal basis.
The trace\index{trace} has the following properties:
\begin{enumerate}
 \item if $A^{\dagger}=A$ then ${\rm Tr} A$ is real;
 \item ${\rm Tr}(\alpha A) = \alpha {\rm Tr} A$;
 \item ${\rm Tr}(A+B) = {\rm Tr}A+{\rm Tr}B$;
 \item ${\rm Tr}(AB)={\rm Tr}(BA)$, the cyclic 
	property.\index{trace!cyclic property}
\end{enumerate}
These properties are easily proved using the knowledge that the trace as 
defined in Eq.~(\ref{trace}) is independent of the basis $\{|\psi_i\rangle\}$.

\section{Tensor product spaces}\index{tensor product!space}

A tensor product\index{tensor product} of two operators $A$ and $B$ is defined 
as follows:
\begin{equation}
 (A\otimes B) (|\psi_i\rangle_1\otimes|\phi_j\rangle_2) = 
 (A|\psi_i\rangle_1)\otimes(B|\phi_j\rangle_2)\; ,
\end{equation}
which is equivalent to $(A\otimes B)(C\otimes D) = (AC\otimes BD)$. In other 
words, every operator sticks to its own Hilbert space.\index{Hilbert space} It 
should be noted, 
however, that {\em not every operator on ${\hilbert}_1\otimes{\hilbert}_2$ is 
of the form} $A\otimes B$. The fact that this is not the 
case is also of fundamental importance to quantum information 
\index{quantum!information} theory, as we 
shall see in the remainder of this thesis (see also appendix \ref{app:povm}).

Other properties of the tensor product of operators are \cite{hilgevoord93}
\begin{enumerate}
 \item $A\otimes 0 = 0\otimes B =0$,
 \item ${\unity}\otimes{\unity} = {\unity}$,
 \item $(A_1+A_2)\otimes B = (A_1\otimes B) + (A_2\otimes B)$,
 \item $\alpha A \otimes \beta B = \alpha\beta (A\otimes B)$,
 \item $(A\otimes B)^{-1} = A^{-1}\otimes B^{-1}$,
 \item $(A\otimes B)^{\dagger} = A^{\dagger}\otimes B^{\dagger}$,
 \item ${\rm Tr}(A\otimes B) = {\rm Tr} A \cdot {\rm Tr} B$.
\end{enumerate}
For notational brevity the tensor product symbol $\otimes$ is often omitted,
yielding, e.g., $|\psi\rangle\otimes|\phi\rangle = |\psi\rangle|\phi\rangle =
|\psi,\phi\rangle$. When this abbreviated notation is used, one should always
remember which state or operator is defined on which Hilbert space.

\section{Projection operators}\index{operator!projection}

So far, we have only considered tensor products\index{tensor product} of 
Hilbert spaces.\index{Hilbert space} However, there is also an operation 
`$\oplus$', called the {\em direct sum}\index{direct!sum} of two vector 
spaces.\index{vector space} The direct sum of two vector spaces 
${\mathcal{W}}={\mathcal{V}}_1\oplus{\mathcal{V}}_2$ is again a vector space, 
and ${\mathcal{V}}_1$ and ${\mathcal{V}}_2$ are called its {\em subspaces}. 
\index{subspace} We can write a vector in 
$\mathcal{W}$ as $\mathbf{w}=(\mathbf{v}_1,\mathbf{v}_2)$, where $\mathbf{v}_1$
and $\mathbf{v}_2$ are vectors in the respective subspaces ${\mathcal{V}}_1$ 
and ${\mathcal{V}}_2$. The subspaces are linear if
\begin{equation}
 \alpha(\mathbf{u}_1,\mathbf{u}_2) + \beta(\mathbf{v}_1,\mathbf{v}_2) = 
(\alpha\mathbf{u}_1+\beta\mathbf{v}_1,\alpha\mathbf{u}_2+\beta\mathbf{v}_2)\; .
\end{equation}
Here, I will only consider linear subspaces.\index{subspace!linear}

Suppose we have a Hilbert space which can be written as the direct sum of
two subspaces. These are again Hilbert spaces:
\begin{equation}
 {\hilbert} = {\hilbert}_1 \oplus {\hilbert}_2\; .
\end{equation}
A state $|\psi\rangle$ in $\hilbert$ can then be written as 
\begin{equation}
 |\psi\rangle = \alpha |\psi_1\rangle + \beta |\psi_2\rangle\; ,
\end{equation}
where $|\psi_1\rangle$ and $|\psi_2\rangle$ are restricted to their respective 
subspaces and $|\alpha|^2 + |\beta|^2 =1$. We can define an operator $P_1$ 
which yields 
\begin{equation}
 P_1 |\psi\rangle = P_1 \left( \alpha |\psi_1\rangle +
 \beta |\psi_2\rangle\right) = \alpha |\psi_1\rangle\; .
\end{equation}
In other words, $P_1$ projects the state $|\psi\rangle$ onto the linear 
subspace $\hilbert_1$. $P_1$ is said to be a {\em projection operator} or 
{\em projector}\index{projector} \cite{hilgevoord93}. An operator is a 
projection operator if and only if 
\begin{equation}\label{projector}
 P^2 = P = P^{\dagger}\; .
\end{equation}
Projection operators have the following properties:
\begin{enumerate}
 \item two projection operators $P$ and $Q$ are called {\em orthogonal} 
	projections if and only if $[P,Q]=0$, they project onto linearly 
	independent subspaces;
 \item the sum of two (orthogonal) projectors is again a projector;
 \item the sum over all orthogonal projectors in $\hilbert$ is the identity 
	operator $\unity$;
 \item the orthocomplement of a projector $P$ in $\hilbert$ is given by 
	$\unity-P$;
 \item the eigenvalues of a projector are 1 and 0.
\end{enumerate} 
When viewed as measurement outcomes\index{measurement!outcome} (see postulate 
3), the eigenvalues\index{eigen!-values} of a projection 
operator\index{operator!projection} indicate whether the state is in the 
subspace\index{subspace} spanned by $P$ or not.

It is easily checked that for any state $|\psi\rangle$ in $\hilbert$ a 
projector on the subspace spanned by $|\psi\rangle$ can be written as
\begin{equation}
 P_{|\psi\rangle} = |\psi\rangle\langle\psi|\; .
\end{equation}
When $\{|\psi_i\rangle\}$ is a complete orthonormal 
basis\index{orthonormal basis} of $\hilbert$, the 
identity operator can then be written as
\begin{equation}
 \unity = \sum_i P_{|\psi_i\rangle} = \sum_i|\psi_i\rangle\langle\psi_i|\; .
\end{equation}
This is called the {\em completeness relation}.\index{completeness relation} 
An operator $A$ with eigenvalues $\alpha_i$ whose eigenvectors are given by 
the basis $\{|\psi_i\rangle\}$ can then be written as 
\begin{equation}
 A = \sum_i \alpha_i |\psi_i\rangle\langle\psi_i|\; .
\end{equation}
This is sometimes called the {\em spectral 
decomposition}\index{spectral decomposition} of $A$. In general, 
when the eigenvectors of an operator $A$ are {\em not} given by this basis, $A$
can be written as
\begin{equation}
 A = \sum_{ij} \alpha_{ij} |\psi_i\rangle\langle\psi_j|\; .
\end{equation}
When $A$ is Hermitian\index{operator!Hermitian} ($A^{\dagger}=A$), we have 
$\alpha_{ij}=\alpha_{ji}^*$.

\newpage
\thispagestyle{empty}

\chapter{States, Operators and Maps}\label{app:povm}

In this appendix I summarise some background knowledge about states,
operators and maps in the context of quantum mechanics. This knowledge is
important for the understanding of the Peres-Horodecki partial transpose
criterion\index{partial transpose!criterion} for the separability 
\index{separability} of bi-partite\index{bi-partite system} density 
matrices,\index{density matrix} and it also 
lays the foundations for the definition of positive operator valued measures. 
\index{POVM} For this appendix I am indebted to professor Rajiah Simon, who 
guided me through Hilbert space.

\section{Single systems}

Suppose we have a physical system which is described by a set of accessible 
states\index{state} $\{|\phi_j\rangle\}$. The superposition 
principle\index{superposition!principle} and the linearity of 
quantum mechanics imply that this set spans a Hilbert 
space\index{Hilbert space} $\hilbert$ of 
dimension $d$. This is a complex vector space\index{vector space!complex} with 
an orthonormal basis\index{orthonormal basis}
\begin{equation}\nonumber
 \begin{pmatrix}
  1 \\
  0 \\
  \vdots \\
  0  
 \end{pmatrix},
 \begin{pmatrix}
  0 \\
  1 \\
  \vdots \\
  0  
 \end{pmatrix}, \ldots,~
 \begin{pmatrix}
  0 \\
  0 \\
  \vdots \\
  1  
 \end{pmatrix}.
\end{equation}
We can define a set of linear operators\index{operator!linear} $\{ A_k\}$ on $
\hilbert$, the elements of which transform one state to another:
\begin{equation}
 A : |\phi\rangle \longrightarrow |\phi'\rangle\; ,
\end{equation}
where $|\phi'\rangle$ is again a state in $\hilbert$. Hermiticity of $A$ 
($A^{\dagger}=A$) implies that all the eigenvalues\index{eigen!-values} of $A$ 
are real. When for all $|\phi\rangle$ we have $\langle\phi|A|\phi\rangle\geq 
0$, $A$ is {\em non-negative}. A non-negative 
operator\index{operator!non-negative} with trace 1 (${\rm Tr}[A]=1$) is 
called a {\em density operator},\index{operator!density} usually denoted by 
$\rho$. Note that all non-negative operators are also 
Hermitian,\index{operator!Hermitian} and have non-negative eigenvalues.
A choice of a basis in $\hilbert$ puts Hermitian non-negative operators into
a one-to-one correspondence with Hermitian non-negative matrices.

The set of all linear operators $\{ A_k\}$ in turn define a Hilbert space 
$\hilbert\otimes\hilbert$ of dimension $d^2$, one orthonormal 
basis\index{orthonormal basis} of which 
can be written as
\begin{equation}
 \begin{pmatrix}
  1 & 0 & \ldots & 0 \\
  0 & \ddots & &\\
  \vdots & & & \vdots\\
  0 & 0 & \ldots & 0 
 \end{pmatrix},
 \begin{pmatrix}
  0 & 1 & \ldots & 0 \\
  0 & \ddots & &\\
  \vdots & & & \vdots\\
  0 & 0 & \ldots & 0 
 \end{pmatrix}, \ldots,~
 \begin{pmatrix}
  0 & 0 & \ldots & 0 \\
  0 & \ddots & &\\
  \vdots & & & \vdots\\
  0 & 0 & \ldots & 1 
 \end{pmatrix}.
\end{equation}

We can now define an even higher set of objects called {\em maps},\index{map} 
denoted by $\{ {\mathcal{L}}\}$, the elements of which linearly transform the 
set of linear operators\index{operator!linear} into itself:
\begin{equation}
 {\mathcal{L}} : A \longrightarrow A' = {\mathcal{L}} (A)\; .
\end{equation}
These linear maps\index{map!linear} are sometimes called {\em super}-operators.
\index{operator!super-} They are operators on the Hilbert 
space\index{Hilbert space} ${\hilbert}\otimes{\hilbert}$, but we give 
them a different name to avoid confusion. A simple example of a map corresponds
to a unitary transformation\index{transformation!unitary} $U : A\rightarrow 
A'=U^{\dagger}AU$. The corresponding map ${\mathcal{L}}_U$ may be written as 
${\mathcal{L}}_U = U^{\dagger}\otimes U$. The set of all maps thus constitutes 
a Hilbert space ${\hilbert}^{\otimes 4}$ of dimension $d^4$. The concept of 
non-negative operators\index{operator!non-negative} lead us to define {\em 
positive} maps.\index{map!positive}

\begin{description}
 \item[Definition:] a map $\mathcal{L}$ is called {\em positive} if for every 
	non-negative operator $A$ the operator $A'={\mathcal{L}}(A)$ is again 
	a non-negative operator.
\end{description}
When ${\rm Tr}[A']={\rm Tr}[A]$ for every $A$, the map $\mathcal{L}$ is a 
{\em trace-preserving} map.\index{map!trace-preserving} Trace-preserving 
positive maps are important in quantum mechanics, since they transform the set 
of density operators\index{operator!density} to itself. This property may lead 
one to expect that these maps correspond to physical processes or 
symmetries.\index{symmetry} It is an interesting aspect of quantum mechanics 
that {\em not all} positive maps can be associated with physical processes. 
This subtle fact becomes important when we consider composite systems.

\section{Composite systems}\index{composite system}

Suppose we have {\em two} systems 1 and 2 with respective accessible states
$\{|\phi^{(1)}_j\rangle\}$ and $\{|\psi^{(2)}_k\rangle\}$. These states span 
two
Hilbert spaces ${\hilbert}^{(1)}$ and ${\hilbert}^{(2)}$ with dimensions $d_1$
and $d_2$ respectively. The accessible states of the composite system can be 
written on the basis of the tensor product\index{tensor product} of the states 
$|\phi^{(1)}_j\rangle\otimes|\psi^{(2)}_k\rangle$, generating a Hilbert space 
${\hilbert}^{(1)}\otimes{\hilbert}^{(2)}$ of dimension $d_1\times d_2$. 
Similarly, the set of linear operators\index{operator!linear} $\{ A_j\}$ on 
${\hilbert}^{(1)}\otimes{\hilbert}^{(2)}$ generates a Hilbert space of 
dimension $(d_1\times d_2)^2$, and the set of maps generates a Hilbert space
of dimension $(d_1\times d_2)^4$.

Consider a map ${\mathcal{L}}_1$, defined for subsystem 1. When this map is 
positive (and trace-preserving) it transforms density operators of the 
subsystem to density operators. When system 1 is part of a composite system 
$1+2$, we want to know when a positive map of system 1 (leaving system 2 
unchanged) would transform a density operator\index{operator!density} defined 
on the composite system\index{composite system} again into a density operator. 
In other words, we ask when the {\em extended}\index{map!extended} 
map ${\mathcal{L}}_{12} = {\mathcal{L}}_{1}\otimes{\unity}_2$, with 
${\unity}_2$ the identity map of system 2, is again positive. 
\begin{description}
 \item[Definition:] a map ${\mathcal{L}}_{1}$ is called {\em completely 
	positive}\index{map!completely positive} if all its extensions are 
	positive.
\end{description}

There exist maps which are positive, but not completely positive. One such 
map is the transpose.\index{transpose} Take, for example, the 
singlet\index{state!singlet} state of a two-level bi-partite 
system\index{bi-partite system} (written in the computational 
basis):\index{computational basis}\index{Hilbert space}
\begin{equation}
 |\Psi\rangle = \frac{1}{\sqrt{2}} \left( |0,1\rangle - |1,0\rangle \right)\; .
\end{equation}
The density operator of this state can be written as
\begin{equation}\label{app:sing}
 \rho = |\Psi\rangle\langle\Psi| = \frac{1}{2} \left( |0,1\rangle\langle 0,1|
 - |0,1\rangle\langle 1,0| - |1,0\rangle\langle 0,1| + 
 |1,0\rangle\langle 1,0| \right)\; .
\end{equation}
The transpose of a general density operator for a single system in this 
notation is given by
\begin{multline}
 T : a |0\rangle\langle 0| + b |0\rangle\langle 1| + c |1\rangle\langle 0| + 
 d |1\rangle\langle 1| \cr \longrightarrow\quad a |0\rangle\langle 0| + b 
 |1\rangle\langle 0| + c |0\rangle\langle 1| + d |1\rangle\langle 1|\; ,
\end{multline}
that is, we {\em exchange} the entries of the bras and kets. This is a 
positive map.\index{map!positive} The extended transpose (or {\em partial} 
transpose)\index{partial transpose} on a 
compound system $PT = T_1\otimes{\unity}_2$, however, is {\em not} positive. 
To see this, apply the extended transpose to the density operator given in 
Eq.~(\ref{app:sing}), we obtain  
\begin{equation}
 PT : \rho \longrightarrow \rho' = \frac{1}{2} \left( |0,1\rangle\langle 0,1|
 - |0,0\rangle\langle 1,1| - |1,1\rangle\langle 0,0| + 
 |1,0\rangle\langle 1,0| \right)\; .
\end{equation}
If the eigenvalues\index{eigen!-values} of $\rho'$ are non-negative, 
$\rho'$ is again a density operator.\index{operator!non-negative} In order to 
find the eigenvalues of this operator we write $\rho'$
in matrix representation on the computational basis:
\begin{equation}
 \rho' = \frac{1}{2} 
 \begin{pmatrix}
   0 & 0 & 0 & -1 \cr
   0 & 1 & 0 &  0 \cr
   0 & 0 & 1 &  0 \cr
  -1 & 0 & 0 &  0
 \end{pmatrix}\; .
\end{equation}
It is easily found that this matrix has eigenvalues 1 (with multiplicity 3)
and $-1$. Therefore, $\rho'$ is not a density operator, and $T$, although
positive, is not a {\em completely} positive 
map.\index{map!completely positive}

I will now present an important class of completely positive maps. Consider 
the general map 
\begin{equation}
 {\mathcal{L}} : A \longrightarrow A' = {\mathcal{L}} (A)\; ,
\end{equation}
with $A$ and $A'$ linear operators\index{operator!linear} on the system 
Hilbert space.\index{Hilbert space} An important special case of such a 
map\index{map} is given by
\begin{equation}\label{effect}
 {\mathcal{L}} : A \longrightarrow A' = \sum_k\lambda_k B_k A B^{\dagger}_k\; ,
\end{equation}
where the $B_k$'s are again linear operators. In particular, we
can define a family of such maps, as given by Eq.~(\ref{family}). When 
$\lambda_k\geq 0$ for all $k$, the map in Eq.~(\ref{effect}) is again a 
positive map.\index{map!positive} To prove this statement, note that 
\begin{equation}
 \langle\phi|B_k A B^{\dagger}_k |\phi\rangle = \langle B^{\dagger}_k\phi| A 
 |B^{\dagger}_k\phi\rangle \equiv \langle\phi'|A|\phi'\rangle \geq 0
\end{equation}
for all $B_k$ and $|\phi\rangle$ if $A$ is non-negative. We then have 
\begin{equation}
 \langle\phi|A'|\phi\rangle = \sum_k \lambda_k \langle\phi|B_k A 
 B^{\dagger}_k|\phi\rangle = \sum_k \lambda_k \langle\phi'_k|A|\phi'_k\rangle
 \; .
\end{equation}
The right-hand side of this equation is positive for all $B_k$'s, 
$\lambda_k\geq 0$'s and non-negative operators\index{operator!non-negative} 
$A$. Hence $A'$ is a non-negative operator and $\mathcal{L}$ is positive. 

Furthermore, when $\lambda_k\geq 0$ such an $\mathcal{L}$ is {\em completely} 
positive.\index{map!completely positive} To prove this statement, let 
$\mathcal{L}$ be a map on system 1 (henceforth denoted by ${\mathcal{L}}_1$) 
and consider a second system 2. Recall that ${\mathcal{L}}_1$ is completely 
positive if all its extensions\index{map!extended} 
${\mathcal{L}}_1\otimes{\unity}_2$ are positive. I will now show that this is 
the case.

Define the extension ${\mathcal{L}}_{12}={\mathcal{L}}_1\otimes{\unity}_2$.
System 2 can have arbitrary dimension, and may itself be composite. We thus 
have to show that ${\mathcal{L}}_{12}$ is positive. Let $A_{12}$ be a 
non-negative operator on the composite system $1+2$:
\begin{equation}
 A_{12} = \sum_{j,k;l,m} a_{jk,lm} |\phi_j\rangle_1|\psi_k\rangle_2
 \langle\psi_m|_1\langle\phi_l|\; ,
\end{equation}
with $a_{jk,lm}=a^*_{lm,jk}$, and define the operator $B_i$ on system 1 as
\begin{equation}
 B_i = \sum_{p,q} b^i_{pq} |\phi_p\rangle\langle\phi_q|\; .
\end{equation}
The map ${\mathcal{L}}_1\otimes{\unity}_2$ is then given by the transformation
\begin{eqnarray}
 A_{12} \rightarrow A_{12}' &=& \sum_i \lambda_i 
 (B_i\otimes{\unity}_2) A_{12} (B_i\otimes{\unity}_2)^{\dagger} \cr
 &=& \sum_i \lambda_i \sum_{p,q;r,s} \sum_{j,k;l,m} a_{jk,lm} b^i_{pq}
 |\phi_p\rangle_1\langle\phi_q|\phi_j\rangle_1 |\psi_k\rangle_2\langle\psi_m|
 _1\langle\phi_l|\phi_r\rangle_1\langle\phi_s| {b^i_{sr}}^* \cr
 &=& \sum_i \lambda_i \sum_{p,s} \sum_{j,k;l,m} a_{jk,lm} b_{pj}
 |\phi_p\rangle_1 |\psi_k\rangle_2\langle\psi_m| _1\langle\phi_s| 
 {b^i_{sl}}^*\cr
 &=& \sum_i \lambda_i \sum_{j,k;l,m} a_{jk,lm}
 |\phi_j'\rangle_1 |\psi_k\rangle_2\langle\psi_m| _1\langle\phi_l'|\; ,
\end{eqnarray}
where I defined $|\phi_j'\rangle = \sum_p b_{pj}|\phi_p\rangle$. Since this is 
a convex sum\index{convex sum} over non-negative 
operators,\index{operator!non-negative} the resulting operator is again
non-negative and $\mathcal{L}$ is completely 
positive.\index{map!completely positive} This completes the proof. 

The fact that positive but not completely positive maps on a subsystem do not 
necessarily transform
density operators on the composite system to density operators can be 
exploited to detect (or witness) quantum entanglement.\index{entanglement} 
This is the subject of the next section.

\section{Partial transpose criterion}\index{partial transpose!criterion}

In quantum information theory,\index{quantum!information} it is important to 
know whether a composite system,\index{composite system} characterised by a 
density operator\index{operator!density} $\rho$ is 
separable\index{separability} or not. One way to test this is to use the 
Peres-Horodecki partial transpose criterion \cite{peres96,horodecki96}.

A density operator $\rho$ of two systems $1+2$ is separable if and only if 
it can be written as
\begin{equation}\label{app:rhosep}
 \rho = \sum_k p_k \rho^{(1)}_k \otimes \rho^{(2)}_k\; ,
\end{equation}
with $p_k > 0$ and $\sum_k p_k =1$. The density operator $\rho^{(j)}_k$ is
defined on system $j=1,2$. Consider again the transpose\index{transpose} of 
an operator $A$:
\begin{equation}
 T : A \longrightarrow A^T\; .
\end{equation} 
As we have seen, this is a trace-preserving\index{map!trace-preserving} 
positive,\index{map!positive} but not completely\index{map!completely positive}
positive map. We extended this map to the {\em partial transpose} 
$PT\equiv T_1 \otimes{\unity}_2$. The partial transpose is not positive on 
the composite system.

Under the partial transpose, the separable density operator from 
Eq.~(\ref{app:rhosep}) will transform according to 
\begin{equation}
 PT : \rho \longrightarrow \rho' = \sum_k p_k \left(\rho^{(1)}_k\right)^T 
 \otimes \rho^{(2)}_k\; .
\end{equation}
However, $T$ is positive and $(\rho^{(1)}_k)^T$ is again a density operator.
Therefore $\rho'$ is another (separable) density operator. Now look at the 
eigenvalues\index{eigen!-values} of $\rho'$. Clearly, if $\rho$ is separable, 
then $\rho'$ has positive eigenvalues.\index{operator!non-negative} Therefore, 
if $\rho'$ has one or more 
{\em negative} eigenvalues, the original density operator $\rho$ must have 
been entangled.\index{entanglement} 
This is the Peres-Horodecki partial transpose criterion. Clearly, it is only
a necessary condition for separability.

It has been proved \cite{horodecki96} that for the Hilbert 
spaces\index{Hilbert space} 
${\hilbert}_2\otimes{\hilbert}_2$ and ${\hilbert}_2\otimes{\hilbert}_3$ the
partial transpose criterion is both necessary and sufficient. In other words,
$\rho$ is separable {\em if and only if} the eigenvalues of its partial
transpose $\rho'$ are positive. For higher dimensional Hilbert spaces this is 
no longer true. In that case there can exist density operators which are {\em 
not} separable, but for which the eigenvalues of $\rho'$ are non-negative. 
Such states are said to exhibit {\em bound} 
entanglement\index{entanglement!bound} 
\cite{horodecki97,horodecki98}. It is generally believed that this form of 
entanglement cannot be purified.\index{purification}

\section{projection operator valued measures}\index{POVM}

Let us now return to the case of a single system and the states, operators
and maps defined on it. Consider a projection 
operator\index{operator!projection} $P$ defined by
\begin{equation}
 P^{\dagger} = P \qquad\mbox{and}\qquad P^2 = P\; .
\end{equation}
In terms of the states $\{ |\phi_j\rangle\}$ this operator can be written as 
\begin{equation}
 P_{|\phi\rangle} = |\phi\rangle\langle\phi|\; .
\end{equation}

Suppose we have a set of projection operators $\{ P_{|\mu\rangle}\}$, with 
the states $|\mu\rangle$ not necessarily orthogonal. We can define a {\em 
generalised} projection operator $\hat{E}_{\nu}$ as a weighted measure over 
this set:
\begin{equation}
 \hat{E}_{\nu} = \sum_{\mu} \lambda^{\nu}_{\mu} P_{|\mu\rangle}\; ,
\end{equation}
with $\lambda^{\nu}_{\mu}>0$ and 
\begin{equation}
 \sum_{\nu} \hat{E}_{\nu} = \unity\; .
\end{equation}
The operator $E_k$ is called a {\em projection operator valued measure} or 
POVM for short \cite{kraus83}.

This can be generalised further by observing that 
\begin{equation}
 \lambda^{\nu}_{\mu} P_{|\mu\rangle} = \alpha^{\nu}_{\mu} 
 |\mu\rangle\langle\nu|\nu\rangle\langle\mu| (\alpha^{\nu}_{\mu})^*\; ,
\end{equation}
with $|\nu\rangle\in\{|\mu\rangle\}$ and $|\alpha^{\nu}_{\mu}|^2 =
\lambda^{\nu}_{\mu}$. When we define the operator $\mathcal{A}$:
\begin{equation}
 {\mathcal{A}} \equiv \sum_{\mu,\nu} \alpha^{\nu}_{\mu} |\mu\rangle\langle\nu|
\end{equation}
we can write the POVM as 
\begin{equation}
 E_{\nu} = \sum_{\mu}{\mathcal{A}}_{\mu\nu}{\mathcal{A}}_{\mu\nu}^{\dagger}\; .
\end{equation}

\chapter{Elementary Group Theory}\label{app:group}

In this appendix I give some background theory of Lie groups. I am indebted 
to the book by De Wit and Smith \cite{dewit86}, which gives a good and concise 
exposition of the subject. Further Lie group theory in particle physics is 
presented in Halzen and Martin \cite{halzen84}. For group theory in quantum 
mechanics, see also Chaichian and Hagedorn \cite{chaichian98}. For a formal 
treatment of Lie groups, see Gilmore \cite{gilmore94}.

A set $\mathsf{G}$ is called a {\em group}\index{group} when $\mathsf{G}$ 
satisfies the following requirements:
\begin{itemize}
 \item There exists a multiplication rule\index{group!multiplication} $(\cdot)$
	such that for every two
	elements $g_1$ and $g_2$ of the group, their product $g_1\cdot g_2$
	is again an element of the group;
 \item the multiplication rule is associative,\index{group!associativity} 
	i.e., $g_1 \cdot (g_2 \cdot g_3)
	= (g_1 \cdot g_2) \cdot g_3$ for all $g_1,g_2,g_3\in{\mathsf{G}}$;
 \item there exists an element $e\in{\mathsf{G}}$, called the unit element, 
	\index{group!unit element}
	for which the product $e\cdot g = g\cdot e = g$, with $g$ any element 
	of $\mathsf{G}$;
 \item for every $g\in{\mathsf{G}}$ there exists an element $g^{-1}\in
	{\mathsf{G}}$, called the inverse element\index{group!inverse element} 
	of $g$, such that $g\cdot g^{-1}=g^{-1}\cdot g = e$.
\end{itemize}
When $g_1 \cdot g_2 = g_2\cdot g_1$, the group is called {\em 
Abelian}.\index{group!Abelian} In
other words, the elements of $\mathsf{G}$ {\em commute}.\index{commutation} 
A subset $\mathsf{H}$ of $\mathsf{G}$ is called a {\em 
subgroup}\index{subgroup} of $\mathsf{G}$ if the group 
requirements above hold for $\mathsf{H}$. This is written as 
${\mathsf{H}}\subset{\mathsf{G}}$. 

\section{Lie groups}

If a group $\mathsf{G}$ has a finite number of elements, this number is called 
the {\em order}\index{group!order} of $\mathsf{G}$. For finite groups, see 
e.g., Serre \cite{serre67}. When the group has an infinite number of 
elements, the group can be either continuous\index{group!continuous} or 
discontinuous.\index{group!discontinuous} In the context 
of this thesis I am mostly interested in continuous groups. If the elements 
of a continuous group $\mathsf{G}$ depend analytically on a (finite) set of 
parameters\index{group!parameter} ($g=g(\vec{\xi})$), we speak of a {\em Lie 
group}\footnote{After 
the Norwegian mathematician Marius Sophus Lie (1842--1899).}. The {\em 
dimension} of the Lie group is given by the number of independent parameters:
if $\vec\xi=(\xi_1,\ldots,\xi_N)$, we have $\dim{\mathsf{G}}=N$. The 
$N$-dimensional space generated by the parameters is called {\em parameter 
space}.\index{parameter space}

Let $\mathsf{G}$ be a one-dimensional Lie group with elements $g(\xi)$. We can 
always choose the parametrisation\index{parametrisation} such that 
\cite{dewit86}
\begin{equation}
 g(\xi) g(\xi') = g(\xi + \xi')
\end{equation}
with
\begin{equation}
 g(0) \equiv e \qquad\mbox{and}\qquad (g(\xi))^{-1} = g(-\xi)\; .
\end{equation}

If we interpret the group elements as operators (acting on other group 
elements), the unit element\index{group!unit element} is the identity 
operator: $e = {\unity}$. In the neighbourhood of the identity,\index{identity}
a group element\index{group!element} can thus be written as an expansion
\begin{equation}
 g(\xi) = g(0) + \xi T + O(\xi^2)\; ,
\end{equation}
where $T$ is some operator. 

Since $\mathsf{G}$ is a continuous group,\index{group!continuous} we can write 
the transformation 
from the identity [written as $g(0)$] to $g(\xi)$ in terms of $n$ small steps
$g(\xi/n)$:
\begin{equation}
 g(\xi) = \left[ g(\xi/n) \right]^n\; .
\end{equation}
If I now take the limit of $n\rightarrow\infty$ the higher-order terms vanish
and we obtain
\begin{equation}
 g(\xi) = \lim_{n\rightarrow\infty} \left[ g(0) + \frac{\xi T}{n} + \ldots 
 \right]^n = \lim_{n\rightarrow\infty} \left[ {\unity} + \frac{\xi T}{n} 
 \right]^n \equiv \exp\left[\xi T\right]\; .
\end{equation}
The operator $T$ is said to be the {\em generator}\index{group!generator} of 
the group $\mathsf{G}$
because it generates the elements of $\mathsf{G}$. This can be generalised 
immediately to $N$-dimensional Lie groups, yielding 
\begin{equation}
 g(\vec\xi) = \exp\left[\sum_{i=1}^N \xi_i T_i\right]\; ,
\end{equation}
where the $\xi_i$'s are the independent parameters\index{group!parameter} of 
the group and the $T_i$'s
the generators. There are as many different generators as there are parameters.

In terms of the generators, the group 
multiplication\index{group!multiplication} can be written as
\begin{equation}
 g(\vec\xi) \cdot g(\vec\xi') = \exp\left[\sum_{i=1}^N \xi_i T_i\right]
 \exp\left[\sum_{i=1}^N \xi_i' T_i\right]\; .
\end{equation}
The right-hand side can be expressed as the argument of a single exponent 
by means of the Baker-Campbell-Hausdorff\index{Baker-Campbell-Hausdorff} 
formula:
\begin{equation}\label{app:BCH}
 \exp\left[\sum_{i=1}^N \xi_i T_i\right] \exp\left[\sum_{i=1}^N \xi_i' 
 T_i\right] = \exp\left[\sum_{i=1}^N (\xi_i + \xi_i') T_i + \frac{1}{2} 
 \sum_{i,j=1}^N \xi_i\xi'_j [T_i,T_j] + \ldots\right]\; ,
\end{equation}
where $[T_i,T_j]$ denotes the commutator\index{commutation} between $T_i$ and 
$T_j$, and the dots
indicate a series of terms with higher-order commutators of the $T$'s (like, 
for example $[T_i,[T_j,T_k]]$). This series does not necessarily terminate.

I started with the condition that $\mathsf{G}$ is a group,\index{group} which 
implies that
the right-hand side of Eq.\ (\ref{app:BCH}) is again an element of $\mathsf{G}$
and thus can be written as $e^{\sum_{i=1}^N \xi_i'' T_i}$. In turn, this 
means that the generators are closed under commutation:
\begin{equation}\label{app:algebra}
 [T_i,T_j] = \sum_{k=1}^N c_{ij}^k T_k\; ,
\end{equation}
with the (complex) numbers $c_{ij}^k$ the so-called {\em structure constants}.
\index{structure constants} To see that this equation must hold, suppose that 
the argument of the right-hand side of Eq.\ (\ref{app:BCH}) {\em does not} 
imply Eq.~(\ref{app:algebra}). There is then a commutator $[T_l,T_m]$ which 
cannot be written as a sum over the generators: $[T_l,T_m] = X$. Repeated 
commutators should then cancel $X$, because the right-hand side of 
Eq.~(\ref{app:BCH}) is a group element. This can only happen when repeated 
commutators yield $X$. By definition, $X$ is then a member of the set of 
generators. This contradicts our assumption.

The structure constants $c_{ij}^k$ define a so-called Lie {\em 
algebra}.\index{algebra} They obey the Jacobi identity for structure constants:
\begin{equation}
 \sum_k c_{ij}^k c_{kl}^m + c_{jl}^k c_{ki}^m + c_{li}^k c_{kj}^m = 0\; .
\end{equation}
This is easily proved using the Jacobi identity for any three operators
$A$, $B$ and $C$:
\begin{equation}
 [[A,B],C] + [[B,C],A] + [[C,A],B] = 0\; .
\end{equation}

\section{Representations}

When we have a set of matrices $M_i$ with $i=1,\ldots,N$, and the commutation
relations between these matrices are given by
\begin{equation}
  [M_i,M_j] = \sum_{k=1}^N c_{ij}^k M_k\; ,
\end{equation}
then this set of matrices is said to form a {\em 
representation}\index{representation} of the Lie algebra\index{algebra} 
defined in Eq.\ (\ref{app:algebra}). When these matrices are
multiplied by $\theta_i$ and exponentiated, they define a representation of 
the group $\mathsf{G}$, denoted by $D({\mathsf{G}})$:
\begin{equation}
 D(g(\vec\theta)) = \exp\left[\sum_{i=1}^N \theta_i M_i \right]\; .
\end{equation}
The matrices $M_i$ form a {\em basis} of the representation. For $N$ matrices
the representation is said to be 
$N$-dimensional. If there exists a non-trivial subspace\index{subspace} of 
$\mathcal{V}$ spanned by the basis $\{ M_i\}$ (i.e., a subspace other than 
$\mathbf{0}$ and $\mathcal{V}$ itself) which is invariant under the group 
transformations, the representation is called {\em 
reducible}.\index{representation!reducible} If no such invariant subspace 
exists, the representation is {\em 
irreducible}.\index{representation!irreducible} The 
theory of representations is important for many applications in physics.

\section{Examples of Lie groups}

One of the most important Lie groups in quantum mechanics must be the group 
of $2\times 2$ unitary matrices. This group is called 
$SU(2)$.\index{group!$SU(2)$} The corresponding Lie algebra 
$su(2)$\index{algebra!$su(2)$} is given by three 
generators\index{group!generator} ($i,j,k\in\{ x,y,z\}$):
\begin{equation}
 [J_i,J_j] = \sum_k i\epsilon_{ij}^k J_k\; ,
\end{equation}
where $\epsilon_{ij}^k$ are the entries of the Levi-Civita 
tensor\index{Levi-Civita tensor} of rank three (entries with even permutations 
of the indices are 1, odd permutations give $-1$, and repeated indices give 0).
The generators are given by $J_i = \frac{1}{2}\sigma_i$, with
\begin{equation}
 \sigma_x = 
 \begin{pmatrix}
  0 & 1 \cr
  1 & 0 
 \end{pmatrix}\; ,\qquad
 \sigma_y = 
 \begin{pmatrix}
  0 & -i \cr
  i & 0 
 \end{pmatrix}\; ,\qquad
 \sigma_z = 
 \begin{pmatrix}
  1 & 0 \cr
  0 & -1 
 \end{pmatrix}\; ,
\end{equation}
the so-called Pauli matrices.\index{Pauli matrices} Representations of this 
group\index{group} are used in the description of angular momentum, spin and 
iso-spin, as well as in quantum optics (see appendix \ref{app:forms} for a 
relation between Lie algebras\index{algebra} and optical devices). If the 
parameters\index{group!parameter} are given by $\xi_x$, $\xi_y$ and 
$\xi_z$, a general $SU(2)$ group element in the fundamental (two-dimensional)
representation can be written as ($\xi\equiv\sqrt{\xi^2_x +\xi^2_y +\xi^2_z}$)
\begin{equation}\label{app:element}
 g(\vec\xi) = \cos\mbox{$\frac{1}{2}$}\xi
 \begin{pmatrix}
  1 & 0 \cr
  0 & 1
 \end{pmatrix} 
 + \frac{i\sin\frac{1}{2}\xi}{\xi}
 \begin{pmatrix}
  \xi_z & \xi_x - i\xi_y \cr
  \xi_x + i\xi_y & -\xi_z
 \end{pmatrix} \; .
\end{equation}
Since the group elements\index{group!element} depend periodically on $\xi$, we 
have $\xi = \sqrt{\xi^2_x +\xi^2_y +\xi^2_z}\leq 2\pi$, which means that the
parameter space\index{parameter space} of $SU(2)$\index{group!$SU(2)$} is {\em 
compact};\index{parameter space!compact} it can be restricted to a sphere with 
radius $2\pi$.

The group $SU(2)$ is closely related to the group $SO(3)$,\index{group!$SO(3)$}
the group of orthogonal $3\times 3$ matrices, better known as the {\em 
rotation group}\index{rotation group}\index{group!rotation} in three 
dimensions. 

We know that a rotation [in $SO(3)$] over $2\pi$ is equal to the identity. 
However, when $\xi=2\pi$ in Eq.\ (\ref{app:element}), we see that the $SU(2)$ 
group element is equal to $-{\unity}$. This behaviour is the reason why 
$SU(2)$, rather than $SO(3)$, is used to describe particles with spin. After 
all, spin $\frac{1}{2}$ particles need a rotation over $4\pi$ in order to 
return to their original state. The group $SU(2)$ is called the {\em covering 
group}\index{covering group}\index{group!covering} of $SO(3)$.

Another Lie group\index{group} which is important in the context of this 
thesis is the group $SU(1,1)$.\index{group!$SU(1,1)$} Its Lie 
algebra\index{algebra} is given by
\begin{equation}
 [J_x,J_y] = -i J_z \; , \qquad [J_y,J_z] = i J_x \; , \qquad 
 [J_z,J_x] = i J_y \; ,
\end{equation} 
and the elements of the fundamental (two-dimensional) representation are 
generated by the matrices $J_i = \frac{1}{2}\rho_i$, with
\begin{equation}
 \rho_x = 
 \begin{pmatrix}
  0 & -1 \cr
  1 & 0
 \end{pmatrix}, \qquad
 \rho_y = 
 \begin{pmatrix}
  0 & i \cr
  i & 0
 \end{pmatrix}, \qquad
 \rho_z = 
 \begin{pmatrix}
  1 & 0 \cr
  0 & -1
 \end{pmatrix}.
\end{equation}
The group elements can be written in the fundamental representation as
\begin{equation}
 g(\vec\zeta) = \cosh\mbox{$\frac{1}{2}$}\zeta
 \begin{pmatrix}
  1 & 0 \cr
  0 & 1 
 \end{pmatrix} 
 + \sinh\mbox{$\frac{1}{2}$}\zeta
 \begin{pmatrix}
  0 & \zeta_1 - i\zeta_2 \cr
  \zeta_1 + i\zeta_2 & 0 
 \end{pmatrix}\; ,
\end{equation}
with $\vec\zeta=(\zeta_1,\zeta_2,\zeta_3)$ and $\zeta\equiv\sqrt{\zeta_3^2
-\zeta_2^2-\zeta_1^2}$. There is no periodicity in $\zeta$, and the 
parameter space is therefore not compact. In quantum optics, this group is 
associated with {\em squeezing}.\index{squeezer}

\newpage
\thispagestyle{empty}

\chapter{Bilinear and Quadratic Forms}\label{app:forms}

In many problems in quantum optics we are faced with a unitary evolution $U$ 
due to a Hermitian operator ${\mathcal{H}}$ (with $U=\exp[-it{\mathcal{H}}/
\hbar]$), generally an interaction Hamiltonian. For computational simplicity 
we often wish that this evolution is in normal ordered 
form.\index{normal ordering} This form (or, 
consequently, the corresponding Baker-Campbell-Hausdorff 
formula)\index{Baker-Campbell-Hausdorff} is usually 
very complicated, if it exists at all. In this appendix I present two 
important classes of operators for which the normal ordered form of 
the unitary evolution can be derived. 

I first define the so-called {\em bilinear}\index{form!bilinear} and {\em 
quadratic forms}\index{form!quadratic} for 
the creation and annihilation operators. We will present the normal ordering 
for evolutions generated by the Hermitian operators which can be written in 
terms of these bilinear and quadratic forms.

Suppose we have two vectors $\vec{x} = (x_1,\ldots,x_n)$ and $\vec{y} = 
(y_1,\ldots,y_n)$. The scalar product between these vectors is denoted by 
$(\vec{x},\vec{y})$. Furthermore, let $C$ be an $n\times n$ matrix.
With $C$ we associate the quadratic form \cite{erdely53}
\begin{equation}\label{quadratic}
 \phi(\vec{x},\vec{x}) = (\vec{x},C\vec{x})\; ,
\end{equation}
and the bilinear form 
\begin{equation}\label{app:bilinear}
 \phi(\vec{x},\vec{y}) = (\vec{x},C\vec{y})\; .
\end{equation}

We have made no assumptions about the nature of the vector components, and it
is possible to define bilinear and quadratic forms in terms of creation and 
annihilation operators $\hat{a}^{\dagger}$ and $\hat{a}$. These operators obey 
the well-known commutation relations
\begin{equation}
 [\hat{a}_i,\hat{a}^{\dagger}_j] = \delta_{ij} \quad\text{and}\quad 
 [\hat{a}_i,\hat{a}_j] = [\hat{a}^{\dagger}_i,\hat{a}^{\dagger}_j] = 0\; .
\end{equation}
The quadratic form now reads $(\vec{a},C\vec{a})$ or $(\vec{a}^{\dagger},C
\vec{a}^{\dagger})$, with $\vec{a}=(\hat{a}_1,\ldots,\hat{a}_n)$ and 
$\vec{a}^{\dagger}=(\hat{a}_1^{\dagger},\ldots,\hat{a}_n^{\dagger})$. The 
bilinear form can be chosen many ways (according to Eq.\ (\ref{app:bilinear})),
but for our present purposes I write it as $(\vec{a}^{\dagger},C\vec{a})$. 

In quantum optics, the bilinear and quadratic form of creation and annihilation
operators occurs very often. Take, for instance, the interaction Hamiltonian 
for the beam-splitter in modes $a_1$ and $a_2$:
\begin{equation}\label{ihbil}
 {\mathcal{H}}_I = \kappa\;\hat{a}_1^{\dagger}\hat{a}_2 + \kappa^* \;
 \hat{a}_2^{\dagger} \hat{a}_1 \; ,
\end{equation}
where $\kappa$ is a coupling constant. This Hamiltonian is a bilinear form 
\index{form!bilinear} which may be written as
\begin{equation}\label{explbil}
 (\vec{a}^{\dagger},C\vec{a}) = (\hat{a}_1^{\dagger},\hat{a}_2^{\dagger})
 \begin{pmatrix}
  0 & \kappa \cr \kappa^* & 0
 \end{pmatrix}
 \begin{pmatrix}
  \hat{a}_1 \cr \hat{a}_2
 \end{pmatrix}\; .
\end{equation}

Another example is the interaction Hamiltonian due to parametric 
down-conversion in two modes $a_1$ and $a_2$. This is a sum of two quadratic 
forms\index{form!quadratic} (one for the creation and one for the 
annihilation operators):
\begin{equation}\label{ihqua}
 {\mathcal{H}}_I = \nu\; \hat{a}_1^{\dagger}\hat{a}_2^{\dagger} +
 \nu^*\; \hat{a}_2\hat{a}_1 \; ,
\end{equation}
where $\nu$ is a coupling constant. In symmetric form, the quadratic form of 
the creation operators reads
\begin{equation}
 (\vec{a}^{\dagger},C\vec{a}^{\dagger}) = \frac{1}{2} (\hat{a}_1^{\dagger},
 \hat{a}_2^{\dagger})
 \begin{pmatrix}
  0 & \nu \cr \nu & 0
 \end{pmatrix}
 \begin{pmatrix}
  \hat{a}_1^{\dagger} \cr \hat{a}_2^{\dagger}
 \end{pmatrix}\; .
\end{equation}

Usually, these interaction Hamiltonians are exponentiated to generate the 
unitary evolution of a system, and studying the behaviour of the bilinear and 
quadratic forms might simplify our computational task. In particular, we would 
like to find the normal ordered form\index{normal ordering} of 
$\exp[-it{{\mathcal{H}}_I}/\hbar]$, where 
${\mathcal{H}}_I$ is given by Eq.\ (\ref{ihbil}) or Eq.\ (\ref{ihqua}).

In the next two sections I will establish relations between the bilinear and 
quadratic forms and the Lie algebras\index{Lie algebra|see{algebra}} 
of\index{Lie group|see{group}} 
$SU(2)$\index{group!$SU(2)$} and $SU(1,1)$\index{group!$SU(1,1)$} respectively.
The two resulting theorems place restrictions on the matrix $C$ in the 
bilinear and quadratic forms. 

\section{Bilinear Forms and {\it SU}(2)}

In quantum optics, linear unitary operations like beam-splitters, half- and 
quar\-ter-wave plates, phase-shifters, polarisation rotations, etc.\  all 
preserve the number of photons. When we write these operations as $U = \exp[-it
{\mathcal{H}}/\hbar]$, with ${\mathcal{H}}$ some Hermitian operator (an 
interaction Hamiltonian), it is clear that every term in ${\mathcal{H}}$
should be a product of an equal number of creation and annihilation operators
(i.e., for every photon which is created, another will be destroyed). The 
lowest order interaction Hamiltonian which satisfies this requirement has a
bilinear form (see, e.g., Eq.\ (\ref{ihbil})). Furthermore, the resulting 
unitary operations form representations\index{representation!of $SU(2)$} of 
the group $SU(2)$, and I therefore study the relation between bilinear forms 
and $SU(2)$.

Define $K_-(\Lambda)$ to be 
\begin{equation}\label{lmin}
 K_-(\Lambda) = \sum_{ij} \hat{a}_i^{\dagger} \lambda_{ij} \hat{a}_j
 \equiv (\vec{a}^{\dagger},\Lambda\vec{a})\; ,
\end{equation}
a bilinear form. Let $K_+(\Lambda)$ be the adjoint of $K_-(\Lambda)$:
\begin{equation}\label{lplus}
 K_+(\Lambda) = K_-^{\dagger}(\Lambda) = (\vec{a}^{\dagger}\Lambda,\vec{a})\; .
\end{equation}

I can now define a third operator $K_0(\Lambda)$ in such a way that the three
operators generate an $su(2)$\index{algebra!$su(2)$} 
algebra\footnote{Traditionally, the {\em group}
is denoted by capital letters, e.g., $SU(2)$, whereas the corresponding {\em
algebra} is written with lowercase letters, e.g., $su(2)$.}:
\begin{equation}\label{lnul}
 K_0 = -\frac{1}{2} [K_-,K_+]\; .
\end{equation}
Using Eqs.\ (\ref{lmin}) and (\ref{lplus}), and normal ordering the 
commutator in Eq.\ (\ref{lnul}) yields
\begin{equation}\label{knul}
 K_0 (\Lambda) = -\frac{1}{2} \sum_{ij} \hat{a}^{\dagger}_i [\Lambda,
 \Lambda^{\dagger}]_{ij} \hat{a}_j = K_0^{\dagger} (\Lambda)\; .
\end{equation}
With $M \equiv [\Lambda,\Lambda^{\dagger}]$ this can be written as
\begin{equation}
 K_0 (M) = -\frac{1}{2} \sum_{ij} \hat{a}^{\dagger}_i \mu_{ij} \hat{a}_j = 
 -\frac{1}{2} (\vec{a}^{\dagger},M\vec{a})\; .
\end{equation}

The last commutation relation which, together with Eq.\ (\ref{lnul}) and 
$K_- = K_+^{\dagger}$, $K_0 = K_0^{\dagger}$ constitutes the $su(2)$ algebra is
\begin{equation}
 [K_0,K_{\pm}] = \pm K_{\pm}\; .
\end{equation}
This equation places a constraint on the allowed matrices $\Lambda$ (all the 
other commutation relations so far have not placed any restrictions on the form
of $\Lambda$). Using Eqs.\ (\ref{lmin}), (\ref{lplus}) and (\ref{knul}) yields
\begin{eqnarray}
 [ K_0, K_- ] = - K_- &\Leftrightarrow& \Lambda = \frac{1}{2}
 [ M,\Lambda] \cr
 [ K_0, K_+ ] = K_+ &\Leftrightarrow& \Lambda^{\dagger} = 
 \frac{1}{2} [ \Lambda^{\dagger},M ] \; .
\end{eqnarray}

This can be summarised in the following theorem:
\begin{description}
 \item[Theorem 1] Consider a bilinear operator of the form $K(\Lambda) = 
	(\vec{a}^{\dagger},\Lambda\vec{a})$. $K$ and $K^{\dagger}$ define a 
	third operator $K_0 = -\frac{1}{2}[K,K^{\dagger}]$. These operators are
	generators\index{algebra!generators of $su(2)$} of an $su(2)$ algebra 
	if and only if $\Lambda = \frac{1}{2} 
	[M,\Lambda]$, with $M = [\Lambda,\Lambda^{\dagger}]$.
\end{description}

Suppose that the interaction Hamiltonian can be written as
\begin{equation}
 {\mathcal{H}}_I = \kappa \; K_+ (\Lambda) + \kappa^* \; K_- (\Lambda) \; ,
\end{equation}
with $\kappa$ again a coupling constant. Note that this is now a {\em sum} of 
two bilinear forms,\index{form!bilinear} unlike in Eq.\ (\ref{explbil}). Since 
$K_+$ and $K_-$ 
generate an $su(2)$ algebra, we know what the normal ordering for the unitary 
evolution $U$ associated with this interaction Hamiltonian is ($\tau = -i
\kappa t/\hbar$ and $\hat{\tau}$ is the unit vector in the direction of $\tau$)
\cite{truax85}:
\begin{eqnarray}
 U(\tau) &=& \exp[\tau K_+ (\Lambda) - \tau^* K_- (\Lambda)] \cr
 &=& \exp[\hat{\tau}\tan|\tau| K_+] \exp[-2\ln\cos|\tau| K_0]\rule{0pt}{10pt}
 \exp[-\hat{\tau}^* \tan|\tau| K_-]\rule{0pt}{10pt}\; .
\end{eqnarray}
This is the Baker-Campbell-Hausdorff 
formula\index{Baker-Campbell-Hausdorff!$su(2)$} for $su(2)$ \cite{gilmore94}.

\section{Quadratic Forms and {\em SU}(1,1)}

In quantum optics, squeezers\index{squeezer} and 
down-converters\index{down-converter} are described by interaction
Hamiltonians which are quadratic in the creation and annihilation operators.
These Hamiltonians generate unitary transformations which do not conserve
the photon number. In particular, these transformations can be viewed as 
photon sources.

Write the unitary evolution of these sources as $U = \exp[-it{\mathcal{H}}/
\hbar]$, with ${\mathcal{H}}$ a Hermitian operator (the interaction 
Hamiltonian). Here, every term in ${\mathcal{H}}$ is proportional to a product
of either two creation or two annihilation operators, i.e., ${\mathcal{H}}$ is 
proportional to a sum of quadratic forms. Subsequently we know that squeezing 
and parametric down-conversion are representations of the group $SU(1,1)$, 
\index{group!$SU(1,1)$}
and I therefore study the relation between this group and quadratic forms.

Let $L_-(\Lambda)$ be a quadratic form:
\begin{equation}
 L_-(\Lambda) = \frac{1}{2} \sum_{ij} \hat{a}_i \lambda_{ij} \hat{a}_j \equiv
 \frac{1}{2} (\vec{a},\Lambda\vec{a})\; .
\end{equation}
Since the annihilation operators commute, it is clear that $\Lambda$ can 
always be chosen symmetric. The adjoint of $L_-(\Lambda)$ is given by
\begin{equation}
 L_+(\Lambda) = L_-^{\dagger}(\Lambda) = \frac{1}{2} \sum_{ij}
 \hat{a}^{\dagger}_i \lambda_{ij}^* \hat{a}^{\dagger}_j \equiv \frac{1}{2}
 (\vec{a}^{\dagger}\Lambda^{\dagger},\vec{a}^{\dagger})\; .
\end{equation}

When we want to construct an $su(1,1)$ algebra\index{algebra!$su(1,1)$} with 
these operators we need to 
show that there exists an operator $L_0(\Lambda)$ which satisfies the 
commutation relations
\begin{equation}\label{su11}
 [L_-,L_+] = 2 L_0 \quad\text{and}\quad [L_0,L_{\pm}] = \pm L_{\pm}\; .
\end{equation}
The first relation in Eq.\ (\ref{su11}) defines $L_0(\Lambda)$:
\begin{equation}
 L_0(\Lambda) = \frac{1}{2} \left\{ (\Lambda\vec{a})^{\dagger} \cdot
 (\Lambda\vec{a}) + \frac{1}{2}\text{Tr}(\Lambda^{\dagger}\Lambda) \right\}
 = L_0^{\dagger} (\Lambda)\; .
\end{equation}
The second relation in Eq.\ (\ref{su11}) places a constraint on $\Lambda$:
\begin{eqnarray}
 [L_0,L_+] = L_+ &\Leftrightarrow& \Lambda^{\dagger} = 
 \Lambda^{\dagger} \Lambda \Lambda^{\dagger}\cr
 [L_0,L_-] = - L_- &\Leftrightarrow& \Lambda = \Lambda 
 \Lambda^{\dagger} \Lambda\; .
\end{eqnarray}
The matrix $\Lambda$ is unitary if it is invertible.

I can now formulate these results in terms of a theorem:
\begin{description}
 \item[Theorem 2] Consider a quadratic operator of the form $L(\Lambda) = 
	(\vec{a},\Lambda\vec{a})$. $L$ and $L^{\dagger}$ define a third 
	operator $L_0 = \frac{1}{2}[L,L^{\dagger}]$. These operators are 
	generators\index{algebra!generators of $su(1,1)$} of an $su(1,1)$ 
	algebra if and only if $\Lambda^{\dagger} = 
	\Lambda^{\dagger} \Lambda \Lambda^{\dagger}$. Such a $\Lambda$ is 
	unitary if and only if it is invertible.
\end{description}

Suppose that the interaction Hamiltonian can be written as
\begin{equation}
 {\mathcal{H}}_I = \nu \; L_+ (\Lambda) + \nu^* \; L_- (\Lambda) \; ,
\end{equation}
with $\nu$ the coupling constant. Since $L_+$ and $L_-$ generate an $su(1,1)$ 
algebra,\index{algebra!$su(1,1)$} we know what the normal 
ordering\index{normal ordering} for the unitary evolution $U$
associated with this interaction Hamiltonian is ($\tau = -i\nu t/\hbar$) 
\cite{truax85}:
\begin{eqnarray}
 U(\tau) &=& \exp[\tau L_+ (\Lambda) - \tau^* L_- (\Lambda)] \cr
 &=& \exp[\hat{\tau}\tanh|\tau| L_+] \exp[-2\ln\cosh|\tau| L_0]\rule{0pt}{10pt}
 \exp[-\hat{\tau}^* \tanh|\tau| L_-]\rule{0pt}{10pt}\; .\quad
\end{eqnarray}
This is the Baker-Campbell-Hausdorff 
formula\index{Baker-Campbell-Hausdorff!$su(1,1)$} for $su(1,1)$.

\newpage
\thispagestyle{empty}

\chapter{Transformation properties of maximal entanglement}\label{app:trans}

In this appendix I will show\footnote{This is not new material, it is included
here for reasons of completeness.} that any maximally entangled 
state\index{state!maximally entangled} can be 
transformed into any other maximally entangled state by means of a unitary 
transformation on only one of the subsystems. I will treat this in a formal 
way by considering an arbitrary maximally entangled state of two $N$-level 
systems in the Schmidt decomposition:\index{Schmidt decomposition}
\begin{equation}\label{ap:max}
 |\psi\rangle = \frac{1}{\sqrt{N}} \sum_{j=1}^N e^{i\phi_j} |n_j,m_j\rangle\; ,
\end{equation}
that is, a state with equal amplitudes on all possible branches. There always 
exist two orthonormal bases\index{orthonormal basis} $\{ |n_i\rangle \}$ and 
$\{ |m_i\rangle \}$ such
that Eq.\ (\ref{ap:max}) can be written this way, by virtue of the definition 
for maximal entanglement. Clearly, we can obtain {\em any} 
maximally entangled state by applying the (bi-local) unitary transformation 
$U_1\otimes U_2$. That is, each maximally entangled state can be 
transformed into any other by a pair of local unitary transformations on each 
of the subsystems. We will now show that any two maximally entangled states 
$|\psi\rangle$ and $|\psi'\rangle$ are connected by a local unitary 
transformation on one subsystem alone:
\begin{equation}
 |\psi\rangle = U\otimes{\unity} |\psi'\rangle = {\unity}\otimes U'|\psi'
 \rangle\; .
\end{equation}

First, I will prove that any transformation $U_1 \otimes U_2$ on a {\em 
particular} maximally entangled state $|\phi\rangle$ can be written as 
$V\otimes\unity |\phi\rangle$, where $V = U_1 U_2^T$. To this end I will
give the proofs for two theorems. Take the special maximally entangled state
\begin{equation}\label{phi}
 |\phi\rangle = \frac{1}{\sqrt{N}} \sum_{j=1}^N |n_j, m_j\rangle\; .
\end{equation}

\begin{description}
 \item[Theorem 1:] For any state $|\phi\rangle$ given by Eq.\ (\ref{phi}) and
 any unitary operator $U$ we have
 \begin{equation}\label{t1}
  U \otimes U^* |\phi\rangle = |\phi\rangle\; .
 \end{equation}
 \item[Proof:] Using the completeness relation
\begin{equation}
 \sum_{k=1}^N |n_k\rangle\langle n_k| = {\unity}
\end{equation}
on both subsystems we have
\begin{equation}
 U \otimes U^* |\phi\rangle = \sum_{k,l=1}^N |n_k,m_l\rangle\langle n_k,m_l| 
 \left( U\otimes\unity \right) \left( \unity\otimes U^* \right) 
 |\phi\rangle\; .
\end{equation}
By writing out $|\phi\rangle$ explicitly according to Eq.\ (\ref{phi}) we 
obtain 
\begin{eqnarray}
 U \otimes U^* |\phi\rangle &=& \frac{1}{\sqrt{N}} \sum_{j,k,l=1}^N U_{kj}
 U_{lj}^* |n_k,m_l\rangle\cr
 &=& \frac{1}{\sqrt{N}} \sum_{k,l=1}^N (UU^{\dagger})_{kl} |n_k,m_l\rangle\cr
 &=& \frac{1}{\sqrt{N}} \sum_{k,l=1}^N \delta_{kl} |n_k,m_l\rangle\; ,
\end{eqnarray}
which is just $|\phi\rangle$. \hfill$\square$
\item[Theorem 2:] Every unitary transformation $U_1 \otimes U_2$ acting on 
 	the state $|\phi\rangle$ given by Eq.\ (\ref{phi}) is equivalent to a 
	transformation $V\otimes\unity$ acting on $|\phi\rangle$, where 
	$V = U_1 U_2^T$. 
 \item[Proof:] The equality
\begin{equation}
 \left( \unity\otimes U^T \right) \left( U\otimes U^* \right) |\phi\rangle =
 \left( U\otimes\unity \right) |\phi\rangle\; ,
\end{equation}
together with theorem 1 immediately gives us
\begin{eqnarray}\label{p1}
 U\otimes\unity |\phi\rangle &=& \unity\otimes U^T |\phi\rangle 
 \qquad\mbox{and}\cr
 U^T \otimes\unity |\phi\rangle &=& \unity\otimes U |\phi\rangle
 \phantom{\Bigg|}\; .
\end{eqnarray}
From Eq.\ (\ref{p1}) we obtain
\begin{eqnarray}
 U_1 \otimes U_2 |\phi\rangle &=& \left( U_1 \otimes\unity \right) 
 \left( \unity\otimes U_2 \right) |\phi\rangle\cr
 &=& \left( U_1 \otimes\unity \right) \left( U_2^T \otimes\unity \right) 
  |\phi\rangle \phantom{\Bigg|} \cr
 &=& U_1 U_2^T \otimes\unity |\phi\rangle\; .
\end{eqnarray}
Similarly,
\begin{equation}
 U_1 \otimes U_2 |\phi\rangle = \unity\otimes U_1^T U_2 |\phi\rangle\; .
\end{equation}
We therefore obtain that $U_1 \otimes U_2 |\phi\rangle$ is equal to 
$V\otimes\unity|\phi\rangle$ with $V = U_1 U_2^T$, and similarly that it is 
equal to $\unity\otimes V' |\phi\rangle$ with $V' = U_1^T U_2$. \hfill$\square$
\end{description}

Since every maximally entangled state\index{state!maximally entangled} can 
be obtained by applying $U_1 \otimes U_2$ to $|\phi\rangle$, two maximally 
entangled states $|\psi\rangle$ and $|\psi'\rangle$ can be transformed into 
any other by choosing 
\begin{eqnarray}
 |\psi\rangle  &=& U\otimes\unity\; |\phi\rangle \cr
 |\psi'\rangle &=& V\otimes\unity\; |\phi\rangle \phantom{\Bigg|}\; ,
\end{eqnarray}
which gives
\begin{equation}
 |\psi\rangle = UV^{\dagger}\otimes\unity\; |\psi'\rangle\; .
\end{equation}
Thus each maximally entangled two-system state can be obtained from any other 
by means of a local unitary transformation on one subsystem alone. 

\newpage
\thispagestyle{empty}

\chapter{Statistical Distance}\label{app:distance}

In this appendix I review the concept of the statistical 
distance.\index{distance!statistical} It is first 
and foremost a concept form {\em classical} probability 
theory,\index{probability!classical theory} which has been 
extended to quantum theory by Wootters \cite{wootters81}, Hilgevoord and 
Uffink \cite{hilgevoord91} and Braunstein and Caves \cite{braunstein94}.

Suppose we have a vase containing red, blue and green marbles in some 
proportion. We can draw a marble from the vase and register its colour. When 
the proportion of red, blue and green marbles is known, we can predict that we 
will draw a red marble with some probability $p_{\rm red}$. In this situation 
the probability quantifies our uncertainty of {\em prediction}.
\index{uncertainty!of prediction}

Alternatively, we might be in a different situation where the proportion of 
red, blue and green marbles is {\em not} known. When we draw a marble, it 
gives us extra knowledge which can be used to estimate the proportion of
marbles. We do not know for certain what the proportion is until we have drawn 
all the marbles from the vase, but every new draw will yield extra information 
about the proportion. The number of drawn red, blue and green marbles estimates
the probability distribution of drawing red, blue and green marbles, and the 
uncertainty after a number of draws is the uncertainty of {\em inference}. 
\index{uncertainty!of inference}
There are therefore two kinds of uncertainty: one associated with the 
prediction of the outcome of a stochastic process and one associated with the 
inference of a probability distribution based on a set of outcomes 
\cite{hilgevoord91}.

\begin{figure}[t]
  \begin{center}
  \begin{psfrags}
     \psfrag{pred}{$p_{\rm red}$}
     \psfrag{pgreen}{$p_{\rm green}$}
     \psfrag{pblue}{$p_{\rm blue}$}
     \psfrag{x}{$\bullet$} 
     \epsfxsize=8in
     \epsfbox[-130 30 570 255]{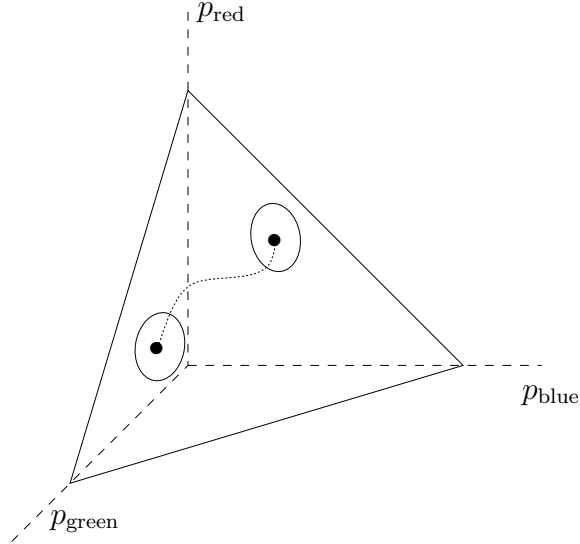}
  \end{psfrags}
  \end{center}
  \caption{The probability simplex corresponding to three possible  
	outcomes `red', `green' and `blue'. The two dots correspond to 
	normalised probability distributions. Their uncertainty regions after 
	$N$ trials is depicted by the circle around the dots. The distance 
	between the two distributions is the shortest path in the simplex, 
	measured in units of the typical statistical fluctuation.}
  \label{fig:app.dis.1}
\end{figure}

The {\em statistical distance} quantifies the distinguishability 
\index{distribution!distinguishability} of two 
probability distributions,\index{distribution!probability} and is therefore 
closely related to the uncertainty 
of inference. Since it is a distance, it obeys the four well-known 
requirements \cite{buskes97}:
\begin{enumerate}
 \item A distance\index{distance} $s(x,y)$ between two points $x$ and $y$ is 
	positive;
 \item $s(x,y)=0$ if and only if $x=y$;
 \item the distance is symmetric: $s(x,y)=s(y,x)$;
 \item the distance obeys the triangle inequality:\index{triangle inequality} 
	$s(x,z) \leq s(x,y) + s(y,z)$.
\end{enumerate}
The points $x$, $y$ and $z$ are elements of some (continuous) space. In the 
case of the statistical distance, these points are probability distributions, 
which are elements of the so-called {\em probability simplex}
\index{probability!simplex} (see figure \ref{fig:app.dis.1}). 

The distance\index{distance} function in a space (in this case the simplex)
\index{probability!simplex} is defined by the so-called {\em 
metric}.\index{metric} The metric $g$ is a real symmetric matrix 
which obeys $\sum_k g_{jk}\cdot g^{kl}=\delta_j^l$, where $\delta_j^l$ is the 
Kronecker delta. Furthermore, it transforms {\em covariant}
\index{covariance} vectors $x_j$ to {\em contravariant}
\index{contravariant} vectors $x^j$, distinguished by lower and upper indices 
respectively\footnote{To avoid confusion, I will {\em not} use Einstein's 
summation convention.}:
\begin{equation}
 x^j = \sum_k g^{jk} x_k \qquad\mbox{and}\qquad x_j = \sum_k g_{jk} x^k\; .
\end{equation}
The contraction $\sum_j x_j y^j$ yields a scalar (which is invariant under all 
transformations).

In general, an (incremental) distance $ds$ on the simplex separating points 
$p^j$ and $p^j + dp^j$ can be written as a quadratic form\index{form!quadratic}
\begin{equation}
 ds^2 = \sum_{jk} g_{jk} dp^j dp^k\; .
\end{equation}
This is a scalar which is invariant under all coordinate transformations. 
The $dp^j$ are the components of the incremental tangent vector $\vec{dp}$ 
along the shortest path between the two probability distributions
\index{distribution!probability} in the 
probability simplex. We now aim to find the metric $g$ of the simplex.

To this end, we define the {\em dual}\index{dual} $A$ to $\vec{dp}$, i.e., 
every component $A_j$ is paired with the component $dp^j$:
\begin{equation}\label{app:mean}
 \langle A\rangle \equiv \sum_j A_j p^j\; ,
\end{equation}
where $\langle A\rangle$ can be interpreted as the mean value of $A$. In order 
to find the metric,\index{metric} we look at the two-point correlation function
\index{two-point correlation function} of $A$:
\begin{equation}
 \langle A^2 \rangle = \sum_{jk} A_j A_k g^{jk} = \sum_j A^2_j p^j\; . 
\end{equation}
The last equality is obtained by using Eq.~(\ref{app:mean}). From this we 
immediately obtain the contravariant form\index{form!contravariant} of the 
metric:
\begin{equation}
 g^{jk} = p^j \delta^{jk}\; ,
\end{equation}
with $\delta^{jk}$ the Kronecker delta. Since $\sum_k g_{jk}\cdot g^{kl} =
\delta_j^l$, the covariant metric is $g_{jk}=\delta_{jk}/p^j$ and the 
statistical distance\index{distance!statistical} becomes
\begin{equation}\label{app:statdist}
 ds^2 = \sum_j \frac{dp^j dp^j}{p^j} \equiv \sum_j \frac{(dp^j)^2}{p^j}\; .
\end{equation}
This is the incremental statistical distance used in chapter \ref{chap2}. When
we make the substitution $p^j = r_j^2$, we find the Euclidean 
distance\index{distance!Euclidean} $ds^2 = 4 \sum_j dr_j^2$.

Note that $ds^2$ tends to infinity when one of the probabilities $p_j$ equals 
zero. This is expected since a probability distribution
\index{distribution!probability} $\vec{p}^{(1)}$ with 
$p_j=0$ is perfectly distinguishable\index{distribution!distinguishability} 
from a distribution $\vec{p}^{(2)}$ with 
$p_j\neq 0$: one outcome corresponding to $p_j$ will immediately tell us that 
we have the probability distribution $\vec{p}^{(2)}$.

In order to find the statistical distance between two well separated 
probability distributions, we have to integrate Eq.~(\ref{app:statdist}). 
Following Wootters \cite{wootters81}, we find
\begin{equation}
 s(\vec{p}^{(1)},\vec{p}^{(2)}) = \arccos\left( \sum_j\sqrt{p^{(1)}_j
 p^{(2)}_j}\right)\; .
\end{equation}
In other words, the statistical distance is the {\em angle} between two 
vectors with coordinates $\sqrt{p^{(1)}_j}$ and $\sqrt{p^{(2)}_j}$. Wootters
\cite{wootters81} proved that this distance measure is the {\em only} 
Riemannian distance\index{distance!Riemannian} measure in a Hilbert space, 
\index{Hilbert space} which is
invariant under all transformations. It should be noted that
we only assumed classical probability theory in our derivation, which makes
the appearance of probability amplitudes\index{probability!amplitude} even 
more surprising.

An alternative way to arrive at the statistical distance is by using the 
Gaussian distribution\index{distribution!Gaussian} for the observed 
frequencies $f_j$ in a large number ($N$) of trials \cite{reif65,braunstein94}:
\begin{equation}
 \rho (f_1,\ldots)\propto\exp\left[ -\frac{N}{2}\sum_j\frac{(f_j - p_j)^2}{p_j}
 \right]\; .
\end{equation}
Two probability distributions\index{distribution!probability} $p^{(1)}$ and 
$p^{(1)}$ can then be distinguished if and only if the Gaussian function 
$\exp\left[ -\frac{N}{2}\sum_j
\frac{(p^{(1)}_j - p^{(2)}_j)^2}{p^{(1)}_j} \right]$ is small. In other words,
if $p^{(1)}_j - p^{(2)}_j\equiv dp_j$ we need
\begin{equation}
 \frac{N}{2} \sum_j \frac{dp_j^2}{p_j} \gtrsim 1\; ,
\end{equation}
or 
\begin{equation}
 ds \gtrsim \frac{1}{\sqrt{N}}\; ,
\end{equation}
which is consistent with Eq.~(\ref{stdstN}).

\newpage 
\thispagestyle{empty}

\chapter{Multi-Dimensional Hermite Polynomials}\label{app:mdhp}

In this appendix I will give the background of multi-dimensional Hermite
polynomials.\index{MDHP|(} Early introductions to the subject were presented 
by P.\ Appell 
and J.\ Kamp\'e de F\'eriet \cite{appell26}, and in the Bateman Manuscript 
Project \cite{erdely53}. M.M.\ Mizrahi \cite{mizrahi75} and M.\ Klauderer 
\cite{klauderer93} further developed the mathematical theory, and in the 
context of quantum optics multi-dimensional Hermite polynomials have been 
applied by V.V.\ Dodonov, V.I.\ Man'ko, O.V.\ Man'ko, V.V.\ Semjonov, A.\ 
Vourdas and R.M.\ Weiner \cite{dodonov84,dodonov94,dodonov96,vourdas87}.

\section{Ordinary Hermite Polynomials}

First, let me revisit the case of the ordinary Hermite 
polynomials,\index{Hermite polynomial} which are 
known to physicists as (part of) the eigenfunctions\index{eigen!-function} of 
the linear harmonic\index{harmonic oscillator}
oscillator in quantum mechanics (see, for example Merzbacher 
\cite{merzbacher98}).

The definition of the Hermite polynomials can be obtained by the construction 
of a so-called {\em generating function}\index{generating function|(} $G(x,s)$:
\begin{equation}\label{genfun}
 G(x,s) = e^{x^2 - (s-x)^2} = \sum_{n=0}^{\infty} \frac{H_n(x)}{n!} s^n \; .
\end{equation}
The last equality will give rise to our definition of the Hermite polynomials 
$H_n (x)$. In order to arrive at this definition we use Taylors expansion:
\index{Taylor expansion}
\begin{equation}
 f(x+s) = \left[ 1 + s\frac{d}{dx} + \frac{s^2}{2!}\left(\frac{d}{dx}\right)^2 
 + \cdots \right] f(x) = e^{s\frac{d}{dx}} f(x)\; ,
\end{equation}
where $\frac{d}{dx}$ denotes the derivative taken with respect to $x$. The 
second equality collects the derivatives in the exponential function 
$\exp[s\frac{d}{dx}]$. Using this relation we write the generating function as
\begin{equation}
 G(x,s) = e^{x^2} e^{-s\frac{d}{dx}} e^{-x^2}\; .
\end{equation}
By expanding the exponential $\exp[-s\frac{d}{dx}]$ and comparing with 
Eq.~(\ref{genfun}) we obtain the definition of the Hermite polynomials 
$H_n (x)$:
\begin{equation}\label{defher}
 H_n (x) = (-1)^n e^{x^2} \frac{d^n}{dx^n} e^{-x^2}\; .
\end{equation}
Every $H_n (x)$ is a polynomial with $n$ real roots and traditionally 
normalised in such a way that the leading term $x^n$ has pre-factor $2^n$. 

There are several relations connecting Hermite polynomials. For instance, the 
Hermite polynomials\index{Hermite polynomial} obey the orthogonality 
relation:\index{orthogonality!relation}
\begin{eqnarray}\label{ortho}
 \int_{-\infty}^{+\infty}e^{-x^2} |H_n (x)|^2\; dx &=& 2^n n!\sqrt{\pi}\; . \cr
 \int_{-\infty}^{+\infty} e^{-x^2} H_n (x) H_m (x)\; dx &=& 0 \qquad\mbox{for}
	\quad n \neq m\; .
\end{eqnarray}
This relation ensures that the eigenfunctions of the harmonic oscillator are 
orthonormal.

Furthermore, there are two types of recursion 
relations\index{recursion relation} connecting Hermite 
polynomials of different order. From the generating function in Eq.\ 
(\ref{genfun}) it is relatively straight\-forward to derive the recursion 
relations
\begin{eqnarray}
 \frac{d}{dx} H_n (x) &=& 2n H_{n-1} (x)\; , \label{rec1} \\
 H_{n+1} (x) - 2x H_n (x) + 2n H_{n-1} (x) &=& 0\; . \label{rec2}
\end{eqnarray}
Combining these two relations yields a second-order homogeneous differential 
equation called the {\em Hermite equation}:\index{Hermite equation}
\begin{equation}\label{hereq}
 \frac{d^2}{dx^2} H_n (x) - 2x \frac{d}{dx} H_n (x) + 2n H_n (x) = 0\; .
\end{equation}

\section{Real Multi-Dimensional Hermite Polynomials}

The ordinary Hermite polynomials are functions of one variable $x$. The 
obvious way to generalise this is taking $x$ to be a vector $\vec{x} =
(x_1\ldots,x_N)$ in an $N$-dimensional vector space. The generating function 
of the multi-dimensional Hermite polynomial (henceforth called {\sc MDHP}) 
then has to change accordingly: $G(x,s) \rightarrow G(\vec{x},\vec{s})$. 

However, rather than replacing $s^2$ by $(\vec{s},\vec{s})$ and $sx$ by 
$(\vec{s},\vec{x})$ (where we denote the inner product of two vectors 
$\vec{a}$ and $\vec{b}$ by $(\vec{a},\vec{b})$), we take the generating 
function to be \cite{erdely53}
\begin{equation}\label{mdhpgenfun}
 G_A(\vec{x},\vec{s}) = \exp\left[ (\vec{s},A\vec{x}) - \frac{1}{2}
 (\vec{s},A\vec{s})\right] \; ,
\end{equation}
where $A$ is a positive definite $N\times N$ matrix, called the {\em defining
matrix}. We can always choose $A$ symmetric. The reason we choose this 
generating function is that we now also include cross-terms $s_i s_j$ and 
$s_i x_j$. Without these cross-terms the 
generalisation would be trivial. When we define $\vec{n}$ as an $N$-tuple 
$(n_1,\ldots,n_N)$ with $n_i$ a non-negative integer, $G_A(\vec{x},\vec{s})$ 
generates the multi-dimensional Hermite polynomials:
\begin{equation}\label{genher}
 G_A(\vec{x},\vec{s}) = \sum_{\vec{n}} \frac{s_1^{n_1}}{n_1!} \cdots 
 \frac{s_N^{n_N}}{n_N!}\; H^A_{\vec{n}} (\vec{x})\; .
\end{equation}
In this equation, $\sum_{\vec{n}}$ means the sum over all possible $N$-tuples
$\vec{n}$.

The generating function leads to the following definition of the real 
multi-dimensional Hermite polynomial:
\begin{equation}
 H^A_{\vec{n}} (\vec{x}) = (-1)^{\sum_i n_i} e^{\frac{1}{2}(\vec{x},A\vec{x})}
 \; \frac{\partial^{\sum_i n_i}}{\partial x_1^{n_1} \cdots \partial x_N^{n_N}}
 \; e^{-\frac{1}{2}(\vec{x},A\vec{x})}\; .
\end{equation}
This definition is derived analogous to the one-dimensional case, which was 
presented above. 

\section{Reduction theorem}\index{reduction theorem}

In order to simplify the derivation of the orthogonality and recursion 
relations for the real {\sc MDHP}'s, I derived a {\em Reduction Theorem}:
\begin{description}
 \item[Reduction Theorem:] For any real $N$-dimensional generating function
  $G_A(\vec{x},\vec{s})$ with positive definite defining matrix $A$ there 
  exists a linear transformation $T$ which transforms $G_A$ into a product of
  $N-M$ generating functions of one-dimensional Hermite polynomials:
  \begin{equation}\label{reduc}
    G_A(\vec{x},\vec{s}) ~\longrightarrow_T~ \prod_{i=1}^{N-M} G(z_i,v_i)\; ,
  \end{equation}
  where $M$ is the number of zero eigenvalues of $A$.
 \item[Proof:] This theorem is proved by explicit construction of $T$. The 
  transformation has two parts: an orthogonal transformation and a rescaling. 
  The $N\times N$ matrix $A$ is real and symmetric. It can therefore be 
  diagonalised by an orthogonal matrix $O$ \cite{gantmacher59}: 
  \begin{equation}
    A = O^T \Lambda O
  \end{equation}
  and 
  \begin{equation}
    (\vec{s},A\vec{x}) = (\vec{s},O \Lambda O^T\vec{x}) =
     (O^T\vec{s},\Lambda O^T\vec{x}) \equiv (\vec{u},\Lambda\vec{y})\; .
  \end{equation}
  This last term can be written as $\sum_i \lambda_i u_i y_i$. The generating 
  function of a real {\sc MDHP} then transforms as
  \begin{equation}
    G_A(\vec{x},\vec{s}) ~\longrightarrow_O~ G_{\Lambda} (\vec{y},\vec{u}) =
     \exp\left[ \sum_i \lambda_i \left( u_i y_i - \frac{1}{2} u_i^2 \right) 
     \right]\; .
  \end{equation}
  We can now rescale the transformed coordinates $\vec{y}$ and $\vec{u}$:
  \begin{equation}
    v_i = \sqrt{\frac{\lambda_i}{2}} u_i \qquad\mbox{and}\qquad z_i = 
     \sqrt{\frac{\lambda_i}{2}} y_i\; .
  \end{equation}
  This {\em rescaled} transformation of the generating function of a real 
  {\sc MDHP} then gives
  \begin{equation}
    G_A(\vec{x},\vec{s}) ~\longrightarrow_T~ G(\vec{z},\vec{v}) = 
     \prod_{i=1}^{N-M} \exp\left[ 2z_i v_i - v_i^2 \right] \; ,
  \end{equation} 
  where $M$ is the number of zero eigenvalues of $A$. This is the 
  transformation $T$ whose existence we had to prove. \hfill $\square$
\end{description}

Since the new variables in Eq.\ (\ref{reduc}) are linearly independent, the 
reduced generating function trivially generates the ordinary Hermite 
polynomials. We can now derive the orthogonality relation of the real {\sc 
MDHP}'s. 

\section{Orthogonality relation}\index{orthogonality!relation}

The Reduction Theorem\index{reduction theorem} yields the diagonalised form:
\begin{eqnarray}\label{diaort}
 && \prod_{i=1}^{N-M} \int_{-\infty}^{+\infty} e^{-z_i^2} | H^A_{n_i}
 (z_i)|^2\; dz_i = \sqrt{\pi^{N-M}} \prod_{i=1}^{N-M} 2^{n_i} n_i! \cr 
 && \prod_{i=1}^{N-M} \int_{-\infty}^{+\infty} e^{-z_i^2} H^A_{n_i} (z_i) 
 H^A_{m_i} (z_i)\; dz_i = 0 \qquad\mbox{if}~\exists~ n_i \neq m_i \; .
\end{eqnarray}

Since the Jacobian\index{Jacobian} $J$ of an orthogonal transformation $O$ is 
equal to 1, I
omit it here. We can now transform Eq.\ (\ref{diaort}) back to the 
non-diagonalised case. This yields
\begin{equation}
 \int_{-\infty}^{+\infty} e^{-\frac{1}{2}(\vec{x},A\vec{x})} H^A_{\vec{n}} 
 (\vec{x}) H^A_{\vec{m}} (\vec{x})\; d\vec{x} = 0\; .
\end{equation}
The orthogonality relations for the real {\sc MDHP}'s are then:
\begin{eqnarray}
 && \int_{-\infty}^{+\infty} e^{-\frac{1}{2}(\vec{x},A\vec{x})} |H^A_{\vec{n}} 
 (\vec{x})|^2\; d\vec{x} = \pi^{N/2} \prod_{i=1}^N 2^{n_i} n_i! \cr 
 && \int_{-\infty}^{+\infty} e^{-\frac{1}{2}(\vec{x},A\vec{x})} H^A_{\vec{n}} 
 (\vec{x}) H^A_{\vec{m}} (\vec{x})\; d\vec{x} = 0 \qquad\mbox{for}~ 
 \vec{n}\neq\vec{m}\; .
\end{eqnarray}

\section{Recursion relations}\index{recursion relation}

There are two classes of recursion relations for the real multi-dimensional
Hermite polynomials. First, we present the differential recursion relations,
which form a generalisation of Eq.\ (\ref{rec1}). Subsequently, we present 
the type of recursion relations which form a generalisation of Eq.\ 
(\ref{rec2}). 

In the generalised form of the Hermite polynomials, we wish to evaluate the 
derivative $\partial_{x_i} H_{\vec{n}}^A(\vec{x})$. We proceed again from the
generating function\index{generating function|)} $G(\vec{x},\vec{s})$:
\begin{equation}\label{derrec1}
 \frac{\partial}{\partial x_i} G(\vec{x},\vec{s}) = \frac{\partial}{\partial
 x_i} e^{-\frac{1}{2} (\vec{s}, A\vec{s})+(\vec{s},A\vec{x})} = \sum_{j=1}^N 
 A_{ij} s_j G(\vec{x},\vec{s})\; .
\end{equation}
Furthermore, from Eq.\ (\ref{genher}) we obtain
\begin{equation}\label{derrec2}
 \frac{\partial}{\partial x_i} G(\vec{x},\vec{s}) = \sum_{\vec{n}} 
 \frac{s_1^{n_1}}{n_1!}\cdots \frac{s_N^{n_N}}{n_N!} \frac{\partial 
 H_{\vec{n}}^A (\vec{x})}{\partial x_i}\; .
\end{equation}

Expanding the right-hand side of Eq.\ (\ref{derrec1}) into Hermite polynomials
\index{MDHP|)}
and equating it with the right-hand side in Eq.\ (\ref{derrec2}) yields
\begin{equation}
 \sum_{\vec{n}} \frac{s_1^{n_1}}{n_1!} \cdots \frac{s_N^{n_N}}{n_N!} 
 \frac{\partial H_{\vec{n}}^A (\vec{x})}{\partial x_i} = \sum_{\vec{n}} 
 \sum_{j=1}^N A_{ij} n_j \frac{s_1^{n_1}}{n_1!} \cdots \frac{s_N^{n_N}}{n_N!}
 H_{\vec{n}-e_j}^A (\vec{x})\; ,
\end{equation}
where $\vec{n}-e_j$ denotes the vector $\vec{n}$ with $n_j$ replaced by
$n_j-1$. Comparing the terms with equal powers in $n_k$ yields the generalised
differential recursion relation
\begin{equation}\label{realrec1}
 \frac{\partial H_{\vec{n}}^A(\vec{x})}{\partial x_i} = \sum_{j=1}^N A_{ij} n_j
 H_{\vec{n}-e_j}^A (\vec{x})\; .
\end{equation} 
This relation is easily generalised for multiple derivatives on 
$H_{\vec{n}}^A (\vec{x})$.

The second recursion relation is given by
\begin{equation}\label{realrec2}
 H^A_{\vec{n}+e_i}(\vec{x}) - \sum_{j=1}^N A_{ij} x_j H^A_{\vec{n}}(\vec{x})
 + \sum_{j=1}^N A_{ij} n_j H^A_{\vec{n}-e_j}(\vec{x}) = 0\; .
\end{equation}
This relation can be proved by taking the derivative to $x_i$ and 
using the recursion relation (\ref{realrec1}). 

\newpage
\thispagestyle{empty}

\chapter{Mathematica Code for Teleportation Modelling}\label{app:math}

In this appendix, I present the {\sc Mathematica} code I used to derive the 
results of chapter \ref{teleportation}.

{\footnotesize
\begin{verbatim}
Krondelta [j_, k_] := If[ j == k, Return [ 1 ], Return [ 0 ] ]

f [ ] := 1
f [0] := 0
f [y___, 1, x___ ] := f[y,x]

f[x___, y_ + z_, w___] := f[x, y, w] + f[x, z, w]

(* 
   Let's define annihilation ops for polarisation k = "x" or "y"
   on mode j = 1, 2, 3, 4, or "a", "u", "v", "d" as a[k,j]
   and the creation ops as ad[k,j]
*)

f[ x___, n_  a[k_ ,j_ ], w___ ] := n f [ x,  a[k, j ], w ] 
f[ x___, n_ ad[k_ ,j_ ], w___ ] := n f [ x, ad[k, j ], w ] 

(*
   The normal ordering rule preserves commutator algebra.
*)

normOrder := f[ x___, a[k_, j_ ], ad[kk_, jj_ ], w___ ] :> 
             f[ x, ad[kk, jj ], a[k, j ], w ] +
             Krondelta [ k, kk ] Krondelta [ j, jj ] f[x, w]

f[ x___, n_. f[y___], w___] := n f[ x, y, w ] 

(*
   To what order do we expand the exponential.
*)

expandExp1 := myExp1[ x__ ] :> g[ x ] + g[ x, x ]/2 

expandExp2 := myExp2[ x__ ] :> g[ 1 ] + g[ x ] 

cutEnd := { g [ x_ ] :> ExpandAll[ x ],  
	    g [ y___, x_] :> f[ g [ y ], ExpandAll [ x ] ] }

(* Calculate the creation and annihilation results *)

numReduce := { f[ x___, ad[k_, ll_], aa__ ket[k_, ll_, n_ ] ] :>
                        Sqrt[n+1] f[ x, aa ket[ k, ll, n+1 ] ],
               f[ x___, a[k_, ll_ ], aa__ ket[k_, ll_, n_ ] ] :>
                        Sqrt[n] f[ x, aa ket[ k, ll, n-1 ] ] ,

(* and the adjoint  *)

               f[ bra[k_, ll_, n_ ] aa__, ad[k_, ll_], x___ ] :>
                        Sqrt[n] f[ aa bra[ k, ll, n-1 ], x ],
               f[ bra[k_, ll_, n_ ] aa__, a[k_, ll_ ], x___ ] :>
                        Sqrt[n+1] f[ aa bra[ k, ll, n+1 ], x ] 
	     }

(* The concise Form *)

consiseForm :=  f[ket["x", h3_, o_] ket["y", h3_, p_]] *
                f[bra["x", bh3_, bo_] bra["y", bh3_, bp_]] :>
       Infix[{" | ",o,", ",p," > < ",bo,", ",bp," |"}, "" ]

(*
   The beam-splitter has five entries: the first two are the input modes,
   the third and the fourth are the respective output modes and the fifth
   entry gives the beam-splitters coefficient.
*)

beamSplitter[ am_, bm_, cm_, dm_, eta_ ] := 
  { ad[ "x", am ] :> Sqrt[eta] ad[ "x", cm ] + Sqrt[1 - eta] ad[ "x", dm ],
     a[ "x", am ] :> Sqrt[eta]  a[ "x", cm ] + Sqrt[1 - eta]  a[ "x", dm ],
    ad[ "x", bm ] :> Sqrt[1 - eta] ad[ "x", cm ] - Sqrt[eta] ad[ "x", dm ],
     a[ "x", bm ] :> Sqrt[1 - eta]  a[ "x", cm ] - Sqrt[eta]  a[ "x", dm ],
    ad[ "y", am ] :> Sqrt[eta] ad[ "y", cm ] + Sqrt[1 - eta] ad[ "y", dm ],
     a[ "y", am ] :> Sqrt[eta]  a[ "y", cm ] + Sqrt[1 - eta]  a[ "y", dm ],
    ad[ "y", bm ] :> Sqrt[1 - eta] ad[ "y", cm ] - Sqrt[eta] ad[ "y", dm ],
     a[ "y", bm ] :> Sqrt[1 - eta]  a[ "y", cm ] - Sqrt[eta]  a[ "y", dm ],
     ket[ "x", am, n_ ] :> ket[ "x", cm, n ],  
     ket[ "y", am, n_ ] :> ket[ "y", cm, n ],  
     ket[ "x", bm, n_ ] :> ket[ "x", dm, n ],  
     ket[ "y", bm, n_ ] :> ket[ "y", dm, n ]  
  }

(*
   The polarisation filter performs a rotation over an angle theta (the
   second entry) on mode am (the first entry). The two directions of
   polarisation are called "x" and "y".
*)

polarizeFilter[ am_, theta_, aam_ ] := {
    ad["x", am] :>   Cos[theta] ad["x", aam ] + Sin[theta] ad["y", aam ],
     a["x", am] :>   Cos[theta]  a["x", aam ] + Sin[theta]  a["y", aam ], 
    ad["y", am] :> - Sin[theta] ad["x", aam ] + Cos[theta] ad["y", aam ],
     a["y", am] :> - Sin[theta]  a["x", aam ] + Cos[theta]  a["y", aam ], 
    ket[ "x", am, n_ ] :> ket[ "x", aam, n ], 
    ket[ "y", am, m_ ] :> ket[ "y", aam, m ] 
  }

(* The takeAdjoint rule changes kets into bras.  *)

takeAdjoint := ket[ k_, l_, n_ ] :> bra[ k, l, n ]
	       
(*
   Polarisation insensitive detector. It assumes detectors cannot distinguish
   between a pulse containing one or more photons. The POVM acts on mode l and
   loss is Sqrt[1-efficiency^2]. The perfect detector therefore corresponds to
   loss=0. Loss is the AMPLITUDE loss.
*)

povMeasure[ b_, loss_ ] := f[ z__ ket[ "x", b, n_] ket[ "y", b, m_ ]] :>
  Sqrt[1-loss^( 2(n + m) )] f[z ket[ "x", b, n ] ket[ "y", b, m ]]


povAngle[ mode_, loss_ ] := { 
   f[ z__ ket["x", mode, n_ ]] :> loss^n f[ z ket["x", mode, n]],
   f[ z__ ket["y", mode, m_ ]] :> loss^m f[ z ket["y", mode, m]] 
			    }

povHit[pol_, mode_, loss_ ] := 
  f[ z__ ket[pol, mode, m_ ]] :> Sqrt[1-loss^(2m)] f[ z ket[pol, mode, m]] 

povMiss[pol_, mode_, loss_ ] := 
  f[ z__ ket[pol, mode, m_ ]] :> loss^m f[ z ket[pol, mode, m]] 

(* partialTrace[ mode_ ] takes the partial trace of mode "mode".  *)

partialTrace[ mode_ ] := { 
	f[l__ bra[ "y", mode, n_ ] u__ ] f[w__ ket[ "y", mode, m_ ] v__ ] :> 
 		f[l u] f[w v] Krondelta[ n, m ],
	f[l__ bra[ "x", mode, n_ ] u__ ] f[w__ ket[ "x", mode, m_ ] v__ ] :> 
		f[l u] f[w v] Krondelta[ n, m ]
			 }

xTrace[ mode_ ] := { 
	f[l__ bra[ "x", mode, n_ ] u__ ] f[w__ ket[ "x", mode, m_ ] v__ ] :> 
		f[l u] f[w v] Krondelta[ n, m ]
			 }

yTrace[ mode_ ] := { 
	f[l__ bra[ "y", mode, n_ ] u__ ] f[w__ ket[ "y", mode, m_ ] v__ ] :> 
		f[l u] f[w v] Krondelta[ n, m ]
			 }

(*
   In the procedure "myCalc", ketval is assigned the function corresponding 
   to the unitary transformation of creating EPR-pairs. After expansion to a
   certain order the beam-splitter on modes "b" and "c" is applied, the 
   polarisation rotation on mode "a" is performed and the creation operators 
   are calculated. Then the Hermitian conjugate is computed. With this we
   can define the density operator (densval). But first we apply the POVM on
   "u" and "v". After expanding the density operator we take the partial 
   traces of "a", "u" and "v", which gives us the output mode "d".
*)

(*
   In the line "ketval = f[ myExp [ ...", tau corresponds to the normal 
   ordered function tau/|tau| tanh(|tau|), and NOT the tau due to the 
   Hamiltonian.
*)

myCalc := Block[ {ketval, braval, densval, myval},
  ketval = f [ myExp1 [ tau f [ ad[ "x", "a"], ad [ "y", "b"] ] -
                  tau f [ ad[ "y", "a"], ad [ "x", "b"] ] ],
	       myExp2 [ tau f [ ad[ "x", "c"], ad [ "y", "d"] ] -
                  tau f [ ad[ "y", "c"], ad [ "x", "d"] ] ],  
                  ket[ "x", "a", 0 ] ket [ "x", "b", 0] *
                  ket[ "y", "a", 0 ] ket [ "y", "b", 0] *
                  ket[ "x", "c", 0 ] ket [ "x", "d", 0] *
                  ket[ "y", "c", 0 ] ket [ "y", "d", 0] ];

  ketval = ketval /. beamSplitter[ "b", "c", "u", "v", 1/2 ];  
  ketval = ketval /. polarizeFilter["a", theta, "a1" ];
  Print[ "<< beamSplitters and polarizeFilter >>" ];

  ketval = ketval /. expandExp1; 
  ketval = ExpandAll[ ketval /. cutEnd ];
  ketval = ketval //. numReduce;
  ketval = ExpandAll[ ketval /. cutEnd ];
  ketval = ketval //. numReduce;
  Print[ "<< expandExp1 >>" ];

  ketval = ketval /. expandExp2;
  ketval = ExpandAll[ ketval /. cutEnd ];
  ketval = ketval //. numReduce;
  Print[ "<< expandExp2 >>" ];

  Save[ "ketval.m", ketval ];

  ketval = ketval /. povMeasure[ "u", loss1 ];
  ketval = ketval /. povMeasure[ "v", loss2 ];

  ketval = ketval /. povHit [ "x", "a1", loss3];		     
  ketval = ketval /. povMiss [ "y", "a1", loss3]; 		     

  Print[ "<< POVM's >>" ];

  ketval = ExpandAll[ ketval ];

  Print[ "<< expansion >>" ];

  braval = ketval /. takeAdjoint;

  ketlen = Length [ ketval ];
  bralen = Length [ braval ];
  Print[ "<< ketlen = ", ketlen, " >>" ];

  densval = Sum [ 
       Print [ N[ 100. jj / ketlen, 3] , " %" ]; 
       Sum [ 
        myval = ketval [[jj]] braval [[kk]];
	myval = myval /. {tau^6 -> 0}; 
        myval = myval //. partialTrace[ "u" ];
        myval = myval //. partialTrace[ "v" ];
	myval = myval //. partialTrace[ "a1"],	 		     
          { kk, 1, bralen } 
           ],
          { jj, 1, ketlen }
                ];

  Save[ "densval4cas1.m", densval ];

  densval = Collect [ densval, tau ];
  densval = densval /. consiseForm;
  densval = Simplify[ densval, TimeConstraint -> Infinity ]
]
\end{verbatim}
}

\newpage
\thispagestyle{empty}

\chapter{Genetic Algorithms}\label{app:genetic}

In this appendix I review genetic algorithms.\index{algorithm!genetic} In the 
first section I present
the basics behind these algorithms \cite{michalewicz92}, and in the second 
section I describe the so-called {\em differential evolution}-approach
\index{evolution!differential} by Price
and Storn \cite{price97}. It was this method I used in chapter 
\ref{lithography}. The fortran code of this application is given in the last 
section.

\section{Genetic algorithms}

Genetic algorithms can be used to find the best solution to a given problem.
As the name already suggests, it is based on `natural selection'
\index{natural selection} over several generations\index{generation} of a 
`population'\index{population} of solutions to the problem. It works as 
follows.

Suppose we have a problem with a set of possible solutions. This set generally
spans a high dimensional solution space. For instance, when the solutions to a
particular problem are given by $x(\theta;a,b,c)=a\cos\theta + b\sin\theta + 
c$ (with $\theta$ its variable and $a$, $b$ and $c$ constants), the solution 
space is a three-dimensional space spanned by the vectors $(a,b,c)$. In 
addition, we have a selection criterion which gives us a measure of the 
`fitness'\index{fitness} of a solution. For example, we might define the 
fitness of a solution as a distance measure\index{distance!measure} between a 
function $y$ and a solution $x$. The smaller 
this distance, the fitter the solution. One such fitness measure for the 
example above may be given by 
\begin{equation}
 f(a,b,c) = \int_0^{2\pi} |y(\theta)-x(\theta;a,b,c)|^2 d\theta\; .
\end{equation}
In the genetic representation,\index{representation!genetic} the numbers $a$, 
$b$ and $c$ are the {\em genes}\index{gene} of a particular solution.

In any genetic algorithm, we first select a population of $n$ solutions
\begin{equation}
 P(t_0) = \{ x_1^{t_0},\ldots,x_n^{t_0} \}\; ,
\end{equation}
where $P$ is the population (taken at the initial time $t_0$) and $x_i^{t_0}$
a candidate solution to the problem. The solutions are evaluated using the 
fitness
measure, yielding a measure set $F(t_0) = \{ f_1^{t_0},\ldots,f_n^{t_0} \}$, 
where $f_i^{t_0}$ is a number associated with the fitness of solution 
$x_i^{t_0}$. 

Depending on the details of our problem, we are looking for the smallest or
the largest number $f_i^{t_0}$. Suppose better fitness means a smaller $f$, 
then we choose the solution $x_k^{t_0}$ corresponding to the smallest 
$f_k^{t_0}$ in our measure set. Let $x_k^{t_0}$ be the best solution for this 
population.\index{population} It will be stored in the memory. This memory 
slot is reserved for the best solution, and it will be updated if some 
solution $x_l^t$ from a later generation\index{generation} outperforms 
$x_k^{t_0}$.

The next step is the crucial step of genetic 
algorithms.\index{algorithm!genetic} The old generation,
the population $P(t_0)$, will now determine the next generation $P(t)$ of
solutions. Low fitness\index{fitness!measure} solutions from $P(t_0)$, 
however, will be discarded: evolution\index{evolution} has destined them to 
die. Thus only the fittest individuals from a 
population will generate a new population: they are making babies.

Just as in the offspring of real populations, the children inherit the traits 
of their parents. But they differ too. In biology, organisms often produce 
genetically different offspring by using {\em crossover},\index{crossover} in 
which genes\index{gene} of 
the parents are mixed. In a population of solutions to a mathematical problem, 
we can also introduce crossover. For example, when two solutions $x_1^{t_0}$ 
and $x_2^{t_0}$ are determined by the vectors
\begin{equation}\nonumber
 x_1^{t_0} = (a_1, b_1, c_1, d_1, e_1) \qquad\mbox{and}\qquad
 x_2^{t_0} = (a_2, b_2, c_2, d_2, e_2)\; ,
\end{equation}
crossover can produce a child-solution $x_j^t = (a_1, b_2, c_2, d_1, e_1)$,
where `genes' $a$, $d$ and $e$ are taken from parent 1 and `genes' $b$ and 
$c$ from parent 2.

\begin{figure}[t]
  \begin{center}
  \begin{psfrags}
     \psfrag{BS}{BF}
     \psfrag{Environment}{Crossover \&} 
     \psfrag{Mutations}{Mutations}
     \psfrag{P(t)}{$P(t)$}
     \psfrag{P(t+1)}{$P(t+1)$}
     \epsfxsize=8in
     \epsfbox[-100 40 800 180]{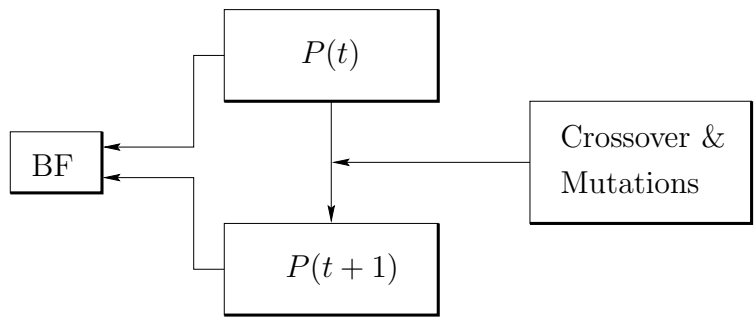}
  \end{psfrags}
  \end{center}
  \caption{Flowchart for genetic algorithms. At time $t$ the fitness of the 
	members of a population $P(t)$ is evaluated according to some 
	criterion. The best fitting member (BF) of $P(t)$ is recorded. 
	Subsequently,
	a new population (the next generation) $P(t+1)$ is formed from $P(t)$.
	In addition, crossover and mutations diversify the next generation.
	This generation is again tested for the best fitting member, which is 
	recorded as the fittest if it defeats the previous fittest.}
  \label{fig:app.gen.1}
\end{figure}

A second mechanism for inducing changes in subsequent generations is {\em 
mutation}.\index{mutation} In real life, background radioactivity or free 
radicals induce 
changes in the DNA structure which will affect future generations. Most of the 
time these changes are a setback and will be eliminated in the next 
generation, but once in a while it increases an individual's fitness. 

Using crossover and mutation, a new population is formed. This is the next 
generation. The population size is usually held constant, but this is not 
necessary. The new population\index{population} is again evaluated, yielding a 
measure set\index{fitness!measure} $F(t) 
= \{ f_1^t,\ldots,f_n^t \}$. The fittest solution of this 
generation\index{generation} is chosen,
and will replace the previous one if it is better.

Next, the process of offspring generation using crossover\index{crossover} and 
mutation\index{mutation} is 
repeated to generate a population $P(t+1)$, the members of which are again 
tested for their fitness, and so on. This process terminates after a given 
number of generations (see also figure \ref{fig:app.gen.1}).

Since less fit members of the parent population do not make children, the
genetic algorithm\index{algorithm!genetic} does not conduct a random search in 
the solution space. Also,
since the best overall solution is recorded, it is not necessarily a member of 
the final population. For example, looking for artistic and scientific traits 
in the human population, The genetic algorithm would probably select Leonardo 
da Vinci, even though he died in 1519.

In short, an optimisation algorithm\index{algorithm!optimisation} is a {\em 
genetic algorithm} if it meets the following criteria:
\begin{enumerate}
 \item The problem must allow a genetic 
	representation\index{representation!genetic} for potential solutions.
	For example, a vector has a genetic representation, in which the 
	entries correspond to genes.\index{gene} As we have seen, crossover 
	exchanges these genes.
 \item An initial population has to be created and a mechanism for producing 
	the next generation must be given. 
 \item A fitness measure\index{fitness!measure} has to be defined in order to 
	guide the evolution. It
	plays the role of the environment in the sense that it induces 
	`natural selection'.
 \item The algorithm needs a crossover and mutation mechanisms to allow the
	generations to evolve.
 \item Finally, the algorithm needs parameters like population size, number of 
	generations, mutation probabilities, etc. 
\end{enumerate} 

In the next section, I will take a closer look at differential evolution. 

\section{Differential evolution}\index{evolution!differential}

The main difference between genetic algorithms and differential evolution lies
in the parent-child relationship. In genetic algorithms described in the 
previous section, two\footnote{Or more: why let biology restrict this 
mathematical protocol?} parents pass their genes on to a child by means
of {\em uniform} crossover.\index{crossover!uniform} This means that all 
parents have equal probability
to pass on their genes to their children (note the distinction with the {\em 
unequal} probability for members of a population of having offspring at all).

In differential evolution,\index{evolution!differential} however, a fitter 
parent has a higher probability of passing on its genes\index{gene} to the 
child. The child is thus more closely related to its fitter parent, and is 
likely to have a good fitness rating.\index{fitness!measure} This 
accommodates a more directed evolution, in which successful branches are 
biased \cite{price97}.

\section{Fortran code for lithography}

I used a genetic algorithm to optimise one-dimensional quantum lithography
\index{quantum lithography}
used in the creation of a trench function. I have omitted the fitness function 
because it is quite lengthy. It can easily be generated using {\sc 
Mathematica}.

{\footnotesize
\begin{verbatim}
        program genetic
c       uses a GENETIC search algorithm
        implicit real*8 (a-h, o-z)
        integer time
        real RAN
        external time, RAN
 
c       number of parameters to fit
        parameter (n = 21)
c       maximum number of generations
        parameter ( gen_max = 1000 )
c       population size
        parameter ( NP = n*10 )
c       scaling mutation parameter
        parameter ( Fscale = 0.5 )
c       recombination parameter
        parameter ( CR = 0.1 )
        real*8 x1(n, NP), x2(n, NP), trial(n), cost(NP), psmallest(n)
c
c       set the random seed
        iseed = time()
c       iseed = 950015448
c
c       initialization
        do 800 i = 1, NP
          do 700 j = 1, n-1
            trial(j) = 2.0d0*RAN(iseed) - 1.0d0
            x1(j,i) = trial(j)
700       continue
c       initialise the exposure time parameter
          trial(21) = 1.0d-2*RAN(iseed)
          x1(21,i) = trial(21)
          cost(i) = f( n, trial )
c         write(*,*) i, trial, cost(i)
800     continue
c       initialise `smallest'
        jsmallest = 1
        smallest = cost(1)*10
 
c       halt after `gen_max' generations
        do 2000 jgen = 1, gen_max
c       loop through the population
          do 1800 i = 1, NP
c       mutate and recombine.
c       randomly generate three *different* vectors from each other and `i'
1001        ia = 1.0 + NP*RAN(iseed)
            if ( ia.eq.i ) goto 1001
1002        ib = 1.0 + NP*RAN(iseed)
            if ( (ib.eq.i) .or. (ib.eq.ia) ) goto 1002
1003        ic = 1.0 + NP*RAN(iseed)
            if ( (ic.eq.i) .or. (ic.eq.ia) .or. (ic.eq.ib) ) goto 1003
c       randomly pick the first parameter
            j = 1.0 + RAN(iseed)*n
c       load n parameters into trial; perform n - 1 binomial trials
            do 1300 k = 1, n
              if ( (RAN(iseed).le.CR) .or. (k.eq.n) ) then
c       source for `trial(j)' is a random vector plus weighted differential..
                trial(j) = x1(j,ic)+Fscale*( x1(j,ia) - x1(j,ib) )
              else
c       ... or the trial parameter comes from `x1(j,i)' itself.
                trial(j) = x1(j,i)
              end if
c       get the next `j' modulo n
              j = j + 1
              if ( j.gt.n ) j = 1
c       last parameter `k=n' comes from noisy random vector.
1300        continue
c       evaluate/select.
c       score this trial
            score = f ( n, trial )
            if ( score.le.cost(i) ) then
              do 1400 j = 1, n
c       move trial to secondary vector (for next generation) ..
                x2(j,i) = trial(j)
1400          continue
              cost(i) = score
            else
              do 1450 j = 1, n
c       ... or place the old population member there
                x2(j,i) = x1(j,i)
1450          continue
            end if
1800      continue
 
c       end of population, swap arrays; move x2 onto x1 for next round
          do 1500 i = 1, NP
            do 1490 j = 1, n
              x1(j,i) = x2(j,i)
1490        continue
1500      continue
 
c       keep a record of progress so far
          do 1900 j = 1, NP
            if ( cost(j).lt. smallest ) then
              smallest = cost(j)
              jsmallest = j
              do 1600 kkk = 1, n
                psmallest(kkk) = x1(kkk,jsmallest)
1600	      continue
            end if
c           write(*,*) jgen, j, cost(j)
1900      continue
c       display the progress each generation
          write(*,*) " gen", jgen, "     score=", float(smallest)
 
2000    continue
        xnorm = 0.0d0
        do 2050 i = 1, n
          xnorm = xnorm + psmallest(i)**2
2050    continue
        xnorm = dsqrt(xnorm)
        do 2100 i = 1, n
          psmallest(i) = psmallest(i) / xnorm
2100    continue
        write(*,*) " parameters:"

        stop
        end

        real*8  function f( n, trial )
        implicit real*8 (a-h, o-z)
        integer n
        real*8 trial(n)
        xnorm = 0.0d0
        do 100 i = 1, n-1
          xnorm = xnorm + trial(i)**2
100     continue
        xnorm = dsqrt ( xnorm )
c       renormalise the trials
        do 200 i = 1, n-1
          trial(i) = trial(i) / xnorm
200     continue

c       f = fitness function to be minimised
\end{verbatim}
}

{\small %\renewcommand{\biblabel}{\oldstylenums\theenumiv}

\newpage
\thispagestyle{empty} }

\newpage\addcontentsline{toc}{chapter}{$\quad\,\,$Index}

{\footnotesize \printindex }

\end{document}